\documentclass[twocolumn]{aastex62}
\pdfoutput=1

\usepackage{amsmath}
\usepackage{bm}
\usepackage{graphics}
\usepackage{booktabs}

\newcommand{\msol}{M$_\odot$}

\graphicspath{{./}{figures/}}

\received{XXXXXX}
\revised{XXXXXX}
\accepted{\today}

\shorttitle{}
\shortauthors{Raaijmakers et al.}

\begin{document}

\title{The Challenges Ahead for Multimessenger Analyses of Gravitational Waves and Kilonova: a Case Study on GW190425}

\newcommand{\GRAPPA}{\affiliation{GRAPPA, Anton Pannekoek Institute for Astronomy and Institute of High-Energy Physics, University of Amsterdam, Science Park 904, 1098 XH Amsterdam, The Netherlands}}
\newcommand{\Nikhef}{\affiliation{Nikhef, Science Park 105, 1098 XG Amsterdam, The Netherlands}}
\newcommand{\UNH}{\affiliation {Department of Physics, University of New Hampshire, 9 Library Way, Durham NH 03824, USA}}
\newcommand{\UAlberta}{\affiliation {Department of Physics, University of Alberta, Edmonton, AB T6G 2E1, Canada}}
\newcommand{\caltech}{\affiliation{Astronomy Department, California Institute of Technology, Pasadena, CA 91125}}
\newcommand{\tokyo}{\affiliation{Research Center for the Early Universe, Graduate School of Science, University of Tokyo, Bunkyo-ku, Tokyo 113-0033, Japan}}
\newcommand{\utrecht}{\affiliation{
Institute for Theoretical Physics,
Utrecht University, Princetonplein 5, 3584 CC Utrecht, The Netherlands}}
\newcommand{\stockholm}{\affiliation{The Oskar Klein Centre, Department of Astronomy, Stockholm University, AlbaNova, SE-106 91 Stockholm, Sweden}
}
\newcommand{\stockholmphysics}{\affiliation{The Oskar Klein Centre, Department of Physics, Stockholm University, AlbaNova, SE-106 91 Stockholm, Sweden}
}
\correspondingauthor{G.~Raaijmakers}
\email{G.Raaijmakers@uva.nl}

\author[0000-0002-9397-786X]{Geert Raaijmakers} \GRAPPA

\author[0000-0001-6573-7773]{Samaya Nissanke} \GRAPPA \Nikhef

\author[0000-0003-4617-4738]{Francois Foucart} \UNH

\author[0000-0002-5619-4938]{Mansi M. Kasliwal}

\affil{Division of Physics, Mathematics, and Astronomy, California Institute of Technology, Pasadena, CA 91125, USA}

\author[0000-0002-8255-5127]{Mattia Bulla} \stockholm

\author[0000-0003-4619-339X]{Rodrigo Fern\'andez} \UAlberta

\author{Amelia Henkel} \UNH

\author{Tanja Hinderer} \GRAPPA \utrecht
 
 \author{Kenta Hotokezaka} \tokyo
 
\author{Kamilė Lukošiūtė} \GRAPPA

\author[0000-0002-1661-2138]{Tejaswi Venumadhav} 
\affiliation{Department of Physics, University of California at Santa Barbara, Santa Barbara, California 93106, USA}
\affiliation{School of Natural Sciences, Institute for Advanced Study, 1 Einstein Drive, Princeton, New Jersey 08540, USA}
\affiliation{International Centre for Theoretical Sciences, Tata Institute of Fundamental Research, Bangalore 560089, India}

\author[0000-0002-7686-3334]{Sarah Antier}
\affiliation{Universit\'e de Paris, CNRS, Astroparticule et Cosmologie, F-75013 Paris, France}

\author[0000-0002-8262-2924]{Michael W. Coughlin}
\affiliation{School of Physics and Astronomy, University of Minnesota, Minneapolis, Minnesota 55455, USA}

\author[0000-0003-2374-307X]{Tim Dietrich}
\affiliation{Institute of Physics and Astronomy, University of Potsdam, Karl-Liebknecht-Str. 24/25, 14476, Potsdam, Germany}
\affiliation{Max Planck Institute for Gravitational Physics (Albert Einstein Institute), Am M\"uhlenberg 1, Potsdam 14476, Germany}

\author{Thomas D. P. Edwards} \stockholmphysics \GRAPPA

\begin{abstract}
In recent years, there have been significant advances in multi-messenger astronomy due to the discovery of the first, and so far only confirmed, gravitational wave event with a simultaneous electromagnetic (EM) counterpart, as well as improvements in numerical simulations, gravitational wave (GW) detectors, and transient astronomy. This has led to the exciting possibility of performing joint analyses of the GW and EM data, providing additional constraints on fundamental properties of the binary progenitor and merger remnant. Here, we present a new Bayesian framework that allows inference of these properties, while taking into account the systematic modeling uncertainties that arise when mapping from GW binary progenitor properties to photometric light curves. We extend the relative binning method presented in \citet{Zackay18} to include extrinsic GW parameters for fast analysis of the GW signal. The focus of our EM framework is on light curves arising from r-process nucleosynthesis in the ejected material during and after merger, the so called kilonova, and particularly on black hole - neutron star systems. As a case study, we examine the recent detection of GW190425, where the primary object is consistent with being either a black hole (BH) or a neutron star (NS).  We show quantitatively how improved mapping between binary progenitor and outflow properties, and/or an increase in EM data quantity and quality are required in order to break degeneracies in the fundamental source parameters.
\end{abstract}


\section{Introduction} \label{sec:intro}
The first gravitational-wave detection of two merging compact objects, during the first observing run of Advanced LIGO, gave us a new direct probe into the fundamental physics governing such objects \citep{GW150914}. In the subsequent observing runs of LIGO and Virgo, tens of binary black hole (BBH) mergers, and several binary neutron star (BNS) and black hole - neutron star (BHNS) mergers have been reported, allowing for population level studies \citep{GWTC1, GWTC2}.
So far, however, the BNS merger GW170817 \citep{GW170817discovery} (combined signal-to-noise ratio (SNR) of 32.4) has been the only\footnote{Leaving out the unconfirmed detection of a candidate electromagnetic counterpart to a binary black hole merger \citep{Graham20}.} event for which electromagnetic (EM) signatures were detected across the frequency spectrum \citep[e.g.,][]{LVCmultimessenger,GW170817grb, GW170817swope, Kasliwal17, Hallinan17}. The extensive follow-up campaign of the counterpart resulted in additional information on many aspects of the merger, ranging from the dense matter equation of state (EOS) of NSs \citep[e.g.,][]{Radice18, Raaijmakers20, Tews20} to measurements of the Hubble constant \citep[e.g.,][]{GW170817H0, LVChubble, Hotokezaka19a, Mukherjee19, Howlett20, Nicolaou20, Dietrich20, Dhawan20}. 

Recent GW events potentially involving NSs --- GW190425 \citep{GW190425discovery}, GW190426 \citep{GWTC2} and GW190814 \citep{GW190814} --- have all led to similar large-scale searches for an EM counterpart \citep{Coughlin19b, Goldstein19, Andreoni20, GRANDMA_o3, SWIFT_o3, GOTO_o3}.
These searches have not resulted in any successful detections so far, though the absence of a counterpart can add weak additional constraints on the binary parameters of the system \citep{Coughlin20a, Coughlin20b, Anand2020, ENGRAVE_190814}. The absence of a detection could be due to the low SNR of some of these events, e.g. GW190425 with a combined SNR of 12.9, producing much larger skymaps compared to GW170817. 

One of the counterparts to a BHNS or BNS merger is the so-called kilonova or macronova \citep{Li98, Kulkarni05}, a thermal ultraviolet-optical-infrared transient (UVOIR) arising from r-process nucleosynthesis \citep{Burbidge57, Cameron57} in the neutron rich material that is ejected during and after the merger \citep{Lattimer74, Rosswog98, Metzger10}. If a kilonova is detected, critical questions remain over whether one could improve upon the constraints of the binary parameters and possibly uncover the ambiguous nature of one of the binary components in e.g., GW170817 \citep[see, e.g.,][]{Hinderer19, Coughlin19c} and GW190425 \citep[see, e.g.,][]{Barbieri20b, Most20, Kyutoku20}.

There have been a number of studies performing a multimessenger analysis of GW170817 and the associated kilonova, most of which focus on the dense matter EOS, using either a limit on
the minimum amount of ejected mass derived from the UVOIR light curves \citep[e.g.,][]{Radice18, Radice19, Hinderer19, Raaijmakers20, Capano20} or a joint Bayesian modeling framework \citep[e.g.,][]{Coughlin18, Coughlin19, Dietrich20, Breschi21b}. The methods in \citet{Coughlin19} and \citet{Dietrich20} allow for additional uncertainty in the analysis, attributed to the uncertainty in the modeling of ejected material, by adding an unknown component to the kilonova. There have also been more general studies developing a Bayesian framework for multimessenger analysis, focusing on several parameters describing the binary progenitor system \citep[e.g.,][]{Barbieri19a, Breschi21a, Nicholl21}. \citet{Coughlin19}, \citet{Dietrich20} and \citet{Barbieri19a} also consider the short gamma-ray burst (GRB) signal \citep[e.g.,][]{Goldstein17} associated with compact object mergers, which can help constrain the inclination angle of the system \citep[e.g.,][]{Ryan20}. \citet{Barbieri19a} furthermore consider the radio remnant due to the kilonova and GRB afterglow \citep[see e.g.][]{Hotokezaka16}. However, \citet{Barbieri19a} do not include uncertainties in their ejecta modeling or light curve computation when performing a joint analysis of the GW and EM data.

More recently, there have been two new Bayesian analyses frameworks developed by \citet{Breschi21b} and \citet{Nicholl21}. In \citet{Breschi21b}, the authors use a three-component model with angular dependence, first presented in \citet{Perego17}, and perform a joint analysis of GW170817 and the associated kilonova. In \citet{Nicholl21}, the authors use the public light curve generation code in \texttt{MOSFiT} \citep{Guillochon18} and divide the ejecta into two components. Additionally they allow for a third component arising from shock-heated material by a GRB jet. They then perform a similar joint analysis of GW170817 and the associated kilonova, focusing on extracting the dense matter EOS. 

In this work, we will present a new multimessenger framework to analyse gravitational waves and the associated UVOIR light curves jointly. We leave the inclusion of the GRB and radio counterpart to future work, and furthermore will not include any interaction between the kilonova material and the jet \citep[see e.g.][]{Kasliwal17, Piro18,Nativi21,Klion21}. We will use the analytical formulae, fitted to numerical simulation of BNS and BHNS, derived in \citet{Kruger20} to connect outflow properties to binary progenitor properties. We incorporate modeling uncertainties by estimating the errors on these formulae and by using a new relation that estimates the amount of material that is ejected from the disk surrounding the merger remnant, which is 
assumed to be a BH. Because of this assumption, our framework is only appropriate for BNS systems with high total masses and for BHNS systems; BNS systems with lower total masses might lead to a NS or hypermassive NS remnant which will significantly alter the light curve properties \citep[see, e.g.,][]{Kawaguchi20}. The paper is structured as follows; in Section \ref{sec:ejecta}, we will discuss how the properties of the binary progenitor are connected to the properties of the outflows during and after the merger; in Section \ref{sec:lightcurve}, we will discuss the kilonova model that is used to compute UVOIR light curves from these outflow properties; in Section \ref{sec:multimessenger}, we will test our analysis framework on a low SNR GW190425-like merger with simulated GW and EM data, and discuss the implications in Section \ref{sec:discussion}. We note that the model presented in this work includes a more careful consideration of the errors for the BHNS scenario than the BNS scenario, which we leave to future work.

\begin{figure*}[t!]
\centering
\includegraphics[width=\textwidth]{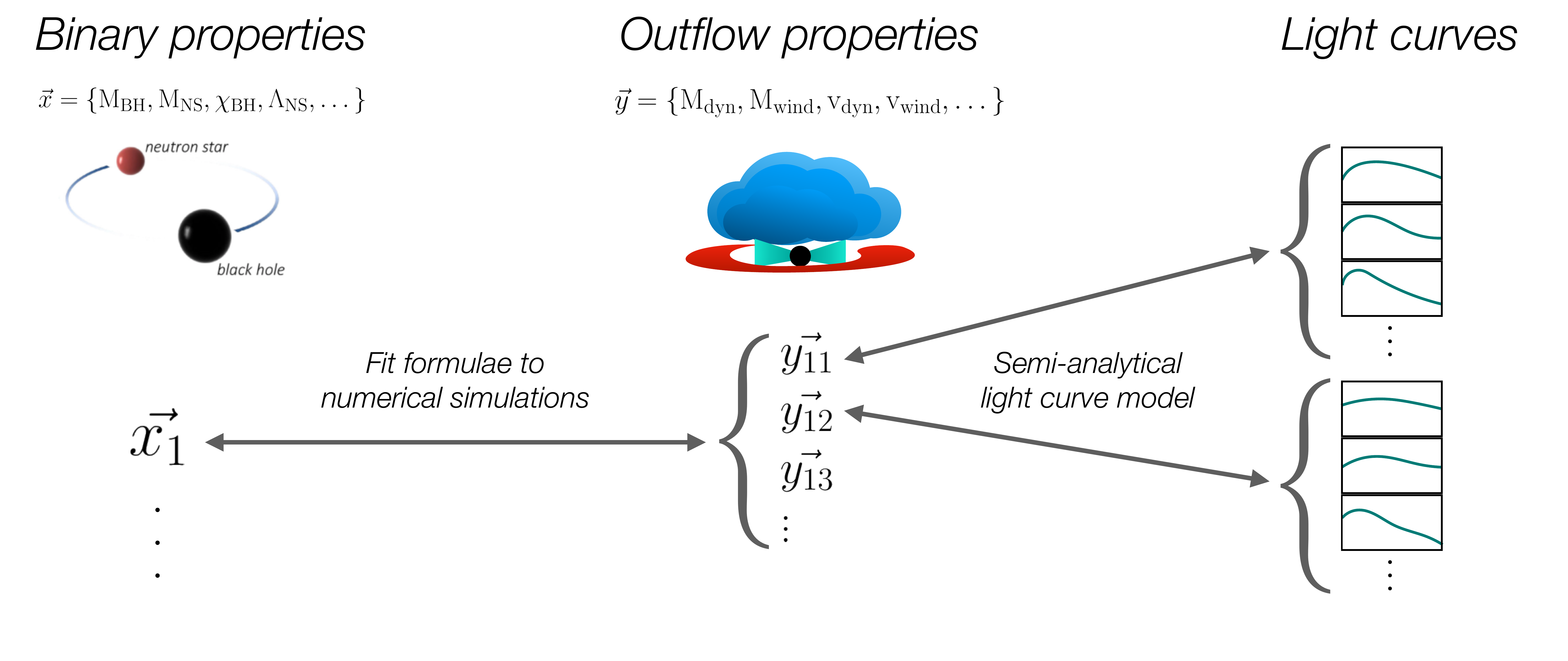}
\caption{Schematic drawing of how binary progenitor properties connect to outflow properties and UVOIR light curves. If we consider a sample of binary properties $\vec{x}$, obtained from, e.g., parameter estimation on the GW strain data, then each point in this sample ($\vec{x}_1, \vec{x}_2 \dots$) maps to outflow properties $\vec{y}$ via analytical formulae calibrated to numerical simulations. Because of the uncertainties in both the numerical simulations and in these formulae, one sample in $\vec{x}$ gives a range of possible outflow properties. From each of these points in $\vec{y}$, light curves can be computed through semi-analytical models, where again the uncertainty in these models leads to multiple light curves for a given point in $\vec{y}$.}
\label{fig:fig1}
\end{figure*}

\section{Connecting binary progenitor to outflow properties}
\label{sec:ejecta}
In order to build a multimessenger analysis framework, we start by connecting the fundamental source properties of the binary progenitor to parameters that describe average properties of the outflows that occur during and after the merger (see also Figure \ref{fig:fig1}). Both the connecting relations and the source properties of interest differ between a BNS and a BHNS system. For a BNS system, the parameters of interest are the masses of the two neutron stars and their tidal deformabilities \citep[see, e.g.,][]{Hinderer10}. In this work, the spin of the neutron star is assumed to be small when computing outflow properties and thus will not impact the kilonova. This assumption is based on the spin distribution of Galactic BNSs, which indicates that the magnitude of the dimensionless spin parameter $\chi = c I \omega/(G M^2) \leq 0.05$ \citep{Zhu18}. In general, however, the spin of the neutron star may have a significant effect on the outflows \citep[see, e.g.][]{Most19, East19, Chaurasia20}. For a BHNS system, the parameters of interest for the neutron star are again the  mass and the tidal deformability. Black holes, however, are thought to have a tidal deformability of zero in general relativity \citep[see e.g.][]{Binnington09, Chia20, Chirenti20}. Contrary to neutron stars, however, the spins of black holes in binaries can be as high as $|\chi_{\rm BH}| \lesssim 0.9$ \citep{GWTC2} and therefore have a much bigger impact on the outflow properties.

The ejecta outflows that contribute to the kilonova also differ between BNS and BHNS systems. In a BNS system, outflows on a dynamical timescale, hereafter called dynamical ejecta, consist of unbound material due to the tidal forces that are exerted on the neutron stars and shock heated material during the merger \citep[see e.g.][and references therein]{Radice18a}. For a BHNS system, the dynamical ejecta is assumed to solely consist of tidally ejected material. A second component that contributes to the kilonova are outflows on longer timescales emerging from the disk surrounding the merger remnant, hereafter, disk wind ejecta.

For both components, the most important parameters describing the outflows are mass, velocity, and opacity. In this Section, we will relate these parameters to the fundamental source properties by employing analytical formulae that are fitted to numerical simulations. We will estimate appropriate uncertainties that account for the unknown modeling errors that go into these simulations, which are larger for the disk wind ejecta. We note that for BNS systems there are several simplifying assumptions applied here which may not accurately represent the complexity of the whole range of possible BNS systems, which we will leave for future work. 

Throughout this section, we will use the quasi-universal relation from \citet{Yagi17} to compute the compactness of the neutron star, $C_{\rm{NS}} = GM_{\rm{NS}}/(c^2 R_{\rm{NS}})$ (where $G$ is the Gravitational constant, $c$ is the speed of light, $M_{\rm NS}$ and $R_{\rm NS}$ the mass and radius of the NS respectively), as a function of its tidal deformability, $\Lambda_{\rm{NS}}$:
\begin{equation}
C_{\rm{NS}} = 0.371 - 0.0391\log(\Lambda_{\rm{NS}}) + 0.001056\log(\Lambda_{\rm{NS}})^2\,.
\end{equation}
While the accuracy of this relation is on the order of $10\%$ for nucleonic EOS, we will assume it to hold true as it evades the need to directly integrate an EOS to obtain a radius. 

\subsection{Dynamical ejecta}
\label{subsec:dynej}
In the case of a BNS system, the mass of the dynamical ejecta is estimated using the formula derived in \citet{Kruger20}:
\begin{eqnarray}
    \frac{M^{\rm BNS}_{\rm dyn}}{10^{-3}M_\odot} = \left(\frac{a}{C_1} + b \frac{M_2^n}{M_1^n} + c C_1\right) M_1 + (1 \leftrightarrow{} 2)\,,
    \label{eq:nsns_ejecta}
\end{eqnarray}
where the coefficients are $a=-9.3335, b=114.17, c=-337.56$, and $n=1.5465$. The subscripts in this equation, and throughout this work, follow the convention that $M_1 > M_2$. We assume the maximum error to be the $2\sigma$ standard deviation obtained from the residuals between the fitting formula and the numerical simulations, which corresponds to an error of $\pm 0.008$ M$_{\odot}$.

When we assume that the progenitor is a BHNS system (with a mass ratio of $Q = 1/q = M_{\rm{BH}}/M_{\rm{NS}}$), we use the following equation (\citet{Kruger20}, based on models by \citet{Kawaguchi16}) to estimate the dynamical ejecta: 
\begin{align}
    \frac{M_{\rm dyn}^{\rm BHNS}}{M_{\rm NS}^b}
        & = a_1 Q^{n_1} \frac{1-2C_{\rm NS}}{C_{\rm NS}} - a_2 Q^{n_2} \frac{R_{\rm ISCO}}{M_{\rm BH}} + a_4\,.
    \label{eq:bhns_ejecta}
\end{align}
Here, M$_{\rm{NS}}^b$ is the baryonic mass of the NS and $R_{\rm{ISCO}}$ is the radius of the innermost stable circular orbit around a black hole of mass $M_{\rm{BH}}$ \citep{Bardeen1972}.
The first can be approximated, following \citet{Lattimer01}, by the simple formula, 
\begin{equation}
M_{\rm{NS}}^b = M_{\rm{NS}} \left( 1 +\frac{0.6 C_{\rm{NS}}}{1 - 0.5C_{\rm{NS}}} \right)\,,
\end{equation}

while the latter is given by, 
\begin{align}
 &   R_{\rm{ISCO}} =3 + Z_2 -\rm{sgn}(\chi_{\rm{BH}}) \sqrt{(3 - Z_1)(3 + Z_1 + 2Z_2)}\,, \\
 & Z_1 = 1 + (1 - \chi_{\rm{BH}}^2)^{1/3}\left[(1 + \chi_{\rm{BH}})^{1/3} + (1 - \chi_{\rm{BH}})^{1/3} \right]\,, \\
 & Z_2 = \sqrt{3\chi_{\rm{BH}}^2 + Z_1^2}\,.
\end{align}

In this scenario, we again assume the error in the mass of the dynamical ejecta to be the $2\sigma$ limit of the numerical error of the fit, i.e. $\Delta$M$_{\rm dyn}^{\rm BHNS} = 0.0094$ \msol. We note, however, that for both the BNS and BHNS scenario, this is an underestimate of the real uncertainty due to the uncertainties in numerical modeling. 

The velocity of the dynamical ejecta and the error in the velocity are assumed to be equal in the BHNS and BNS systems, which we justify by assuming that for the BNS system the dynamical ejecta is dominated by the tidal tail (see next paragraph). The velocity and the error are approximated using the formulae derived in \citet{Foucart17}: 
\begin{align}
& v_{\rm{dyn}} = 0.0149 Q + 0.1493\,, \\
& \Delta v_{\rm{dyn}} = 0.1 v_{\rm{dyn}}\,.
\label{eq:vdyn}
\end{align}

Finally, the composition of the dynamical ejecta is extremely neutron rich, which leads to a large lanthanide-fraction through rapid neutron capture. Since lanthanides have many absorption lines, this will cause the ejecta to be very opaque, and so we will use an effective grey\footnote{Independent of frequency.} opacity for this component between $1$ and $10$ cm$^2$ g$^{-1}$ for both BHNS and BNS systems. However, for (especially symmetric) BNS mergers, the dynamical ejecta includes shock heated ejecta on top of the tidal tail which are predicted to have a lower lanthanide fraction, and thus a lower opacity \citep[see, e.g.,][]{Sekiguchi16, Dietrich17, Tanaka20,Nedora21}. For simplicity, we assume that this component does not contribute significantly compared to the tidal tail. For the case study considered in Section \ref{sec:multimessenger}, this assumption can be justified by the asymmetric masses of the simulated signal. 

\begin{figure}[t!]
\centering
\includegraphics[width=.49\textwidth]{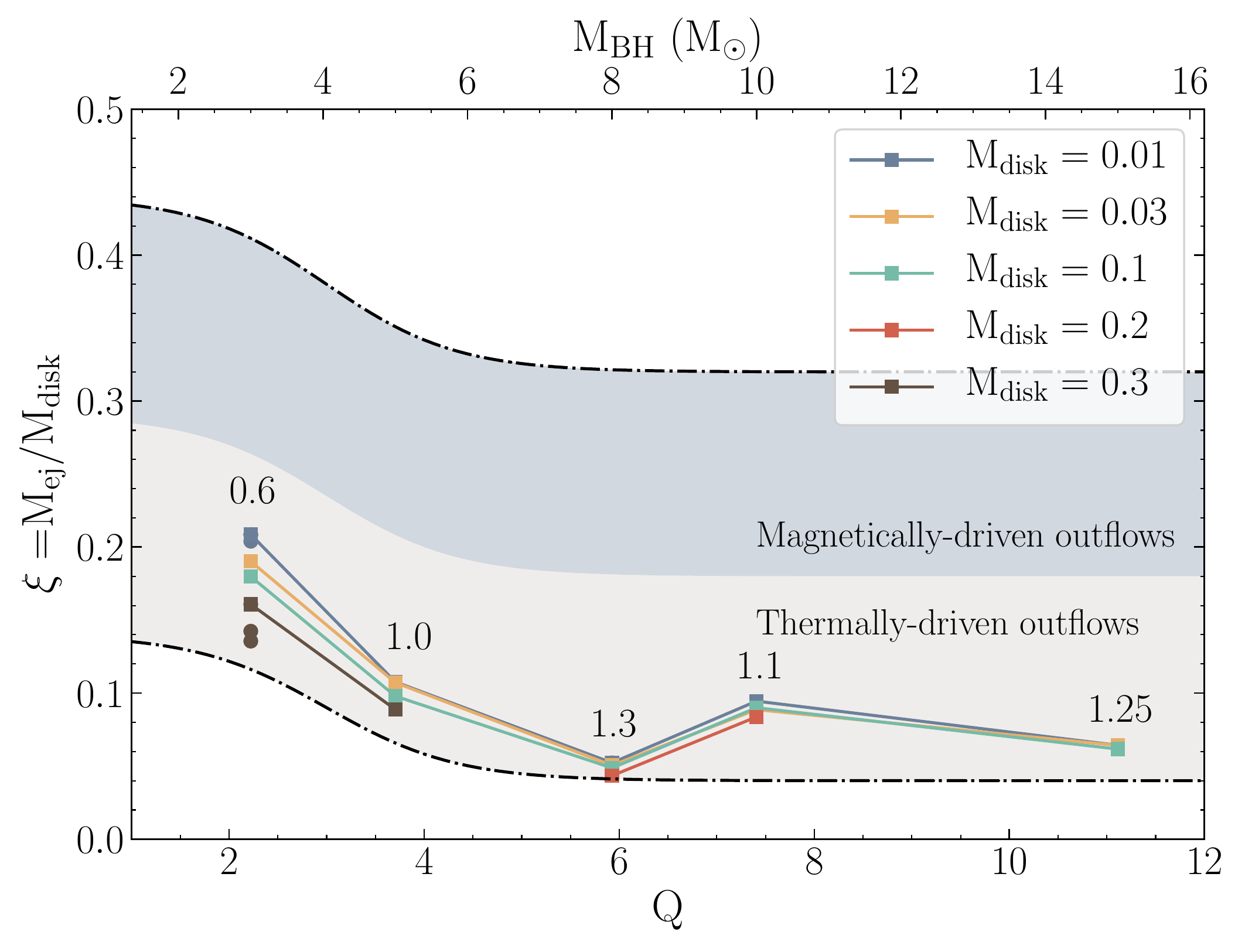}
\caption{Uncertainty in the fraction of the initial disk M$_{\rm disk}$ that is eventually ejected as a function of the mass ratio of the progenitor binary $Q$. The points show the results of the viscous hydrodynamic simulations of BH accretion disks from \citet{Fernandez20}, which
capture the thermally-driven outflow due to angular momentum transport
(i.e., dissipation of MRI turbulence) and nuclear recombination. The numbers in the plot indicate the disk compactness $C_d$ and the top axis is the black hole mass used for the initial setup. As an estimation, we connect to the mass ratio by dividing the black hole mass by a fiducial neutron star mass of $1.35$ \msol. The light-shaded band indicates the uncertainty captured by Equation (\ref{eq:fraction}), while the darker-shaded band shows the possible enhancement in the
fraction of the disk mass ejected through magnetic driving, ranging between $0\%$ and $20\%$ depending on the initial field geometry of the disk (c.f. \citealt{Christie19})}
\label{fig:fig2}
\end{figure}

\subsection{Disk wind ejecta}
\label{sec:disk_wind_ejecta}

The second type of ejecta comes from matter outflows in the accretion disk surrounding the merger remnant. It is important to note that the following discussion only applies when the merger remnant is a BH, which for BNS systems is only the case when the total mass is high enough that the merger remnant undergoes collapse to a BH. 
The first step to estimate the mass of these outflows is to compute the mass of the disk, i.e. the material that is still bound to the merger remnant and supported by rotation against collapse. In the case of a BNS system we use the formula derived in \citet{Kruger20}  to compute: 
\begin{equation}
    M_{\rm disk}^{\rm BNS} = M_2~\mathrm{max} \left\{5\times 10^{-4}, (aC_2 + c)^d\right\}\,.
    \label{eq:nsns_disk}
\end{equation}

For the BHNS scenario, we first compute the remnant mass outside of the black hole after the merger using the fitting formula of \citet{Foucart18}: 
\begin{equation}
    \hat{M}_{\rm{rem}} = \left[ \rm{Max}\left( \alpha \frac{1 - 2C_{\rm{NS}}}{\eta^{1/3}} - \beta\frac{C_{\rm{NS}}}{\eta}, 0 \right) \right]^\delta \,,
    \label{eq:bhns_disk}
\end{equation}
where $\eta$ is the symmetric mass ratio, i.e. $\eta = Q/(1 + Q)^2$. 

From the remnant mass, we can compute the mass of the disk surrounding the black hole by simply subtracting the dynamical ejecta, i.e. $M_{\rm{disk}}^{\rm BHNS} = \hat{M}_{\rm{rem}} - M_{\rm{dyn}}^{\rm BHNS}$. 

A certain fraction of the disk will not be accreted onto the black hole but will instead become gravitationally unbound in what we will call a disk wind. Numerical simulations of a disk surrounding a BH merger remnant indicate that the ejected material from the disk can roughly be divided into two further sub-components based on their velocity. The first thermally-driven component arises from energy dissipation due to angular momentum transport 
(i.e. MRI-driven
turbulence) and nuclear recombination on timescales longer than the dynamical time, which can be modeled well by viscous hydrodynamic simulations \citep[see, e.g.,][]{Fernandez20}. A faster magnetically-driven component appears on shorter timescales in general relativistic magnetohydrodynamic (GRMHD) simulations \citep[see, e.g.,][]{Siegel17, Fernandez19, De20}. 
For the thermally-driven outflow component generated around a BH remnant, \citet{Fernandez20} found an inverse linear dependency of the ejected mass on the compactness of the disk using viscous hydrodynamic simulations. A disk formed closer to the BH (more compact) ejects a smaller fraction of the initial disk mass than an initially more extended disk because a larger fraction of the disk is accreted before the thermal outflow can begin, given its
% higher 
proximity to $R_{\rm{ISCO}}$. We assume that the compactness of the disk is roughly proportional to the mass ratio of the progenitor binary so that we can estimate the fraction of the disk that is ejected thermally as a function of the initial binary properties. In Figure \ref{fig:fig2}, we show the fraction $\xi$ of ejected material $M_{\rm ej}$ over the initial disk mass M$_{\rm disk}$ found in \citet{Fernandez20}. On the top axis, we show the value of the black hole mass used in the simulations, while the numbers in the plot indicate the disk compactness $C_d = (M_{\rm BH}/5$ \msol $) (50 $km$ /R_d)$, where $R_d$ is the radius of the disk. The initial conditions used in \citet{Fernandez20} are inspired by numerical relativity simulations using neutron stars of mass $M_{\rm NS} \sim 1.2-1.4 M_\odot$. Thus, we estimate the mass ratio by simply dividing the mass of the black hole by a fiducial neutron star with mass $1.35$ \msol. To capture the uncertainty in the viscous outflows as a function of the mass ratio, we use the ansatz: 
\begin{equation}
\label{eq:fraction}
    \xi = \frac{M_{\rm ej}}{M_{\rm disk}} = \xi_1 + \frac{\xi_2-\xi_1}{1 + e^{1.5(Q-3)}}\,,
\end{equation}
where for the lower bound $\xi_1 = 0.04$ and $\xi_2 = 0.14$, and for the upper bound, $\xi_1 = 0.32$ and $\xi_2 = 0.44$.
In addition to the thermally-driven outflows, we assume that the magnetically-driven component could contribute anywhere between $0\%$ and $20 \%$ of the initial disk mass (see Figure \ref{fig:fig2}). The large uncertainty can be attributed to the magnetic field strength and configuration at the time of disk formation, with strong poloidal fields ejecting larger fractions of the disk than weak poloidal fields or toroidal fields \citep[see][]{Christie19}.

The velocity of the disk wind ejecta is very uncertain as there are only a few three-dimensional GRMHD simulations that capture the magnetically-driven outflow. The velocity distributions in \citet{Siegel17} and \citet{De20} show that the bulk of the material is centered around $\sim0.1$c, similar to what \citet{Christie19} find for their model with a weak and strong initial poloidal magnetic field, although for the latter, there is a more pronounced high velocity tail in the distribution. 
For their model with a toroidal magnetic field configuration, this high velocity tail is almost absent and the bulk of the material has velocities that are roughly $\leq0.1 c$. In \citet{Fernandez19}, the authors find that the material ejected due to the magnetic field has velocities $\geq 0.1c$, which is expected considering their poloidal magnetic field configuration. The velocity of the thermally-driven outflows is much lower ($\leq 0.1c$), with the bulk of the material around $0.03c- 0.05c$ \citep[see also][]{Fernandez20}. We take the average velocity to be between $0.1c$ and $0.3c$, with a minimum and maximum velocity distribution cutoff above $0.1 v_{\rm wind}$ and below $2 v_{\rm wind}$ respectively (see Table \ref{tab:tab1}).

The composition of the disk outflows is less accurately known than for the dynamical ejecta. The thermal
component is ejected on timescales long enough that it is mostly reprocessed by neutrino emission and absorption. The composition is primarily light $r$-process elements, with variable amounts of lanthanide-rich matter depending on detailed properties of the disk and central object (e.g., \citealt{just2015,martin2015,wu2016,lippuner2017}). The composition of the magnetic component is less well understood given the limited number of long-term GRMHD simulations. The faster overall velocities imply shorter expansion times and less neutrino reprocessing, with a larger expected proportion of lanthanide-rich matter than the thermal component \citep{Siegel17,Fernandez19}. Neutrino absorption has been included in one GRMHD study only, with results of the early disk evolution indicating that it is indeed an important process in regulating the composition \citep{miller2019}. To model the opacity we choose an effective grey opacity between $0.1 - 1.0$ cm$^2$ g$^{-1}$ for the thermal component and between $1 - 10$ cm$^2$ g$^{-1}$ for the magnetic component.

An important note to make here is that in the case of the merger remnant being a hypermassive neutron star (HMNS), both the fraction of the disk that is ejected and the composition change drastically. Due to absorption of neutrinos emitted by the HMNS and absence of a BH mass sink, the disk could be entirely ejected, depending on the lifetime of the HMNS \citep[see, e.g.,][]{metzger2014,fujibayashi2018,fahlman2018,fujibayashi2020, Moesta20}. We do not consider this possibility here, but will leave it for future work.

\begin{figure}[t!]
\centering
\includegraphics[width=.5\textwidth]{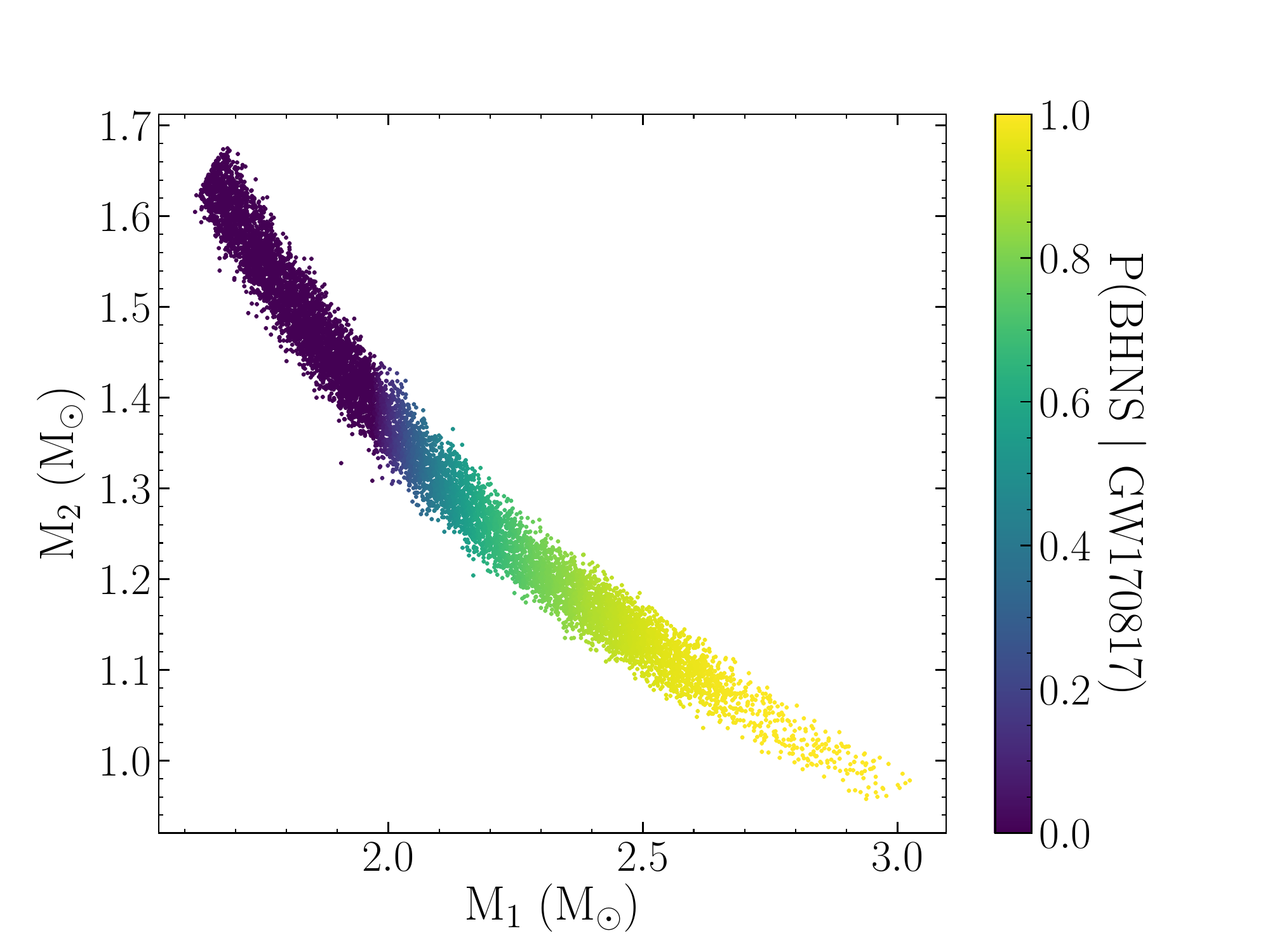}
\caption{Posterior samples of the two source masses in GW190425 for the high-spin priors using the \texttt{IMRPhenomDNRT} waveform model. The samples are  color coded by the probability of GW190425 being a BHNS or a BNS system, given the constraints on the EOS by GW170817 \citep{GW170817eos}.}
\label{fig:fig3}
\end{figure}

\subsection{Ejecta mass distributions GW190425}

\begin{figure*}[t!]
\centering
\includegraphics[width=.48\textwidth]{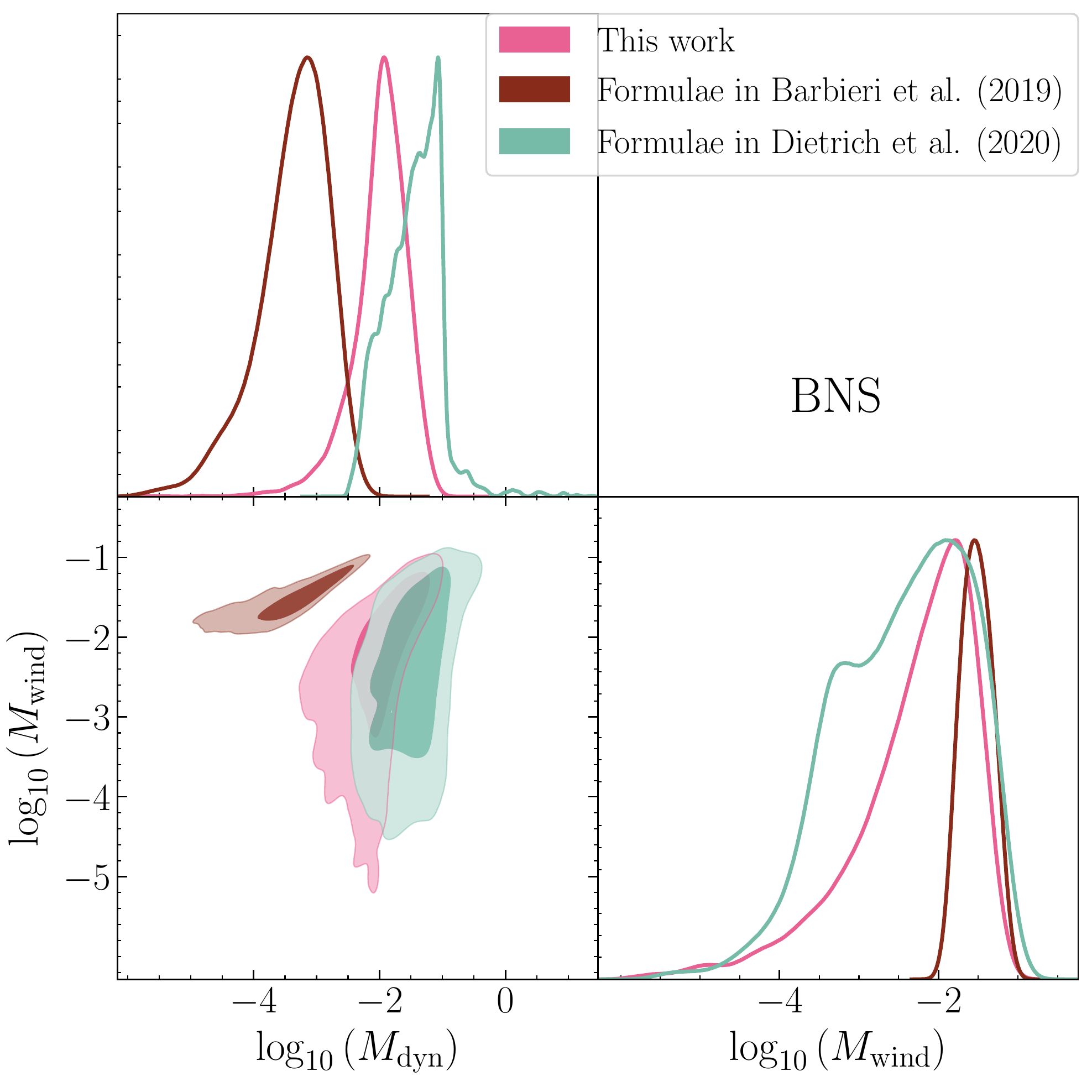}
\includegraphics[width=.48\textwidth]{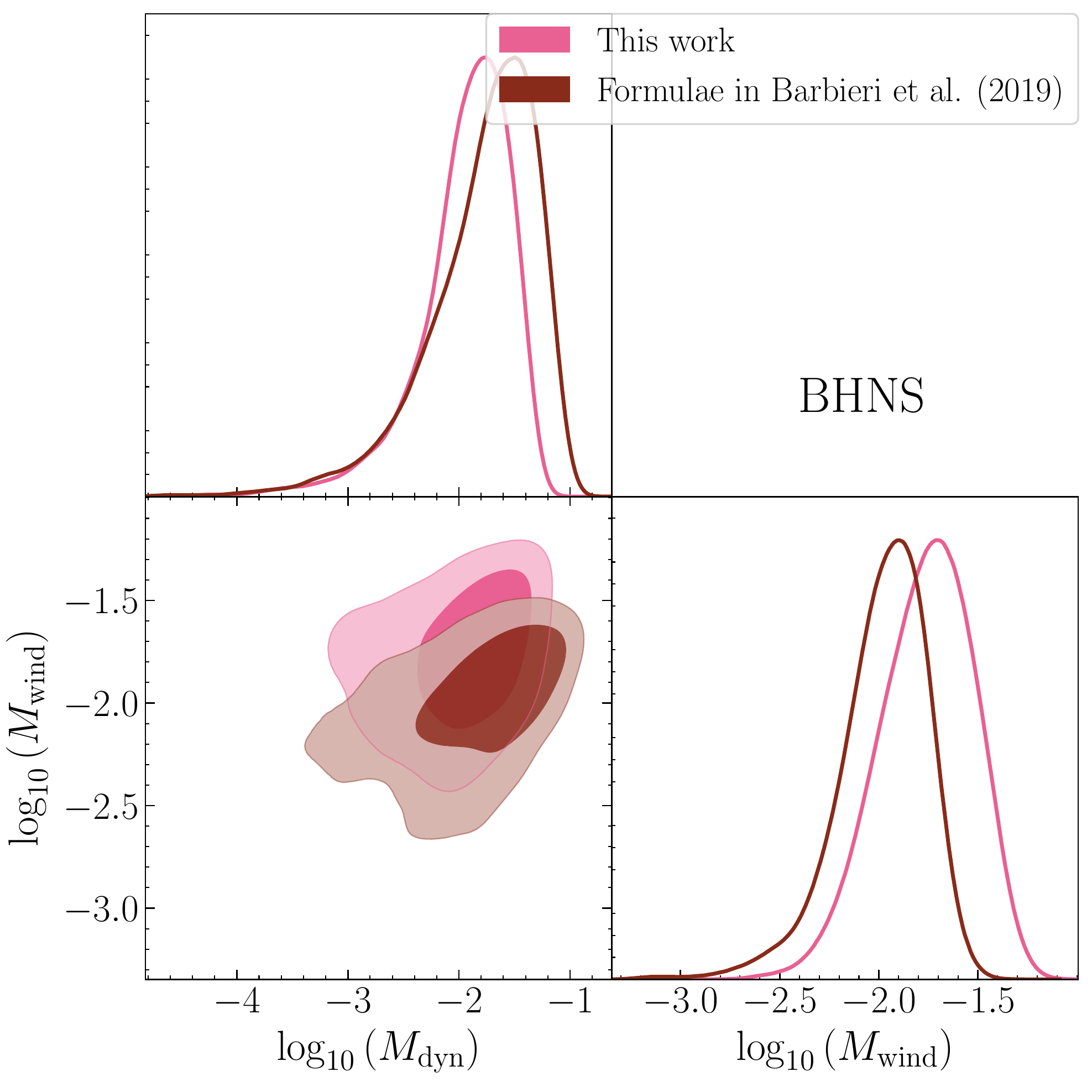}
\caption{Comparison of ejecta masses for the dynamical component and the disk wind component for the BNS scenario (left panel) and the BHNS scenario (right panel). The masses are computed by using different sets of fitting formulae \citep{Coughlin19, Barbieri20} on the high-spin posterior distribution of GW190425 \citep{GW190425discovery}. The shaded regions contain respectively $68\%$ and $95\%$ of all ejecta masses \textit{after} discarding the samples where M$_{\rm dyn}$ or M$_{\rm wind}$ is zero. The fractions of discarded samples are $58\%$, $78\%$, and $30\%$ for the BNS formulae in this work, in the work of \citet{Barbieri19a}, and in the work of \citet{Dietrich20}, respectively. For the BHNS scenario, the fractions of discarded samples are $40\%$ and $25\%$ for the formulae in this work and in the work of \citet{Barbieri19a}, respectively.}
\label{fig:fig4}
\end{figure*}

To better understand the formulae discussed in Section \ref{sec:disk_wind_ejecta}, we apply them to the gravitational wave event GW190425 \citep[][]{GW190425discovery}. GW190425 was detected in the LIGO Livingston and the Virgo detectors. The two component masses, also shown in Figure \ref{fig:fig3}, are consistent with the system being either a BNS or a BHNS, depending on the maximum mass of NSs as determined by the dense matter EOS. 

In order to apply the formulae to GW190425, we randomly draw $\mathcal{O}(10^4)$ samples from the public\footnote{\url{https://dcc.ligo.org/LIGO-P2000223/public}, high-spin prior with waveform model \texttt{IMRPhenomDNRT}} posterior probability density function (PDF) for GW190425. For each draw, we simultaneously take a random EOS sample from the public\footnote{\url{https://dcc.ligo.org/LIGO-P1800115/public}, low-spin prior with waveform model \texttt{IMRPhenomPNRT}} posterior EOS samples of GW170817 \citep{GW170817eos}. Depending on whether the primary mass of our draw is above or below the maximum mass of the chosen EOS, we can thus assume the nature of the primary component of GW190425 is a BH or NS respectively. The dimensionless BH spin, $\chi_{\rm BH}$, needed in case of a BHNS system is part of the sample that is drawn from GW190425. However, the tidal deformability of the two objects in GW190425 was unconstrained, so instead we set the tidal deformability of the NS(s) by evaluating the drawn EOS at the values drawn for the mass(es). We then apply Equations \ref{eq:nsns_ejecta} - \ref{eq:bhns_disk} to compute the corresponding ejecta masses where, for the disk wind ejecta, we randomly pick a value for $\xi$ between the upper and lower bound of Equation (\ref{eq:fraction}) to determine how much of the disk is unbound.

The resulting distributions of ejecta masses are shown in Figure \ref{fig:fig4} in pink, where the different shades illustrate the regions containing $68\%$ and $95\%$ of all values. For comparison, we also show the distributions one would compute when using different sets of fitting formulae. In brown, we show the distributions following the method from \citet{Barbieri19a}, where the authors use fits for the dynamical ejecta and disk mass from \citet{Kawaguchi16} and \citet{Foucart18} respectively in a BHNS merger and from \citet{Radice18b} and \citet{Barbieri19a} respectively in the case of a BNS merger. For the BHNS scenario, there is reasonable agreement between the distribution using the formulae in \citet{Barbieri19a} and the distribution computed in this work. This is due to the fact that the same formula for the disk mass is used, and the formula for the dynamical ejecta in \citet{Kruger20} is an adjusted version of the formula in \citet{Kawaguchi16}. For the BNS scenario, the formula used in \citet{Barbieri19a} predicts a lower dynamical ejecta mass. The predictions for disk wind mass is consistent between \citet{Barbieri19a} and this work.  
In green we also show the distributions obtained from using the same method as in \citet{Dietrich20} for BNS systems. The authors use formulae for the dynamical ejecta mass and disk wind mass derived in \citet{Coughlin19} and \citet{Dietrich20}, respectively. Although the predictions for dynamical ejecta mass are slightly higher than in this work, there is general agreement between the two distributions.

\begin{table*}[]
\caption{All parameters used in the model described in Section \ref{sec:ejecta} and their prior support in the full analysis of the GW signal and it's EM counterpart. The notation $U(a,b)$ here means uniformly drawn between boundaries $a$ and $b$. We also show the adjusted priors for the Case 4 run discussed in Section \ref{subsec:multimessenger}}
\centering
\begin{tabular}{@{}llll@{}}
\toprule
Parameters                             & Description                                        & Prior density and support              &   (Case 4)              \\ \midrule
\multicolumn{4}{c}{\textit{Binary properties}}                                                                                                     \\ \midrule
M$_{1}$ [M$_{\odot}$]                  & Mass of the primary object                         & $\in P(\bf{x}~|~\vec{d}_{GW190425})$                       \\
M$_{2}$ [M$_{\odot}$]                  & Mass of the secondary object                       & $\in P(\bf{x}~|~\vec{d}_{GW190425})$                       \\
$\chi_{1}$                             & Spin parameter of the primary object (BHNS)        & $\in P(\bf{x}~|~\vec{d}_{GW190425})$                       \\
$\Lambda_{1}$                          & Tidal deformability of the primary object  & $\Lambda$(M$_{1}$ ; EOS$_{GW170817}$)                \\
$\Lambda_{2}$                          & Tidal deformability of the secondary object (BNS)       & $\Lambda$(M$_{2}$ ; EOS$_{GW170817}$)                \\ \midrule
\multicolumn{4}{c}{\textit{Ejecta and light curve properties}}                                                                                     \\ \midrule
M$_{\rm{dyn}}$ [M$_{\odot}$]           & Mass of the dynamical ejecta                       &  Eq. (\ref{eq:nsns_ejecta}) (BNS) or (\ref{eq:bhns_ejecta}) (BHNS)  \\
v$_{\rm{dyn}}$ [c]                     & Velocity of the dynamical ejecta                   &          Eq. (\ref{eq:vdyn})                 \\
$v_{\rm{min, dyn}}$ [c]                & Minimum velocity of the dynamical ejecta           & $\sim$ U(0.1, 1.0) v$_{\rm{dyn}}$     &     $\sim$ U(0.7, 0.9) v$_{\rm{dyn}}$           \\
$v_{\rm{max, dyn}}$ [c]                & Maximum velocity of the dynamical ejecta           & $\sim$ U(1.5, 2.5) v$_{\rm{dyn}}$    & $\sim$ U(1.5, 1.7) v$_{\rm{dyn}}$                \\
n$_{\rm{dyn}}$                         & Power law index of density distribution            & $\sim$ U(3.5, 4.5)                 &                  \\
$\kappa_{\rm{dyn}}$ [cm$^2$ g$^{-1}$]  & Effective grey opacity of the dynamical ejecta     & $\sim$ U(1.0, 10.0)         & $\sim$ U(5.0, 8.0)                          \\ \midrule
M$_{\rm{wind}}$ [M$_{\odot}$]          & Mass of the disk wind ejecta                       &  \multicolumn{2}{l}{Eq. (\ref{eq:nsns_disk}) (BNS) or (\ref{eq:bhns_disk}) (BHNS) and Eq. \ref{eq:fraction}      }             \\
v$_{\rm{wind}}$ [c]                    & Velocity of the disk wind ejecta                   & $\sim$ U(0.1, 0.3)         & $\sim$ U(0.12, 0.18)                          \\
$v_{\rm{min, wind}}$ [c]               & Minimum velocity of the disk wind ejecta           & $\sim$ U(0.1, 1.0) v$_{\rm{wind}}$  & $\sim$ U(0.3, 0.5) v$_{\rm{wind}}$                \\
$v_{\rm{max, wind}}$ [c]               & Maximum velocity of the disk wind ejecta           & $\sim$ U(1.5, 2.0) v$_{\rm{wind}}$      & $\sim$ U(1.5, 1.7) v$_{\rm{wind}}$             \\
n$_{\rm{wind}}$                        & Power law index of density distribution            & $\sim$ U(3.5, 4.5)                                   \\
v$_{\kappa}$ [c]                       & Transition velocity between low and high $\kappa$  & $\sim$ U(v$_{\rm{min, wind}}$, v$_{\rm{max, wind}}$) & \\
$\kappa_{\rm{low}}$ [cm$^2$ g$^{-1}$]  & Effective grey opacity for v$ \leq$ v$_{\kappa}$   & $\sim$ U(0.1, 1.0)                 & $\sim$ U(0.1, 0.3)                  \\
$\kappa_{\rm{high}}$ [cm$^2$ g$^{-1}$] & Effective grey opacity for v$ > $v$_{\kappa}$      & $\sim$ U(1.0, 10.0)  & $\sim$ U(6.0, 10.0)\\   \bottomrule                              
\end{tabular}
\label{tab:tab1}
\end{table*}

\begin{figure*}[t!]
\centering
\includegraphics[width=.9\textwidth]{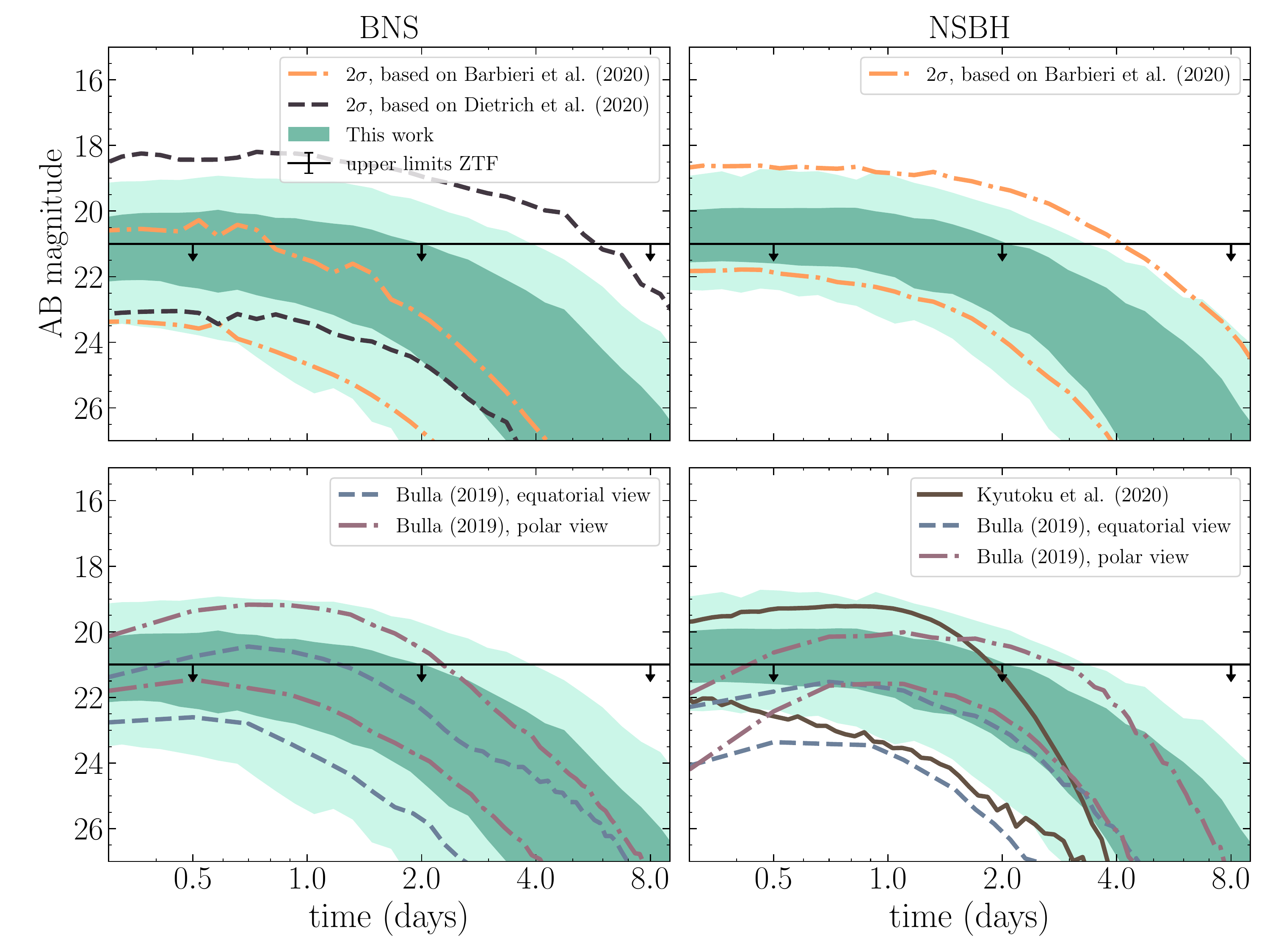}
\caption{The light curves in the g-band from assuming that GW190425 is a BNS system (left panels) or a BHNS system (right panels), given the equation of state constraints from GW170817. The different shaded bands indicate the regions that contain $68\%$ and $95\%$ of all light curves respectively. Also shown in the upper panels are the $95\%$ regions for light curves computed using the code by \citet{Hotokezaka19}, but using the methods in \citet{Barbieri20} (orange dashed-dotted lines) and \citet{Dietrich20} (black dashed line). Note that, contrary to the upper bounds, the lower bounds of the $95\%$ credible regions should not be taken definitively as part of the parameter space consistent with GW190425 results in (almost) no EM radiation. The brown solid lines in the lower right panel are two models from \citet{Kyutoku20}; one for a lanthanide-rich disk outflow and dynamical outflow at 140 Mpc and one for a lanthanide-poor disk outflow and dynamical outflow at 130 Mpc. The blue (grey) dashed lines in the lower panels correspond to an equatorial view (polar view) of light curves computed with \texttt{POSSIS} \citep{Bulla19}, with ejecta masses set by $0.01 \leq$ M$_{\rm dyn}$, M$_{\rm wind} \leq 0.03$ \msol~for BHNS, $0.001 \leq$ M$_{\rm dyn} \leq 0.02$ \msol~and $0.01 \leq$ M$_{\rm wind} \leq 0.03$ \msol~for BNS, and distances in the range $120-200$ Mpc. The black horizontal lines correspond to the upper limits found with ZTF by the GROWTH collaboration, covering $21\%$ integrated probability of the skymap for this event. \citep{GROWTH190425}. }
\label{fig:fig3a}
\end{figure*}

\begin{figure*}[t!]
\centering
\includegraphics[width=.9\textwidth]{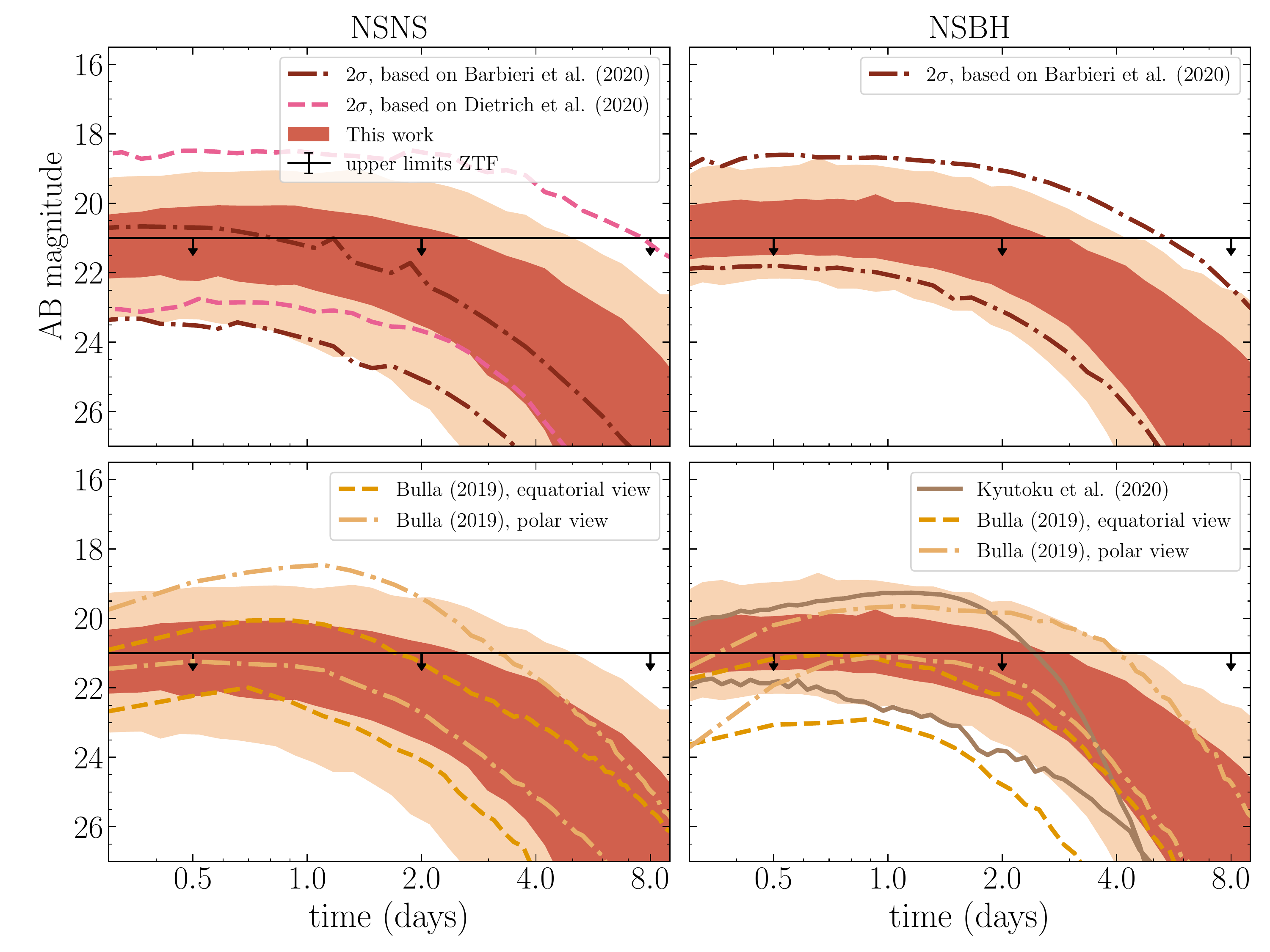}
\caption{Same as Figure \ref{fig:fig3a}, but for the r-band.}
\label{fig:fig3b}
\end{figure*}

\section{Light curve modeling} 
\label{sec:lightcurve}
From the inferred ejecta parameters discussed in the previous section, we can generate the corresponding light curves. To do so, we use the publicly\footnote{\url{https://github.com/hotokezaka/HeatingRate}} available code by \citet{Hotokezaka19}, a semi-analytic model based on the work by \citet{Li98}. In contrast to other semi-analytic light curve models \citep[see, e.g.][]{Waxman19}, both the effect of the time delay between photon production and photon emergence and the gradient in the velocity are  incorporated. We note, however, that the code does not take into account the asymmetric geometry of the outflows \citep[see, e.g.,][for viewing angle dependent light curve models]{Barbieri19a, Bulla19, Kawaguchi20, Zhu20, Heinzel21} which can have a significant impact on the luminosity and color of the observed light curve \citep{Bulla19, Darbha20, Korobkin20}. We also fix, for an increase in computational efficiency,\footnote{The code by \citet{Hotokezaka19} does include a calculation of the thermalization based on the ejecta properties.} the nuclear heating rate to the model by \citet{Korobkin12} (with thermalization efficiency $\epsilon = 0.5$). Again, we recognize that this can lead to additional uncertainty in the kilonova light curve \citep{Barnes20}. 

For each sample of M$_{\rm dyn}$, v$_{\rm dyn}$, and M$_{\rm wind}$ computed from the posterior PDFs of GW190425, we estimate the corresponding light curve. Since the software by \citet{Hotokezaka19} does not only take an average velocity as input but also requires a velocity distribution, which we have chosen to be a power law distribution, we have to consider a few extra parameters. These consist of a power law index $n$, varied between $3.5 - 4.5$, and a minimum and maximum velocity cutoff. We split the calculation into two components, a dynamical component and a post-merger disk wind component. For the dynamical component, the minimum velocity cutoff varies between $0.1$v$_{\rm dyn}$ and v$_{\rm dyn}$, and the maximum velocity cutoff between $1.5$v$_{\rm dyn}$ and $2.5$v$_{\rm dyn}$. For the post-merger disk winds, the average velocity of the ejecta is not connected to the properties of the binary but is taken to be between $0.1c$ and $0.3c$. The minimum velocity cutoff is the same as for the dynamical component, while the maximum velocity cutoff is slightly lower, i.e., between $1.5$v$_{\rm wind}$ and $2.0$v$_{\rm wind}$. 

The final parameters that we need to include are the effective opacities of the ejected material. The dynamically ejected material is very neutron-rich and will, therefore, produce lanthanides through r-process nucleosynthesis, leading to a high opacity.\footnote{Again with the caveat that more symmetric NS mergers will also produce shock heated material with lower opacities (see Section \ref{subsec:dynej}).} We allow the effective grey opacity for this component to be within the range of $1 - 10$ cm$^2$ g$^{-1}$. Since the post-merger disk winds consist of ejecta with varying composition, we assign two effective grey opacities, one in the range $0.1-1.0$ cm$^2$ g$^{-1}$ for $v \leq v_{\kappa}$ and one in the range $1 - 10$ cm$^2$ g$^{-1}$ for $v>v_{\kappa}$. The transition velocity $v_{\kappa}$ is randomly drawn between the minimum and maximum of the velocity distribution. Table \ref{tab:tab1} shows a summary of all parameters and their allowed ranges. 

The final light curves are then obtained by adding the two components, where estimates for different photometric bands rely on the assumption of blackbody emission.\footnote{The temperature is estimated through the bolometric luminosity $L_{\rm bol}$ and the radius of the photosphere $r_{\rm ph}$ as $T = L_{\rm bol}/(4\pi \sigma r^2_{\rm ph})$.} The uncertainty in the distance is also included by randomly drawing a distance measurement from the posterior PDF of GW190425. In Figures \ref{fig:fig3a} and \ref{fig:fig3b}, we show the light curves corresponding to posterior PDF distributions of GW190425 in the g-band and r-band, for both the BNS scenario (left panels) and the BHNS (scenario). The dark and light shaded regions indicate the regions where $68\%$ and $95\%$ of the light curves are contained respectively. The black horizontal line shows the upper limits obtained by the GROWTH collaboration using ZTF \citep{GROWTH190425}.

As a comparison, and to better assess the current modeling uncertainties, we also include light curves obtained with our code when the input parameters are derived from other methods (upper panels). First, we compute light curves from the distribution of ejecta masses and velocities following the methods outlined in \citet{Coughlin19} with updated fitting formulae from \citet{Dietrich20}, which are tailored to BNS systems. In \citet{Coughlin19} the authors also use a two-component model, divided into dynamical and disk wind ejecta. They relate the dynamical ejecta mass to the total ejected mass in the first component by asserting the proportionality M$_{\rm{ej,1}} = \frac{1}{\alpha}$ M$_{\rm{dyn}}$, where $0.01 \leq \alpha \leq 1$. For the second component, the mass is assumed to be a fraction of the disk mass M$_{\rm{ej,2}} = \xi$ M$_{\rm{disk}}$, where $0 \leq \xi \leq 0.5$. The velocity of the second component and the lanthanide fraction of both components are not related to any binary parameters, but considered as free parameters in the ranges $0.03 \leq$ v$_{\rm{disk}} \leq 0.3$ and $-9 \leq \log_{10}(X_{\rm{lan}}) \leq -1$. The black dashed (pink dashed) lines in Figure \ref{fig:fig3a} (\ref{fig:fig3a}) illustrate the region where $95\%$ of the light curves are contained. Since the parameters $\alpha$ and $\xi$ allow for a larger range of ejecta masses, the resulting kilonova light curves are relatively broad, but consistent with the light curves presented in this paper.  

Second, we show the light curves computed with our code using the ejecta properties obtained by following the methods in \citet{Barbieri19a}. The authors use a similar approach to the methods presented in this paper, though with a three-component model; a dynamical component, a neutrino-driven disk wind component, and a viscous (secular, corresponding to `thermally-driven' in our framework, Equation \ref{eq:fraction}) component. From the disk mass, they estimate that for a BHNS system, $1\%$ is ejected through neutrino-driven winds while $20\%$ is ejected through viscous processes, both with a velocity of $0.1$c. For a BNS system, $5\%$ is ejected through neutrino-driven winds and $20\%$ through viscous processes with velocities of $0.067$c and $0.04$c respectively. 
An effective grey opacity is used, which is set to $15$ cm$^2$ g$^{-1}$ for the dynamical ejecta and to $1$ and $5$ cm$^2$ g$^{-1}$ for respectively the wind and viscous ejecta in a BHNS scenario. For BNSs, only the opacity of the disk wind is lowered to $0.5$ cm$^2$ g$^{-1}$. The authors do include an angle dependence in their light curve computations but this is not included here. The effect of including angle dependencies is larger for the BHNS scenario, where the geometry of the dynamical ejecta is more likely to be confined to the equatorial plane and not in all azimuthal directions. As a result the kilonova appears brighter when we observe in the polar direction and fainter when we observe in the equatorial plane. The yellow (black) dot-dashed lines in Figure \ref{fig:fig3a} (\ref{fig:fig3b}) again indicate the region where $95\%$ of all light curves are contained. For the BHNS scenario, these regions are comparable to our results, mostly due to the fact that the same underlying fitting formula for the disk mass is used \citep[see][]{Foucart17}. Their method does, however, predict fainter light curves for the BNS scenario. Although Figure \ref{fig:fig4} suggests that the disk wind mass is comparable to our results, there is a large fraction of samples not shown, where the dynamical ejecta mass is zero and the disk wind mass is very low, that dominate the resulting region containing $95\%$ of all light curves. Another factor that contributes to the differences in the resulting light curves are the higher opacities chosen for dynamical and disk wind ejecta. 

We also show light curves that are computed using different approaches to generate light curves. In the case of a BHNS system, we show the light curves computed by \citet{Kyutoku20} (shown in their Figure 1). The authors use a Monte-Carlo radiation-transfer code \citep[see][]{Tanaka13, Kawaguchi20} to estimate light curves from a representative disk mass and dynamical mass outflow. Instead of using an effective grey opacity, as in this work, the opacity is time-dependent and derived with updated atomic structure calculations of r-process elements \citep{Tanaka20}. The variation in their light curves is a combination of different viewing angles, composition, and distances. The lower limit shown here corresponds to their equatorial view of a lanthanide-rich disk outflow and dynamical ejecta at 140 Mpc, while the upper limit corresponds to a polar view of a lanthanide-poor disk outflow with dynamical ejecta at 130 Mpc. 

Lastly, we show light curves computed with \texttt{POSSIS} \citep{Bulla19}, a three dimensional Monte-Carlo code that models the radiation transport in kilonovae, taking into account the geometry of the merger outflows and the viewing angle. For faster light curve generation the code does not solve the full radiative transfer equation but rather uses time- and wavelength dependent opacities as direct input. For the BHNS case, we use the grid presented in \cite{Anand2020} and restrict to ejecta masses that are consistent with the distribution of ejecta masses computed in Section \ref{sec:ejecta}, i.e.  M$_{\rm dyn} = 0.01, 0.02, 0.03$ \msol and M$_{\rm wind} = 0.01, 0.02, 0.03$ \msol. The distance is varied between $120-200$ Mpc, corresponding to the $1\sigma$ range of the luminosity distance measure inferred in \citet{GW190425discovery}. We show the resulting bands for two viewing angles, one in the equatorial and one in the polar plane. Similarly, for the BNS case, we use the grid computed in \citet{Dietrich20} and restrict to M$_{\rm dyn} = 0.001, 0.005, 0.01, 0.02$ \msol and M$_{\rm wind} = 0.01, 0.03$ \msol.

We conclude that although the methods to compute the light curves in Figure \ref{fig:fig3a} and \ref{fig:fig3b} are quite different, there is broad agreement on the photometric magnitude range that is predicted for GW190425. This is due to the large statistical uncertainties in the GW posterior PDF, and specifically the large spread in distance that directly affects the magnitude of the light curve. There is, however, some discrepancy between the light curves computed from the ejecta properties based on the fitting formulae in \citet{Barbieri19a} and the formulae in \citet{Dietrich20} and this work. This is again a result of the lower predicted $M_{\rm dyn}$ and $M_{\rm wind}$ in this particular part of parameter space using the formulae in \citet{Radice18b} and \citet{Barbieri19a}. When comparing the light curve ranges between BHNS and BNS it is not clear that an EM detection would be able to distinguish the nature of the system, but this warrants a more detailed study which we will leave for future work.

\begin{figure*}[t!]
\centering
\includegraphics[width=.95\textwidth]{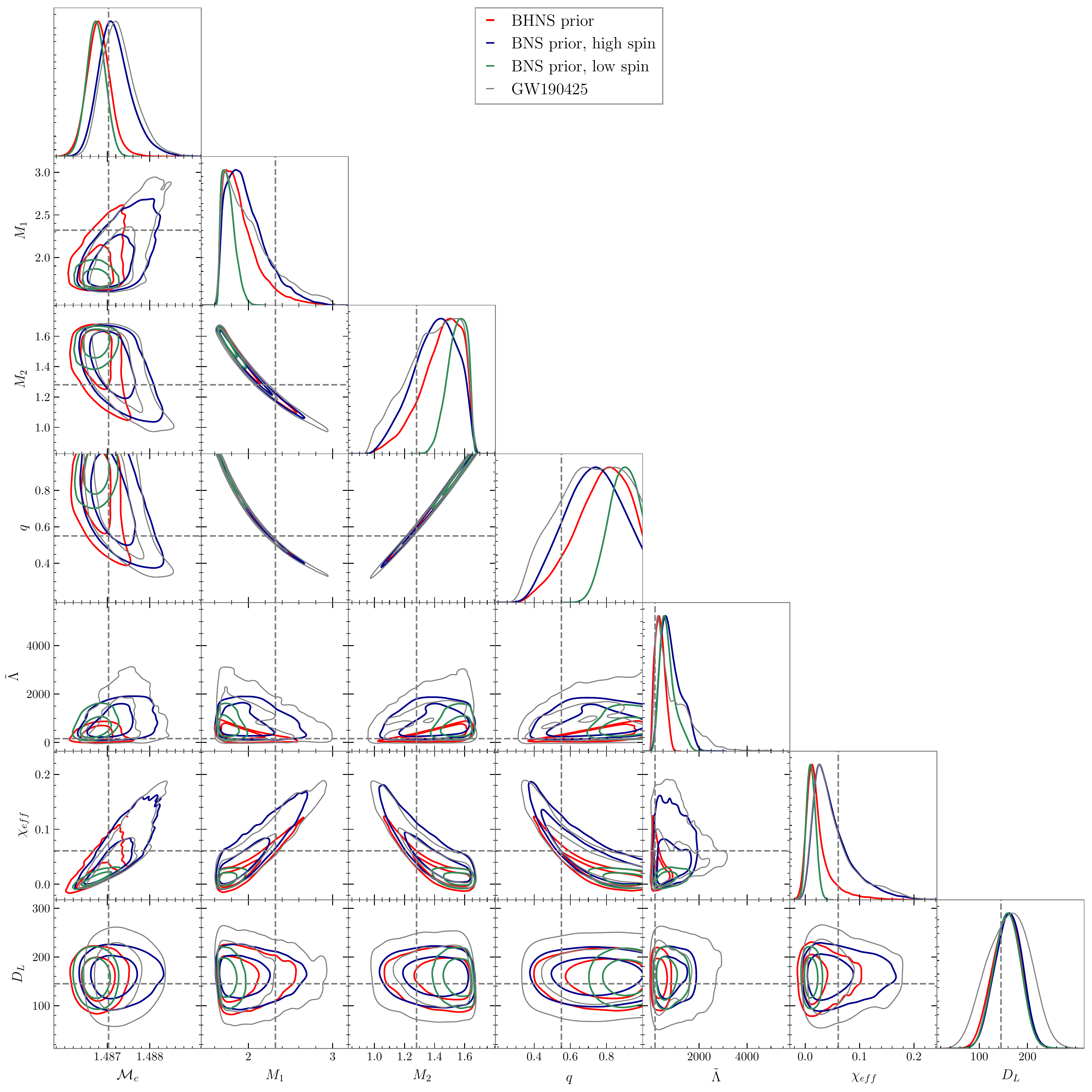}
\caption{Posterior distributions of binary properties inferred from a gravitational wave signal. The grey contours correspond to the posteriors obtained in \citet{GW190425discovery} with a high spin prior. The colored contours are the posteriors obtained with relative binning from our simulated signal. The red, blue and green colors indicate a BHNS prior, a BNS prior with high spins and a BNS prior with low spins respectively. The dashed grey lines indicate the injected values.}
\label{fig:fig5}
\end{figure*}

\begin{figure}[h!]
\centering
\includegraphics[width=.45\textwidth]{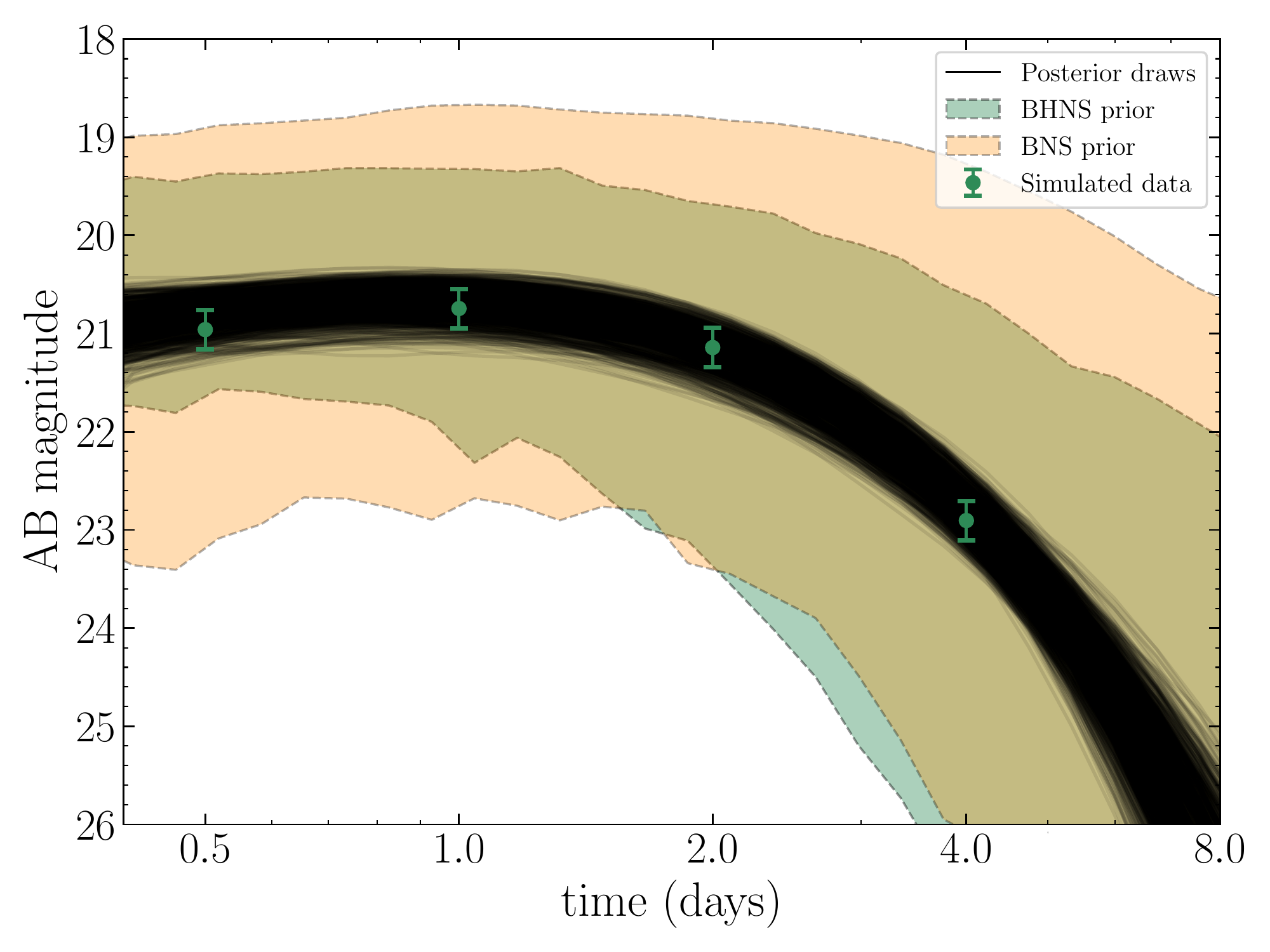}
\caption{The green points indicate the EM data points in the g-band used for the analysis in Section \ref{sec:multimessenger}, while the shaded bands indicate the $95\%$ credible regions for the priors, assuming either a BHNS or a BNS system. In black we show a selection of posterior draws from the posterior distribution obtained in Case 2 (see main text Section \ref{subsec:multimessenger}).}
\label{fig:fig6}
\end{figure}

\section{Multimessenger parameter estimation}
\label{sec:multimessenger}
\subsection{Gravitation wave analysis}
\label{subsec:gwanalysis}
For fast analysis of the GW signal, we use the relative binning method introduced by \citet{Zackay18}, which exploits the fact that the difference between gravitational waveforms with nonzero posterior probability density in the frequency domain can be described by a smoothly varying perturbation. By using the ratio of a gravitational waveform with some fiducial waveform close to where the likelihood peaks, the number of frequency points where the waveform is evaluated reduces to $\mathcal{O}(10^2)$, compared to $\mathcal{O}(10^7)$ for traditional GW parameter inference. However, \citet{Zackay18} did not include any extrinsic geometric parameters \citep[see, e.g.,][]{GWdataguide} in their method so here we have extended the relative binning method used in our code to include parameters such as distance, sky location, inclination, and polarization \citep[see also][for a similar approach]{Finstad20}. 

\begin{table}[]

\caption{Source properties of the simulated gravitational wave signal.}
\centering
\begin{tabular}{ll}
\toprule
Parameter                                        & Injected value \\ \midrule
$\mathcal{M}_c$ (source frame)   & 1.441    \msol    \\
$\mathcal{M}_c$ (detector frame) & 1.4871   \msol      \\
M$_{1}$ (source frame) & 2.25 \msol \\
M$_{2}$ (source frame) & 1.24 \msol \\
q                                                & 0.55           \\
$\Lambda_1$                                      & 0              \\
$\Lambda_2$                                      & 1000          \\
$\chi_{1,z}$                                     & 0.1            \\
$\chi_{2,z}$                                     & -0.02          \\
Distance                                    & 145 Mpc          \\ \bottomrule
\end{tabular}
\label{tab:sourceparams}
\end{table}

We simulate a waveform similar to GW190425 (see Table \ref{tab:sourceparams} for the source parameters), using the \texttt{IMRPhenomD\_NRTidalv2} model \citep{Dietrich19}, and inject this into the LIGO Livingston and Virgo detectors with an SNR of 11.6 and 4.7 respectively. We note that the mass of the primary object, M$_1$, is within the range of possible NS masses \citep[see e.g.][]{Raaijmakers20}, but by setting the tidal deformability to be  $\Lambda_1 = 0$, we set this object to be a BH.\footnote{We note that there exist now publicly available waveform models dedicated to BHNS signals that were not available when this work started \citep[see, e.g.,][]{Matas20, Thompson20}.} Furthermore, it is important to mention that the spin of the BH is outside of the range of the low-spin prior, which was chosen as a test case to see how the typically employed prior choice of LVC analyses might biases the parameter estimation if the type of the progenitor system is not certainly determined. To generate the noise, we use a realistic Power Spectral Density (PSD)\footnote{https://dcc.ligo.org/LIGO-T2000012/public} from O3a and assume it to be stationary and Gaussian. We then compute the summary data necessary for relative binning and set up our likelihood function. If we assume the noise in the detectors to be Gaussian and defined by a power spectrum $S(f)$, we can write the likelihood as,

\begin{align}
    {\cal L}(\bm{d}_{\rm GW} &~|~ \bm{\theta}_{\rm GW})   \propto \\
    & \exp \left[-\frac{1}{2}\langle \bm{d}_{\rm GW} - \bm{h} (\bm{\theta}_{\rm GW}) ~|~ \bm{d}_{\rm GW} - \bm{h} (\bm{\theta}_{\rm GW})\rangle \right], \nonumber
\end{align}

where $\bm{d}_{\rm GW}$ is the GW strain data, $\bm{h}(\bm{\theta}_{\rm GW})$ a gravitational waveform defined by the vector set of model parameters $\bm{\theta}_{\rm GW}$, and

\begin{equation}
    \langle \bm{a}~|~\bm{b} \rangle = 4 \Re \int_{f_{\rm min}}^{f_{\rm max}} \frac{\tilde{a}^{*}(f) \tilde{b}(f)}{S(f)} df.
\end{equation}
We use Bayes' theorem to express,
\begin{equation}
\mathcal{P}\left(\bm{\theta}_{\rm GW} ~|~ \bm{d}_{\rm GW}\right) \propto \pi \left(\bm{\theta}_{\rm GW}\right) \mathcal{L}\left(\bm{d}_{\rm GW} ~|~ \bm{\theta}_{\rm GW} \right),
\end{equation}

where $\mathcal{P}\left(\bm{\theta}_{\rm GW} ~|~ \bm{d}_{\rm GW}\right)$ is the posterior distribution on the  gravitational wave parameters given the data, and $\pi \left(\bm{\theta}_{\rm GW}\right)$ is the prior distribution on these parameters.
To sample from the posterior PDF distribution, we use the nested sampling algorithm \texttt{MultiNest} \citep{Feroz09, Feroz13, Buchner14}. Instead of sampling in component masses, we sample in the parameters chirp mass and mass ratio to speed up the convergence of our posterior distribution. We choose a uniform prior in both, within the range $\mathcal{M}_{c} \in [1.477, 1.497]$ and $q = 1/Q \in [0.2, 1]$. For the tidal deformability of the two components, we sample uniformly as well within the range $\Lambda_{1,2} \in [0, 5000]$ when assuming a BNS system, while for our BHNS prior we set $\Lambda_1 = 0$. The prior for the two dimensionless spins is uniform in magnitude, while the orientation is isotropically distributed, although here we only consider spins that are aligned with the orbital angular momentum to speed up the parameter inference. When assuming a BNS system, we either choose a low spin, $|\chi| \in [0, 0.05]$, or a high spin prior, $|\chi| \in [0, 0.89]$. For the BHNS prior, the spin of the heavier object is uniform over $|\chi| \in [0, 0.89]$, while the lighter object is uniform over $|\chi| \in [0, 0.05]$. The prior for the luminosity distance is uniform in comoving volume, such that $\pi(D_L) \propto D_L^2$, with bounds chosen to be far from where the posterior probability is non-negligible. The priors for the right ascension and declination are taken uniformly across the sky, and the priors for the inclination and polarization are uniformly distributed in angle from 0 to 2$\pi$ or $\pi$, respectively. We show the posterior PDF distributions of all three prior choices in Figure \ref{fig:fig5} and include the posterior distribution of GW190425 for the high spin prior case in \citet{GW190425discovery} for comparison, which is the most agnostic case when the nature of the binary is not known.

Figure~\ref{fig:fig5} shows the effects of changing the priors on the spins as seen by comparing the green and blue curves. When comparing the red to the other colored curves, it also shows the effects of the assumption on the nature of the binary, which involves different priors on tidal deformabilities (for the BH it is set to zero) as well as on the spins (high spin for the BH, low spin for the NS). As expected, the luminosity distance measured from the GW amplitude is largely unaffected by changes to the priors on parameters that mainly impact the phasing. We note that even with the BHNS priors corresponding to the injected signal, the true values of the parameters are not correctly recovered due to the small SNR. We also note that the BHNS assumption actually does not yield results close to the injected values for any of the parameters except for the tidal deformability. Instead, for the masses and spin parameter the high-spin BNS priors lead to results closest to the true values. 
As expected, the mass and spin measurements are most impacted by the spin priors, as seen by comparing the blue and green curves. This is a known degeneracy due to the fact that effects of the mass ratio and spins first enter the phasing only separated by half of a post-Newtonian order and thus accumulate similarly with the GW frequency. Our results also show that the PDF for tidal deformability is most strongly dependent on the prior assumptions on the type of binary system.  For reference, the results from \citet{GW190425discovery} with a high spin prior are also shown as the grey contours, which are in good agreement with our corresponding simulated results depicted by the blue curves.  Note that the GW signal considered here has a low SNR, similar to GW190425. Therefore the following discussion on additional constraining power from EM observations will only apply to such low-SNR signals. High-SNR signals, and signals detected across a broader frequency range by third generation GW detectors, already provide tighter constraints on binary parameters, which affect the additional constraining power of EM observations. 

\begin{figure*}[t!]
\centering
\includegraphics[width=.49\textwidth]{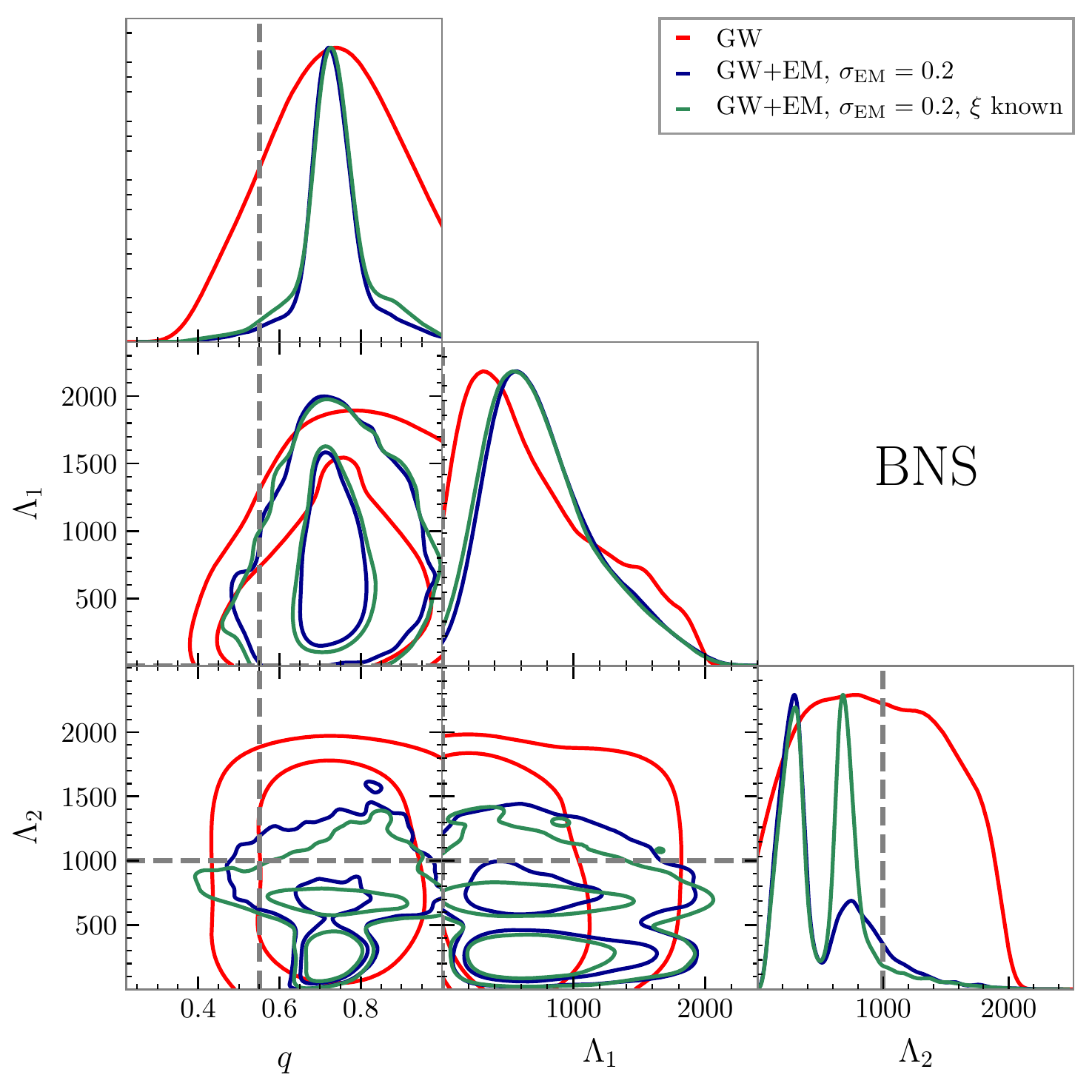}
\includegraphics[width=.49\textwidth]{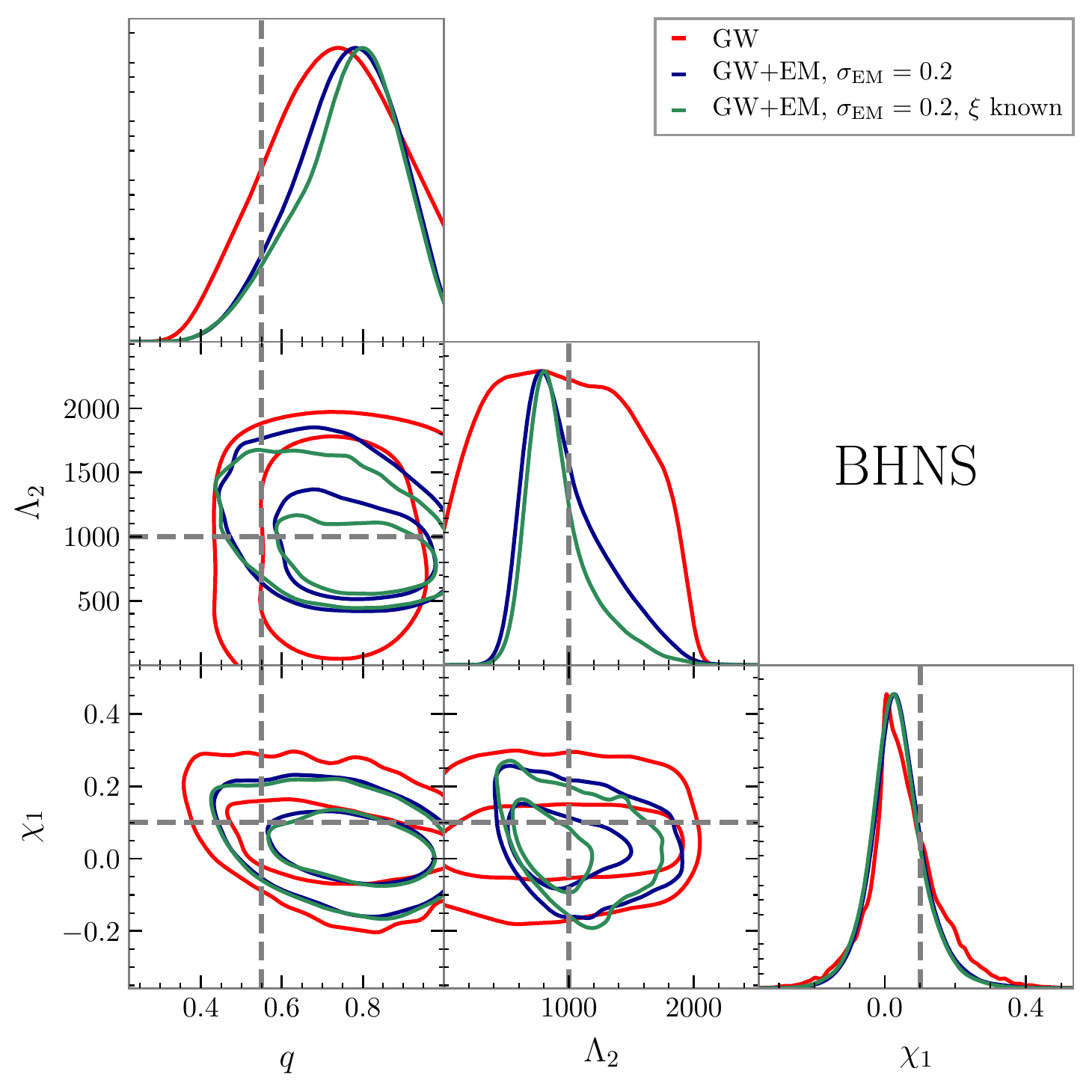}
\caption{Posterior distribution on binary parameters given the GW data (red) using the high-spin BNS prior, and when including an EM counterpart with either unknown parameter $\xi$ (blue) or known (green). The grey dashed lines indicate the values of the binary properties used to simulate the GW and EM data (note that since $\Lambda_1 = 0$ the underlying value is not visible here) .}
\label{fig:fig7}
\end{figure*}

\subsection{Multimessenger analysis} 
\label{subsec:multimessenger}
In order to test how an EM counterpart to a GW event can constrain the properties of the merging binary, we combine our EM model defined in Section \ref{sec:ejecta} and \ref{sec:lightcurve} with the GW analysis mentioned above in Section \ref{subsec:gwanalysis}. Taking the values of our simulated GW event, we compute the corresponding light curve by using Equations (\ref{eq:nsns_ejecta})-(\ref{eq:bhns_disk}), and additionally setting $v_{\rm{min, dyn}} = 0.8v_{\rm dyn}$, $v_{\rm{max, dyn}} = 1.5v_{\rm dyn}$, $\kappa_{\rm dyn} =  5$ cm$^2$ g$^{-1}$ for the dynamically ejected component and $v_{\rm wind} = 0.15$c, $v_{\rm{min, wind}} = 0.4v_{\rm wind}$, $v_{\rm{max, wind}} = 1.5v_{\rm wind}$, $\kappa_{\rm low} = 0.1$ cm$^2$ g$^{-1}$, $\kappa_{\rm high} = 8$ cm$^2$ g$^{-1}$, and $v_{\kappa} = 0.18$c for the disk wind ejecta. For the definition of these parameters, we refer back to Section \ref{sec:lightcurve} or Table \ref{tab:tab1}. We evaluate this light curve at $t = [0.5, 1, 2, 4]$ days in the g- and r-band to obtain our EM dataset (see Figure \ref{fig:fig6}).

Under the assumption that the EM dataset is independent of the GW dataset, we can separate each likelihood and express the posterior PDF distribution as

\begin{align}
\label{eq:posterior}
\mathcal{P}(\bm{\theta}~|~ \bm{d}) &\propto \pi(\bm{\theta}) \mathcal{L}(\bm{d} ~|~ \bm{\theta}) \\
& \propto \pi(\bm{\theta}) \mathcal{L}(\bm{d}_{\rm GW} ~|~ \bm{\theta}_{\rm GW,EM}) \mathcal{L} (\bm{d}_{\rm EM} ~|~ \bm{\theta}_{\rm EM})\,.
\end{align}
The vector $\bm{\theta}$ now contains both the parameters needed for the GW waveform generation that affect the light curves, $\bm{\theta}_{\rm GW,EM}$, and the parameters to compute an EM light curve, $\bm{\theta}_{\rm EM}$. For the GW likelihood, we assume that it is proportional to the posterior distribution obtained from the GW analysis described in the previous section: 
\begin{equation}
\label{eq:gwlikelihood}
    \mathcal{L}(\bm{d}_{\rm GW} ~|~ \bm{\theta}_{\rm GW,EM}) \propto \int \mathcal{P}\left(\bm{\theta} ~|~ \bm{d}_{\rm GW}\right) d\bm{\theta}_{\rm nuisance}\,,
\end{equation}
We marginalize over all parameters in the GW analysis that are not connected to the light curve model ($\bm{\theta}_{\rm nuisance}$) so that the vector $\bm{\theta}_{\rm GW,EM}$ only contains $\{ \mathcal{M}_{c}, q, \chi_{BH}, \Lambda_{NS}, D\}$ and $\{ \mathcal{M}_{c}, q, \Lambda_1, \Lambda_2 , D\}$ for a BHNS and BNS system, respectively. One implication of Equation (\ref{eq:gwlikelihood}) is that the prior distribution on these parameters is already incorporated into the likelihood, which we need to take into account when choosing our priors. % \samaya{cite Kenta H0 paper for appendix}.
Since some of the parameters in $\bm{\theta}_{\rm EM}$ depend on the parameters $\bm{\theta}_{\rm GW,EM}$ through Equations \ref{eq:nsns_ejecta} - \ref{eq:bhns_disk}, we can rewrite Equation \ref{eq:posterior} using a hierarchical approach: 
\begin{align}
    \mathcal{P}(\bm{\theta}~|~ \bm{d}) \propto ~& \pi(\bm{\theta}_{\rm GW,EM})~ \pi(\bm{\theta}_{\rm EM} ~|~\bm{\theta}_{\rm GW,EM}) \\ 
   & \times \mathcal{L}(\bm{d}_{\rm GW} ~|~ \bm{\theta}_{\rm GW,EM})~ \mathcal{L} (\bm{d}_{\rm EM} ~|~ \bm{\theta}_{\rm EM}) \nonumber \,.
\end{align}

Lastly, we assume a Gaussian likelihood for the EM data: 
\begin{equation}
    \mathcal{L}(\bm{d}_{\rm EM} ~|~ \bm{\theta}_{\rm EM}) \propto \exp{\left(-\frac{(\bm{d}_{\rm EM} - L(\bm{\theta}_{\rm EM}))^2}{2 \sigma_{\rm EM}^2}\right)},
\end{equation}
where $\bm{d}_{\rm EM}$ are the photometric measurements of the light curve and $L(\bm{\theta}_{\rm EM})$ the computed light curve using the model presented in this work. Initially we set $\sigma_{\rm EM} = 0.2$ mag. However, to investigate how the results might depend on future improvements on the uncertainty in the photometric measurements, we also consider a hypothetical case where $\sigma_{\rm EM} = 0.01$ mag.

For our prior distributions, we choose a uniform prior over the parameters $\bm{\theta}_{\rm GW,EM}$ in $\pi(\bm{\theta}_{\rm GW,EM})$ to minimize the effect of imposing prior information twice. For the parameters $\bm{\theta}_{\rm EM}$ that do not depend on $\bm{\theta}_{\rm GW,EM}$, we use uniform priors in the ranges described in Table \ref{tab:tab1}. For parameters in the distribution $\bm{\theta}_{\rm EM}$ that do depend on $\bm{\theta}_{\rm GW,EM}$, i.e. the dynamical ejecta mass and velocity and the disk wind mass, we take a uniform distribution in the uncertainty range defined for that parameter according to Equations \ref{eq:nsns_ejecta} - \ref{eq:bhns_disk}. We assume that the detection of the EM counterpart leads to an identification of the host galaxy, which we simplify to fixing the distance to the true value $D = 145$ Mpc, both when computing the light curve magnitude and when evaluating $\mathcal{L}(\bm{d}_{\rm GW} ~|~ \bm{\theta}_{\rm GW})$. In Figure \ref{fig:fig6}, we show the $90\%$ credible region of the prior light curves in the g-band for both the BHNS and BNS scenario, together with the simulated EM data. 

The posterior PDF distribution is again sampled with \texttt{MultiNest} and shown for the parameters of interest in Figure \ref{fig:fig7} when using the high spin BNS prior for the gravitational wave data and both the BNS prior (left panel) and BHNS prior (right panel) for the EM data. Three different analyses are shown where: (i) only the GW data is considered; (ii) both GW and EM data are considered with $\sigma_{\rm EM} = 0.2$; and (iii) both GW and EM data are considered with $\sigma_{\rm EM} = 0.2$ \textit{and} $\xi$ is fixed. Hereafter we refer to these analyses as Case 1-3. 

For the posterior distributions using the BNS prior for the EM data, we first note that including EM information (Case 2) puts significant constraints on the mass ratio of the system. This can be attributed to the strong correlation between the mass ratio and M$_{\rm dyn}^{\rm BNS}$, where lower and higher values of $q$ lead to too much or too little dynamical ejecta respectively. The tidal deformability is not well constrained for the primary object, since this only has a small effect on the dynamical ejecta. However, for the secondary object the tidal deformability is the main parameter that relates to the disk winds and is therefore much more constrained. The double peak structure is the result of the correlation with both the disk wind and dynamical ejecta velocity. A higher tidal deformability leads to a higher disk wind mass with low velocity, but to be consistent with the early EM data, this requires a higher dynamical ejecta velocity. On the other hand a lower tidal deformability leads to a lower disk wind mass with high velocity, and thus requires a lower dynamical ejecta velocity. Fixing the $\xi$ parameter (Case 3) only strengthens the double peak structure in $\Lambda_2$, as the higher values of $\Lambda_2$ are more consistent with the now better measured disk mass.

\begin{figure}[t!]
\centering
\includegraphics[width=.48\textwidth]{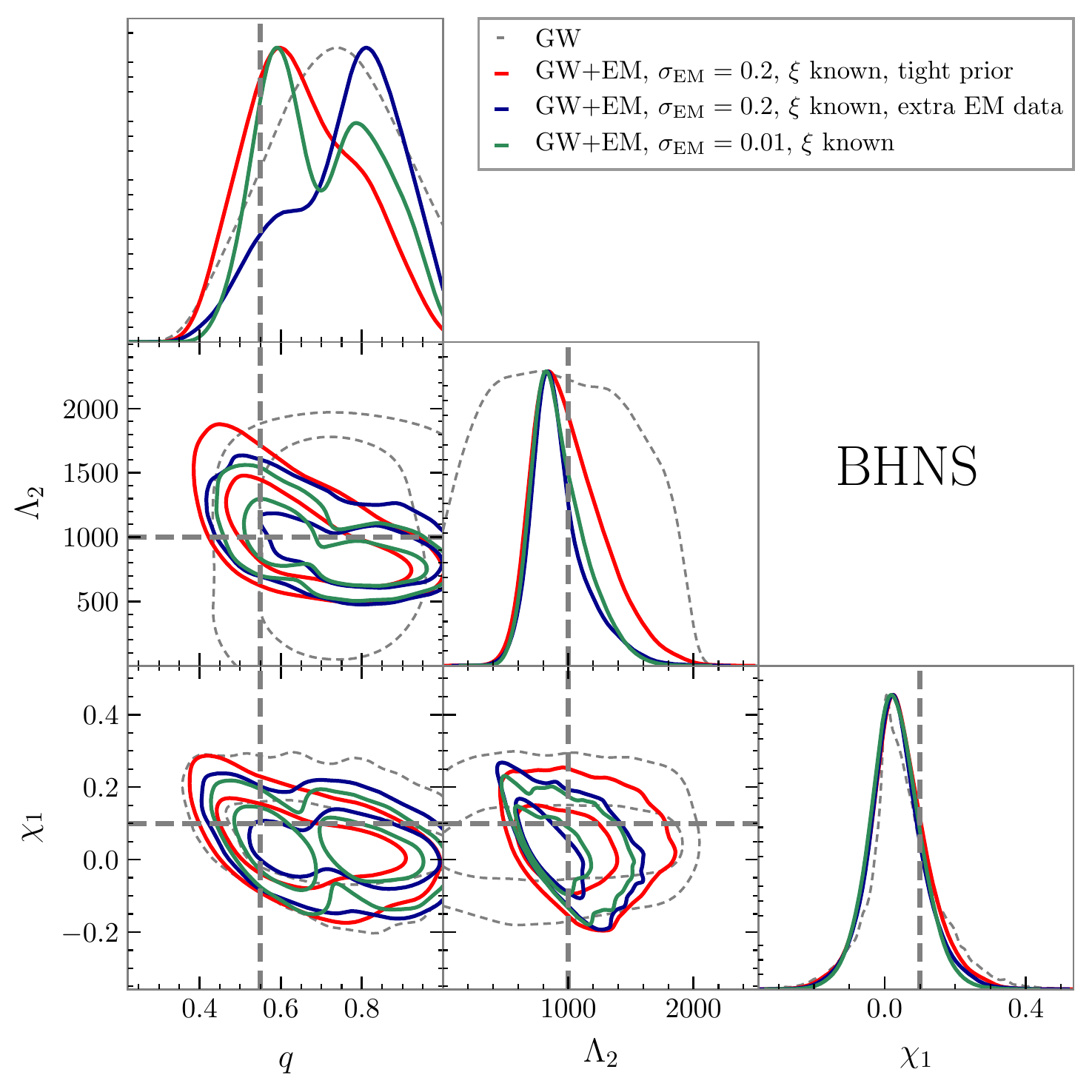}
\caption{Same as Figure \ref{fig:fig7}, but now showing posterior distributions where we decrease the prior volume for the EM parameters (see Table \ref{tab:tab1}) (red), add EM data in the i- and z-band (blue) and decrease the uncertainty in the EM data points to $\sigma_{\rm EM} = 0.01$ (green). The dashed, grey lines indicate the GW posterior for the high-spin BNS prior.}
\label{fig:fig8}
\end{figure}

In the right panel of Figure \ref{fig:fig7}, where a BHNS prior is used to analyze the EM data, we note that for Case 2 and 3, where $\sigma_{\rm EM} = 0.2$, there are only slightly better constraints on the mass ratio than for Case 1, due to the degeneracies between parameters in $\bm{\theta}_{\rm EM}$. The most significant improvement is made in the $\Lambda_2$ parameter. This is the result of both M$_{\rm disk}$ and M$_{\rm dyn}$ depending similarly on $\Lambda_2$, meaning that low values will result in too little ejected total mass, while high values will lead to too much ejected mass to be consistent with the EM data. Lastly, we note that the EM data considered here has very little additional constraining power on the dimensionless spin of the black hole.

To investigate whether we can improve on the constraints on the binary parameters, we study three more cases where the BHNS prior on the EM data is used. These are: (i) a scenario where we put a tighter prior on all parameters in $\bm{\theta}_{\rm EM}$ (see Table \ref{tab:tab1}); (ii) a scenario where on top of the EM data in the g- and r-band we also have data in the i- and z-band at the same epochs; and (iii) a scenario where the EM data have $\sigma_{\rm EM} = 0.01$. Hereafter, these are referred to as Case 4-6. Case 4 could be seen as a hypothetical future scenario where there is a better understanding of the relation between the binary properties and all outflow properties on top of the parameter $\xi$. Case 5 and 6 could help answer the question of whether the constraints improve more from taking data in other photometric bands or decreasing the uncertainty in the photometric points one already has. 

\begin{figure*}[t!]
\centering
\includegraphics[width=.95\textwidth]{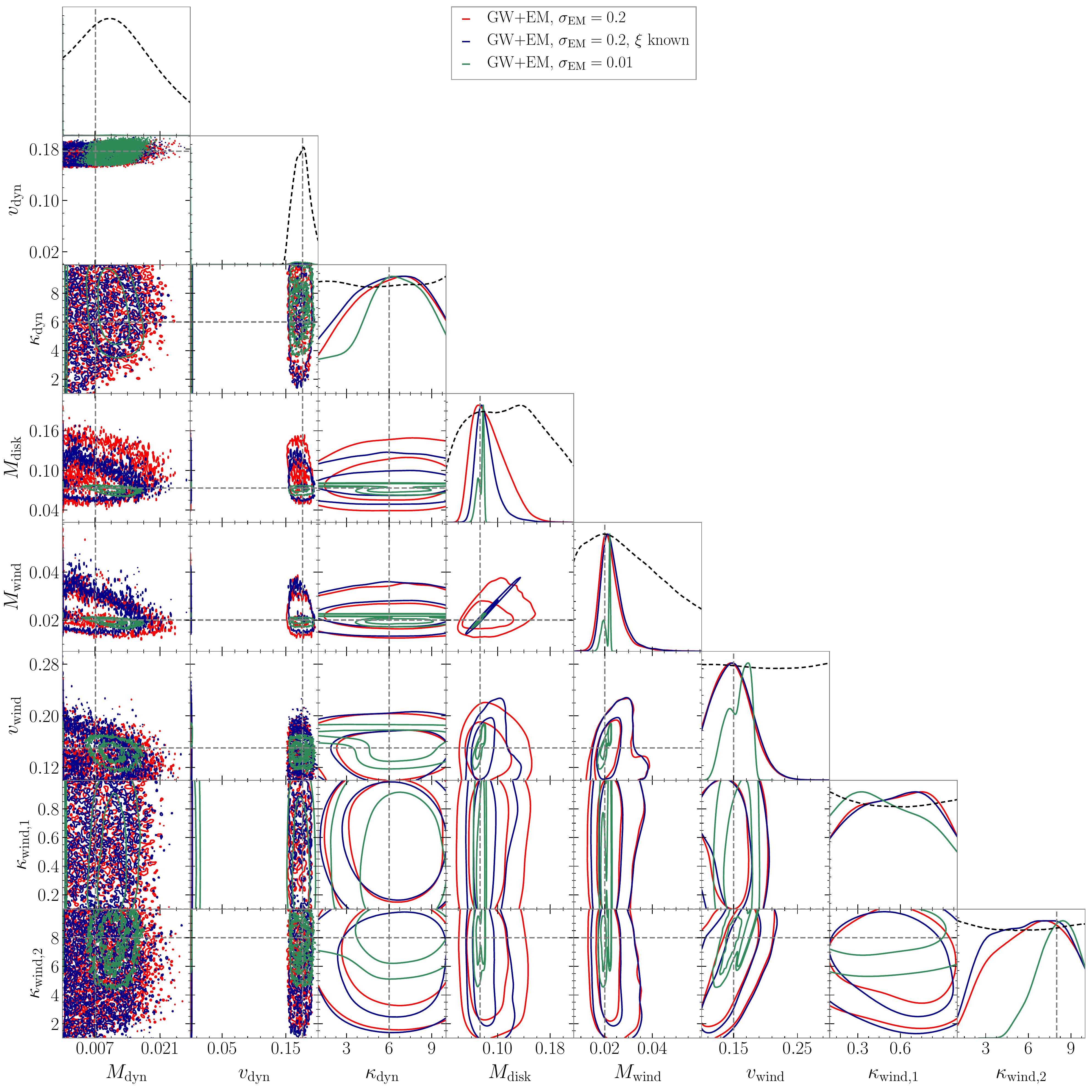}
\caption{Posterior distributions for outflow properties for Case 1, 2 and 6. The grey dashed lines indicate the injected values of outflow properties used to generate the EM data. We note that for Case 1 and 2 (red and blue lines) the dynamical ejecta mass and velocity are mostly consistent with being zero.}
\label{fig:fig9}
\end{figure*}

The posterior distributions corresponding to Case 4-6 are shown in Figure \ref{fig:fig8}. For all cases, the constraints on the binary parameters are tighter than in Case 2, although for Case 5, where data in more photometric bands are considered, the peak of the posterior still has more support for near-equal mass ratios away from the injected parameter value. This is a result of both the posterior for the GW data alone having more support for near-equal mass ratios (see Figure \ref{fig:fig5}) and the system considered here having a small amount of dynamical ejecta compared to disk wind ejecta. That means that the EM data can be well-fitted by models with only disk wind ejecta. This is illustrated in Figure \ref{fig:fig9}, where the posterior distributions on the outflow properties show very little support for dynamical ejecta larger than zero. Such models, with small dynamical ejecta compared to disk wind ejecta, are associated with more equal mass systems, which is why there is more posterior support for larger values of $q$. For Case 4 and 6, the data is less well fitted by models with only disk wind ejecta, causing the true underlying values to be within the $68\%$ credible regions. For Case 4, this is due to some of the degeneracy between the dynamical and disk wind ejecta being alleviated while for Case 6, the small error on the EM data allows for the small amount of dynamical ejecta to be distinguished from the disk wind ejecta.

\begin{figure}[t!]
\centering
\includegraphics[width=.49\textwidth]{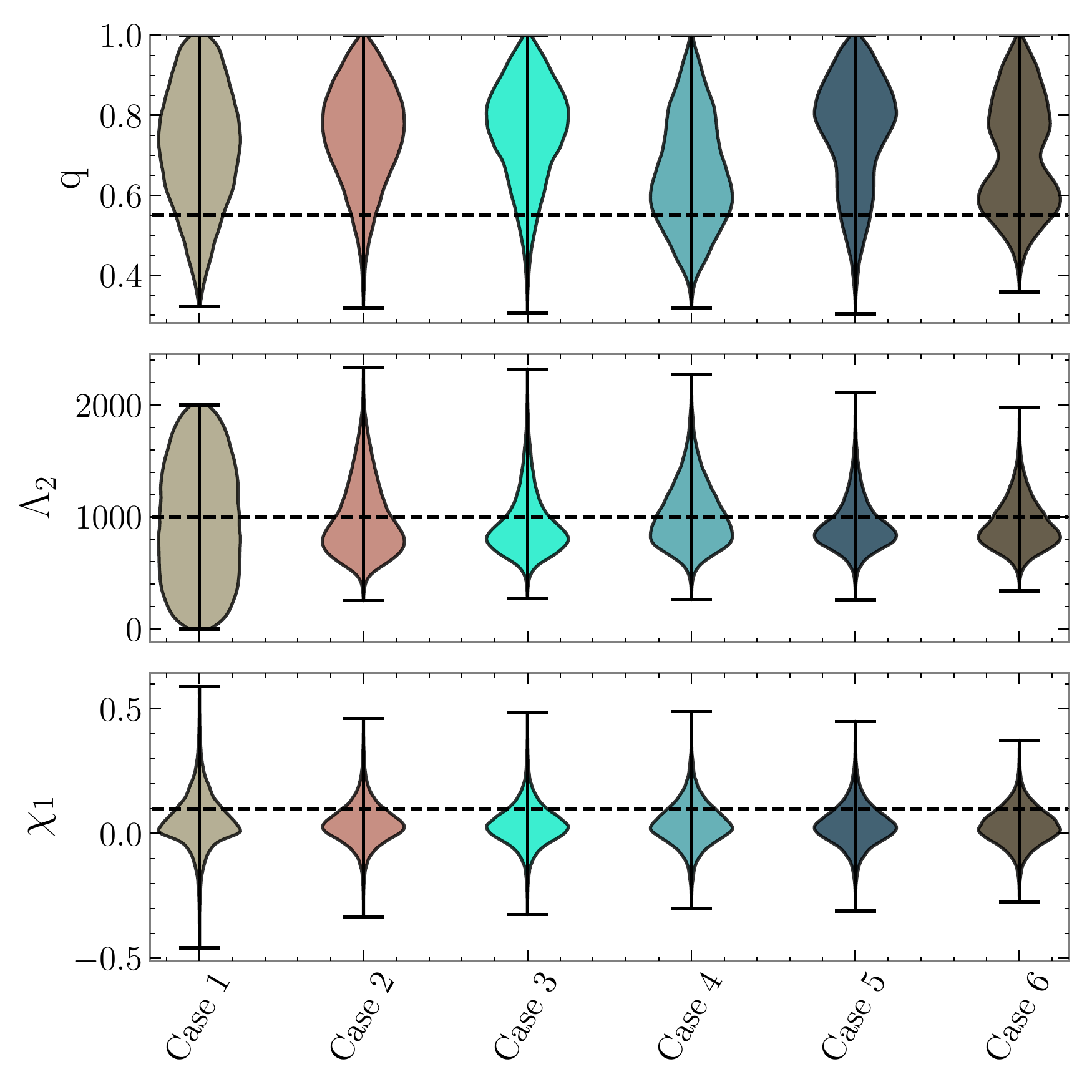}
\caption{Marginalized one-dimensional posterior distributions for the three binary parameters that connect to outflow properties. We show all six cases considered here, described in the main text of Section \ref{subsec:multimessenger}, and conclude that with EM observations (Cases 2 - 5) the constraining power increases when comparing to using only GW data (Case 1).}
\label{fig:fig10}
\end{figure}

\section{Discussion}
\label{sec:discussion}
Based on the results presented in Figures \ref{fig:fig7}, \ref{fig:fig8}, and \ref{fig:fig10}, we conclude that a detection of an EM counterpart in the future can lead to improved constraints on the binary progenitor parameters for systems similar to the low-SNR, BHNS/BNS system considered here. However, even in this idealized scenario where the EM data can be exactly modeled within this work's framework, the high-dimensionality of the problem and the existing degeneracies between parameters impact how well one can measure the fundamental properties of the progenitor. In this Section, we will first briefly summarize our results and then discuss the important caveats and limiting assumptions of our work.

On one hand, when modeling the EM lightcurve, if we assume that the GW190425-like signal (Section \ref{sec:multimessenger}) is a BHNS, the injected parameter values are within the $95\%$ credible region for all posterior distributions (right panel of Figure \ref{fig:fig7} and Figure \ref{fig:fig8}). However, the posterior distributions on the mass ratio have more support for near-equal mass ratios when only considering GW data (case 1), which does not change when adding EM data (Cases 2,3 and 5). This is because the small amount of dynamical ejecta in our simulated model allows for the EM data to be well fitted with models that have enough disk wind ejecta, i.e., models with higher values of $q$. This is resolved in Cases 4 and 6, where the posterior distribution on the mass ratio peaks around the injected value (see also Figure \ref{fig:fig10}). In all cases, the constraints on the tidal deformability improve the most, while constraints on the spin of the BH improve the least, compared to Case 1 (only GW data). This implies that currently the addition of EM data to GW190425-like events, where the tidal deformability is unconstrained \citep{GW190425discovery}, could give valuable information on the dense matter EOS. From the results of Cases 5 and 6, it is not yet clear whether it is preferred to take more data in different photometric bands or to improve the accuracy of the datapoints, which would need a more detailed study of the full parameter space and is left to future work. As expected, the constraining power, however, does increase in both cases. 

On the other hand, when modeling the EM light curves, if we assume that the GW190425-like signal is a BNS (Section \ref{sec:multimessenger}), the posterior distributions suggest that the injected binary parameters are not always within the $68\%$ credible region. Although this work focuses on the BNS framework in less detail, we note that parameter inference is highly dependent on the assumption of the nature of the system. GRMHD simulations by \citet{Most20} show that high-mass BNS systems lead to a fast dynamical ejecta component, whereas this component is absent for systems with similar parameters bar the primary object being a BH. In principle, this could help distinguish the nature of the system. However, if such a signal is not detected, both the GW data and EM data might look very similar (as shown in Figures \ref{fig:fig3a} and \ref{fig:fig3b}). In this case, one needs to be careful when interpreting results from parameter estimation analyses that assume the nature of the system.

 Furthermore, there are simplifying assumptions in the framework, detailed below, that could have a significant impact on multimessenger parameter inference, and need to be investigated in more detail in future work. 
 
 Firstly, throughout this work, we assume that ejecta outflows are spherically symmetric, which is a poor assumption especially for BHNS systems \citep[see, e.g.,][]{Kawaguchi16,Foucart17}. Depending on the viewing angle with which the system is observed, the light curve can be brighter or fainter in different photometric bands \citep{Bulla19, Darbha20, Korobkin20}.

 Secondly, information on the viewing angle is potentially obtainable from the observation of a relativistic jet, which can improve parameter estimation on systems with asymmetric outflow geometries \citep[see, e.g.,][]{Heinzel21}. Currently, the relativistic jet that is launched in some mergers \citep[see, e.g.][]{Ruiz21} is not taken into account in the model presented here. Neither is the interaction between the jet and the dynamical and disk wind ejecta which can alter the kilonova light curve \citep[][]{Nativi21,Klion21,Nicholl21}. 
 
 Thirdly, although we have made the simplifying assumption that the nuclear heating rate is fixed to a specific model derived by \citet{Korobkin12}, there exist multiple models for nuclear heating rates \citep[see, e.g.,][]{Hotokezaka19, Kasen19}. The uncertainty in the nuclear physics input can lead to up to one order of magnitude uncertainty in the bolometric luminosity \citep{Zhu20}. 

Fourthly, we note that the fitting formulae presented in Section \ref{sec:ejecta} do not cover the entire parameter space that our parameter inference spans (Section \ref{sec:multimessenger}). This means that the results presented here are dependent on interpolations and extrapolations of existing numerical simulations, which might be subject to change when more simulations are available in the (near) future. One could also increase the number of free parameters in these fitting formulae, by making use of numerical simulations where, e.g., the spin of the neutron star \citep{Ruiz20} or the orbital eccentricity of the binary \citep{Papenfort18} is taken into account.

Fifthly, we stress that the framework developed here is mostly focused on BHNS systems, with an approximate applicability to high-mass BNS systems when there is no short- or long-lived merger remnant. 

Finally, a more extensive parameter study is needed to investigate how EM observations of a larger diversity of binary mergers can constrain fundamental binary parameters. In order to better understand modelling errors, our kilonova model cross-comparison should be extended to incorporate parameter inference \citep[see, e.g.][]{Dietrich20, Heinzel21}. 

In conclusion, EM observations add constraining power to the GW data for the low SNR, BHNS system considered in this work. Our results suggest that this constraining power will improve when numerical models of merger outflows, telescope sensitivity and GW detectors progress \citep[see, e.g.,][]{Cowperthwaite19, Chornock19, Palmese19}. Currently, significant effort is underway into adding more sophisticated treatments of microphysical processes to numerical models, as well as increasing the coverage of the parameter space studied with these models \citep{baiotti16, Shibata19, Dietrich20a, Ciolfi20, Foucart20, Ruiz21a}. The Vera Rubin Observatory, an optical, wide-field telescope expected to begin operations in 2021 \citep{LSST}, will have a large impact on taking high-accuracy EM data and detecting GW events at larger distances, while telescopes as ZTF \citep{ZTF} and (near-future) dedicated GW follow-up telescopes such as BlackGEM \citep{BlackGEM}, GOTO \citep{GOTO} will increase the number of detected EM counterparts. Concurrently, next generation GW detectors, such as the Einstein Telescope \citep{EinsteinTelescope} and the Cosmic Explorer \citep{CosmicExplorer}, will increase the SNR of GW detections and widen the frequency range in comparison with LIGO/Virgo, allowing higher-precision measurements of binary properties. 

\acknowledgments 
We thank Kyohei Kawaguchi for providing us with their light curves and Andrew Williamson, Matthew Liska, Doosoo Yoon, Koushik Chatterjee, Philipp Moesta, Om Salafia,, Masaomi Tanaka and  Kenta Kuichi for useful discussions. We thank Barbara Patricelli and Leo Singer for their careful reading of the paper and their comments.  We are very grateful to the the LIGO Scientific, Virgo and KAGRA Collaborations for public access to their data products of GW170817, GW190425. We also thank the GROWTH collaboration for public access to their observational data products.  GR, SMN, TH, KL are grateful for financial support from the Nederlandse Organisatie voor Wetenschappelijk Onderzoek (NWO) through the Projectruimte and VIDI grants (Nissanke). TH also acknowledges funding from the NWO sectorplan. FF and AH gratefully acknowledge support from NASA through grant number 80NSSC18K0565, from the DOE through Early Career Award DE-SC0020435, and from the NSF through grant number PHY-1806278. MMK acknowledges the GROWTH project funded by the National Science Foundation under PIRE Grant No 1545949. MMK acknowledges generous support from the David and Lucille Packard Foundation. MB acknowledges support from the Swedish Research Council (Reg. no. 2020-03330).
RF acknowledges support from the Natural Sciences and Engineering Research
Council of Canada (NSERC) through Discovery Grant RGPIN-2017-04286, and from the
Faculty of Science at the University of Alberta. TV acknowledge support by the John Bahcall Fellowship at the Institute for Advanced Study and by the National Science Foundation under Grant No. 2012086. M.C. acknowledges support from the National Science Foundation with grant number PHY-2010970. T.E. acknowledges support by the Vetenskapsr{\aa}det (Swedish Research Council) through contract No.  638-2013-8993 and the Oskar Klein Centre for Cosmoparticle Physics.

\software{Python/C~language~\citep{python2007}, NumPy~\citep{Numpy2011}, Cython~\citep{cython2011}, SciPy~\citep{Scipy}, MPI~\citep{MPI}, MPI for Python~\citep{mpi4py}, Matplotlib~\citep{Hunter:2007,matplotlibv2}, IPython~\citep{IPython2007}, Jupyter~\citep{Kluyver:2016aa}, \textsc{MultiNest}~\citep{Feroz09}, \textsc{PyMultiNest}~\citep{Buchner14}, GetDist \citep[][]{GetDist}.}

\bibliography{references}

\begin{thebibliography}{}
\expandafter\ifx\csname natexlab\endcsname\relax\def\natexlab#1{#1}\fi
\providecommand{\url}[1]{\href{#1}{#1}}

\bibitem[{{Abbott} {et~al.}(2016){Abbott}, {Abbott}, {Abbott}, {Abernathy},
  {Acernese}, {Ackley}, {Adams}, {Adams}, {Addesso}, {Adhikari}, {Adya},
  {Affeldt}, {Agathos}, {Agatsuma}, {Aggarwal}, {Aguiar}, {Aiello}, {Ain},
  {Ajith}, {Allen}, {Allocca}, {Altin}, {Anderson}, {Anderson}, {Arai},
  {Arain}, {Araya}, {Arceneaux}, {Areeda}, {Arnaud}, {Arun}, {Ascenzi},
  {Ashton}, {Ast}, {Aston}, {Astone}, {Aufmuth}, {Aulbert}, {Babak}, {Bacon},
  {Bader}, {Baker}, {Baldaccini}, {Ballardin}, {Ballmer}, {Barayoga},
  {Barclay}, {Barish}, {Barker}, {Barone}, {Barr}, {Barsotti}, {Barsuglia},
  {Barta}, {Bartlett}, {Barton}, {Bartos}, {Bassiri}, {Basti}, {Batch},
  {Baune}, {Bavigadda}, {Bazzan}, {Behnke}, {Bejger}, {Belczynski}, {Bell},
  {Bell}, {Berger}, {Bergman}, {Bergmann}, {Berry}, {Bersanetti}, {Bertolini},
  {Betzwieser}, {Bhagwat}, {Bhandare}, {Bilenko}, {Billingsley}, {Birch},
  {Birney}, {Birnholtz}, {Biscans}, {Bisht}, {Bitossi}, {Biwer}, {Bizouard},
  {Blackburn}, {Blair}, {Blair}, {Blair}, {Bloemen}, {Bock}, {Bodiya}, {Boer},
  {Bogaert}, {Bogan}, {Bohe}, {Bojtos}, {Bond}, {Bondu}, {Bonnand}, {Boom},
  {Bork}, {Boschi}, {Bose}, {Bouffanais}, {Bozzi}, {Bradaschia}, {Brady},
  {Braginsky}, {Branchesi}, {Brau}, {Briant}, {Brillet}, {Brinkmann},
  {Brisson}, {Brockill}, {Brooks}, {Brown}, {Brown}, {Brown}, {Buchanan},
  {Buikema}, {Bulik}, {Bulten}, {Buonanno}, {Buskulic}, {Buy}, {Byer},
  {Cabero}, {Cadonati}, {Cagnoli}, {Cahillane}, {Bustillo}, {Callister},
  {Calloni}, {Camp}, {Cannon}, {Cao}, {Capano}, {Capocasa}, {Carbognani},
  {Caride}, {Casanueva Diaz}, {Casentini}, {Caudill}, {Cavagli{\`a}},
  {Cavalier}, {Cavalieri}, {Cella}, {Cepeda}, {Baiardi}, {Cerretani},
  {Cesarini}, {Chakraborty}, {Chalermsongsak}, {Chamberlin}, {Chan}, {Chao},
  {Charlton}, {Chassande-Mottin}, {Chen}, {Chen}, {Cheng}, {Chincarini},
  {Chiummo}, {Cho}, {Cho}, {Chow}, {Christensen}, {Chu}, {Chua}, {Chung},
  {Ciani}, {Clara}, {Clark}, {Cleva}, {Coccia}, {Cohadon}, {Colla}, {Collette},
  {Cominsky}, {Constancio}, {Conte}, {Conti}, {Cook}, {Corbitt}, {Cornish},
  {Corsi}, {Cortese}, {Costa}, {Coughlin}, {Coughlin}, {Coulon}, {Countryman},
  {Couvares}, {Cowan}, {Coward}, {Cowart}, {Coyne}, {Coyne}, {Craig},
  {Creighton}, {Creighton}, {Cripe}, {Crowder}, {Cruise}, {Cumming},
  {Cunningham}, {Cuoco}, {Dal Canton}, {Danilishin}, {D'Antonio}, {Danzmann},
  {Darman}, {Da Silva Costa}, {Dattilo}, {Dave}, {Daveloza}, {Davier},
  {Davies}, {Daw}, {Day}, {De}, {DeBra}, {Debreczeni}, {Degallaix}, {De
  Laurentis}, {Del{\'e}glise}, {Del Pozzo}, {Denker}, {Dent}, {Dereli},
  {Dergachev}, {DeRosa}, {De Rosa}, {DeSalvo}, {Dhurandhar}, {D{\'\i}az}, {Di
  Fiore}, {Di Giovanni}, {Di Lieto}, {Di Pace}, {Di Palma}, {Di Virgilio},
  {Dojcinoski}, {Dolique}, {Donovan}, {Dooley}, {Doravari}, {Douglas},
  {Downes}, {Drago}, {Drever}, {Driggers}, {Du}, {Ducrot}, {Dwyer}, {Edo},
  {Edwards}, {Effler}, {Eggenstein}, {Ehrens}, {Eichholz}, {Eikenberry},
  {Engels}, {Essick}, {Etzel}, {Evans}, {Evans}, {Everett}, {Factourovich},
  {Fafone}, {Fair}, {Fairhurst}, {Fan}, {Fang}, {Farinon}, {Farr}, {Farr},
  {Favata}, {Fays}, {Fehrmann}, {Fejer}, {Feldbaum}, {Ferrante}, {Ferreira},
  {Ferrini}, {Fidecaro}, {Finn}, {Fiori}, {Fiorucci}, {Fisher}, {Flaminio},
  {Fletcher}, {Fong}, {Fournier}, {Franco}, {Frasca}, {Frasconi}, {Frede},
  {Frei}, {Freise}, {Frey}, {Frey}, {Fricke}, {Fritschel}, {Frolov}, {Fulda},
  {Fyffe}, {Gabbard}, {Gair}, {Gammaitoni}, {Gaonkar}, {Garufi}, {Gatto},
  {Gaur}, {Gehrels}, {Gemme}, {Gendre}, {Genin}, {Gennai}, {George}, {Gergely},
  {Germain}, {Ghosh}, {Ghosh}, {Ghosh}, {Giaime}, {Giardina}, {Giazotto},
  {Gill}, {Glaefke}, {Gleason}, {Goetz}, {Goetz}, {Gondan}, {Gonz{\'a}lez},
  {Castro}, {Gopakumar}, {Gordon}, {Gorodetsky}, {Gossan}, {Gosselin},
  {Gouaty}, {Graef}, {Graff}, {Granata}, {Grant}, {Gras}, {Gray}, {Greco},
  {Green}, {Greenhalgh}, {Groot}, {Grote}, {Grunewald}, {Guidi}, {Guo},
  {Gupta}, {Gupta}, {Gushwa}, {Gustafson}, {Gustafson}, {Hacker}, {Hall},
  {Hall}, {Hammond}, {Haney}, {Hanke}, {Hanks}, {Hanna}, {Hannam}, {Hanson},
  {Hardwick}, {Harms}, {Harry}, {Harry}, {Hart}, {Hartman}, {Haster},
  {Haughian}, {Healy}, {Heefner}, {Heidmann}, {Heintze}, {Heinzel}, {Heitmann},
  {Hello}, {Hemming}, {Hendry}, {Heng}, {Hennig}, {Heptonstall}, {Heurs},
  {Hild}, {Hoak}, {Hodge}, {Hofman}, {Hollitt}, {Holt}, {Holz}, {Hopkins},
  {Hosken}, {Hough}, {Houston}, {Howell}, {Hu}, {Huang}, {Huerta}, {Huet},
  {Hughey}, {Husa}, {Huttner}, {Huynh-Dinh}, {Idrisy}, {Indik}, {Ingram},
  {Inta}, {Isa}, {Isac}, {Isi}, {Islas}, {Isogai}, {Iyer}, {Izumi}, {Jacobson},
  {Jacqmin}, {Jang}, {Jani}, {Jaranowski}, {Jawahar}, {Jim{\'e}nez-Forteza},
  {Johnson}, {Johnson-McDaniel}, {Jones}, {Jones}, {Jonker}, {Ju}, {Haris},
  {Kalaghatgi}, {Kalogera}, {Kandhasamy}, {Kang}, {Kanner}, {Karki},
  {Kasprzack}, {Katsavounidis}, {Katzman}, {Kaufer}, {Kaur}, {Kawabe},
  {Kawazoe}, {K{\'e}f{\'e}lian}, {Kehl}, {Keitel}, {Kelley}, {Kells},
  {Kennedy}, {Keppel}, {Key}, {Khalaidovski}, {Khalili}, {Khan}, {Khan},
  {Khan}, {Khazanov}, {Kijbunchoo}, {Kim}, {Kim}, {Kim}, {Kim}, {Kim}, {Kim},
  {King}, {King}, {Kinzel}, {Kissel}, {Kleybolte}, {Klimenko}, {Koehlenbeck},
  {Kokeyama}, {Koley}, {Kondrashov}, {Kontos}, {Koranda}, {Korobko}, {Korth},
  {Kowalska}, {Kozak}, {Kringel}, {Krishnan}, {Kr{\'o}lak}, {Krueger}, {Kuehn},
  {Kumar}, {Kumar}, {Kuo}, {Kutynia}, {Kwee}, {Lackey}, {Landry}, {Lange},
  {Lantz}, {Lasky}, {Lazzarini}, {Lazzaro}, {Leaci}, {Leavey}, {Lebigot},
  {Lee}, {Lee}, {Lee}, {Lee}, {Lenon}, {Leonardi}, {Leong}, {Leroy},
  {Letendre}, {Levin}, {Levine}, {Li}, {Libson}, {Littenberg}, {Lockerbie},
  {Logue}, {Lombardi}, {London}, {Lord}, {Lorenzini}, {Loriette}, {Lormand},
  {Losurdo}, {Lough}, {Lousto}, {Lovelace}, {L{\"u}ck}, {Lundgren}, {Luo},
  {Lynch}, {Ma}, {MacDonald}, {Machenschalk}, {MacInnis}, {Macleod},
  {Maga{\~n}a-Sandoval}, {Magee}, {Mageswaran}, {Majorana}, {Maksimovic},
  {Malvezzi}, {Man}, {Mandel}, {Mandic}, {Mangano}, {Mansell}, {Manske},
  {Mantovani}, {Marchesoni}, {Marion}, {M{\'a}rka}, {M{\'a}rka}, {Markosyan},
  {Maros}, {Martelli}, {Martellini}, {Martin}, {Martin}, {Martynov}, {Marx},
  {Mason}, {Masserot}, {Massinger}, {Masso-Reid}, {Matichard}, {Matone},
  {Mavalvala}, {Mazumder}, {Mazzolo}, {McCarthy}, {McClelland}, {McCormick},
  {McGuire}, {McIntyre}, {McIver}, {McManus}, {McWilliams}, {Meacher},
  {Meadors}, {Meidam}, {Melatos}, {Mendell}, {Mendoza-Gandara}, {Mercer},
  {Merilh}, {Merzougui}, {Meshkov}, {Messenger}, {Messick}, {Meyers},
  {Mezzani}, {Miao}, {Michel}, {Middleton}, {Mikhailov}, {Milano}, {Miller},
  {Millhouse}, {Minenkov}, {Ming}, {Mirshekari}, {Mishra}, {Mitra},
  {Mitrofanov}, {Mitselmakher}, {Mittleman}, {Moggi}, {Mohan}, {Mohapatra},
  {Montani}, {Moore}, {Moore}, {Moraru}, {Moreno}, {Morriss}, {Mossavi},
  {Mours}, {Mow-Lowry}, {Mueller}, {Mueller}, {Muir}, {Mukherjee}, {Mukherjee},
  {Mukherjee}, {Mukund}, {Mullavey}, {Munch}, {Murphy}, {Murray}, {Mytidis},
  {Nardecchia}, {Naticchioni}, {Nayak}, {Necula}, {Nedkova}, {Nelemans},
  {Neri}, {Neunzert}, {Newton}, {Nguyen}, {Nielsen}, {Nissanke}, {Nitz},
  {Nocera}, {Nolting}, {Normandin}, {Nuttall}, {Oberling}, {Ochsner}, {O'Dell},
  {Oelker}, {Ogin}, {Oh}, {Oh}, {Ohme}, {Oliver}, {Oppermann}, {Oram},
  {O'Reilly}, {O'Shaughnessy}, {Ott}, {Ottaway}, {Ottens}, {Overmier}, {Owen},
  {Pai}, {Pai}, {Palamos}, {Palashov}, {Palomba}, {Pal-Singh}, {Pan}, {Pan},
  {Pankow}, {Pannarale}, {Pant}, {Paoletti}, {Paoli}, {Papa}, {Paris},
  {Parker}, {Pascucci}, {Pasqualetti}, {Passaquieti}, {Passuello},
  {Patricelli}, {Patrick}, {Pearlstone}, {Pedraza}, {Pedurand}, {Pekowsky},
  {Pele}, {Penn}, {Perreca}, {Pfeiffer}, {Phelps}, {Piccinni}, {Pichot},
  {Pickenpack}, {Piergiovanni}, {Pierro}, {Pillant}, {Pinard}, {Pinto},
  {Pitkin}, {Poeld}, {Poggiani}, {Popolizio}, {Post}, {Powell}, {Prasad},
  {Predoi}, {Premachandra}, {Prestegard}, {Price}, {Prijatelj}, {Principe},
  {Privitera}, {Prix}, {Prodi}, {Prokhorov}, {Puncken}, {Punturo}, {Puppo},
  {P{\"u}rrer}, {Qi}, {Qin}, {Quetschke}, {Quintero}, {Quitzow-James}, {Raab},
  {Rabeling}, {Radkins}, {Raffai}, {Raja}, {Rakhmanov}, {Ramet}, {Rapagnani},
  {Raymond}, {Razzano}, {Re}, {Read}, {Reed}, {Regimbau}, {Rei}, {Reid},
  {Reitze}, {Rew}, {Reyes}, {Ricci}, {Riles}, {Robertson}, {Robie}, {Robinet},
  {Rocchi}, {Rolland}, {Rollins}, {Roma}, {Romano}, {Romano}, {Romanov},
  {Romie}, {Rosi{\'n}ska}, {Rowan}, {R{\"u}diger}, {Ruggi}, {Ryan}, {Sachdev},
  {Sadecki}, {Sadeghian}, {Salconi}, {Saleem}, {Salemi}, {Samajdar}, {Sammut},
  {Sampson}, {Sanchez}, {Sandberg}, {Sandeen}, {Sanders}, {Sanders},
  {Sassolas}, {Sathyaprakash}, {Saulson}, {Sauter}, {Savage}, {Sawadsky},
  {Schale}, {Schilling}, {Schmidt}, {Schmidt}, {Schnabel}, {Schofield},
  {Sch{\"o}nbeck}, {Schreiber}, {Schuette}, {Schutz}, {Scott}, {Scott},
  {Sellers}, {Sengupta}, {Sentenac}, {Sequino}, {Sergeev}, {Serna},
  {Setyawati}, {Sevigny}, {Shaddock}, {Shaffer}, {Shah}, {Shahriar}, {Shaltev},
  {Shao}, {Shapiro}, {Shawhan}, {Sheperd}, {Shoemaker}, {Shoemaker}, {Siellez},
  {Siemens}, {Sigg}, {Silva}, {Simakov}, {Singer}, {Singer}, {Singh}, {Singh},
  {Singhal}, {Sintes}, {Slagmolen}, {Smith}, {Smith}, {Smith}, {Smith}, {Son},
  {Sorazu}, {Sorrentino}, {Souradeep}, {Srivastava}, {Staley}, {Steinke},
  {Steinlechner}, {Steinlechner}, {Steinmeyer}, {Stephens}, {Stevenson},
  {Stone}, {Strain}, {Straniero}, {Stratta}, {Strauss}, {Strigin}, {Sturani},
  {Stuver}, {Summerscales}, {Sun}, {Sutton}, {Swinkels}, {Szczepa{\'n}czyk},
  {Tacca}, {Talukder}, {Tanner}, {T{\'a}pai}, {Tarabrin}, {Taracchini},
  {Taylor}, {Theeg}, {Thirugnanasambandam}, {Thomas}, {Thomas}, {Thomas},
  {Thorne}, {Thorne}, {Thrane}, {Tiwari}, {Tiwari}, {Tokmakov}, {Tomlinson},
  {Tonelli}, {Torres}, {Torrie}, {T{\"o}yr{\"a}}, {Travasso}, {Traylor},
  {Trifir{\`o}}, {Tringali}, {Trozzo}, {Tse}, {Turconi}, {Tuyenbayev},
  {Ugolini}, {Unnikrishnan}, {Urban}, {Usman}, {Vahlbruch}, {Vajente},
  {Valdes}, {Vallisneri}, {van Bakel}, {van Beuzekom}, {van den Brand}, {Van
  Den Broeck}, {Vander-Hyde}, {van der Schaaf}, {van Heijningen}, {van Veggel},
  {Vardaro}, {Vass}, {Vas{\'u}th}, {Vaulin}, {Vecchio}, {Vedovato}, {Veitch},
  {Veitch}, {Venkateswara}, {Verkindt}, {Vetrano}, {Vicer{\'e}}, {Vinciguerra},
  {Vine}, {Vinet}, {Vitale}, {Vo}, {Vocca}, {Vorvick}, {Voss}, {Vousden},
  {Vyatchanin}, {Wade}, {Wade}, {Wade}, {Waldman}, {Walker}, {Wallace},
  {Walsh}, {Wang}, {Wang}, {Wang}, {Wang}, {Wang}, {Ward}, {Ward}, {Warner},
  {Was}, {Weaver}, {Wei}, {Weinert}, {Weinstein}, {Weiss}, {Welborn}, {Wen},
  {We{\ss}els}, {Westphal}, {Wette}, {Whelan}, {Whitcomb}, {White}, {Whiting},
  {Wiesner}, {Wilkinson}, {Willems}, {Williams}, {Williams}, {Williamson},
  {Willis}, {Willke}, {Wimmer}, {Winkelmann}, {Winkler}, {Wipf}, {Wiseman},
  {Wittel}, {Woan}, {Worden}, {Wright}, {Wu}, {Yablon}, {Yakushin}, {Yam},
  {Yamamoto}, {Yancey}, {Yap}, {Yu}, {Yvert}, {Zadro{\.Z}ny}, {Zangrando},
  {Zanolin}, {Zendri}, {Zevin}, {Zhang}, {Zhang}, {Zhang}, {Zhang}, {Zhao},
  {Zhou}, {Zhou}, {Zhu}, {Zucker}, {Zuraw}, {Zweizig}, {LIGO Scientific
  Collaboration}, \& {Virgo Collaboration}}]{GW150914}
{Abbott}, B.~P., {Abbott}, R., {Abbott}, T.~D., {et~al.} 2016, \prl, 116,
  061102

\bibitem[{{Abbott} {et~al.}(2017{\natexlab{a}}){Abbott}, {Abbott}, {Abbott},
  {Acernese}, {Ackley}, {Adams}, {Adams}, {Addesso}, {Adhikari}, {Adya},
  {Affeldt}, {Afrough}, {Agarwal}, {Agathos}, {Agatsuma}, {Aggarwal}, {Aguiar},
  {Aiello}, {Ain}, {Ajith}, {Allen}, {Allen}, {Allocca}, {Altin}, {Amato},
  {Ananyeva}, {Anderson}, {Anderson}, {Angelova}, {Antier}, {Appert}, {Arai},
  {Araya}, {Areeda}, {Arnaud}, {Arun}, {Ascenzi}, {Ashton}, {Ast}, {Aston},
  {Astone}, {Atallah}, {Aufmuth}, {Aulbert}, {AultONeal}, {Austin},
  {Avila-Alvarez}, {Babak}, {Bacon}, {Bader}, {Bae}, {Bailes}, {Baker},
  {Baldaccini}, {Ballardin}, {Ballmer}, {Banagiri}, {Barayoga}, {Barclay},
  {Barish}, {Barker}, {Barkett}, {Barone}, {Barr}, {Barsotti}, {Barsuglia},
  {Barta}, {Barthelmy}, {Bartlett}, {Bartos}, {Bassiri}, {Basti}, {Batch},
  {Bawaj}, {Bayley}, {Bazzan}, {B{\'e}csy}, {Beer}, {Bejger}, {Belahcene},
  {Bell}, {Berger}, {Bergmann}, {Bernuzzi}, {Bero}, {Berry}, {Bersanetti},
  {Bertolini}, {Betzwieser}, {Bhagwat}, {Bhandare}, {Bilenko}, {Billingsley},
  {Billman}, {Birch}, {Birney}, {Birnholtz}, {Biscans}, {Biscoveanu}, {Bisht},
  {Bitossi}, {Biwer}, {Bizouard}, {Blackburn}, {Blackman}, {Blair}, {Blair},
  {Blair}, {Bloemen}, {Bock}, {Bode}, {Boer}, {Bogaert}, {Bohe}, {Bondu},
  {Bonilla}, {Bonnand}, {Boom}, {Bork}, {Boschi}, {Bose}, {Bossie},
  {Bouffanais}, {Bozzi}, {Bradaschia}, {Brady}, {Branchesi}, {Brau}, {Briant},
  {Brillet}, {Brinkmann}, {Brisson}, {Brockill}, {Broida}, {Brooks}, {Brown},
  {Brown}, {Brunett}, {Buchanan}, {Buikema}, {Bulik}, {Bulten}, {Buonanno},
  {Buskulic}, {Buy}, {Byer}, {Cabero}, {Cadonati}, {Cagnoli}, {Cahillane},
  {Calder{\'o}n Bustillo}, {Callister}, {Calloni}, {Camp}, {Canepa},
  {Canizares}, {Cannon}, {Cao}, {Cao}, {Capano}, {Capocasa}, {Carbognani},
  {Caride}, {Carney}, {Carullo}, {Casanueva Diaz}, {Casentini}, {Caudill},
  {Cavagli{\`a}}, {Cavalier}, {Cavalieri}, {Cella}, {Cepeda},
  {Cerd{\'a}-Dur{\'a}n}, {Cerretani}, {Cesarini}, {Chamberlin}, {Chan}, {Chao},
  {Charlton}, {Chase}, {Chassande-Mottin}, {Chatterjee}, {Chatziioannou},
  {Cheeseboro}, {Chen}, {Chen}, {Chen}, {Cheng}, {Chia}, {Chincarini},
  {Chiummo}, {Chmiel}, {Cho}, {Cho}, {Chow}, {Christensen}, {Chu}, {Chua},
  {Chua}, {Chung}, {Chung}, {Ciani}, {Ciolfi}, {Cirelli}, {Cirone}, {Clara},
  {Clark}, {Clearwater}, {Cleva}, {Cocchieri}, {Coccia}, {Cohadon}, {Cohen},
  {Colla}, {Collette}, {Cominsky}, {Constancio}, {Conti}, {Cooper}, {Corban},
  {Corbitt}, {Cordero-Carri{\'o}n}, {Corley}, {Cornish}, {Corsi}, {Cortese},
  {Costa}, {Coughlin}, {Coughlin}, {Coulon}, {Countryman}, {Couvares}, {Covas},
  {Cowan}, {Coward}, {Cowart}, {Coyne}, {Coyne}, {Creighton}, {Creighton},
  {Cripe}, {Crowder}, {Cullen}, {Cumming}, {Cunningham}, {Cuoco}, {Dal Canton},
  {D{\'a}lya}, {Danilishin}, {D'Antonio}, {Danzmann}, {Dasgupta}, {Da Silva
  Costa}, {Dattilo}, {Dave}, {Davier}, {Davis}, {Daw}, {Day}, {De}, {DeBra},
  {Degallaix}, {De Laurentis}, {Del{\'e}glise}, {Del Pozzo}, {Demos}, {Denker},
  {Dent}, {De Pietri}, {Dergachev}, {De Rosa}, {DeRosa}, {De Rossi}, {DeSalvo},
  {de Varona}, {Devenson}, {Dhurandhar}, {D{\'\i}az}, {Dietrich}, {Di Fiore},
  {Di Giovanni}, {Di Girolamo}, {Di Lieto}, {Di Pace}, {Di Palma}, {Di Renzo},
  {Doctor}, {Dolique}, {Donovan}, {Dooley}, {Doravari}, {Dorrington},
  {Douglas}, {Dovale {\'A}lvarez}, {Downes}, {Drago}, {Dreissigacker},
  {Driggers}, {Du}, {Ducrot}, {Dudi}, {Dupej}, {Dwyer}, {Edo}, {Edwards},
  {Effler}, {Eggenstein}, {Ehrens}, {Eichholz}, {Eikenberry}, {Eisenstein},
  {Essick}, {Estevez}, {Etienne}, {Etzel}, {Evans}, {Evans}, {Factourovich},
  {Fafone}, {Fair}, {Fairhurst}, {Fan}, {Farinon}, {Farr}, {Farr},
  {Fauchon-Jones}, {Favata}, {Fays}, {Fee}, {Fehrmann}, {Feicht}, {Fejer},
  {Fernandez-Galiana}, {Ferrante}, {Ferreira}, {Ferrini}, {Fidecaro},
  {Finstad}, {Fiori}, {Fiorucci}, {Fishbach}, {Fisher}, {Fitz-Axen},
  {Flaminio}, {Fletcher}, {Fong}, {Font}, {Forsyth}, {Forsyth}, {Fournier},
  {Frasca}, {Frasconi}, {Frei}, {Freise}, {Frey}, {Frey}, {Fries}, {Fritschel},
  {Frolov}, {Fulda}, {Fyffe}, {Gabbard}, {Gadre}, {Gaebel}, {Gair},
  {Gammaitoni}, {Ganija}, {Gaonkar}, {Garcia-Quiros}, {Garufi}, {Gateley},
  {Gaudio}, {Gaur}, {Gayathri}, {Gehrels}, {Gemme}, {Genin}, {Gennai},
  {George}, {George}, {Gergely}, {Germain}, {Ghonge}, {Ghosh}, {Ghosh},
  {Ghosh}, {Giaime}, {Giardina}, {Giazotto}, {Gill}, {Glover}, {Goetz},
  {Goetz}, {Gomes}, {Goncharov}, {Gonz{\'a}lez}, {Gonzalez Castro},
  {Gopakumar}, {Gorodetsky}, {Gossan}, {Gosselin}, {Gouaty}, {Grado}, {Graef},
  {Granata}, {Grant}, {Gras}, {Gray}, {Greco}, {Green}, {Gretarsson}, {Groot},
  {Grote}, {Grunewald}, {Gruning}, {Guidi}, {Guo}, {Gupta}, {Gupta}, {Gushwa},
  {Gustafson}, {Gustafson}, {Halim}, {Hall}, {Hall}, {Hamilton}, {Hammond},
  {Haney}, {Hanke}, {Hanks}, {Hanna}, {Hannam}, {Hannuksela}, {Hanson},
  {Hardwick}, {Harms}, {Harry}, {Harry}, {Hart}, {Haster}, {Haughian}, {Healy},
  {Heidmann}, {Heintze}, {Heitmann}, {Hello}, {Hemming}, {Hendry}, {Heng},
  {Hennig}, {Heptonstall}, {Heurs}, {Hild}, {Hinderer}, {Ho}, {Hoak}, {Hofman},
  {Holt}, {Holz}, {Hopkins}, {Horst}, {Hough}, {Houston}, {Howell}, {Hreibi},
  {Hu}, {Huerta}, {Huet}, {Hughey}, {Husa}, {Huttner}, {Huynh-Dinh}, {Indik},
  {Inta}, {Intini}, {Isa}, {Isac}, {Isi}, {Iyer}, {Izumi}, {Jacqmin}, {Jani},
  {Jaranowski}, {Jawahar}, {Jim{\'e}nez-Forteza}, {Johnson},
  {Johnson-McDaniel}, {Jones}, {Jones}, {Jonker}, {Ju}, {Junker}, {Kalaghatgi},
  {Kalogera}, {Kamai}, {Kandhasamy}, {Kang}, {Kanner}, {Kapadia}, {Karki},
  {Karvinen}, {Kasprzack}, {Kastaun}, {Katolik}, {Katsavounidis}, {Katzman},
  {Kaufer}, {Kawabe}, {K{\'e}f{\'e}lian}, {Keitel}, {Kemball}, {Kennedy},
  {Kent}, {Key}, {Khalili}, {Khan}, {Khan}, {Khan}, {Khazanov}, {Kijbunchoo},
  {Kim}, {Kim}, {Kim}, {Kim}, {Kim}, {Kim}, {Kimbrell}, {King}, {King},
  {Kinley-Hanlon}, {Kirchhoff}, {Kissel}, {Kleybolte}, {Klimenko}, {Knowles},
  {Koch}, {Koehlenbeck}, {Koley}, {Kondrashov}, {Kontos}, {Korobko}, {Korth},
  {Kowalska}, {Kozak}, {Kr{\"a}mer}, {Kringel}, {Krishnan}, {Kr{\'o}lak},
  {Kuehn}, {Kumar}, {Kumar}, {Kumar}, {Kuo}, {Kutynia}, {Kwang}, {Lackey},
  {Lai}, {Landry}, {Lang}, {Lange}, {Lantz}, {Lanza}, {Larson},
  {Lartaux-Vollard}, {Lasky}, {Laxen}, {Lazzarini}, {Lazzaro}, {Leaci},
  {Leavey}, {Lee}, {Lee}, {Lee}, {Lee}, {Lee}, {Lehmann}, {Lenon}, {Leon},
  {Leonardi}, {Leroy}, {Letendre}, {Levin}, {Li}, {Linker}, {Littenberg},
  {Liu}, {Liu}, {Lo}, {Lockerbie}, {London}, {Lord}, {Lorenzini}, {Loriette},
  {Lormand}, {Losurdo}, {Lough}, {Lousto}, {Lovelace}, {L{\"u}ck}, {Lumaca},
  {Lundgren}, {Lynch}, {Ma}, {Macas}, {Macfoy}, {Machenschalk}, {MacInnis},
  {Macleod}, {Maga{\~n}a Hernandez}, {Maga{\~n}a-Sandoval}, {Maga{\~n}a
  Zertuche}, {Magee}, {Majorana}, {Maksimovic}, {Man}, {Mandic}, {Mangano},
  {Mansell}, {Manske}, {Mantovani}, {Marchesoni}, {Marion}, {M{\'a}rka},
  {M{\'a}rka}, {Markakis}, {Markosyan}, {Markowitz}, {Maros}, {Marquina},
  {Marsh}, {Martelli}, {Martellini}, {Martin}, {Martin}, {Martynov}, {Marx},
  {Mason}, {Massera}, {Masserot}, {Massinger}, {Masso-Reid}, {Mastrogiovanni},
  {Matas}, {Matichard}, {Matone}, {Mavalvala}, {Mazumder}, {McCarthy},
  {McClelland}, {McCormick}, {McCuller}, {McGuire}, {McIntyre}, {McIver},
  {McManus}, {McNeill}, {McRae}, {McWilliams}, {Meacher}, {Meadors}, {Mehmet},
  {Meidam}, {Mejuto-Villa}, {Melatos}, {Mendell}, {Mercer}, {Merilh},
  {Merzougui}, {Meshkov}, {Messenger}, {Messick}, {Metzdorff}, {Meyers},
  {Miao}, {Michel}, {Middleton}, {Mikhailov}, {Milano}, {Miller}, {Miller},
  {Miller}, {Millhouse}, {Milovich-Goff}, {Minazzoli}, {Minenkov}, {Ming},
  {Mishra}, {Mitra}, {Mitrofanov}, {Mitselmakher}, {Mittleman}, {Moffa},
  {Moggi}, {Mogushi}, {Mohan}, {Mohapatra}, {Molina}, {Montani}, {Moore},
  {Moraru}, {Moreno}, {Morisaki}, {Morriss}, {Mours}, {Mow-Lowry}, {Mueller},
  {Muir}, {Mukherjee}, {Mukherjee}, {Mukherjee}, {Mukund}, {Mullavey}, {Munch},
  {Mu{\~n}iz}, {Muratore}, {Murray}, {Nagar}, {Napier}, {Nardecchia},
  {Naticchioni}, {Nayak}, {Neilson}, {Nelemans}, {Nelson}, {Nery}, {Neunzert},
  {Nevin}, {Newport}, {Newton}, {Ng}, {Nguyen}, {Nguyen}, {Nichols}, {Nielsen},
  {Nissanke}, {Nitz}, {Noack}, {Nocera}, {Nolting}, {North}, {Nuttall},
  {Oberling}, {O'Dea}, {Ogin}, {Oh}, {Oh}, {Ohme}, {Okada}, {Oliver},
  {Oppermann}, {Oram}, {O'Reilly}, {Ormiston}, {Ortega}, {O'Shaughnessy},
  {Ossokine}, {Ottaway}, {Overmier}, {Owen}, {Pace}, {Page}, {Page}, {Pai},
  {Pai}, {Palamos}, {Palashov}, {Palomba}, {Pal-Singh}, {Pan}, {Pan}, {Pang},
  {Pang}, {Pankow}, {Pannarale}, {Pant}, {Paoletti}, {Paoli}, {Papa}, {Parida},
  {Parker}, {Pascucci}, {Pasqualetti}, {Passaquieti}, {Passuello}, {Patil},
  {Patricelli}, {Pearlstone}, {Pedraza}, {Pedurand}, {Pekowsky}, {Pele},
  {Penn}, {Perez}, {Perreca}, {Perri}, {Pfeiffer}, {Phelps}, {Piccinni},
  {Pichot}, {Piergiovanni}, {Pierro}, {Pillant}, {Pinard}, {Pinto}, {Pirello},
  {Pitkin}, {Poe}, {Poggiani}, {Popolizio}, {Porter}, {Post}, {Powell},
  {Prasad}, {Pratt}, {Pratten}, {Predoi}, {Prestegard}, {Prijatelj},
  {Principe}, {Privitera}, {Prix}, {Prodi}, {Prokhorov}, {Puncken}, {Punturo},
  {Puppo}, {P{\"u}rrer}, {Qi}, {Quetschke}, {Quintero}, {Quitzow-James},
  {Raab}, {Rabeling}, {Radkins}, {Raffai}, {Raja}, {Rajan}, {Rajbhandari},
  {Rakhmanov}, {Ramirez}, {Ramos-Buades}, {Rapagnani}, {Raymond}, {Razzano},
  {Read}, {Regimbau}, {Rei}, {Reid}, {Reitze}, {Ren}, {Reyes}, {Ricci},
  {Ricker}, {Rieger}, {Riles}, {Rizzo}, {Robertson}, {Robie}, {Robinet},
  {Rocchi}, {Rolland}, {Rollins}, {Roma}, {Romano}, {Romano}, {Romel}, {Romie},
  {Rosi{\'n}ska}, {Ross}, {Rowan}, {R{\"u}diger}, {Ruggi}, {Rutins}, {Ryan},
  {Sachdev}, {Sadecki}, {Sadeghian}, {Sakellariadou}, {Salconi}, {Saleem},
  {Salemi}, {Samajdar}, {Sammut}, {Sampson}, {Sanchez}, {Sanchez},
  {Sanchis-Gual}, {Sandberg}, {Sanders}, {Sassolas}, {Sathyaprakash},
  {Saulson}, {Sauter}, {Savage}, {Sawadsky}, {Schale}, {Scheel}, {Scheuer},
  {Schmidt}, {Schmidt}, {Schnabel}, {Schofield}, {Sch{\"o}nbeck}, {Schreiber},
  {Schuette}, {Schulte}, {Schutz}, {Schwalbe}, {Scott}, {Scott}, {Seidel},
  {Sellers}, {Sengupta}, {Sentenac}, {Sequino}, {Sergeev}, {Shaddock},
  {Shaffer}, {Shah}, {Shahriar}, {Shaner}, {Shao}, {Shapiro}, {Shawhan},
  {Sheperd}, {Shoemaker}, {Shoemaker}, {Siellez}, {Siemens}, {Sieniawska},
  {Sigg}, {Silva}, {Singer}, {Singh}, {Singhal}, {Sintes}, {Slagmolen},
  {Smith}, {Smith}, {Smith}, {Somala}, {Son}, {Sonnenberg}, {Sorazu},
  {Sorrentino}, {Souradeep}, {Spencer}, {Srivastava}, {Staats}, {Staley},
  {Steinke}, {Steinlechner}, {Steinlechner}, {Steinmeyer}, {Stevenson},
  {Stone}, {Stops}, {Strain}, {Stratta}, {Strigin}, {Strunk}, {Sturani},
  {Stuver}, {Summerscales}, {Sun}, {Sunil}, {Suresh}, {Sutton}, {Swinkels},
  {Szczepa{\'n}czyk}, {Tacca}, {Tait}, {Talbot}, {Talukder}, {Tanner},
  {T{\'a}pai}, {Taracchini}, {Tasson}, {Taylor}, {Taylor}, {Tewari}, {Theeg},
  {Thies}, {Thomas}, {Thomas}, {Thomas}, {Thorne}, {Thorne}, {Thrane},
  {Tiwari}, {Tiwari}, {Tokmakov}, {Toland}, {Tonelli}, {Tornasi},
  {Torres-Forn{\'e}}, {Torrie}, {T{\"o}yr{\"a}}, {Travasso}, {Traylor},
  {Trinastic}, {Tringali}, {Trozzo}, {Tsang}, {Tse}, {Tso}, {Tsukada}, {Tsuna},
  {Tuyenbayev}, {Ueno}, {Ugolini}, {Unnikrishnan}, {Urban}, {Usman},
  {Vahlbruch}, {Vajente}, {Valdes}, {Vallisneri}, {van Bakel}, {van Beuzekom},
  {van den Brand}, {Van Den Broeck}, {Vander-Hyde}, {van der Schaaf}, {van
  Heijningen}, {van Veggel}, {Vardaro}, {Varma}, {Vass}, {Vas{\'u}th},
  {Vecchio}, {Vedovato}, {Veitch}, {Veitch}, {Venkateswara}, {Venugopalan},
  {Verkindt}, {Vetrano}, {Vicer{\'e}}, {Viets}, {Vinciguerra}, {Vine}, {Vinet},
  {Vitale}, {Vo}, {Vocca}, {Vorvick}, {Vyatchanin}, {Wade}, {Wade}, {Wade},
  {Walet}, {Walker}, {Wallace}, {Walsh}, {Wang}, {Wang}, {Wang}, {Wang},
  {Wang}, {Ward}, {Warner}, {Was}, {Watchi}, {Weaver}, {Wei}, {Weinert},
  {Weinstein}, {Weiss}, {Wen}, {Wessel}, {We{\ss}els}, {Westerweck},
  {Westphal}, {Wette}, {Whelan}, {Whitcomb}, {Whiting}, {Whittle}, {Wilken},
  {Williams}, {Williams}, {Williamson}, {Willis}, {Willke}, {Wimmer},
  {Winkler}, {Wipf}, {Wittel}, {Woan}, {Woehler}, {Wofford}, {Wong}, {Worden},
  {Wright}, {Wu}, {Wysocki}, {Xiao}, {Yamamoto}, {Yancey}, {Yang}, {Yap},
  {Yazback}, {Yu}, {Yu}, {Yvert}, {Zadro{\.Z}ny}, {Zanolin}, {Zelenova},
  {Zendri}, {Zevin}, {Zhang}, {Zhang}, {Zhang}, {Zhang}, {Zhao}, {Zhou},
  {Zhou}, {Zhu}, {Zhu}, {Zimmerman}, {Zucker}, {Zweizig}, {LIGO Scientific
  Collaboration}, \& {Virgo Collaboration}}]{GW170817discovery}
---. 2017{\natexlab{a}}, \prl, 119, 161101

\bibitem[{{Abbott} {et~al.}(2017{\natexlab{b}}){Abbott}, {Abbott}, {Abbott},
  {Acernese}, {Ackley}, {Adams}, {Adams}, {Addesso}, {Adhikari}, {Adya},
  {Affeldt}, {Afrough}, {Agarwal}, {Agathos}, {Agatsuma}, {Aggarwal}, {Aguiar},
  {Aiello}, {Ain}, {Ajith}, {Allen}, {Allen}, {Allocca}, {Altin}, {Amato},
  {Ananyeva}, {Anderson}, {Anderson}, {Angelova}, {Antier}, {Appert}, {Arai},
  {Araya}, {Areeda}, {Arnaud}, {Arun}, {Ascenzi}, {Ashton}, {Ast}, {Aston},
  {Astone}, {Atallah}, {Aufmuth}, {Aulbert}, {AultONeal}, {Austin},
  {Avila-Alvarez}, {Babak}, {Bacon}, {Bader}, {Bae}, {Baker}, {Baldaccini},
  {Ballardin}, {Ballmer}, {Banagiri}, {Barayoga}, {Barclay}, {Barish},
  {Barker}, {Barkett}, {Barone}, {Barr}, {Barsotti}, {Barsuglia}, {Barta},
  {Barthelmy}, {Bartlett}, {Bartos}, {Bassiri}, {Basti}, {Batch}, {Bawaj},
  {Bayley}, {Bazzan}, {B{\'e}csy}, {Beer}, {Bejger}, {Belahcene}, {Bell},
  {Berger}, {Bergmann}, {Bero}, {Berry}, {Bersanetti}, {Bertolini},
  {Betzwieser}, {Bhagwat}, {Bhandare}, {Bilenko}, {Billingsley}, {Billman},
  {Birch}, {Birney}, {Birnholtz}, {Biscans}, {Biscoveanu}, {Bisht}, {Bitossi},
  {Biwer}, {Bizouard}, {Blackburn}, {Blackman}, {Blair}, {Blair}, {Blair},
  {Bloemen}, {Bock}, {Bode}, {Boer}, {Bogaert}, {Bohe}, {Bondu}, {Bonilla},
  {Bonnand}, {Boom}, {Bork}, {Boschi}, {Bose}, {Bossie}, {Bouffanais}, {Bozzi},
  {Bradaschia}, {Brady}, {Branchesi}, {Brau}, {Briant}, {Brillet}, {Brinkmann},
  {Brisson}, {Brockill}, {Broida}, {Brooks}, {Brown}, {Brown}, {Brunett},
  {Buchanan}, {Buikema}, {Bulik}, {Bulten}, {Buonanno}, {Buskulic}, {Buy},
  {Byer}, {Cabero}, {Cadonati}, {Cagnoli}, {Cahillane}, {Calder{\'o}n
  Bustillo}, {Callister}, {Calloni}, {Camp}, {Canepa}, {Canizares}, {Cannon},
  {Cao}, {Cao}, {Capano}, {Capocasa}, {Carbognani}, {Caride}, {Carney},
  {Casanueva Diaz}, {Casentini}, {Caudill}, {Cavagli{\`a}}, {Cavalier},
  {Cavalieri}, {Cella}, {Cepeda}, {Cerd{\'a}-Dur{\'a}n}, {Cerretani},
  {Cesarini}, {Chamberlin}, {Chan}, {Chao}, {Charlton}, {Chase},
  {Chassande-Mottin}, {Chatterjee}, {Chatziioannou}, {Cheeseboro}, {Chen},
  {Chen}, {Chen}, {Cheng}, {Chia}, {Chincarini}, {Chiummo}, {Chmiel}, {Cho},
  {Cho}, {Chow}, {Christensen}, {Chu}, {Chua}, {Chua}, {Chung}, {Chung},
  {Ciani}, {Ciolfi}, {Cirelli}, {Cirone}, {Clara}, {Clark}, {Clearwater},
  {Cleva}, {Cocchieri}, {Coccia}, {Cohadon}, {Cohen}, {Colla}, {Collette},
  {Cominsky}, {Constancio}, {Conti}, {Cooper}, {Corban}, {Corbitt},
  {Cordero-Carri{\'o}n}, {Corley}, {Cornish}, {Corsi}, {Cortese}, {Costa},
  {Coughlin}, {Coughlin}, {Coulon}, {Countryman}, {Couvares}, {Covas}, {Cowan},
  {Coward}, {Cowart}, {Coyne}, {Coyne}, {Creighton}, {Creighton}, {Cripe},
  {Crowder}, {Cullen}, {Cumming}, {Cunningham}, {Cuoco}, {Dal Canton},
  {D{\'a}lya}, {Danilishin}, {D'Antonio}, {Danzmann}, {Dasgupta}, {Da Silva
  Costa}, {Dattilo}, {Dave}, {Davier}, {Davis}, {Daw}, {Day}, {De}, {DeBra},
  {Degallaix}, {De Laurentis}, {Del{\'e}glise}, {Del Pozzo}, {Demos}, {Denker},
  {Dent}, {De Pietri}, {Dergachev}, {De Rosa}, {DeRosa}, {De Rossi}, {DeSalvo},
  {de Varona}, {Devenson}, {Dhurandhar}, {D{\'\i}az}, {Di Fiore}, {Di
  Giovanni}, {Di Girolamo}, {Di Lieto}, {Di Pace}, {Di Palma}, {Di Renzo},
  {Doctor}, {Dolique}, {Donovan}, {Dooley}, {Doravari}, {Dorrington},
  {Douglas}, {Dovale {\'A}lvarez}, {Downes}, {Drago}, {Dreissigacker},
  {Driggers}, {Du}, {Ducrot}, {Dupej}, {Dwyer}, {Edo}, {Edwards}, {Effler},
  {Ehrens}, {Eichholz}, {Eikenberry}, {Eisenstein}, {Essick}, {Estevez},
  {Etienne}, {Etzel}, {Evans}, {Evans}, {Factourovich}, {Fafone}, {Fair},
  {Fairhurst}, {Fan}, {Farinon}, {Farr}, {Farr}, {Fauchon-Jones}, {Favata},
  {Fays}, {Fee}, {Fehrmann}, {Feicht}, {Fejer}, {Fernandez-Galiana},
  {Ferrante}, {Ferreira}, {Ferrini}, {Fidecaro}, {Finstad}, {Fiori},
  {Fiorucci}, {Fishbach}, {Fisher}, {Fitz-Axen}, {Flaminio}, {Fletcher},
  {Fong}, {Font}, {Forsyth}, {Forsyth}, {Fournier}, {Frasca}, {Frasconi},
  {Frei}, {Freise}, {Frey}, {Frey}, {Fries}, {Fritschel}, {Frolov}, {Fulda},
  {Fyffe}, {Gabbard}, {Gadre}, {Gaebel}, {Gair}, {Gammaitoni}, {Ganija},
  {Gaonkar}, {Garcia-Quiros}, {Garufi}, {Gateley}, {Gaudio}, {Gaur},
  {Gayathri}, {Gehrels}, {Gemme}, {Genin}, {Gennai}, {George}, {George},
  {Gergely}, {Germain}, {Ghonge}, {Ghosh}, {Ghosh}, {Ghosh}, {Giaime},
  {Giardina}, {Giazotto}, {Gill}, {Glover}, {Goetz}, {Goetz}, {Gomes},
  {Goncharov}, {Gonz{\'a}lez}, {Gonzalez Castro}, {Gopakumar}, {Gorodetsky},
  {Gossan}, {Gosselin}, {Gouaty}, {Grado}, {Graef}, {Granata}, {Grant}, {Gras},
  {Gray}, {Greco}, {Green}, {Gretarsson}, {Griswold}, {Groot}, {Grote},
  {Grunewald}, {Gruning}, {Guidi}, {Guo}, {Gupta}, {Gupta}, {Gushwa},
  {Gustafson}, {Gustafson}, {Halim}, {Hall}, {Hall}, {Hamilton}, {Hammond},
  {Haney}, {Hanke}, {Hanks}, {Hanna}, {Hannam}, {Hannuksela}, {Hanson},
  {Hardwick}, {Harms}, {Harry}, {Harry}, {Hart}, {Haster}, {Haughian}, {Healy},
  {Heidmann}, {Heintze}, {Heitmann}, {Hello}, {Hemming}, {Hendry}, {Heng},
  {Hennig}, {Heptonstall}, {Heurs}, {Hild}, {Hinderer}, {Hoak}, {Hofman},
  {Holt}, {Holz}, {Hopkins}, {Horst}, {Hough}, {Houston}, {Howell}, {Hreibi},
  {Hu}, {Huerta}, {Huet}, {Hughey}, {Husa}, {Huttner}, {Huynh-Dinh}, {Indik},
  {Inta}, {Intini}, {Isa}, {Isac}, {Isi}, {Iyer}, {Izumi}, {Jacqmin}, {Jani},
  {Jaranowski}, {Jawahar}, {Jim{\'e}nez-Forteza}, {Johnson}, {Jones}, {Jones},
  {Jonker}, {Ju}, {Junker}, {Kalaghatgi}, {Kalogera}, {Kamai}, {Kandhasamy},
  {Kang}, {Kanner}, {Kapadia}, {Karki}, {Karvinen}, {Kasprzack}, {Katolik},
  {Katsavounidis}, {Katzman}, {Kaufer}, {Kawabe}, {K{\'e}f{\'e}lian}, {Keitel},
  {Kemball}, {Kennedy}, {Kent}, {Key}, {Khalili}, {Khan}, {Khan}, {Khan},
  {Khazanov}, {Kijbunchoo}, {Kim}, {Kim}, {Kim}, {Kim}, {Kim}, {Kim},
  {Kimbrell}, {King}, {King}, {Kinley-Hanlon}, {Kirchhoff}, {Kissel},
  {Kleybolte}, {Klimenko}, {Knowles}, {Koch}, {Koehlenbeck}, {Koley},
  {Kondrashov}, {Kontos}, {Korobko}, {Korth}, {Kowalska}, {Kozak},
  {Kr{\"a}mer}, {Kringel}, {Krishnan}, {Kr{\'o}lak}, {Kuehn}, {Kumar}, {Kumar},
  {Kumar}, {Kuo}, {Kutynia}, {Kwang}, {Lackey}, {Lai}, {Landry}, {Lang},
  {Lange}, {Lantz}, {Lanza}, {Larson}, {Lartaux-Vollard}, {Lasky}, {Laxen},
  {Lazzarini}, {Lazzaro}, {Leaci}, {Leavey}, {Lee}, {Lee}, {Lee}, {Lee}, {Lee},
  {Lehmann}, {Lenon}, {Leonardi}, {Leroy}, {Letendre}, {Levin}, {Li}, {Linker},
  {Littenberg}, {Liu}, {Lo}, {Lockerbie}, {London}, {Lord}, {Lorenzini},
  {Loriette}, {Lormand}, {Losurdo}, {Lough}, {Lousto}, {Lovelace}, {L{\"u}ck},
  {Lumaca}, {Lundgren}, {Lynch}, {Ma}, {Macas}, {Macfoy}, {Machenschalk},
  {MacInnis}, {Macleod}, {Maga{\~n}a Hernandez}, {Maga{\~n}a-Sandoval},
  {Maga{\~n}a Zertuche}, {Magee}, {Majorana}, {Maksimovic}, {Man}, {Mandic},
  {Mangano}, {Mansell}, {Manske}, {Mantovani}, {Marchesoni}, {Marion},
  {M{\'a}rka}, {M{\'a}rka}, {Markakis}, {Markosyan}, {Markowitz}, {Maros},
  {Marquina}, {Marsh}, {Martelli}, {Martellini}, {Martin}, {Martin},
  {Martynov}, {Mason}, {Massera}, {Masserot}, {Massinger}, {Masso-Reid},
  {Mastrogiovanni}, {Matas}, {Matichard}, {Matone}, {Mavalvala}, {Mazumder},
  {McCarthy}, {McClelland}, {McCormick}, {McCuller}, {McGuire}, {McIntyre},
  {McIver}, {McManus}, {McNeill}, {McRae}, {McWilliams}, {Meacher}, {Meadors},
  {Mehmet}, {Meidam}, {Mejuto-Villa}, {Melatos}, {Mendell}, {Mercer}, {Merilh},
  {Merzougui}, {Meshkov}, {Messenger}, {Messick}, {Metzdorff}, {Meyers},
  {Miao}, {Michel}, {Middleton}, {Mikhailov}, {Milano}, {Miller}, {Miller},
  {Miller}, {Millhouse}, {Milovich-Goff}, {Minazzoli}, {Minenkov}, {Ming},
  {Mishra}, {Mitra}, {Mitrofanov}, {Mitselmakher}, {Mittleman}, {Moffa},
  {Moggi}, {Mogushi}, {Mohan}, {Mohapatra}, {Montani}, {Moore}, {Moraru},
  {Moreno}, {Morriss}, {Mours}, {Mow-Lowry}, {Mueller}, {Muir}, {Mukherjee},
  {Mukherjee}, {Mukherjee}, {Mukund}, {Mullavey}, {Munch}, {Mu{\~n}iz},
  {Muratore}, {Murray}, {Napier}, {Nardecchia}, {Naticchioni}, {Nayak},
  {Neilson}, {Nelemans}, {Nelson}, {Nery}, {Neunzert}, {Nevin}, {Newport},
  {Newton}, {Ng}, {Nguyen}, {Nguyen}, {Nichols}, {Nielsen}, {Nissanke}, {Nitz},
  {Noack}, {Nocera}, {Nolting}, {North}, {Nuttall}, {Oberling}, {O'Dea},
  {Ogin}, {Oh}, {Oh}, {Ohme}, {Okada}, {Oliver}, {Oppermann}, {Oram},
  {O'Reilly}, {Ormiston}, {Ortega}, {O'Shaughnessy}, {Ossokine}, {Ottaway},
  {Overmier}, {Owen}, {Pace}, {Page}, {Page}, {Pai}, {Pai}, {Palamos},
  {Palashov}, {Palomba}, {Pal-Singh}, {Pan}, {Pan}, {Pang}, {Pang}, {Pankow},
  {Pannarale}, {Pant}, {Paoletti}, {Paoli}, {Papa}, {Parida}, {Parker},
  {Pascucci}, {Pasqualetti}, {Passaquieti}, {Passuello}, {Patil}, {Patricelli},
  {Pearlstone}, {Pedraza}, {Pedurand}, {Pekowsky}, {Pele}, {Penn}, {Perez},
  {Perreca}, {Perri}, {Pfeiffer}, {Phelps}, {Piccinni}, {Pichot},
  {Piergiovanni}, {Pierro}, {Pillant}, {Pinard}, {Pinto}, {Pirello}, {Pitkin},
  {Poe}, {Poggiani}, {Popolizio}, {Porter}, {Post}, {Powell}, {Prasad},
  {Pratt}, {Pratten}, {Predoi}, {Prestegard}, {Price}, {Prijatelj}, {Principe},
  {Privitera}, {Prodi}, {Prokhorov}, {Puncken}, {Punturo}, {Puppo},
  {P{\"u}rrer}, {Qi}, {Quetschke}, {Quintero}, {Quitzow-James}, {Raab},
  {Rabeling}, {Radkins}, {Raffai}, {Raja}, {Rajan}, {Rajbhandari}, {Rakhmanov},
  {Ramirez}, {Ramos-Buades}, {Rapagnani}, {Raymond}, {Razzano}, {Read},
  {Regimbau}, {Rei}, {Reid}, {Reitze}, {Ren}, {Reyes}, {Ricci}, {Ricker},
  {Rieger}, {Riles}, {Rizzo}, {Robertson}, {Robie}, {Robinet}, {Rocchi},
  {Rolland}, {Rollins}, {Roma}, {Romano}, {Romel}, {Romie}, {Rosi{\'n}ska},
  {Ross}, {Rowan}, {R{\"u}diger}, {Ruggi}, {Rutins}, {Ryan}, {Sachdev},
  {Sadecki}, {Sadeghian}, {Sakellariadou}, {Salconi}, {Saleem}, {Salemi},
  {Samajdar}, {Sammut}, {Sampson}, {Sanchez}, {Sanchez}, {Sanchis-Gual},
  {Sandberg}, {Sanders}, {Sassolas}, {Sathyaprakash}, {Saulson}, {Sauter},
  {Savage}, {Sawadsky}, {Schale}, {Scheel}, {Scheuer}, {Schmidt}, {Schmidt},
  {Schnabel}, {Schofield}, {Sch{\"o}nbeck}, {Schreiber}, {Schuette}, {Schulte},
  {Schutz}, {Schwalbe}, {Scott}, {Scott}, {Seidel}, {Sellers}, {Sengupta},
  {Sentenac}, {Sequino}, {Sergeev}, {Shaddock}, {Shaffer}, {Shah}, {Shahriar},
  {Shaner}, {Shao}, {Shapiro}, {Shawhan}, {Sheperd}, {Shoemaker}, {Shoemaker},
  {Siellez}, {Siemens}, {Sieniawska}, {Sigg}, {Silva}, {Singer}, {Singh},
  {Singhal}, {Sintes}, {Slagmolen}, {Smith}, {Smith}, {Smith}, {Somala}, {Son},
  {Sonnenberg}, {Sorazu}, {Sorrentino}, {Souradeep}, {Spencer}, {Srivastava},
  {Staats}, {Staley}, {Steinke}, {Steinlechner}, {Steinlechner}, {Steinmeyer},
  {Stevenson}, {Stone}, {Stops}, {Strain}, {Stratta}, {Strigin}, {Strunk},
  {Sturani}, {Stuver}, {Summerscales}, {Sun}, {Sunil}, {Suresh}, {Sutton},
  {Swinkels}, {Szczepa{\'n}czyk}, {Tacca}, {Tait}, {Talbot}, {Talukder},
  {Tanner}, {T{\'a}pai}, {Taracchini}, {Tasson}, {Taylor}, {Taylor}, {Tewari},
  {Theeg}, {Thies}, {Thomas}, {Thomas}, {Thomas}, {Thorne}, {Thorne}, {Thrane},
  {Tiwari}, {Tiwari}, {Tokmakov}, {Toland}, {Tonelli}, {Tornasi},
  {Torres-Forn{\'e}}, {Torrie}, {T{\"o}yr{\"a}}, {Travasso}, {Traylor},
  {Trinastic}, {Tringali}, {Trozzo}, {Tsang}, {Tse}, {Tso}, {Tsukada}, {Tsuna},
  {Tuyenbayev}, {Ueno}, {Ugolini}, {Unnikrishnan}, {Urban}, {Usman},
  {Vahlbruch}, {Vajente}, {Valdes}, {van Bakel}, {van Beuzekom}, {van den
  Brand}, {Van Den Broeck}, {Vander-Hyde}, {van der Schaaf}, {van Heijningen},
  {van Veggel}, {Vardaro}, {Varma}, {Vass}, {Vas{\'u}th}, {Vecchio},
  {Vedovato}, {Veitch}, {Veitch}, {Venkateswara}, {Venugopalan}, {Verkindt},
  {Vetrano}, {Vicer{\'e}}, {Viets}, {Vinciguerra}, {Vine}, {Vinet}, {Vitale},
  {Vo}, {Vocca}, {Vorvick}, {Vyatchanin}, {Wade}, {Wade}, {Wade}, {Walet},
  {Walker}, {Wallace}, {Walsh}, {Wang}, {Wang}, {Wang}, {Wang}, {Wang}, {Ward},
  {Warner}, {Was}, {Watchi}, {Weaver}, {Wei}, {Weinert}, {Weinstein}, {Weiss},
  {Wen}, {Wessel}, {Wessels}, {Westerweck}, {Westphal}, {Wette}, {Whelan},
  {Whitcomb}, {Whiting}, {Whittle}, {Wilken}, {Williams}, {Williams},
  {Williamson}, {Willis}, {Willke}, {Wimmer}, {Winkler}, {Wipf}, {Wittel},
  {Woan}, {Woehler}, {Wofford}, {Wong}, {Worden}, {Wright}, {Wu}, {Wysocki},
  {Xiao}, {Yamamoto}, {Yancey}, {Yang}, {Yap}, {Yazback}, {Yu}, {Yu}, {Yvert},
  {Zadro{\.z}ny}, {Zanolin}, {Zelenova}, {Zendri}, {Zevin}, {Zhang}, {Zhang},
  {Zhang}, {Zhang}, {Zhao}, {Zhou}, {Zhou}, {Zhu}, {Zhu}, {Zimmerman},
  {Zucker}, {Zweizig}, {LIGO Scientific Collaboration}, {Virgo Collaboration},
  {Wilson-Hodge}, {Bissaldi}, {Blackburn}, {Briggs}, {Burns}, {Cleveland},
  {Connaughton}, {Gibby}, {Giles}, {Goldstein}, {Hamburg}, {Jenke}, {Hui},
  {Kippen}, {Kocevski}, {McBreen}, {Meegan}, {Paciesas}, {Poolakkil}, {Preece},
  {Racusin}, {Roberts}, {Stanbro}, {Veres}, {von Kienlin}, {GBM}, {Savchenko},
  {Ferrigno}, {Kuulkers}, {Bazzano}, {Bozzo}, {Brandt}, {Chenevez},
  {Courvoisier}, {Diehl}, {Domingo}, {Hanlon}, {Jourdain}, {Laurent}, {Lebrun},
  {Lutovinov}, {Martin-Carrillo}, {Mereghetti}, {Natalucci}, {Rodi}, {Roques},
  {Sunyaev}, {Ubertini}, {INTEGRAL}, {Aartsen}, {Ackermann}, {Adams},
  {Aguilar}, {Ahlers}, {Ahrens}, {Samarai}, {Altmann}, {Andeen}, {Anderson},
  {Ansseau}, {Anton}, {Arg{\"u}elles}, {Auffenberg}, {Axani}, {Bagherpour},
  {Bai}, {Barron}, {Barwick}, {Baum}, {Bay}, {Beatty}, {Becker Tjus},
  {Bernardini}, {Besson}, {Binder}, {Bindig}, {Blaufuss}, {Blot}, {Bohm},
  {B{\"o}rner}, {Bos}, {Bose}, {B{\"o}ser}, {Botner}, {Bourbeau}, {Bourbeau},
  {Bradascio}, {Braun}, {Brayeur}, {Brenzke}, {Bretz}, {Bron},
  {Brostean-Kaiser}, {Burgman}, {Carver}, {Casey}, {Casier}, {Cheung},
  {Chirkin}, {Christov}, {Clark}, {Classen}, {Coenders}, {Collin}, {Conrad},
  {Cowen}, {Cross}, {Day}, {de Andr{\'e}}, {De Clercq}, {DeLaunay},
  {Dembinski}, {De Ridder}, {Desiati}, {de Vries}, {de Wasseige}, {de With},
  {DeYoung}, {D{\'\i}az-V{\'e}lez}, {di Lorenzo}, {Dujmovic}, {Dumm},
  {Dunkman}, {Dvorak}, {Eberhardt}, {Ehrhardt}, {Eichmann}, {Eller}, {Evenson},
  {Fahey}, {Fazely}, {Felde}, {Filimonov}, {Finley}, {Flis}, {Franckowiak},
  {Friedman}, {Fuchs}, {Gaisser}, {Gallagher}, {Gerhardt}, {Ghorbani}, {Giang},
  {Glauch}, {Gl{\"u}senkamp}, {Goldschmidt}, {Gonzalez}, {Grant}, {Griffith},
  {Haack}, {Hallgren}, {Halzen}, {Hanson}, {Hebecker}, {Heereman}, {Helbing},
  {Hellauer}, {Hickford}, {Hignight}, {Hill}, {Hoffman}, {Hoffmann},
  {Hokanson-Fasig}, {Hoshina}, {Huang}, {Huber}, {Hultqvist}, {H{\"u}nnefeld},
  {In}, {Ishihara}, {Jacobi}, {Japaridze}, {Jeong}, {Jero}, {Jones},
  {Kalaczynski}, {Kang}, {Kappes}, {Karg}, {Karle}, {Kauer}, {Keivani},
  {Kelley}, {Kheirandish}, {Kim}, {Kim}, {Kintscher}, {Kiryluk}, {Kittler},
  {Klein}, {Kohnen}, {Koirala}, {Kolanoski}, {K{\"o}pke}, {Kopper}, {Kopper},
  {Koschinsky}, {Koskinen}, {Kowalski}, {Krings}, {Kroll}, {Kr{\"u}ckl},
  {Kunnen}, {Kunwar}, {Kurahashi}, {Kuwabara}, {Kyriacou}, {Labare},
  {Lanfranchi}, {Larson}, {Lauber}, {Lesiak-Bzdak}, {Leuermann}, {Liu}, {Lu},
  {L{\"u}nemann}, {Luszczak}, {Madsen}, {Maggi}, {Mahn}, {Mancina}, {Maruyama},
  {Mase}, {Maunu}, {McNally}, {Meagher}, {Medici}, {Meier}, {Menne}, {Merino},
  {Meures}, {Miarecki}, {Micallef}, {Moment{\'e}}, {Montaruli}, {Moore},
  {Moulai}, {Nahnhauer}, {Nakarmi}, {Naumann}, {Neer}, {Niederhausen},
  {Nowicki}, {Nygren}, {Obertacke Pollmann}, {Olivas}, {O'Murchadha},
  {Palczewski}, {Pandya}, {Pankova}, {Peiffer}, {Pepper}, {P{\'e}rez de los
  Heros}, {Pieloth}, {Pinat}, {Price}, {Przybylski}, {Raab}, {R{\"a}del},
  {Rameez}, {Rawlins}, {Rea}, {Reimann}, {Relethford}, {Relich}, {Resconi},
  {Rhode}, {Richman}, {Robertson}, {Rongen}, {Rott}, {Ruhe}, {Ryckbosch},
  {Rysewyk}, {S{\"a}lzer}, {Sanchez Herrera}, {Sandrock}, {Sandroos},
  {Santander}, {Sarkar}, {Sarkar}, {Satalecka}, {Schlunder}, {Schmidt},
  {Schneider}, {Schoenen}, {Sch{\"o}neberg}, {Schumacher}, {Seckel},
  {Seunarine}, {Soedingrekso}, {Soldin}, {Song}, {Spiczak}, {Spiering},
  {Stachurska}, {Stamatikos}, {Stanev}, {Stasik}, {Stettner}, {Steuer},
  {Stezelberger}, {Stokstad}, {St{\"o}ssl}, {Strotjohann}, {Stuttard},
  {Sullivan}, {Sutherland}, {Taboada}, {Tatar}, {Tenholt}, {Ter-Antonyan},
  {Terliuk}, {Te{\v{s}}i{\'c}}, {Tilav}, {Toale}, {Tobin}, {Toscano}, {Tosi},
  {Tselengidou}, {Tung}, {Turcati}, {Turley}, {Ty}, {Unger}, {Usner},
  {Vandenbroucke}, {Van Driessche}, {van Eijndhoven}, {Vanheule}, {van Santen},
  {Vehring}, {Vogel}, {Vraeghe}, {Walck}, {Wallace}, {Wallraff}, {Wandler},
  {Wandkowsky}, {Waza}, {Weaver}, {Weiss}, {Wendt}, {Werthebach}, {Whelan},
  {Wiebe}, {Wiebusch}, {Wille}, {Williams}, {Wills}, {Wolf}, {Wood}, {Woolsey},
  {Woschnagg}, {Xu}, {Xu}, {Xu}, {Yanez}, {Yodh}, {Yoshida}, {Yuan}, {Zoll},
  {IceCube Collaboration}, {Balasubramanian}, {Mate}, {Bhalerao},
  {Bhattacharya}, {Vibhute}, {Dewangan}, {Rao}, {Vadawale}, {AstroSat Cadmium
  Zinc Telluride Imager Team}, {Svinkin}, {Hurley}, {Aptekar}, {Frederiks},
  {Golenetskii}, {Kozlova}, {Lysenko}, {Oleynik}, {Tsvetkova}, {Ulanov},
  {Cline}, {IPN Collaboration}, {Li}, {Xiong}, {Zhang}, {Lu}, {Song}, {Cao},
  {Chang}, {Chen}, {Chen}, {Chen}, {Chen}, {Chen}, {Chen}, {Cui}, {Cui},
  {Deng}, {Dong}, {Du}, {Fu}, {Gao}, {Gao}, {Gao}, {Ge}, {Gu}, {Guan}, {Guo},
  {Han}, {Hu}, {Huang}, {Huo}, {Jia}, {Jiang}, {Jiang}, {Jin}, {Jin}, {Li},
  {Li}, {Li}, {Li}, {Li}, {Li}, {Li}, {Li}, {Li}, {Li}, {Li}, {Liang}, {Liao},
  {Liu}, {Liu}, {Liu}, {Liu}, {Liu}, {Liu}, {Liu}, {Lu}, {Lu}, {Luo}, {Ma},
  {Meng}, {Nang}, {Nie}, {Ou}, {Qu}, {Sai}, {Sun}, {Tan}, {Tao}, {Tao}, {Tuo},
  {Wang}, {Wang}, {Wang}, {Wang}, {Wang}, {Wen}, {Wu}, {Wu}, {Xiao}, {Xu},
  {Xu}, {Yan}, {Yang}, {Yang}, {Yang}, {Zhang}, {Zhang}, {Zhang}, {Zhang},
  {Zhang}, {Zhang}, {Zhang}, {Zhang}, {Zhang}, {Zhang}, {Zhang}, {Zhang},
  {Zhang}, {Zhang}, {Zhang}, {Zhang}, {Zhang}, {Zhang}, {Zhao}, {Zhao}, {Zhao},
  {Zheng}, {Zhu}, {Zhu}, {Zou}, {Insight-HXMT Collaboration}, {Albert},
  {Andr{\'e}}, {Anghinolfi}, {Ardid}, {Aubert}, {Aublin}, {Avgitas}, {Baret},
  {Barrios-Mart{\'\i}}, {Basa}, {Belhorma}, {Bertin}, {Biagi}, {Bormuth},
  {Bourret}, {Bouwhuis}, {Br{\^a}nza{\c{s}}}, {Bruijn}, {Brunner}, {Busto},
  {Capone}, {Caramete}, {Carr}, {Celli}, {Cherkaoui El Moursli}, {Chiarusi},
  {Circella}, {Coelho}, {Coleiro}, {Coniglione}, {Costantini}, {Coyle},
  {Creusot}, {D{\'\i}az}, {Deschamps}, {De Bonis}, {Distefano}, {Di Palma},
  {Domi}, {Donzaud}, {Dornic}, {Drouhin}, {Eberl}, {El Bojaddaini}, {El
  Khayati}, {Els{\"a}sser}, {Enzenh{\"o}fer}, {Ettahiri}, {Fassi}, {Felis},
  {Fusco}, {Gay}, {Giordano}, {Glotin}, {Gr{\'e}goire}, {Ruiz}, {Graf},
  {Hallmann}, {van Haren}, {Heijboer}, {Hello}, {Hern{\'a}ndez-Rey},
  {H{\"o}ssl}, {Hofest{\"a}dt}, {Hugon}, {Illuminati}, {James}, {de Jong},
  {Jongen}, {Kadler}, {Kalekin}, {Katz}, {Kiessling}, {Kouchner}, {Kreter},
  {Kreykenbohm}, {Kulikovskiy}, {Lachaud}, {Lahmann}, {Lef{\`e}vre}, {Leonora},
  {Lotze}, {Loucatos}, {Marcelin}, {Margiotta}, {Marinelli},
  {Mart{\'\i}nez-Mora}, {Mele}, {Melis}, {Michael}, {Migliozzi}, {Moussa},
  {Navas}, {Nezri}, {Organokov}, {P{\u{a}}v{\u{a}}la{\c{s}}}, {Pellegrino},
  {Perrina}, {Piattelli}, {Popa}, {Pradier}, {Quinn}, {Racca}, {Riccobene},
  {S{\'a}nchez-Losa}, {Salda{\~n}a}, {Salvadori}, {Samtleben}, {Sanguineti},
  {Sapienza}, {Sieger}, {Spurio}, {Stolarczyk}, {Taiuti}, {Tayalati},
  {Trovato}, {Turpin}, {T{\"o}nnis}, {Vallage}, {Van Elewyck}, {Versari},
  {Vivolo}, {Vizzoca}, {Wilms}, {Zornoza}, {Z{\'u}{\~n}iga}, {ANTARES
  Collaboration}, {Beardmore}, {Breeveld}, {Burrows}, {Cenko}, {Cusumano},
  {D'A{\`\i}}, {de Pasquale}, {Emery}, {Evans}, {Giommi}, {Gronwall}, {Kennea},
  {Krimm}, {Kuin}, {Lien}, {Marshall}, {Melandri}, {Nousek}, {Oates},
  {Osborne}, {Pagani}, {Page}, {Palmer}, {Perri}, {Siegel}, {Sbarufatti},
  {Tagliaferri}, {Tohuvavohu}, {Swift Collaboration}, {Tavani}, {Verrecchia},
  {Bulgarelli}, {Evangelista}, {Pacciani}, {Feroci}, {Pittori}, {Giuliani},
  {Del Monte}, {Donnarumma}, {Argan}, {Trois}, {Ursi}, {Cardillo}, {Piano},
  {Longo}, {Lucarelli}, {Munar-Adrover}, {Fuschino}, {Labanti}, {Marisaldi},
  {Minervini}, {Fioretti}, {Parmiggiani}, {Gianotti}, {Trifoglio}, {Di Persio},
  {Antonelli}, {Barbiellini}, {Caraveo}, {Cattaneo}, {Costa}, {Colafrancesco},
  {D'Amico}, {Ferrari}, {Morselli}, {Paoletti}, {Picozza}, {Pilia}, {Rappoldi},
  {Soffitta}, {Vercellone}, {AGILE Team}, {Foley}, {Coulter}, {Kilpatrick},
  {Drout}, {Piro}, {Shappee}, {Siebert}, {Simon}, {Ulloa}, {Kasen}, {Madore},
  {Murguia-Berthier}, {Pan}, {Prochaska}, {Ramirez-Ruiz}, {Rest},
  {Rojas-Bravo}, {1M2H Team}, {Berger}, {Soares-Santos}, {Annis}, {Alexander},
  {Allam}, {Balbinot}, {Blanchard}, {Brout}, {Butler}, {Chornock}, {Cook},
  {Cowperthwaite}, {Diehl}, {Drlica-Wagner}, {Drout}, {Durret}, {Eftekhari},
  {Finley}, {Fong}, {Frieman}, {Fryer}, {Garc{\'\i}a-Bellido}, {Gruendl},
  {Hartley}, {Herner}, {Kessler}, {Lin}, {Lopes}, {Louren{\c{c}}o}, {Margutti},
  {Marshall}, {Matheson}, {Medina}, {Metzger}, {Mu{\~n}oz}, {Muir}, {Nicholl},
  {Nugent}, {Palmese}, {Paz-Chinch{\'o}n}, {Quataert}, {Sako}, {Sauseda},
  {Schlegel}, {Scolnic}, {Secco}, {Smith}, {Sobreira}, {Villar}, {Vivas},
  {Wester}, {Williams}, {Yanny}, {Zenteno}, {Zhang}, {Abbott}, {Banerji},
  {Bechtol}, {Benoit-L{\'e}vy}, {Bertin}, {Brooks}, {Buckley-Geer}, {Burke},
  {Capozzi}, {Carnero Rosell}, {Carrasco Kind}, {Castander}, {Crocce}, {Cunha},
  {D'Andrea}, {da Costa}, {Davis}, {DePoy}, {Desai}, {Dietrich}, {Eifler},
  {Fernandez}, {Flaugher}, {Fosalba}, {Gaztanaga}, {Gerdes}, {Giannantonio},
  {Goldstein}, {Gruen}, {Gschwend}, {Gutierrez}, {Honscheid}, {James},
  {Jeltema}, {Johnson}, {Johnson}, {Kent}, {Krause}, {Kron}, {Kuehn}, {Lahav},
  {Lima}, {Maia}, {March}, {Martini}, {McMahon}, {Menanteau}, {Miller},
  {Miquel}, {Mohr}, {Nichol}, {Ogando}, {Plazas}, {Romer}, {Roodman}, {Rykoff},
  {Sanchez}, {Scarpine}, {Schindler}, {Schubnell}, {Sevilla-Noarbe}, {Sheldon},
  {Smith}, {Smith}, {Stebbins}, {Suchyta}, {Swanson}, {Tarle}, {Thomas},
  {Troxel}, {Tucker}, {Vikram}, {Walker}, {Wechsler}, {Weller}, {Carlin},
  {Gill}, {Li}, {Marriner}, {Neilsen}, {Dark Energy Camera GW-EM
  Collaboration}, {DES Collaboration}, {Haislip}, {Kouprianov}, {Reichart},
  {Sand}, {Tartaglia}, {Valenti}, {Yang}, {DLT40 Collaboration}, {Benetti},
  {Brocato}, {Campana}, {Cappellaro}, {Covino}, {D'Avanzo}, {D'Elia}, {Getman},
  {Ghirlanda}, {Ghisellini}, {Limatola}, {Nicastro}, {Palazzi}, {Pian},
  {Piranomonte}, {Possenti}, {Rossi}, {Salafia}, {Tomasella}, {Amati},
  {Antonelli}, {Bernardini}, {Bufano}, {Capaccioli}, {Casella}, {Dadina}, {De
  Cesare}, {Di Paola}, {Giuffrida}, {Giunta}, {Israel}, {Lisi}, {Maiorano},
  {Mapelli}, {Masetti}, {Pescalli}, {Pulone}, {Salvaterra}, {Schipani},
  {Spera}, {Stamerra}, {Stella}, {Testa}, {Turatto}, {Vergani}, {Aresu},
  {Bachetti}, {Buffa}, {Burgay}, {Buttu}, {Caria}, {Carretti}, {Casasola},
  {Castangia}, {Carboni}, {Casu}, {Concu}, {Corongiu}, {Deiana}, {Egron},
  {Fara}, {Gaudiomonte}, {Gusai}, {Ladu}, {Loru}, {Leurini}, {Marongiu},
  {Melis}, {Melis}, {Migoni}, {Milia}, {Navarrini}, {Orlati}, {Ortu}, {Palmas},
  {Pellizzoni}, {Perrodin}, {Pisanu}, {Poppi}, {Righini}, {Saba}, {Serra},
  {Serrau}, {Stagni}, {Surcis}, {Vacca}, {Vargiu}, {Hunt}, {Jin}, {Klose},
  {Kouveliotou}, {Mazzali}, {M{\o}ller}, {Nava}, {Piran}, {Selsing}, {Vergani},
  {Wiersema}, {Toma}, {Higgins}, {Mundell}, {di Serego Alighieri}, {G{\'o}tz},
  {Gao}, {Gomboc}, {Kaper}, {Kobayashi}, {Kopac}, {Mao}, {Starling}, {Steele},
  {van der Horst}, {GRAWITA: GRAvitational Wave Inaf TeAm}, {Acero}, {Atwood},
  {Baldini}, {Barbiellini}, {Bastieri}, {Berenji}, {Bellazzini}, {Bissaldi},
  {Blandford}, {Bloom}, {Bonino}, {Bottacini}, {Bregeon}, {Buehler}, {Buson},
  {Cameron}, {Caputo}, {Caraveo}, {Cavazzuti}, {Chekhtman}, {Cheung}, {Chiang},
  {Ciprini}, {Cohen-Tanugi}, {Cominsky}, {Costantin}, {Cuoco}, {D'Ammando}, {de
  Palma}, {Digel}, {Di Lalla}, {Di Mauro}, {Di Venere}, {Dubois}, {Fegan},
  {Focke}, {Franckowiak}, {Fukazawa}, {Funk}, {Fusco}, {Gargano}, {Gasparrini},
  {Giglietto}, {Giordano}, {Giroletti}, {Glanzman}, {Green}, {Grondin},
  {Guillemot}, {Guiriec}, {Harding}, {Horan}, {J{\'o}hannesson}, {Kamae},
  {Kensei}, {Kuss}, {La Mura}, {Latronico}, {Lemoine-Goumard}, {Longo},
  {Loparco}, {Lovellette}, {Lubrano}, {Magill}, {Maldera}, {Manfreda},
  {Mazziotta}, {McEnery}, {Meyer}, {Michelson}, {Mirabal}, {Monzani},
  {Moretti}, {Morselli}, {Moskalenko}, {Negro}, {Nuss}, {Ojha}, {Omodei},
  {Orienti}, {Orlando}, {Palatiello}, {Paliya}, {Paneque}, {Pesce-Rollins},
  {Piron}, {Porter}, {Principe}, {Rain{\`o}}, {Rando}, {Razzano}, {Razzaque},
  {Reimer}, {Reimer}, {Reposeur}, {Rochester}, {Saz Parkinson}, {Sgr{\`o}},
  {Siskind}, {Spada}, {Spandre}, {Suson}, {Takahashi}, {Tanaka}, {Thayer},
  {Thayer}, {Thompson}, {Tibaldo}, {Torres}, {Torresi}, {Troja}, {Venters},
  {Vianello}, {Zaharijas}, {Fermi Large Area Telescope Collaboration},
  {Allison}, {Bannister}, {Dobie}, {Kaplan}, {Lenc}, {Lynch}, {Murphy},
  {Sadler}, {Australia Telescope Compact Array}, {Hotan}, {James}, {Oslowski},
  {Raja}, {Shannon}, {Whiting}, {Australian SKA Pathfinder}, {Arcavi},
  {Howell}, {McCully}, {Hosseinzadeh}, {Hiramatsu}, {Poznanski}, {Barnes},
  {Zaltzman}, {Vasylyev}, {Maoz}, {Las Cumbres Observatory Group}, {Cooke},
  {Bailes}, {Wolf}, {Deller}, {Lidman}, {Wang}, {Gendre}, {Andreoni}, {Ackley},
  {Pritchard}, {Bessell}, {Chang}, {M{\"o}ller}, {Onken}, {Scalzo},
  {Ridden-Harper}, {Sharp}, {Tucker}, {Farrell}, {Elmer}, {Johnston},
  {Venkatraman Krishnan}, {Keane}, {Green}, {Jameson}, {Hu}, {Ma}, {Sun}, {Wu},
  {Wang}, {Shang}, {Hu}, {Ashley}, {Yuan}, {Li}, {Tao}, {Zhu}, {Zhang},
  {Suntzeff}, {Zhou}, {Yang}, {Orange}, {Morris}, {Cucchiara}, {Giblin},
  {Klotz}, {Staff}, {Thierry}, {Schmidt}, {OzGrav}, {(Deeper}, {Wider},
  {program}, {AST3}, {CAASTRO Collaborations}, {Tanvir}, {Levan}, {Cano}, {de
  Ugarte-Postigo}, {Gonz{\'a}lez-Fern{\'a}ndez}, {Greiner}, {Hjorth}, {Irwin},
  {Kr{\"u}hler}, {Mandel}, {Milvang-Jensen}, {O'Brien}, {Rol}, {Rosetti},
  {Rosswog}, {Rowlinson}, {Steeghs}, {Th{\"o}ne}, {Ulaczyk}, {Watson}, {Bruun},
  {Cutter}, {Figuera Jaimes}, {Fujii}, {Fruchter}, {Gompertz}, {Jakobsson},
  {Hodosan}, {J{\`e}rgensen}, {Kangas}, {Kann}, {Rabus}, {Schr{\o}der},
  {Stanway}, {Wijers}, {VINROUGE Collaboration}, {Lipunov}, {Gorbovskoy},
  {Kornilov}, {Tyurina}, {Balanutsa}, {Kuznetsov}, {Vlasenko}, {Podesta},
  {Lopez}, {Podesta}, {Levato}, {Saffe}, {Mallamaci}, {Budnev}, {Gress},
  {Kuvshinov}, {Gorbunov}, {Vladimirov}, {Zimnukhov}, {Gabovich}, {Yurkov},
  {Sergienko}, {Rebolo}, {Serra-Ricart}, {Tlatov}, {Ishmuhametova}, {MASTER
  Collaboration}, {Abe}, {Aoki}, {Aoki}, {Asakura}, {Baar}, {Barway}, {Bond},
  {Doi}, {Finet}, {Fujiyoshi}, {Furusawa}, {Honda}, {Itoh}, {Kanda},
  {Kawabata}, {Kawabata}, {Kim}, {Koshida}, {Kuroda}, {Lee}, {Liu},
  {Matsubayashi}, {Miyazaki}, {Morihana}, {Morokuma}, {Motohara}, {Murata},
  {Nagai}, {Nagashima}, {Nagayama}, {Nakaoka}, {Nakata}, {Ohsawa}, {Ohshima},
  {Ohta}, {Okita}, {Saito}, {Saito}, {Sako}, {Sekiguchi}, {Sumi}, {Tajitsu},
  {Takahashi}, {Takayama}, {Tamura}, {Tanaka}, {Tanaka}, {Terai}, {Tominaga},
  {Tristram}, {Uemura}, {Utsumi}, {Yamaguchi}, {Yasuda}, {Yoshida}, {Zenko},
  {J-GEM}, {Adams}, {Anupama}, {Bally}, {Barway}, {Bellm}, {Blagorodnova},
  {Cannella}, {Chandra}, {Chatterjee}, {Clarke}, {Cobb}, {Cook}, {Copperwheat},
  {De}, {Emery}, {Feindt}, {Foster}, {Fox}, {Frail}, {Fremling}, {Frohmaier},
  {Garcia}, {Ghosh}, {Giacintucci}, {Goobar}, {Gottlieb}, {Grefenstette},
  {Hallinan}, {Harrison}, {Heida}, {Helou}, {Ho}, {Horesh}, {Hotokezaka}, {Ip},
  {Itoh}, {Jacobs}, {Jencson}, {Kasen}, {Kasliwal}, {Kassim}, {Kim}, {Kiran},
  {Kuin}, {Kulkarni}, {Kupfer}, {Lau}, {Madsen}, {Mazzali}, {Miller},
  {Miyasaka}, {Mooley}, {Myers}, {Nakar}, {Ngeow}, {Nugent}, {Ofek},
  {Palliyaguru}, {Pavana}, {Perley}, {Peters}, {Pike}, {Piran}, {Qi}, {Quimby},
  {Rana}, {Rosswog}, {Rusu}, {Sadler}, {Van Sistine}, {Sollerman}, {Xu}, {Yan},
  {Yatsu}, {Yu}, {Zhang}, {Zhao}, {GROWTH}, {JAGWAR}, {Caltech-NRAO},
  {TTU-NRAO}, {NuSTAR Collaborations}, {Chambers}, {Huber}, {Schultz},
  {Bulger}, {Flewelling}, {Magnier}, {Lowe}, {Wainscoat}, {Waters}, {Willman},
  {Pan-STARRS}, {Ebisawa}, {Hanyu}, {Harita}, {Hashimoto}, {Hidaka}, {Hori},
  {Ishikawa}, {Isobe}, {Iwakiri}, {Kawai}, {Kawai}, {Kawamuro}, {Kawase},
  {Kitaoka}, {Makishima}, {Matsuoka}, {Mihara}, {Morita}, {Morita}, {Nakahira},
  {Nakajima}, {Nakamura}, {Negoro}, {Oda}, {Sakamaki}, {Sasaki}, {Serino},
  {Shidatsu}, {Shimomukai}, {Sugawara}, {Sugita}, {Sugizaki}, {Tachibana},
  {Takao}, {Tanimoto}, {Tomida}, {Tsuboi}, {Tsunemi}, {Ueda}, {Ueno}, {Yamada},
  {Yamaoka}, {Yamauchi}, {Yatabe}, {Yoneyama}, {Yoshii}, {MAXI Team}, {Coward},
  {Crisp}, {Macpherson}, {Andreoni}, {Laugier}, {Noysena}, {Klotz}, {Gendre},
  {Thierry}, {Turpin}, {Consortium}, {Im}, {Choi}, {Kim}, {Yoon}, {Lim}, {Lee},
  {Lee}, {Kim}, {Ko}, {Joe}, {Kwon}, {Kim}, {Lim}, {Choi}, {KU Collaboration},
  {Fynbo}, {Malesani}, {Xu}, {Optical Telescope}, {Smartt}, {Jerkstrand},
  {Kankare}, {Sim}, {Fraser}, {Inserra}, {Maguire}, {Leloudas}, {Magee},
  {Shingles}, {Smith}, {Young}, {Kotak}, {Gal-Yam}, {Lyman}, {Homan},
  {Agliozzo}, {Anderson}, {Angus}, {Ashall}, {Barbarino}, {Bauer}, {Berton},
  {Botticella}, {Bulla}, {Cannizzaro}, {Cartier}, {Cikota}, {Clark}, {De Cia},
  {Della Valle}, {Dennefeld}, {Dessart}, {Dimitriadis}, {Elias-Rosa}, {Firth},
  {Fl{\"o}rs}, {Frohmaier}, {Galbany}, {Gonz{\'a}lez-Gait{\'a}n}, {Gromadzki},
  {Guti{\'e}rrez}, {Hamanowicz}, {Harmanen}, {Heintz}, {Hernandez}, {Hodgkin},
  {Hook}, {Izzo}, {James}, {Jonker}, {Kerzendorf}, {Kostrzewa-Rutkowska},
  {Kromer}, {Kuncarayakti}, {Lawrence}, {Manulis}, {Mattila}, {McBrien},
  {M{\"u}ller}, {Nordin}, {O'Neill}, {Onori}, {Palmerio}, {Pastorello},
  {Patat}, {Pignata}, {Podsiadlowski}, {Razza}, {Reynolds}, {Roy}, {Ruiter},
  {Rybicki}, {Salmon}, {Pumo}, {Prentice}, {Seitenzahl}, {Smith}, {Sollerman},
  {Sullivan}, {Szegedi}, {Taddia}, {Taubenberger}, {Terreran}, {Van Soelen},
  {Vos}, {Walton}, {Wright}, {Wyrzykowski}, {Yaron}, {pre=''(''>ePESSTO},
  {Chen}, {Kr{\"u}hler}, {Schady}, {Wiseman}, {Greiner}, {Rau}, {Schweyer},
  {Klose}, {Nicuesa Guelbenzu}, {GROND}, {Palliyaguru}, {Tech University},
  {Shara}, {Williams}, {Vaisanen}, {Potter}, {Romero Colmenero}, {Crawford},
  {Buckley}, {Mao}, {SALT Group}, {D{\'\i}az}, {Macri}, {Garc{\'\i}a Lambas},
  {Mendes de Oliveira}, {Nilo Castell{\'o}n}, {Ribeiro}, {S{\'a}nchez},
  {Schoenell}, {Abramo}, {Akras}, {Alcaniz}, {Artola}, {Beroiz}, {Bonoli},
  {Cabral}, {Camuccio}, {Chavushyan}, {Coelho}, {Colazo}, {Costa-Duarte},
  {Cuevas Larenas}, {Dom{\'\i}nguez Romero}, {Dultzin}, {Fern{\'a}ndez},
  {Garc{\'\i}a}, {Girardini}, {Gon{\c{c}}alves}, {Gon{\c{c}}alves}, {Gurovich},
  {Jim{\'e}nez-Teja}, {Kanaan}, {Lares}, {Lopes de Oliveira}, {L{\'o}pez-Cruz},
  {Melia}, {Molino}, {Padilla}, {Pe{\~n}uela}, {Placco}, {Qui{\~n}ones},
  {Ram{\'\i}rez Rivera}, {Renzi}, {Riguccini}, {R{\'\i}os-L{\'o}pez},
  {Rodriguez}, {Sampedro}, {Schneiter}, {Sodr{\'e}}, {Starck}, {Torres-Flores},
  {Tornatore}, {Zadro{\.z}ny}, {Castillo}, {TOROS: Transient Robotic
  Observatory of South Collaboration}, {Castro-Tirado}, {Tello}, {Hu}, {Zhang},
  {Cunniffe}, {Castell{\'o}n}, {Hiriart}, {Caballero-Garc{\'\i}a},
  {Jel{\'\i}nek}, {Kub{\'a}nek}, {P{\'e}rez del Pulgar}, {Park}, {Jeong},
  {Castro Cer{\'o}n}, {Pandey}, {Yock}, {Querel}, {Fan}, {Wang}, {BOOTES
  Collaboration}, {Beardsley}, {Brown}, {Crosse}, {Emrich}, {Franzen},
  {Gaensler}, {Horsley}, {Johnston-Hollitt}, {Kenney}, {Morales}, {Pallot},
  {Sokolowski}, {Steele}, {Tingay}, {Trott}, {Walker}, {Wayth}, {Williams},
  {Wu}, {Murchison Widefield Array}, {Yoshida}, {Sakamoto}, {Kawakubo},
  {Yamaoka}, {Takahashi}, {Asaoka}, {Ozawa}, {Torii}, {Shimizu}, {Tamura},
  {Ishizaki}, {Cherry}, {Ricciarini}, {Penacchioni}, {Marrocchesi}, {CALET
  Collaboration}, {Pozanenko}, {Volnova}, {Mazaeva}, {Minaev}, {Krugov},
  {Kusakin}, {Reva}, {Moskvitin}, {Rumyantsev}, {Inasaridze}, {Klunko},
  {Tungalag}, {Schmalz}, {Burhonov}, {IKI-GW Follow-up Collaboration},
  {Abdalla}, {Abramowski}, {Aharonian}, {Ait Benkhali}, {Ang{\"u}ner},
  {Arakawa}, {Arrieta}, {Aubert}, {Backes}, {Balzer}, {Barnard}, {Becherini},
  {Becker Tjus}, {Berge}, {Bernhard}, {Bernl{\"o}hr}, {Blackwell},
  {B{\"o}ttcher}, {Boisson}, {Bolmont}, {Bonnefoy}, {Bordas}, {Bregeon},
  {Brun}, {Brun}, {Bryan}, {B{\"u}chele}, {Bulik}, {Capasso}, {Caroff},
  {Carosi}, {Casanova}, {Cerruti}, {Chakraborty}, {Chaves}, {Chen},
  {Chevalier}, {Colafrancesco}, {Condon}, {Conrad}, {Davids}, {Decock}, {Deil},
  {Devin}, {deWilt}, {Dirson}, {Djannati-Ata{\"\i}}, {Donath}, {O'C. Drury},
  {Dutson}, {Dyks}, {Edwards}, {Egberts}, {Emery}, {Ernenwein}, {Eschbach},
  {Farnier}, {Fegan}, {Fernandes}, {Fiasson}, {Fontaine}, {Funk},
  {F{\"u}ssling}, {Gabici}, {Gallant}, {Garrigoux}, {Gat{\'e}}, {Giavitto},
  {Giebels}, {Glawion}, {Glicenstein}, {Gottschall}, {Grondin}, {Hahn},
  {Haupt}, {Hawkes}, {Heinzelmann}, {Henri}, {Hermann}, {Hinton}, {Hofmann},
  {Hoischen}, {Holch}, {Holler}, {Horns}, {Ivascenko}, {Iwasaki},
  {Jacholkowska}, {Jamrozy}, {Jankowsky}, {Jankowsky}, {Jingo}, {Jouvin},
  {Jung-Richardt}, {Kastendieck}, {Katarzy{\'n}ski}, {Katsuragawa},
  {Kerszberg}, {Khangulyan}, {Kh{\'e}lifi}, {King}, {Klepser}, {Klochkov},
  {Klu{\'z}niak}, {Komin}, {Kosack}, {Krakau}, {Kraus}, {Kr{\"u}ger}, {Laffon},
  {Lamanna}, {Lau}, {Lees}, {Lefaucheur}, {Lemi{\`e}re}, {Lemoine-Goumard},
  {Lenain}, {Leser}, {Lohse}, {Lorentz}, {Liu}, {Lypova}, {Malyshev},
  {Marandon}, {Marcowith}, {Mariaud}, {Marx}, {Maurin}, {Maxted}, {Mayer},
  {Meintjes}, {Meyer}, {Mitchell}, {Moderski}, {Mohamed}, {Mohrmann},
  {Mor{\r{a}}}, {Moulin}, {Murach}, {Nakashima}, {de Naurois}, {Ndiyavala},
  {Niederwanger}, {Niemiec}, {Oakes}, {O'Brien}, {Odaka}, {Ohm}, {Ostrowski},
  {Oya}, {Padovani}, {Panter}, {Parsons}, {Pekeur}, {Pelletier}, {Perennes},
  {Petrucci}, {Peyaud}, {Piel}, {Pita}, {Poireau}, {Poon}, {Prokhorov},
  {Prokoph}, {P{\"u}hlhofer}, {Punch}, {Quirrenbach}, {Raab}, {Rauth},
  {Reimer}, {Reimer}, {Renaud}, {de los Reyes}, {Rieger}, {Rinchiuso},
  {Romoli}, {Rowell}, {Rudak}, {Rulten}, {Sahakian}, {Saito}, {Sanchez},
  {Santangelo}, {Sasaki}, {Schlickeiser}, {Sch{\"u}ssler}, {Schulz},
  {Schwanke}, {Schwemmer}, {Seglar-Arroyo}, {Settimo}, {Seyffert}, {Shafi},
  {Shilon}, {Shiningayamwe}, {Simoni}, {Sol}, {Spanier}, {Spir-Jacob},
  {Stawarz}, {Steenkamp}, {Stegmann}, {Steppa}, {Sushch}, {Takahashi},
  {Tavernet}, {Tavernier}, {Taylor}, {Terrier}, {Tibaldo}, {Tiziani},
  {Tluczykont}, {Trichard}, {Tsirou}, {Tsuji}, {Tuffs}, {Uchiyama}, {van der
  Walt}, {van Eldik}, {van Rensburg}, {van Soelen}, {Vasileiadis}, {Veh},
  {Venter}, {Viana}, {Vincent}, {Vink}, {Voisin}, {V{\"o}lk}, {Vuillaume},
  {Wadiasingh}, {Wagner}, {Wagner}, {Wagner}, {White}, {Wierzcholska},
  {Willmann}, {W{\"o}rnlein}, {Wouters}, {Yang}, {Zaborov}, {Zacharias},
  {Zanin}, {Zdziarski}, {Zech}, {Zefi}, {Ziegler}, {Zorn}, {{\.Z}ywucka},
  {H.~E.~S.~S. Collaboration}, {Fender}, {Broderick}, {Rowlinson}, {Wijers},
  {Stewart}, {ter Veen}, {Shulevski}, {LOFAR Collaboration}, {Kavic},
  {Simonetti}, {League}, {Tsai}, {Obenberger}, {Nathaniel}, {Taylor}, {Dowell},
  {Liebling}, {Estes}, {Lippert}, {Sharma}, {Vincent}, {Farella}, {Wavelength
  Array}, {Abeysekara}, {Albert}, {Alfaro}, {Alvarez}, {Arceo},
  {Arteaga-Vel{\'a}zquez}, {Avila Rojas}, {Ayala Solares}, {Barber}, {Becerra
  Gonzalez}, {Becerril}, {Belmont-Moreno}, {BenZvi}, {Berley}, {Bernal},
  {Braun}, {Brisbois}, {Caballero-Mora}, {Capistr{\'a}n}, {Carrami{\~n}ana},
  {Casanova}, {Castillo}, {Cotti}, {Cotzomi}, {Couti{\~n}o de Le{\'o}n}, {De
  Le{\'o}n}, {De la Fuente}, {Diaz Hernandez}, {Dichiara}, {Dingus},
  {DuVernois}, {D{\'\i}az-V{\'e}lez}, {Ellsworth}, {Engel},
  {Enr{\'\i}quez-Rivera}, {Fiorino}, {Fleischhack}, {Fraija},
  {Garc{\'\i}a-Gonz{\'a}lez}, {Garfias}, {Gerhardt}, {Gonz{\~o}lez Mu{\~n}oz},
  {Gonz{\'a}lez}, {Goodman}, {Hampel-Arias}, {Harding}, {Hernandez},
  {Hernandez-Almada}, {Hona}, {H{\"u}ntemeyer}, {Iriarte}, {Jardin-Blicq},
  {Joshi}, {Kaufmann}, {Kieda}, {Lara}, {Lauer}, {Lennarz}, {Le{\'o}n Vargas},
  {Linnemann}, {Longinotti}, {Raya}, {Luna-Garc{\'\i}a}, {L{\'o}pez-Coto},
  {Malone}, {Marinelli}, {Martinez}, {Martinez-Castellanos},
  {Mart{\'\i}nez-Castro}, {Mart{\'\i}nez-Huerta}, {Matthews},
  {Miranda-Romagnoli}, {Moreno}, {Mostaf{\'a}}, {Nellen}, {Newbold}, {Nisa},
  {Noriega-Papaqui}, {Pelayo}, {Pretz}, {P{\'e}rez-P{\'e}rez}, {Ren}, {Rho},
  {Rivi{\`e}re}, {Rosa-Gonz{\'a}lez}, {Rosenberg}, {Ruiz-Velasco}, {Salazar},
  {Salesa Greus}, {Sandoval}, {Schneider}, {Schoorlemmer}, {Sinnis}, {Smith},
  {Springer}, {Surajbali}, {Tibolla}, {Tollefson}, {Torres}, {Ukwatta},
  {Weisgarber}, {Westerhoff}, {Wisher}, {Wood}, {Yapici}, {Yodh}, {Younk},
  {Zhou}, {{\'A}lvarez}, {HAWC Collaboration}, {Aab}, {Abreu}, {Aglietta},
  {Albuquerque}, {Albury}, {Allekotte}, {Almela}, {Alvarez Castillo},
  {Alvarez-Mu{\~n}iz}, {Anastasi}, {Anchordoqui}, {Andrada}, {Andringa},
  {Aramo}, {Arsene}, {Asorey}, {Assis}, {Avila}, {Badescu}, {Balaceanu},
  {Barbato}, {Barreira Luz}, {Becker}, {Bellido}, {Berat}, {Bertaina},
  {Bertou}, {Biermann}, {Biteau}, {Blaess}, {Blanco}, {Blazek}, {Bleve},
  {Boh{\'a}{\v{c}}ov{\'a}}, {Bonifazi}, {Borodai}, {Botti}, {Brack}, {Brancus},
  {Bretz}, {Bridgeman}, {Briechle}, {Buchholz}, {Bueno}, {Buitink}, {Buscemi},
  {Caballero-Mora}, {Caccianiga}, {Cancio}, {Canfora}, {Caruso}, {Castellina},
  {Catalani}, {Cataldi}, {Cazon}, {Chavez}, {Chinellato}, {Chudoba}, {Clay},
  {Cobos Cerutti}, {Colalillo}, {Coleman}, {Collica}, {Coluccia},
  {Concei{\c{c}}{\~a}o}, {Consolati}, {Contreras}, {Cooper}, {Coutu},
  {Covault}, {Cronin}, {D'Amico}, {Daniel}, {Dasso}, {Daumiller}, {Dawson},
  {Day}, {de Almeida}, {de Jong}, {De Mauro}, {de Mello Neto}, {De Mitri}, {de
  Oliveira}, {de Souza}, {Debatin}, {Deligny}, {D{\'\i}az Castro}, {Diogo},
  {Dobrigkeit}, {D'Olivo}, {Dorosti}, {Dos Anjos}, {Dova}, {Dundovic}, {Ebr},
  {Engel}, {Erdmann}, {Erfani}, {Escobar}, {Espadanal}, {Etchegoyen}, {Falcke},
  {Farmer}, {Farrar}, {Fauth}, {Fazzini}, {Feldbusch}, {Fenu}, {Fick},
  {Figueira}, {Filip{\v{c}}i{\v{c}}}, {Freire}, {Fujii}, {Fuster},
  {Ga{\"\i}or}, {Garc{\'\i}a}, {Gat{\'e}}, {Gemmeke}, {Gherghel-Lascu}, {Ghia},
  {Giaccari}, {Giammarchi}, {Giller}, {G{\l}as}, {Glaser}, {Golup}, {G{\'o}mez
  Berisso}, {G{\'o}mez Vitale}, {Gonz{\'a}lez}, {Gorgi}, {Gottowik}, {Grillo},
  {Grubb}, {Guarino}, {Guedes}, {Halliday}, {Hampel}, {Hansen}, {Harari},
  {Harrison}, {Harvey}, {Haungs}, {Hebbeker}, {Heck}, {Heimann}, {Herve},
  {Hill}, {Hojvat}, {Holt}, {Homola}, {H{\"o}randel}, {Horvath},
  {Hrabovsk{\'y}}, {Huege}, {Hulsman}, {Insolia}, {Isar}, {Jandt}, {Johnsen},
  {Josebachuili}, {Jurysek}, {K{\"a}{\"a}p{\"a}}, {Kampert}, {Keilhauer},
  {Kemmerich}, {Kemp}, {Kieckhafer}, {Klages}, {Kleifges}, {Kleinfeller},
  {Krause}, {Krohm}, {Kuempel}, {Kukec Mezek}, {Kunka}, {Kuotb Awad}, {Lago},
  {LaHurd}, {Lang}, {Lauscher}, {Legumina}, {Leigui de Oliveira},
  {Letessier-Selvon}, {Lhenry-Yvon}, {Link}, {Lo Presti}, {Lopes}, {L{\'o}pez},
  {L{\'o}pez Casado}, {Lorek}, {Luce}, {Lucero}, {Malacari}, {Mallamaci},
  {Mandat}, {Mantsch}, {Mariazzi}, {Maris}, {Marsella}, {Martello}, {Martinez},
  {Mart{\'\i}nez Bravo}, {Mas{\'\i}as Meza}, {Mathes}, {Mathys}, {Matthews},
  {Matthiae}, {Mayotte}, {Mazur}, {Medina}, {Medina-Tanco}, {Melo},
  {Menshikov}, {Merenda}, {Michal}, {Micheletti}, {Middendorf}, {Miramonti},
  {Mitrica}, {Mockler}, {Mollerach}, {Montanet}, {Morello}, {Morlino},
  {M{\"u}ller}, {M{\"u}ller}, {Muller}, {M{\"u}ller}, {Mussa}, {Naranjo},
  {Nguyen}, {Niculescu-Oglinzanu}, {Niechciol}, {Niemietz}, {Niggemann},
  {Nitz}, {Nosek}, {Novotny}, {No{\v{z}}ka}, {N{\'u}{\~n}ez}, {Oikonomou},
  {Olinto}, {Palatka}, {Pallotta}, {Papenbreer}, {Parente}, {Parra}, {Paul},
  {Pech}, {Pedreira}, {P{\c{e}}kala}, {Pe{\~n}a-Rodriguez}, {Pereira},
  {Perlin}, {Perrone}, {Peters}, {Petrera}, {Phuntsok}, {Pierog}, {Pimenta},
  {Pirronello}, {Platino}, {Plum}, {Poh}, {Porowski}, {Prado}, {Privitera},
  {Prouza}, {Quel}, {Querchfeld}, {Quinn}, {Ramos-Pollan}, {Rautenberg},
  {Ravignani}, {Ridky}, {Riehn}, {Risse}, {Ristori}, {Rizi}, {Rodrigues de
  Carvalho}, {Rodriguez Fernandez}, {Rodriguez Rojo}, {Roncoroni}, {Roth},
  {Roulet}, {Rovero}, {Ruehl}, {Saffi}, {Saftoiu}, {Salamida}, {Salazar},
  {Saleh}, {Salina}, {S{\'a}nchez}, {Sanchez-Lucas}, {Santos}, {Santos},
  {Sarazin}, {Sarmento}, {Sarmiento-Cano}, {Sato}, {Schauer}, {Scherini},
  {Schieler}, {Schimp}, {Schmidt}, {Scholten}, {Schov{\'a}nek}, {Schr{\"o}der},
  {Schr{\"o}der}, {Schulz}, {Schumacher}, {Sciutto}, {Segreto}, {Shadkam},
  {Shellard}, {Sigl}, {Silli}, {{\v{S}}m{\'\i}da}, {Snow}, {Sommers},
  {Sonntag}, {Soriano}, {Squartini}, {Stanca}, {Stani{\v{c}}}, {Stasielak},
  {Stassi}, {Stolpovskiy}, {Strafella}, {Streich}, {Suarez},
  {Suarez-Dur{\'a}n}, {Sudholz}, {Suomij{\"a}rvi}, {Supanitsky},
  {{\v{S}}up{\'\i}k}, {Swain}, {Szadkowski}, {Taboada}, {Taborda},
  {Timmermans}, {Todero Peixoto}, {Tomankova}, {Tom{\'e}}, {Torralba Elipe},
  {Travnicek}, {Trini}, {Tueros}, {Ulrich}, {Unger}, {Urban}, {Vald{\'e}s
  Galicia}, {Vali{\~n}o}, {Valore}, {van Aar}, {van Bodegom}, {van den Berg},
  {van Vliet}, {Varela}, {Vargas C{\'a}rdenas}, {V{\'a}zquez}, {Veberi{\v{c}}},
  {Ventura}, {Vergara Quispe}, {Verzi}, {Vicha}, {Villase{\~n}or}, {Vorobiov},
  {Wahlberg}, {Wainberg}, {Walz}, {Watson}, {Weber}, {Weindl}, {Wiede{\'n}ski},
  {Wiencke}, {Wilczy{\'n}ski}, {Wirtz}, {Wittkowski}, {Wundheiler}, {Yang},
  {Yushkov}, {Zas}, {Zavrtanik}, {Zavrtanik}, {Zepeda}, {Zimmermann},
  {Ziolkowski}, {Zong}, {Zuccarello}, {Pierre Auger Collaboration}, {Kim},
  {Schulze}, {Bauer}, {Corral-Santana}, {de Gregorio-Monsalvo},
  {Gonz{\'a}lez-L{\'o}pez}, {Hartmann}, {Ishwara-Chandra}, {Mart{\'\i}n},
  {Mehner}, {Misra}, {Micha{\l}owski}, {Resmi}, {ALMA Collaboration}, {Paragi},
  {Agudo}, {An}, {Beswick}, {Casadio}, {Frey}, {Jonker}, {Kettenis}, {Marcote},
  {Moldon}, {Szomoru}, {van Langevelde}, {Yang}, {Euro VLBI Team}, {Cwiek},
  {Cwiok}, {Czyrkowski}, {Dabrowski}, {Kasprowicz}, {Mankiewicz}, {Nawrocki},
  {Opiela}, {Piotrowski}, {Wrochna}, {Zaremba}, {{\.Z}arnecki}, {Pi of the Sky
  Collaboration}, {Haggard}, {Nynka}, {Ruan}, {Chandra Team at McGill
  University}, {Bland}, {Booler}, {Devillepoix}, {de Gois}, {Hancock}, {Howie},
  {Paxman}, {Sansom}, {Towner}, {Desert Fireball Network}, {Tonry}, {Coughlin},
  {Stubbs}, {Denneau}, {Heinze}, {Stalder}, {Weiland}, {ATLAS}, {Eatough},
  {Kramer}, {Kraus}, {Time Resolution Universe Survey}, {Troja}, {Piro},
  {Becerra Gonz{\'a}lez}, {Butler}, {Fox}, {Khandrika}, {Kutyrev}, {Lee},
  {Ricci}, {Ryan}, {S{\'a}nchez-Ram{\'\i}rez}, {Veilleux}, {Watson},
  {Wieringa}, {Burgess}, {van Eerten}, {Fontes}, {Fryer}, {Korobkin},
  {Wollaeger}, {RIMAS}, {RATIR}, {Camilo}, {Foley}, {Goedhart}, {Makhathini},
  {Oozeer}, {Smirnov}, {Fender}, {Woudt}, \& {South
  Africa/MeerKAT}}]{LVCmultimessenger}
---. 2017{\natexlab{b}}, \apjl, 848, L12

\bibitem[{{Abbott} {et~al.}(2017{\natexlab{c}}){Abbott}, {Abbott}, {Abbott},
  {Acernese}, {Ackley}, {Adams}, {Adams}, {Addesso}, {Adhikari}, {Adya},
  {Affeldt}, {Afrough}, {Agarwal}, {Agathos}, {Agatsuma}, {Aggarwal}, {Aguiar},
  {Aiello}, {Ain}, {Ajith}, {Allen}, {Allen}, {Allocca}, {Aloy}, {Altin},
  {Amato}, {Ananyeva}, {Anderson}, {Anderson}, {Angelova}, {Antier}, {Appert},
  {Arai}, {Araya}, {Areeda}, {Arnaud}, {Arun}, {Ascenzi}, {Ashton}, {Ast},
  {Aston}, {Astone}, {Atallah}, {Aufmuth}, {Aulbert}, {AultONeal}, {Austin},
  {Avila-Alvarez}, {Babak}, {Bacon}, {Bader}, {Bae}, {Baker}, {Baldaccini},
  {Ballardin}, {Ballmer}, {Banagiri}, {Barayoga}, {Barclay}, {Barish},
  {Barker}, {Barkett}, {Barone}, {Barr}, {Barsotti}, {Barsuglia}, {Barta},
  {Bartlett}, {Bartos}, {Bassiri}, {Basti}, {Batch}, {Bawaj}, {Bayley},
  {Bazzan}, {B{\'e}csy}, {Beer}, {Bejger}, {Belahcene}, {Bell}, {Berger},
  {Bergmann}, {Bero}, {Berry}, {Bersanetti}, {Bertolini}, {Betzwieser},
  {Bhagwat}, {Bhandare}, {Bilenko}, {Billingsley}, {Billman}, {Birch},
  {Birney}, {Birnholtz}, {Biscans}, {Biscoveanu}, {Bisht}, {Bitossi}, {Biwer},
  {Bizouard}, {Blackburn}, {Blackman}, {Blair}, {Blair}, {Blair}, {Bloemen},
  {Bock}, {Bode}, {Boer}, {Bogaert}, {Bohe}, {Bondu}, {Bonilla}, {Bonnand},
  {Boom}, {Bork}, {Boschi}, {Bose}, {Bossie}, {Bouffanais}, {Bozzi},
  {Bradaschia}, {Brady}, {Branchesi}, {Brau}, {Briant}, {Brillet}, {Brinkmann},
  {Brisson}, {Brockill}, {Broida}, {Brooks}, {Brown}, {Brown}, {Brunett},
  {Buchanan}, {Buikema}, {Bulik}, {Bulten}, {Buonanno}, {Buskulic}, {Buy},
  {Byer}, {Cabero}, {Cadonati}, {Cagnoli}, {Cahillane}, {Calder{\'o}n
  Bustillo}, {Callister}, {Calloni}, {Camp}, {Canepa}, {Canizares}, {Cannon},
  {Cao}, {Cao}, {Capano}, {Capocasa}, {Carbognani}, {Caride}, {Carney},
  {Casanueva Diaz}, {Casentini}, {Caudill}, {Cavagli{\`a}}, {Cavalier},
  {Cavalieri}, {Cella}, {Cepeda}, {Cerd{\'a}-Dur{\'a}n}, {Cerretani},
  {Cesarini}, {Chamberlin}, {Chan}, {Chao}, {Charlton}, {Chase},
  {Chassande-Mottin}, {Chatterjee}, {Chatziioannou}, {Cheeseboro}, {Chen},
  {Chen}, {Chen}, {Cheng}, {Chia}, {Chincarini}, {Chiummo}, {Chmiel}, {Cho},
  {Cho}, {Chow}, {Christensen}, {Chu}, {Chua}, {Chua}, {Chung}, {Chung},
  {Ciani}, {Ciolfi}, {Cirelli}, {Cirone}, {Clara}, {Clark}, {Clearwater},
  {Cleva}, {Cocchieri}, {Coccia}, {Cohadon}, {Cohen}, {Colla}, {Collette},
  {Cominsky}, {Constancio}, {Conti}, {Cooper}, {Corban}, {Corbitt},
  {Cordero-Carri{\'o}n}, {Corley}, {Cornish}, {Corsi}, {Cortese}, {Costa},
  {Coughlin}, {Coughlin}, {Coulon}, {Countryman}, {Couvares}, {Covas}, {Cowan},
  {Coward}, {Cowart}, {Coyne}, {Coyne}, {Creighton}, {Creighton}, {Cripe},
  {Crowder}, {Cullen}, {Cumming}, {Cunningham}, {Cuoco}, {Dal Canton},
  {D{\'a}lya}, {Danilishin}, {D'Antonio}, {Danzmann}, {Dasgupta}, {Da Silva
  Costa}, {Dattilo}, {Dave}, {Davier}, {Davis}, {Daw}, {Day}, {De}, {DeBra},
  {Degallaix}, {De Laurentis}, {Del{\'e}glise}, {Del Pozzo}, {Demos}, {Denker},
  {Dent}, {De Pietri}, {Dergachev}, {De Rosa}, {DeRosa}, {De Rossi}, {DeSalvo},
  {de Varona}, {Devenson}, {Dhurandhar}, {D{\'\i}az}, {Di Fiore}, {Di
  Giovanni}, {Di Girolamo}, {Di Lieto}, {Di Pace}, {Di Palma}, {Di Renzo},
  {Doctor}, {Dolique}, {Donovan}, {Dooley}, {Doravari}, {Dorrington},
  {Douglas}, {Dovale {\'A}lvarez}, {Downes}, {Drago}, {Dreissigacker},
  {Driggers}, {Du}, {Ducrot}, {Dupej}, {Dwyer}, {Edo}, {Edwards}, {Effler},
  {Eggenstein}, {Ehrens}, {Eichholz}, {Eikenberry}, {Eisenstein}, {Essick},
  {Estevez}, {Etienne}, {Etzel}, {Evans}, {Evans}, {Factourovich}, {Fafone},
  {Fair}, {Fairhurst}, {Fan}, {Farinon}, {Farr}, {Farr}, {Fauchon-Jones},
  {Favata}, {Fays}, {Fee}, {Fehrmann}, {Feicht}, {Fejer}, {Fernandez-Galiana},
  {Ferrante}, {Ferreira}, {Ferrini}, {Fidecaro}, {Finstad}, {Fiori},
  {Fiorucci}, {Fishbach}, {Fisher}, {Fitz-Axen}, {Flaminio}, {Fletcher},
  {Fong}, {Font}, {Forsyth}, {Forsyth}, {Fournier}, {Frasca}, {Frasconi},
  {Frei}, {Freise}, {Frey}, {Frey}, {Fries}, {Fritschel}, {Frolov}, {Fulda},
  {Fyffe}, {Gabbard}, {Gadre}, {Gaebel}, {Gair}, {Gammaitoni}, {Ganija},
  {Gaonkar}, {Garcia-Quiros}, {Garufi}, {Gateley}, {Gaudio}, {Gaur},
  {Gayathri}, {Gehrels}, {Gemme}, {Genin}, {Gennai}, {George}, {George},
  {Gergely}, {Germain}, {Ghonge}, {Ghosh}, {Ghosh}, {Ghosh}, {Giaime},
  {Giardina}, {Giazotto}, {Gill}, {Glover}, {Goetz}, {Goetz}, {Gomes},
  {Goncharov}, {Gonz{\'a}lez}, {Gonzalez Castro}, {Gopakumar}, {Gorodetsky},
  {Gossan}, {Gosselin}, {Gouaty}, {Grado}, {Graef}, {Granata}, {Grant}, {Gras},
  {Gray}, {Greco}, {Green}, {Gretarsson}, {Groot}, {Grote}, {Grunewald},
  {Gruning}, {Guidi}, {Guo}, {Gupta}, {Gupta}, {Gushwa}, {Gustafson},
  {Gustafson}, {Halim}, {Hall}, {Hall}, {Hamilton}, {Hammond}, {Haney},
  {Hanke}, {Hanks}, {Hanna}, {Hannam}, {Hannuksela}, {Hanson}, {Hardwick},
  {Harms}, {Harry}, {Harry}, {Hart}, {Haster}, {Haughian}, {Healy}, {Heidmann},
  {Heintze}, {Heitmann}, {Hello}, {Hemming}, {Hendry}, {Heng}, {Hennig},
  {Heptonstall}, {Heurs}, {Hild}, {Hinderer}, {Hoak}, {Hofman}, {Holt}, {Holz},
  {Hopkins}, {Horst}, {Hough}, {Houston}, {Howell}, {Hreibi}, {Hu}, {Huerta},
  {Huet}, {Hughey}, {Husa}, {Huttner}, {Huynh-Dinh}, {Indik}, {Inta}, {Intini},
  {Isa}, {Isac}, {Isi}, {Iyer}, {Izumi}, {Jacqmin}, {Jani}, {Jaranowski},
  {Jawahar}, {Jim{\'e}nez-Forteza}, {Johnson}, {Johnson-McDaniel}, {Jones},
  {Jones}, {Jonker}, {Ju}, {Junker}, {Kalaghatgi}, {Kalogera}, {Kamai},
  {Kandhasamy}, {Kang}, {Kanner}, {Kapadia}, {Karki}, {Karvinen}, {Kasprzack},
  {Kastaun}, {Katolik}, {Katsavounidis}, {Katzman}, {Kaufer}, {Kawabe},
  {K{\'e}f{\'e}lian}, {Keitel}, {Kemball}, {Kennedy}, {Kent}, {Key}, {Khalili},
  {Khan}, {Khan}, {Khan}, {Khazanov}, {Kijbunchoo}, {Kim}, {Kim}, {Kim}, {Kim},
  {Kim}, {Kim}, {Kimbrell}, {King}, {King}, {Kinley-Hanlon}, {Kirchhoff},
  {Kissel}, {Kleybolte}, {Klimenko}, {Knowles}, {Koch}, {Koehlenbeck}, {Koley},
  {Kondrashov}, {Kontos}, {Korobko}, {Korth}, {Kowalska}, {Kozak},
  {Kr{\"a}mer}, {Kringel}, {Krishnan}, {Kr{\'o}lak}, {Kuehn}, {Kumar}, {Kumar},
  {Kumar}, {Kuo}, {Kutynia}, {Kwang}, {Lackey}, {Lai}, {Landry}, {Lang},
  {Lange}, {Lantz}, {Lanza}, {Lartaux-Vollard}, {Lasky}, {Laxen}, {Lazzarini},
  {Lazzaro}, {Leaci}, {Leavey}, {Lee}, {Lee}, {Lee}, {Lee}, {Lee}, {Lehmann},
  {Lenon}, {Leonardi}, {Leroy}, {Letendre}, {Levin}, {Li}, {Linker},
  {Littenberg}, {Liu}, {Lo}, {Lockerbie}, {London}, {Lord}, {Lorenzini},
  {Loriette}, {Lormand}, {Losurdo}, {Lough}, {Lousto}, {Lovelace}, {L{\"u}ck},
  {Lumaca}, {Lundgren}, {Lynch}, {Ma}, {Macas}, {Macfoy}, {Machenschalk},
  {MacInnis}, {Macleod}, {Maga{\~n}a Hernandez}, {Maga{\~n}a-Sandoval},
  {Maga{\~n}a Zertuche}, {Magee}, {Majorana}, {Maksimovic}, {Man}, {Mandic},
  {Mangano}, {Mansell}, {Manske}, {Mantovani}, {Marchesoni}, {Marion},
  {M{\'a}rka}, {M{\'a}rka}, {Markakis}, {Markosyan}, {Markowitz}, {Maros},
  {Marquina}, {Martelli}, {Martellini}, {Martin}, {Martin}, {Martynov},
  {Mason}, {Massera}, {Masserot}, {Massinger}, {Masso-Reid}, {Mastrogiovanni},
  {Matas}, {Matichard}, {Matone}, {Mavalvala}, {Mazumder}, {McCarthy},
  {McClelland}, {McCormick}, {McCuller}, {McGuire}, {McIntyre}, {McIver},
  {McManus}, {McNeill}, {McRae}, {McWilliams}, {Meacher}, {Meadors}, {Mehmet},
  {Meidam}, {Mejuto-Villa}, {Melatos}, {Mendell}, {Mercer}, {Merilh},
  {Merzougui}, {Meshkov}, {Messenger}, {Messick}, {Metzdorff}, {Meyers},
  {Miao}, {Michel}, {Middleton}, {Mikhailov}, {Milano}, {Miller}, {Miller},
  {Miller}, {Millhouse}, {Milovich-Goff}, {Minazzoli}, {Minenkov}, {Ming},
  {Mishra}, {Mitra}, {Mitrofanov}, {Mitselmakher}, {Mittleman}, {Moffa},
  {Moggi}, {Mogushi}, {Mohan}, {Mohapatra}, {Montani}, {Moore}, {Moraru},
  {Moreno}, {Morriss}, {Mours}, {Mow-Lowry}, {Mueller}, {Muir}, {Mukherjee},
  {Mukherjee}, {Mukherjee}, {Mukund}, {Mullavey}, {Munch}, {Mu{\~n}iz},
  {Muratore}, {Murray}, {Napier}, {Nardecchia}, {Naticchioni}, {Nayak},
  {Neilson}, {Nelemans}, {Nelson}, {Nery}, {Neunzert}, {Nevin}, {Newport},
  {Newton}, {Ng}, {Nguyen}, {Nichols}, {Nielsen}, {Nissanke}, {Nitz}, {Noack},
  {Nocera}, {Nolting}, {North}, {Nuttall}, {Oberling}, {O'Dea}, {Ogin}, {Oh},
  {Oh}, {Ohme}, {Okada}, {Oliver}, {Oppermann}, {Oram}, {O'Reilly}, {Ormiston},
  {Ortega}, {O'Shaughnessy}, {Ossokine}, {Ottaway}, {Overmier}, {Owen}, {Pace},
  {Page}, {Page}, {Pai}, {Pai}, {Palamos}, {Palashov}, {Palomba}, {Pal-Singh},
  {Pan}, {Pan}, {Pang}, {Pang}, {Pankow}, {Pannarale}, {Pant}, {Paoletti},
  {Paoli}, {Papa}, {Parida}, {Parker}, {Pascucci}, {Pasqualetti},
  {Passaquieti}, {Passuello}, {Patil}, {Patricelli}, {Pearlstone}, {Pedraza},
  {Pedurand}, {Pekowsky}, {Pele}, {Penn}, {Perez}, {Perreca}, {Perri},
  {Pfeiffer}, {Phelps}, {Piccinni}, {Pichot}, {Piergiovanni}, {Pierro},
  {Pillant}, {Pinard}, {Pinto}, {Pirello}, {Pitkin}, {Poe}, {Poggiani},
  {Popolizio}, {Porter}, {Post}, {Powell}, {Prasad}, {Pratt}, {Pratten},
  {Predoi}, {Prestegard}, {Prijatelj}, {Principe}, {Privitera}, {Prodi},
  {Prokhorov}, {Puncken}, {Punturo}, {Puppo}, {P{\"u}rrer}, {Qi}, {Quetschke},
  {Quintero}, {Quitzow-James}, {Raab}, {Rabeling}, {Radkins}, {Raffai}, {Raja},
  {Rajan}, {Rajbhandari}, {Rakhmanov}, {Ramirez}, {Ramos-Buades}, {Rapagnani},
  {Raymond}, {Razzano}, {Read}, {Regimbau}, {Rei}, {Reid}, {Reitze}, {Ren},
  {Reyes}, {Ricci}, {Ricker}, {Rieger}, {Riles}, {Rizzo}, {Robertson}, {Robie},
  {Robinet}, {Rocchi}, {Rolland}, {Rollins}, {Roma}, {Romano}, {Romel},
  {Romie}, {Rosi{\'n}ska}, {Ross}, {Rowan}, {R{\"u}diger}, {Ruggi}, {Rutins},
  {Ryan}, {Sachdev}, {Sadecki}, {Sadeghian}, {Sakellariadou}, {Salconi},
  {Saleem}, {Salemi}, {Samajdar}, {Sammut}, {Sampson}, {Sanchez}, {Sanchez},
  {Sanchis-Gual}, {Sandberg}, {Sanders}, {Sassolas}, {Sathyaprakash},
  {Saulson}, {Sauter}, {Savage}, {Sawadsky}, {Schale}, {Scheel}, {Scheuer},
  {Schmidt}, {Schmidt}, {Schnabel}, {Schofield}, {Sch{\"o}nbeck}, {Schreiber},
  {Schuette}, {Schulte}, {Schutz}, {Schwalbe}, {Scott}, {Scott}, {Seidel},
  {Sellers}, {Sengupta}, {Sentenac}, {Sequino}, {Sergeev}, {Shaddock},
  {Shaffer}, {Shah}, {Shahriar}, {Shaner}, {Shao}, {Shapiro}, {Shawhan},
  {Sheperd}, {Shoemaker}, {Shoemaker}, {Siellez}, {Siemens}, {Sieniawska},
  {Sigg}, {Silva}, {Singer}, {Singh}, {Singhal}, {Sintes}, {Slagmolen},
  {Smith}, {Smith}, {Smith}, {Somala}, {Son}, {Sonnenberg}, {Sorazu},
  {Sorrentino}, {Souradeep}, {Spencer}, {Srivastava}, {Staats}, {Staley},
  {Steinke}, {Steinlechner}, {Steinlechner}, {Steinmeyer}, {Stevenson},
  {Stone}, {Stops}, {Strain}, {Stratta}, {Strigin}, {Strunk}, {Sturani},
  {Stuver}, {Summerscales}, {Sun}, {Sunil}, {Suresh}, {Sutton}, {Swinkels},
  {Szczepa{\'n}czyk}, {Tacca}, {Tait}, {Talbot}, {Talukder}, {Tanner},
  {T{\'a}pai}, {Taracchini}, {Tasson}, {Taylor}, {Taylor}, {Tewari}, {Theeg},
  {Thies}, {Thomas}, {Thomas}, {Thomas}, {Thorne}, {Thorne}, {Thrane},
  {Tiwari}, {Tiwari}, {Tokmakov}, {Toland}, {Tonelli}, {Tornasi},
  {Torres-Forn{\'e}}, {Torrie}, {T{\"o}yr{\"a}}, {Travasso}, {Traylor},
  {Trinastic}, {Tringali}, {Trozzo}, {Tsang}, {Tse}, {Tso}, {Tsukada}, {Tsuna},
  {Tuyenbayev}, {Ueno}, {Ugolini}, {Unnikrishnan}, {Urban}, {Usman},
  {Vahlbruch}, {Vajente}, {Valdes}, {van Bakel}, {van Beuzekom}, {van den
  Brand}, {Van Den Broeck}, {Vander-Hyde}, {van der Schaaf}, {van Heijningen},
  {van Veggel}, {Vardaro}, {Varma}, {Vass}, {Vas{\'u}th}, {Vecchio},
  {Vedovato}, {Veitch}, {Veitch}, {Venkateswara}, {Venugopalan}, {Verkindt},
  {Vetrano}, {Vicer{\'e}}, {Viets}, {Vinciguerra}, {Vine}, {Vinet}, {Vitale},
  {Vo}, {Vocca}, {Vorvick}, {Vyatchanin}, {Wade}, {Wade}, {Wade}, {Walet},
  {Walker}, {Wallace}, {Walsh}, {Wang}, {Wang}, {Wang}, {Wang}, {Wang}, {Ward},
  {Warner}, {Was}, {Watchi}, {Weaver}, {Wei}, {Weinert}, {Weinstein}, {Weiss},
  {Wen}, {Wessel}, {We{\ss}els}, {Westerweck}, {Westphal}, {Wette}, {Whelan},
  {Whitcomb}, {Whiting}, {Whittle}, {Wilken}, {Williams}, {Williams},
  {Williamson}, {Willis}, {Willke}, {Wimmer}, {Winkler}, {Wipf}, {Wittel},
  {Woan}, {Woehler}, {Wofford}, {Wong}, {Worden}, {Wright}, {Wu}, {Wysocki},
  {Xiao}, {Yamamoto}, {Yancey}, {Yang}, {Yap}, {Yazback}, {Yu}, {Yu}, {Yvert},
  {Zadro{\.z}ny}, {Zanolin}, {Zelenova}, {Zendri}, {Zevin}, {Zhang}, {Zhang},
  {Zhang}, {Zhang}, {Zhao}, {Zhou}, {Zhou}, {Zhu}, {Zhu}, {Zimmerman},
  {Zucker}, {Zweizig}, {(LIGO Scientific Collaboration}, {Virgo Collaboration},
  {Burns}, {Veres}, {Kocevski}, {Racusin}, {Goldstein}, {Connaughton},
  {Briggs}, {Blackburn}, {Hamburg}, {Hui}, {von Kienlin}, {McEnery}, {Preece},
  {Wilson-Hodge}, {Bissaldi}, {Cleveland}, {Gibby}, {Giles}, {Kippen},
  {McBreen}, {Meegan}, {Paciesas}, {Poolakkil}, {Roberts}, {Stanbro},
  {Gamma-ray Burst Monitor}, {Savchenko}, {Ferrigno}, {Kuulkers}, {Bazzano},
  {Bozzo}, {Brandt}, {Chenevez}, {Courvoisier}, {Diehl}, {Domingo}, {Hanlon},
  {Jourdain}, {Laurent}, {Lebrun}, {Lutovinov}, {Mereghetti}, {Natalucci},
  {Rodi}, {Roques}, {Sunyaev}, {Ubertini}, \& {(INTEGRAL}}]{GW170817grb}
---. 2017{\natexlab{c}}, \apjl, 848, L13

\bibitem[{{Abbott} {et~al.}(2017{\natexlab{d}}){Abbott}, {Abbott}, {Abbott},
  {Acernese}, {Ackley}, {Adams}, {Adams}, {Addesso}, {Adhikari}, {Adya},
  {Affeldt}, {Afrough}, {Agarwal}, {Agathos}, {Agatsuma}, {Aggarwal}, {Aguiar},
  {Aiello}, {Ain}, {Ajith}, {Allen}, {Allen}, {Allocca}, {Altin}, {Amato},
  {Ananyeva}, {Anderson}, {Anderson}, {Angelova}, {Antier}, {Appert}, {Arai},
  {Araya}, {Areeda}, {Arnaud}, {Arun}, {Ascenzi}, {Ashton}, {Ast}, {Aston},
  {Astone}, {Atallah}, {Aufmuth}, {Aulbert}, {Aultoneal}, {Austin},
  {Avila-Alvarez}, {Babak}, {Bacon}, {Bader}, {Bae}, {Baker}, {Baldaccini},
  {Ballardin}, {Ballmer}, {Banagiri}, {Barayoga}, {Barclay}, {Barish},
  {Barker}, {Barkett}, {Barone}, {Barr}, {Barsotti}, {Barsuglia}, {Barta},
  {Bartlett}, {Bartos}, {Bassiri}, {Basti}, {Batch}, {Bawaj}, {Bayley},
  {Bazzan}, {B{\'e}csy}, {Beer}, {Bejger}, {Belahcene}, {Bell}, {Berger},
  {Bergmann}, {Bero}, {Berry}, {Bersanetti}, {Bertolini}, {Betzwieser},
  {Bhagwat}, {Bhandare}, {Bilenko}, {Billingsley}, {Billman}, {Birch},
  {Birney}, {Birnholtz}, {Biscans}, {Biscoveanu}, {Bisht}, {Bitossi}, {Biwer},
  {Bizouard}, {Blackburn}, {Blackman}, {Blair}, {Blair}, {Blair}, {Bloemen},
  {Bock}, {Bode}, {Boer}, {Bogaert}, {Bohe}, {Bondu}, {Bonilla}, {Bonnand},
  {Boom}, {Bork}, {Boschi}, {Bose}, {Bossie}, {Bouffanais}, {Bozzi},
  {Bradaschia}, {Brady}, {Branchesi}, {Brau}, {Briant}, {Brillet}, {Brinkmann},
  {Brisson}, {Brockill}, {Broida}, {Brooks}, {Brown}, {Brown}, {Brunett},
  {Buchanan}, {Buikema}, {Bulik}, {Bulten}, {Buonanno}, {Buskulic}, {Buy},
  {Byer}, {Cabero}, {Cadonati}, {Cagnoli}, {Cahillane}, {Bustillo},
  {Callister}, {Calloni}, {Camp}, {Canepa}, {Canizares}, {Cannon}, {Cao},
  {Cao}, {Capano}, {Capocasa}, {Carbognani}, {Caride}, {Carney}, {Diaz},
  {Casentini}, {Caudill}, {Cavagli{\`a}}, {Cavalier}, {Cavalieri}, {Cella},
  {Cepeda}, {Cerd{\'a}-Dur{\'a}n}, {Cerretani}, {Cesarini}, {Chamberlin},
  {Chan}, {Chao}, {Charlton}, {Chase}, {Chassande-Mottin}, {Chatterjee},
  {Chatziioannou}, {Cheeseboro}, {Chen}, {Chen}, {Chen}, {Cheng}, {Chia},
  {Chincarini}, {Chiummo}, {Chmiel}, {Cho}, {Cho}, {Chow}, {Christensen},
  {Chu}, {Chua}, {Chua}, {Chung}, {Chung}, {Ciani}, {Ciolfi}, {Cirelli},
  {Cirone}, {Clara}, {Clark}, {Clearwater}, {Cleva}, {Cocchieri}, {Coccia},
  {Cohadon}, {Cohen}, {Colla}, {Collette}, {Cominsky}, {Constancio}, {Conti},
  {Cooper}, {Corban}, {Corbitt}, {Cordero-Carri{\'o}n}, {Corley}, {Cornish},
  {Corsi}, {Cortese}, {Costa}, {Coughlin}, {Coughlin}, {Coulon}, {Countryman},
  {Couvares}, {Covas}, {Cowan}, {Coward}, {Cowart}, {Coyne}, {Coyne},
  {Creighton}, {Creighton}, {Cripe}, {Crowder}, {Cullen}, {Cumming},
  {Cunningham}, {Cuoco}, {Dal Canton}, {D{\'a}lya}, {Danilishin}, {D'Antonio},
  {Danzmann}, {Dasgupta}, {da Silva Costa}, {Datrier}, {Dattilo}, {Dave},
  {Davier}, {Davis}, {Daw}, {Day}, {de}, {Debra}, {Degallaix}, {de Laurentis},
  {Del{\'e}glise}, {Del Pozzo}, {Demos}, {Denker}, {Dent}, {de Pietri},
  {Dergachev}, {De Rosa}, {Derosa}, {de Rossi}, {Desalvo}, {de Varona},
  {Devenson}, {Dhurandhar}, {D{\'\i}az}, {di Fiore}, {di Giovanni}, {di
  Girolamo}, {di Lieto}, {di Pace}, {di Palma}, {di Renzo}, {Doctor},
  {Dolique}, {Donovan}, {Dooley}, {Doravari}, {Dorrington}, {Douglas}, {Dovale
  {\'A}lvarez}, {Downes}, {Drago}, {Dreissigacker}, {Driggers}, {Du}, {Ducrot},
  {Dupej}, {Dwyer}, {Edo}, {Edwards}, {Effler}, {Eggenstein}, {Ehrens},
  {Eichholz}, {Eikenberry}, {Eisenstein}, {Essick}, {Estevez}, {Etienne},
  {Etzel}, {Evans}, {Evans}, {Factourovich}, {Fafone}, {Fair}, {Fairhurst},
  {Fan}, {Farinon}, {Farr}, {Farr}, {Fauchon-Jones}, {Favata}, {Fays}, {Fee},
  {Fehrmann}, {Feicht}, {Fejer}, {Fernandez-Galiana}, {Ferrante}, {Ferreira},
  {Ferrini}, {Fidecaro}, {Finstad}, {Fiori}, {Fiorucci}, {Fishbach}, {Fisher},
  {Fitz-Axen}, {Flaminio}, {Fletcher}, {Fong}, {Font}, {Forsyth}, {Forsyth},
  {Fournier}, {Frasca}, {Frasconi}, {Frei}, {Freise}, {Frey}, {Frey}, {Fries},
  {Fritschel}, {Frolov}, {Fulda}, {Fyffe}, {Gabbard}, {Gadre}, {Gaebel},
  {Gair}, {Gammaitoni}, {Ganija}, {Gaonkar}, {Garcia-Quiros}, {Garufi},
  {Gateley}, {Gaudio}, {Gaur}, {Gayathri}, {Gehrels}, {Gemme}, {Genin},
  {Gennai}, {George}, {George}, {Gergely}, {Germain}, {Ghonge}, {Ghosh},
  {Ghosh}, {Ghosh}, {Giaime}, {Giardina}, {Giazotto}, {Gill}, {Glover},
  {Goetz}, {Goetz}, {Gomes}, {Goncharov}, {Gonz{\'a}lez}, {Castro},
  {Gopakumar}, {Gorodetsky}, {Gossan}, {Gosselin}, {Gouaty}, {Grado}, {Graef},
  {Granata}, {Grant}, {Gras}, {Gray}, {Greco}, {Green}, {Gretarsson}, {Groot},
  {Grote}, {Grunewald}, {Gruning}, {Guidi}, {Guo}, {Gupta}, {Gupta}, {Gushwa},
  {Gustafson}, {Gustafson}, {Halim}, {Hall}, {Hall}, {Hamilton}, {Hammond},
  {Haney}, {Hanke}, {Hanks}, {Hanna}, {Hannam}, {Hannuksela}, {Hanson},
  {Hardwick}, {Harms}, {Harry}, {Harry}, {Hart}, {Haster}, {Haughian}, {Healy},
  {Heidmann}, {Heintze}, {Heitmann}, {Hello}, {Hemming}, {Hendry}, {Heng},
  {Hennig}, {Heptonstall}, {Heurs}, {Hild}, {Hinderer}, {Hoak}, {Hofman},
  {Holt}, {Holz}, {Hopkins}, {Horst}, {Hough}, {Houston}, {Howell}, {Hreibi},
  {Hu}, {Huerta}, {Huet}, {Hughey}, {Husa}, {Huttner}, {Huynh-Dinh}, {Indik},
  {Inta}, {Intini}, {Isa}, {Isac}, {Isi}, {Iyer}, {Izumi}, {Jacqmin}, {Jani},
  {Jaranowski}, {Jawahar}, {Jim{\'e}nez-Forteza}, {Johnson}, {Jones}, {Jones},
  {Jonker}, {Ju}, {Junker}, {Kalaghatgi}, {Kalogera}, {Kamai}, {Kandhasamy},
  {Kang}, {Kanner}, {Kapadia}, {Karki}, {Karvinen}, {Kasprzack}, {Katolik},
  {Katsavounidis}, {Katzman}, {Kaufer}, {Kawabe}, {K{\'e}f{\'e}lian}, {Keitel},
  {Kemball}, {Kennedy}, {Kent}, {Key}, {Khalili}, {Khan}, {Khan}, {Khan},
  {Khazanov}, {Kijbunchoo}, {Kim}, {Kim}, {Kim}, {Kim}, {Kim}, {Kim},
  {Kimbrell}, {King}, {King}, {Kinley-Hanlon}, {Kirchhoff}, {Kissel},
  {Kleybolte}, {Klimenko}, {Knowles}, {Koch}, {Koehlenbeck}, {Koley},
  {Kondrashov}, {Kontos}, {Korobko}, {Korth}, {Kowalska}, {Kozak},
  {Kr{\"a}mer}, {Kringel}, {Krishnan}, {Kr{\'o}lak}, {Kuehn}, {Kumar}, {Kumar},
  {Kumar}, {Kuo}, {Kutynia}, {Kwang}, {Lackey}, {Lai}, {Landry}, {Lang},
  {Lange}, {Lantz}, {Lanza}, {Lartaux-Vollard}, {Lasky}, {Laxen}, {Lazzarini},
  {Lazzaro}, {Leaci}, {Leavey}, {Lee}, {Lee}, {Lee}, {Lee}, {Lee}, {Lehmann},
  {Lenon}, {Leonardi}, {Leroy}, {Letendre}, {Levin}, {Li}, {Linker},
  {Littenberg}, {Liu}, {Liu}, {Lo}, {Lockerbie}, {London}, {Lord}, {Lorenzini},
  {Loriette}, {Lormand}, {Losurdo}, {Lough}, {Lousto}, {Lovelace}, {L{\"u}ck},
  {Lumaca}, {Lundgren}, {Lynch}, {Ma}, {Macas}, {Macfoy}, {Machenschalk},
  {Macinnis}, {MacLeod}, {Hernandez}, {Maga{\~n}a-Sandoval}, {Zertuche},
  {Magee}, {Majorana}, {Maksimovic}, {Man}, {Mandic}, {Mangano}, {Mansell},
  {Manske}, {Mantovani}, {Marchesoni}, {Marion}, {M{\'a}rka}, {M{\'a}rka},
  {Markakis}, {Markosyan}, {Markowitz}, {Maros}, {Marquina}, {Martelli},
  {Martellini}, {Martin}, {Martin}, {Martynov}, {Mason}, {Massera}, {Masserot},
  {Massinger}, {Masso-Reid}, {Mastrogiovanni}, {Matas}, {Matichard}, {Matone},
  {Mavalvala}, {Mazumder}, {McCarthy}, {McClelland}, {McCormick}, {McCuller},
  {McGuire}, {McIntyre}, {McIver}, {McManus}, {McNeill}, {McRae}, {McWilliams},
  {Meacher}, {Meadors}, {Mehmet}, {Meidam}, {Mejuto-Villa}, {Melatos},
  {Mendell}, {Mercer}, {Merilh}, {Merzougui}, {Meshkov}, {Messenger},
  {Messick}, {Metzdorff}, {Meyers}, {Miao}, {Michel}, {Middleton}, {Mikhailov},
  {Milano}, {Miller}, {Miller}, {Miller}, {Millhouse}, {Milovich-Goff},
  {Minazzoli}, {Minenkov}, {Ming}, {Mishra}, {Mitra}, {Mitrofanov},
  {Mitselmakher}, {Mittleman}, {Moffa}, {Moggi}, {Mogushi}, {Mohan},
  {Mohapatra}, {Montani}, {Moore}, {Moraru}, {Moreno}, {Morriss}, {Mours},
  {Mow-Lowry}, {Mueller}, {Muir}, {Mukherjee}, {Mukherjee}, {Mukherjee},
  {Mukund}, {Mullavey}, {Munch}, {Mu{\~n}iz}, {Muratore}, {Murray}, {Napier},
  {Nardecchia}, {Naticchioni}, {Nayak}, {Neilson}, {Nelemans}, {Nelson},
  {Nery}, {Neunzert}, {Nevin}, {Newport}, {Newton}, {Ng}, {Nguyen}, {Nichols},
  {Nielsen}, {Nissanke}, {Nitz}, {Noack}, {Nocera}, {Nolting}, {North},
  {Nuttall}, {Oberling}, {O'Dea}, {Ogin}, {Oh}, {Oh}, {Ohme}, {Okada},
  {Oliver}, {Oppermann}, {Oram}, {O'Reilly}, {Ormiston}, {Ortega},
  {O'Shaughnessy}, {Ossokine}, {Ottaway}, {Overmier}, {Owen}, {Pace}, {Page},
  {Page}, {Pai}, {Pai}, {Palamos}, {Palashov}, {Palomba}, {Pal-Singh}, {Pan},
  {Pan}, {Pang}, {Pang}, {Pankow}, {Pannarale}, {Pant}, {Paoletti}, {Paoli},
  {Papa}, {Parida}, {Parker}, {Pascucci}, {Pasqualetti}, {Passaquieti},
  {Passuello}, {Patil}, {Patricelli}, {Pearlstone}, {Pedraza}, {Pedurand},
  {Pekowsky}, {Pele}, {Penn}, {Perez}, {Perreca}, {Perri}, {Pfeiffer},
  {Phelps}, {Piccinni}, {Pichot}, {Piergiovanni}, {Pierro}, {Pillant},
  {Pinard}, {Pinto}, {Pirello}, {Pitkin}, {Poe}, {Poggiani}, {Popolizio},
  {Porter}, {Post}, {Powell}, {Prasad}, {Pratt}, {Pratten}, {Predoi},
  {Prestegard}, {Prijatelj}, {Principe}, {Privitera}, {Prodi}, {Prokhorov},
  {Puncken}, {Punturo}, {Puppo}, {P{\"u}rrer}, {Qi}, {Quetschke}, {Quintero},
  {Quitzow-James}, {Raab}, {Rabeling}, {Radkins}, {Raffai}, {Raja}, {Rajan},
  {Rajbhandari}, {Rakhmanov}, {Ramirez}, {Ramos-Buades}, {Rapagnani},
  {Raymond}, {Razzano}, {Read}, {Regimbau}, {Rei}, {Reid}, {Reitze}, {Ren},
  {Reyes}, {Ricci}, {Ricker}, {Rieger}, {Riles}, {Rizzo}, {Robertson}, {Robie},
  {Robinet}, {Rocchi}, {Rolland}, {Rollins}, {Roma}, {Romano}, {Romano},
  {Romel}, {Romie}, {Rosi{\'n}ska}, {Ross}, {Rowan}, {R{\"u}diger}, {Ruggi},
  {Rutins}, {Ryan}, {Sachdev}, {Sadecki}, {Sadeghian}, {Sakellariadou},
  {Salconi}, {Saleem}, {Salemi}, {Samajdar}, {Sammut}, {Sampson}, {Sanchez},
  {Sanchez}, {Sanchis-Gual}, {Sandberg}, {Sanders}, {Sassolas},
  {Sathyaprakash}, {Saulson}, {Sauter}, {Savage}, {Sawadsky}, {Schale},
  {Scheel}, {Scheuer}, {Schmidt}, {Schmidt}, {Schnabel}, {Schofield},
  {Sch{\"o}nbeck}, {Schreiber}, {Schuette}, {Schulte}, {Schutz}, {Schwalbe},
  {Scott}, {Scott}, {Seidel}, {Sellers}, {Sengupta}, {Sentenac}, {Sequino},
  {Sergeev}, {Shaddock}, {Shaffer}, {Shah}, {Shahriar}, {Shaner}, {Shao},
  {Shapiro}, {Shawhan}, {Sheperd}, {Shoemaker}, {Shoemaker}, {Siellez},
  {Siemens}, {Sieniawska}, {Sigg}, {Silva}, {Singer}, {Singh}, {Singhal},
  {Sintes}, {Slagmolen}, {Smith}, {Smith}, {Smith}, {Somala}, {Son},
  {Sonnenberg}, {Sorazu}, {Sorrentino}, {Souradeep}, {Spencer}, {Srivastava},
  {Staats}, {Staley}, {Steer}, {Steinke}, {Steinlechner}, {Steinlechner},
  {Steinmeyer}, {Stevenson}, {Stone}, {Stops}, {Strain}, {Stratta}, {Strigin},
  {Strunk}, {Sturani}, {Stuver}, {Summerscales}, {Sun}, {Sunil}, {Suresh},
  {Sutton}, {Swinkels}, {Szczepa{\'n}czyk}, {Tacca}, {Tait}, {Talbot},
  {Talukder}, {Tanner}, {T{\'a}pai}, {Taracchini}, {Tasson}, {Taylor},
  {Taylor}, {Tewari}, {Theeg}, {Thies}, {Thomas}, {Thomas}, {Thomas}, {Thorne},
  {Thrane}, {Tiwari}, {Tiwari}, {Tokmakov}, {Toland}, {Tonelli}, {Tornasi},
  {Torres-Forn{\'e}}, {Torrie}, {T{\"o}yr{\"a}}, {Travasso}, {Traylor},
  {Trinastic}, {Tringali}, {Trozzo}, {Tsang}, {Tse}, {Tso}, {Tsukada}, {Tsuna},
  {Tuyenbayev}, {Ueno}, {Ugolini}, {Unnikrishnan}, {Urban}, {Usman},
  {Vahlbruch}, {Vajente}, {Valdes}, {van Bakel}, {van Beuzekom}, {van den
  Brand}, {van den Broeck}, {Vander-Hyde}, {van der Schaaf}, {van Heijningen},
  {van Veggel}, {Vardaro}, {Varma}, {Vass}, {Vas{\'u}th}, {Vecchio},
  {Vedovato}, {Veitch}, {Veitch}, {Venkateswara}, {Venugopalan}, {Verkindt},
  {Vetrano}, {Vicer{\'e}}, {Viets}, {Vinciguerra}, {Vine}, {Vinet}, {Vitale},
  {Vo}, {Vocca}, {Vorvick}, {Vyatchanin}, {Wade}, {Wade}, {Wade}, {Walet},
  {Walker}, {Wallace}, {Walsh}, {Wang}, {Wang}, {Wang}, {Wang}, {Wang}, {Ward},
  {Warner}, {Was}, {Watchi}, {Weaver}, {Wei}, {Weinert}, {Weinstein}, {Weiss},
  {Wen}, {Wessel}, {We{\ss}els}, {Westerweck}, {Westphal}, {Wette}, {Whelan},
  {Whitcomb}, {Whiting}, {Whittle}, {Wilken}, {Williams}, {Williams},
  {Williamson}, {Willis}, {Willke}, {Wimmer}, {Winkler}, {Wipf}, {Wittel},
  {Woan}, {Woehler}, {Wofford}, {Wong}, {Worden}, {Wright}, {Wu}, {Wysocki},
  {Xiao}, {Yamamoto}, {Yancey}, {Yang}, {Yap}, {Yazback}, {Yu}, {Yu}, {Yvert},
  {Zadro{\.z}ny}, {Zanolin}, {Zelenova}, {Zendri}, {Zevin}, {Zhang}, {Zhang},
  {Zhang}, {Zhang}, {Zhao}, {Zhou}, {Zhou}, {Zhu}, {Zhu}, {Zimmerman},
  {Zucker}, {Zweizig}, {Foley}, {Coulter}, {Drout}, {Kasen}, {Kilpatrick},
  {Madore}, {Murguia-Berthier}, {Pan}, {Piro}, {Prochaska}, {Ramirez-Ruiz},
  {Rest}, {Rojas-Bravo}, {Shappee}, {Siebert}, {Simon}, {Ulloa}, {Annis},
  {Soares-Santos}, {Brout}, {Scolnic}, {Diehl}, {Frieman}, {Berger},
  {Alexander}, {Allam}, {Balbinot}, {Blanchard}, {Butler}, {Chornock}, {Cook},
  {Cowperthwaite}, {Drlica-Wagner}, {Drout}, {Durret}, {Eftekhari}, {Finley},
  {Fong}, {Fryer}, {Garc{\'\i}a-Bellido}, {Gill}, {Gruendl}, {Hanna},
  {Hartley}, {Herner}, {Huterer}, {Kasen}, {Kessler}, {Li}, {Lin}, {Lopes},
  {Louren{\c{c}}o}, {Margutti}, {Marriner}, {Marshall}, {Matheson}, {Medina},
  {Metzger}, {Mu{\~n}oz}, {Muir}, {Nicholl}, {Nugent}, {Palmese},
  {Paz-Chinch{\'o}n}, {Quataert}, {Sako}, {Sauseda}, {Schlegel}, {Secco},
  {Smith}, {Sobreira}, {Stebbins}, {Villar}, {Vivas}, {Wester}, {Williams},
  {Yanny}, {Zenteno}, {Abbott}, {Abdalla}, {Bechtol}, {Benoit-L{\'e}vy},
  {Bertin}, {Bridle}, {Brooks}, {Buckley-Geer}, {Burke}, {Rosell}, {Kind},
  {Carretero}, {Castander}, {Cunha}, {D'Andrea}, {da Costa}, {Davis}, {Depoy},
  {Desai}, {Dietrich}, {Estrada}, {Fernandez}, {Flaugher}, {Fosalba},
  {Gaztanaga}, {Gerdes}, {Giannantonio}, {Goldstein}, {Gruen}, {Gutierrez},
  {Hartley}, {Honscheid}, {Jain}, {James}, {Jeltema}, {Johnson}, {Kent},
  {Krause}, {Kron}, {Kuehn}, {Kuhlmann}, {Kuropatkin}, {Lahav}, {Lima}, {Maia},
  {March}, {Miller}, {Miquel}, {Neilsen}, {Nord}, {Ogando}, {Plazas}, {Romer},
  {Roodman}, {Rykoff}, {Sanchez}, {Scarpine}, {Schubnell}, {Sevilla-Noarbe},
  {Smith}, {Smith}, {Suchyta}, {Tarle}, {Thomas}, {Thomas}, {Troxel}, {Tucker},
  {Vikram}, {Walker}, {Weller}, {Zhang}, {Haislip}, {Kouprianov}, {Reichart},
  {Tartaglia}, {Sand}, {Valenti}, {Yang}, {Arcavi}, {Hosseinzadeh}, {Howell},
  {McCully}, {Poznanski}, {Vasylyev}, {Tanvir}, {Levan}, {Hjorth}, {Cano},
  {Copperwheat}, {de Ugarte-Postigo}, {Evans}, {Fynbo},
  {Gonz{\'a}lez-Fern{\'a}ndez}, {Greiner}, {Irwin}, {Lyman}, {Mandel},
  {McMahon}, {Milvang-Jensen}, {O'Brien}, {Osborne}, {Perley}, {Pian},
  {Palazzi}, {Rol}, {Rosetti}, {Rosswog}, {Rowlinson}, {Schulze}, {Steeghs},
  {Th{\"o}ne}, {Ulaczyk}, {Watson}, {Wiersema}, {Lipunov}, {Gorbovskoy},
  {Kornilov}, {Tyurina}, {Balanutsa}, {Vlasenko}, {Gorbunov}, {Podesta},
  {Levato}, {Saffe}, {Buckley}, {Budnev}, {Gress}, {Yurkov}, {Rebolo}, \&
  {Serra-Ricart}}]{GW170817H0}
---. 2017{\natexlab{d}}, \nat, 551, 85

\bibitem[{{Abbott} {et~al.}(2018){Abbott}, {Abbott}, {Abbott}, {Acernese},
  {Ackley}, {Adams}, {Adams}, {Addesso}, {Adhikari}, {Adya}, {Affeldt},
  {Agarwal}, {Agathos}, {Agatsuma}, {Aggarwal}, {Aguiar}, {Aiello}, {Ain},
  {Ajith}, {Allen}, {Allen}, {Allocca}, {Aloy}, {Altin}, {Amato}, {Ananyeva},
  {Anderson}, {Anderson}, {Angelova}, {Antier}, {Appert}, {Arai}, {Araya},
  {Areeda}, {Ar{\`e}ne}, {Arnaud}, {Arun}, {Ascenzi}, {Ashton}, {Ast}, {Aston},
  {Astone}, {Atallah}, {Aubin}, {Aufmuth}, {Aulbert}, {AultONeal}, {Austin},
  {Avila-Alvarez}, {Babak}, {Bacon}, {Badaracco}, {Bader}, {Bae}, {Baker},
  {Baldaccini}, {Ballardin}, {Ballmer}, {Banagiri}, {Barayoga}, {Barclay},
  {Barish}, {Barker}, {Barkett}, {Barnum}, {Barone}, {Barr}, {Barsotti},
  {Barsuglia}, {Barta}, {Bartlett}, {Bartos}, {Bassiri}, {Basti}, {Batch},
  {Bawaj}, {Bayley}, {Bazzan}, {B{\'e}csy}, {Beer}, {Bejger}, {Belahcene},
  {Bell}, {Beniwal}, {Bensch}, {Berger}, {Bergmann}, {Bernuzzi}, {Bero},
  {Berry}, {Bersanetti}, {Bertolini}, {Betzwieser}, {Bhandare}, {Bilenko},
  {Bilgili}, {Billingsley}, {Billman}, {Birch}, {Birney}, {Birnholtz},
  {Biscans}, {Biscoveanu}, {Bisht}, {Bitossi}, {Bizouard}, {Blackburn},
  {Blackman}, {Blair}, {Blair}, {Blair}, {Bloemen}, {Bock}, {Bode}, {Boer},
  {Boetzel}, {Bogaert}, {Bohe}, {Bondu}, {Bonilla}, {Bonnand}, {Booker},
  {Boom}, {Booth}, {Bork}, {Boschi}, {Bose}, {Bossie}, {Bossilkov}, {Bosveld},
  {Bouffanais}, {Bozzi}, {Bradaschia}, {Brady}, {Bramley}, {Branchesi}, {Brau},
  {Briant}, {Brighenti}, {Brillet}, {Brinkmann}, {Brisson}, {Brockill},
  {Brooks}, {Brown}, {Brunett}, {Buchanan}, {Buikema}, {Bulik}, {Bulten},
  {Buonanno}, {Buskulic}, {Buy}, {Byer}, {Cabero}, {Cadonati}, {Cagnoli},
  {Cahillane}, {Calder{\'o}n Bustillo}, {Callister}, {Calloni}, {Camp},
  {Canepa}, {Canizares}, {Cannon}, {Cao}, {Cao}, {Capano}, {Capocasa},
  {Carbognani}, {Caride}, {Carney}, {Carullo}, {Casanueva Diaz}, {Casentini},
  {Caudill}, {Cavagli{\`a}}, {Cavalier}, {Cavalieri}, {Cella}, {Cepeda},
  {Cerd{\'a}-Dur{\'a}n}, {Cerretani}, {Cesarini}, {Chaibi}, {Chamberlin},
  {Chan}, {Chao}, {Charlton}, {Chase}, {Chassande-Mottin}, {Chatterjee},
  {Chatziioannou}, {Cheeseboro}, {Chen}, {Chen}, {Chen}, {Cheng}, {Chia},
  {Chincarini}, {Chiummo}, {Chmiel}, {Cho}, {Cho}, {Chow}, {Christensen},
  {Chu}, {Chua}, {Chua}, {Chung}, {Chung}, {Ciani}, {Ciobanu}, {Ciolfi},
  {Cipriano}, {Cirelli}, {Cirone}, {Clara}, {Clark}, {Clearwater}, {Cleva},
  {Cocchieri}, {Coccia}, {Cohadon}, {Cohen}, {Colla}, {Collette}, {Collins},
  {Cominsky}, {Constancio}, {Conti}, {Cooper}, {Corban}, {Corbitt},
  {Cordero-Carri{\'o}n}, {Corley}, {Cornish}, {Corsi}, {Cortese}, {Costa},
  {Cotesta}, {Coughlin}, {Coughlin}, {Coulon}, {Countryman}, {Couvares},
  {Covas}, {Cowan}, {Coward}, {Cowart}, {Coyne}, {Coyne}, {Creighton},
  {Creighton}, {Cripe}, {Crowder}, {Cullen}, {Cumming}, {Cunningham}, {Cuoco},
  {Canton}, {D{\'a}lya}, {Danilishin}, {D'Antonio}, {Danzmann}, {Dasgupta}, {Da
  Silva Costa}, {Dattilo}, {Dave}, {Davier}, {Davis}, {Daw}, {Day}, {DeBra},
  {Deenadayalan}, {Degallaix}, {De Laurentis}, {Del{\'e}glise}, {Del Pozzo},
  {Demos}, {Denker}, {Dent}, {De Pietri}, {Derby}, {Dergachev}, {De Rosa}, {De
  Rossi}, {DeSalvo}, {de Varona}, {Dhurandhar}, {D{\'\i}az}, {Dietrich}, {Di
  Fiore}, {Di Giovanni}, {Di Girolamo}, {Di Lieto}, {Ding}, {Di Pace}, {Di
  Palma}, {Di Renzo}, {Dmitriev}, {Doctor}, {Dolique}, {Donovan}, {Dooley},
  {Doravari}, {Dorrington}, {Dovale {\'A}lvarez}, {Downes}, {Drago},
  {Dreissigacker}, {Driggers}, {Du}, {Dupej}, {Dwyer}, {Easter}, {Edo},
  {Edwards}, {Effler}, {Eggenstein}, {Ehrens}, {Eichholz}, {Eikenberry},
  {Eisenmann}, {Eisenstein}, {Essick}, {Estelles}, {Estevez}, {Etienne},
  {Etzel}, {Evans}, {Evans}, {Fafone}, {Fair}, {Fairhurst}, {Fan}, {Farinon},
  {Farr}, {Farr}, {Fauchon-Jones}, {Favata}, {Fays}, {Fee}, {Fehrmann},
  {Feicht}, {Fejer}, {Feng}, {Fernandez-Galiana}, {Ferrante}, {Ferreira},
  {Ferrini}, {Fidecaro}, {Fiori}, {Fiorucci}, {Fishbach}, {Fisher}, {Fishner},
  {Fitz-Axen}, {Flaminio}, {Fletcher}, {Fong}, {Font}, {Forsyth}, {Forsyth},
  {Fournier}, {Frasca}, {Frasconi}, {Frei}, {Freise}, {Frey}, {Frey},
  {Fritschel}, {Frolov}, {Fulda}, {Fyffe}, {Gabbard}, {Gadre}, {Gaebel},
  {Gair}, {Gammaitoni}, {Ganija}, {Gaonkar}, {Garcia},
  {Garc{\'\i}a-Quir{\'o}s}, {Garufi}, {Gateley}, {Gaudio}, {Gaur}, {Gayathri},
  {Gemme}, {Genin}, {Gennai}, {George}, {George}, {Gergely}, {Germain},
  {Ghonge}, {Ghosh}, {Ghosh}, {Ghosh}, {Giacomazzo}, {Giaime}, {Giardina},
  {Giazotto}, {Gill}, {Giordano}, {Glover}, {Goetz}, {Goetz}, {Goncharov},
  {Gonz{\'a}lez}, {Gonzalez Castro}, {Gopakumar}, {Gorodetsky}, {Gossan},
  {Gosselin}, {Gouaty}, {Grado}, {Graef}, {Granata}, {Grant}, {Gras}, {Gray},
  {Greco}, {Green}, {Green}, {Gretarsson}, {Groot}, {Grote}, {Grunewald},
  {Gruning}, {Guidi}, {Gulati}, {Guo}, {Gupta}, {Gupta}, {Gushwa}, {Gustafson},
  {Gustafson}, {Halim}, {Hall}, {Hall}, {Hamilton}, {Hamilton}, {Hammond},
  {Haney}, {Hanke}, {Hanks}, {Hanna}, {Hannam}, {Hannuksela}, {Hanson},
  {Hardwick}, {Harms}, {Harry}, {Harry}, {Hart}, {Haster}, {Haughian}, {Healy},
  {Heidmann}, {Heintze}, {Heitmann}, {Hello}, {Hemming}, {Hendry}, {Heng},
  {Hennig}, {Heptonstall}, {Hernandez}, {Heurs}, {Hild}, {Hinderer}, {Ho},
  {Hoak}, {Hochheim}, {Hofman}, {Holland}, {Holt}, {Holz}, {Hopkins}, {Horst},
  {Hough}, {Houston}, {Howell}, {Hreibi}, {Huerta}, {Huet}, {Hughey}, {Hulko},
  {Husa}, {Huttner}, {Huynh-Dinh}, {Iess}, {Indik}, {Ingram}, {Inta}, {Intini},
  {Irwin}, {Isa}, {Isac}, {Isi}, {Iyer}, {Izumi}, {Jacqmin}, {Jani},
  {Jaranowski}, {Johnson}, {Johnson}, {Jones}, {Jones}, {Jonker}, {Ju},
  {Junker}, {Kalaghatgi}, {Kalogera}, {Kamai}, {Kandhasamy}, {Kang}, {Kanner},
  {Kapadia}, {Karki}, {Karvinen}, {Kasprzack}, {Katolik}, {Katsanevas},
  {Katsavounidis}, {Katzman}, {Kaufer}, {Kawabe}, {Keerthana},
  {K{\'e}f{\'e}lian}, {Keitel}, {Kemball}, {Kennedy}, {Key}, {Khalili},
  {Khamesra}, {Khan}, {Khan}, {Khan}, {Khan}, {Khazanov}, {Kijbunchoo}, {Kim},
  {Kim}, {Kim}, {Kim}, {Kim}, {Kim}, {King}, {King}, {Kinley-Hanlon},
  {Kirchhoff}, {Kissel}, {Kleybolte}, {Klimenko}, {Knowles}, {Koch},
  {Koehlenbeck}, {Koley}, {Kondrashov}, {Kontos}, {Korobko}, {Korth},
  {Kowalska}, {Kozak}, {Kr{\"a}mer}, {Kringel}, {Krishnan}, {Kr{\'o}lak},
  {Kuehn}, {Kumar}, {Kumar}, {Kumar}, {Kuo}, {Kutynia}, {Kwang}, {Lackey},
  {Lai}, {Landry}, {Landry}, {Lang}, {Lange}, {Lantz}, {Lanza},
  {Lartaux-Vollard}, {Lasky}, {Laxen}, {Lazzarini}, {Lazzaro}, {Leaci},
  {Leavey}, {Lee}, {Lee}, {Lee}, {Lee}, {Lee}, {Lehmann}, {Lenon}, {Leonardi},
  {Leroy}, {Letendre}, {Levin}, {Li}, {Li}, {Li}, {Linker}, {Littenberg},
  {Liu}, {Liu}, {Lo}, {Lockerbie}, {London}, {Longo}, {Lorenzini}, {Loriette},
  {Lormand}, {Losurdo}, {Lough}, {Lousto}, {Lovelace}, {L{\"u}ck}, {Lumaca},
  {Lundgren}, {Lynch}, {Ma}, {Macas}, {Macfoy}, {Machenschalk}, {MacInnis},
  {Macleod}, {Maga{\~n}a Hernandez}, {Maga{\~n}a-Sandoval}, {Maga{\~n}a
  Zertuche}, {Magee}, {Majorana}, {Maksimovic}, {Man}, {Mandic}, {Mangano},
  {Mansell}, {Manske}, {Mantovani}, {Marchesoni}, {Marion}, {M{\'a}rka},
  {M{\'a}rka}, {Markakis}, {Markosyan}, {Markowitz}, {Maros}, {Marquina},
  {Martelli}, {Martellini}, {Martin}, {Martin}, {Martynov}, {Mason}, {Massera},
  {Masserot}, {Massinger}, {Masso-Reid}, {Mastrogiovanni}, {Matas},
  {Matichard}, {Matone}, {Mavalvala}, {Mazumder}, {McCann}, {McCarthy},
  {McClelland}, {McCormick}, {McCuller}, {McGuire}, {McIver}, {McManus},
  {McRae}, {McWilliams}, {Meacher}, {Meadors}, {Mehmet}, {Meidam},
  {Mejuto-Villa}, {Melatos}, {Mendell}, {Mendoza-Gandara}, {Mercer}, {Mereni},
  {Merilh}, {Merzougui}, {Meshkov}, {Messenger}, {Messick}, {Metzdorff},
  {Meyers}, {Miao}, {Michel}, {Middleton}, {Mikhailov}, {Milano}, {Miller},
  {Miller}, {Miller}, {Miller}, {Millhouse}, {Mills}, {Milovich-Goff},
  {Minazzoli}, {Minenkov}, {Ming}, {Mishra}, {Mitra}, {Mitrofanov},
  {Mitselmakher}, {Mittleman}, {Moffa}, {Mogushi}, {Mohan}, {Mohapatra},
  {Montani}, {Moore}, {Moraru}, {Moreno}, {Morisaki}, {Mours}, {Mow-Lowry},
  {Mueller}, {Muir}, {Mukherjee}, {Mukherjee}, {Mukherjee}, {Mukund},
  {Mullavey}, {Munch}, {Mu{\~n}iz}, {Muratore}, {Murray}, {Nagar}, {Napier},
  {Nardecchia}, {Naticchioni}, {Nayak}, {Neilson}, {Nelemans}, {Nelson},
  {Nery}, {Neunzert}, {Nevin}, {Newport}, {Ng}, {Ng}, {Nguyen}, {Nguyen},
  {Nichols}, {Nielsen}, {Nissanke}, {Nitz}, {Nocera}, {Nolting}, {North},
  {Nuttall}, {Obergaulinger}, {Oberling}, {O'Brien}, {O'Dea}, {Ogin}, {Oh},
  {Oh}, {Ohme}, {Ohta}, {Okada}, {Oliver}, {Oppermann}, {Oram}, {O'Reilly},
  {Ormiston}, {Ortega}, {O'Shaughnessy}, {Ossokine}, {Ottaway}, {Overmier},
  {Owen}, {Pace}, {Pagano}, {Page}, {Page}, {Pai}, {Pai}, {Palamos},
  {Palashov}, {Palomba}, {Pal-Singh}, {Pan}, {Pan}, {Pang}, {Pang}, {Pankow},
  {Pannarale}, {Pant}, {Paoletti}, {Paoli}, {Papa}, {Parida}, {Parker},
  {Pascucci}, {Pasqualetti}, {Passaquieti}, {Passuello}, {Patil}, {Patricelli},
  {Pearlstone}, {Pedersen}, {Pedraza}, {Pedurand}, {Pekowsky}, {Pele}, {Penn},
  {Perego}, {Perez}, {Perreca}, {Perri}, {Pfeiffer}, {Phelps}, {Phukon},
  {Piccinni}, {Pichot}, {Piergiovanni}, {Pierro}, {Pillant}, {Pinard}, {Pinto},
  {Pirello}, {Pitkin}, {Poggiani}, {Popolizio}, {Porter}, {Possenti}, {Post},
  {Powell}, {Prasad}, {Pratt}, {Pratten}, {Predoi}, {Prestegard}, {Principe},
  {Privitera}, {Prodi}, {Prokhorov}, {Puncken}, {Punturo}, {Puppo},
  {P{\"u}rrer}, {Qi}, {Quetschke}, {Quintero}, {Quitzow-James}, {Raab},
  {Rabeling}, {Radkins}, {Raffai}, {Raja}, {Rajan}, {Rajbhandari}, {Rakhmanov},
  {Ramirez}, {Ramos-Buades}, {Rana}, {Rapagnani}, {Raymond}, {Razzano}, {Read},
  {Regimbau}, {Rei}, {Reid}, {Reitze}, {Ren}, {Ricci}, {Ricker},
  {Riemenschneider}, {Riles}, {Rizzo}, {Robertson}, {Robie}, {Robinet},
  {Robson}, {Rocchi}, {Rolland}, {Rollins}, {Roma}, {Romano}, {Romel}, {Romie},
  {Rosi{\'n}ska}, {Ross}, {Rowan}, {R{\"u}diger}, {Ruggi}, {Rutins}, {Ryan},
  {Sachdev}, {Sadecki}, {Sakellariadou}, {Salconi}, {Saleem}, {Salemi},
  {Samajdar}, {Sammut}, {Sampson}, {Sanchez}, {Sanchez}, {Sanchis-Gual},
  {Sandberg}, {Sanders}, {Sarin}, {Sassolas}, {Sathyaprakash}, {Saulson},
  {Sauter}, {Savage}, {Sawadsky}, {Schale}, {Scheel}, {Scheuer}, {Schmidt},
  {Schnabel}, {Schofield}, {Sch{\"o}nbeck}, {Schreiber}, {Schuette}, {Schulte},
  {Schutz}, {Schwalbe}, {Scott}, {Scott}, {Seidel}, {Sellers}, {Sengupta},
  {Sentenac}, {Sequino}, {Sergeev}, {Setyawati}, {Shaddock}, {Shaffer}, {Shah},
  {Shahriar}, {Shaner}, {Shao}, {Shapiro}, {Shawhan}, {Shen}, {Shoemaker},
  {Shoemaker}, {Siellez}, {Siemens}, {Sieniawska}, {Sigg}, {Silva}, {Singer},
  {Singh}, {Singhal}, {Sintes}, {Slagmolen}, {Slaven-Blair}, {Smith}, {Smith},
  {Smith}, {Somala}, {Son}, {Sorazu}, {Sorrentino}, {Souradeep}, {Spencer},
  {Srivastava}, {Staats}, {Steinke}, {Steinlechner}, {Steinlechner},
  {Steinmeyer}, {Steltner}, {Stevenson}, {Stocks}, {Stone}, {Stops}, {Strain},
  {Stratta}, {Strigin}, {Strunk}, {Sturani}, {Stuver}, {Summerscales}, {Sun},
  {Sunil}, {Suresh}, {Sutton}, {Swinkels}, {Szczepa{\'n}czyk}, {Tacca}, {Tait},
  {Talbot}, {Talukder}, {Tanner}, {T{\'a}pai}, {Taracchini}, {Tasson},
  {Taylor}, {Taylor}, {Tewari}, {Theeg}, {Thies}, {Thomas}, {Thomas}, {Thomas},
  {Thorne}, {Thrane}, {Tiwari}, {Tiwari}, {Tokmakov}, {Toland}, {Tonelli},
  {Tornasi}, {Torres-Forn{\'e}}, {Torrie}, {T{\"o}yr{\"a}}, {Travasso},
  {Traylor}, {Trinastic}, {Tringali}, {Trovato}, {Trozzo}, {Tsang}, {Tse},
  {Tso}, {Tsuna}, {Tsukada}, {Tuyenbayev}, {Ueno}, {Ugolini}, {Urban}, {Usman},
  {Vahlbruch}, {Vajente}, {Valdes}, {van Bakel}, {van Beuzekom}, {van den
  Brand}, {Van Den Broeck}, {Vander-Hyde}, {van der Schaaf}, {van Heijningen},
  {van Veggel}, {Vardaro}, {Varma}, {Vass}, {Vas{\'u}th}, {Vecchio},
  {Vedovato}, {Veitch}, {Veitch}, {Venkateswara}, {Venugopalan}, {Verkindt},
  {Vetrano}, {Vicer{\'e}}, {Viets}, {Vinciguerra}, {Vine}, {Vinet}, {Vitale},
  {Vo}, {Vocca}, {Vorvick}, {Vyatchanin}, {Wade}, {Wade}, {Wade}, {Walet},
  {Walker}, {Wallace}, {Walsh}, {Wang}, {Wang}, {Wang}, {Wang}, {Wang}, {Ward},
  {Warner}, {Was}, {Watchi}, {Weaver}, {Wei}, {Weinert}, {Weinstein}, {Weiss},
  {Wellmann}, {Wen}, {Wessel}, {We{\ss}els}, {Westerweck}, {Wette}, {Whelan},
  {Whiting}, {Whittle}, {Wilken}, {Williams}, {Williams}, {Williamson},
  {Willis}, {Willke}, {Wimmer}, {Winkler}, {Wipf}, {Wittel}, {Woan}, {Woehler},
  {Wofford}, {Wong}, {Worden}, {Wright}, {Wu}, {Wysocki}, {Xiao}, {Yam},
  {Yamamoto}, {Yancey}, {Yang}, {Yap}, {Yazback}, {Yu}, {Yu}, {Yvert},
  {Zadro{\.Z}ny}, {Zanolin}, {Zelenova}, {Zendri}, {Zevin}, {Zhang}, {Zhang},
  {Zhang}, {Zhang}, {Zhang}, {Zhao}, {Zhou}, {Zhou}, {Zhu}, {Zhu}, {Zimmerman},
  {Zlochower}, {Zucker}, {Zweizig}, {LIGO Scientific Collaboration}, \& {Virgo
  Collaboration}}]{GW170817eos}
---. 2018, \prl, 121, 161101

\bibitem[{{Abbott} {et~al.}(2019){Abbott}, {Abbott}, {Abbott}, {Abraham},
  {Acernese}, {Ackley}, {Adams}, {Adhikari}, {Adya}, {Affeldt}, {Agathos},
  {Agatsuma}, {Aggarwal}, {Aguiar}, {Aiello}, {Ain}, {Ajith}, {Allen},
  {Allocca}, {Aloy}, {Altin}, {Amato}, {Ananyeva}, {Anderson}, {Anderson},
  {Angelova}, {Antier}, {Appert}, {Arai}, {Araya}, {Areeda}, {Ar{\`e}ne},
  {Arnaud}, {Arun}, {Ascenzi}, {Ashton}, {Aston}, {Astone}, {Aubin}, {Aufmuth},
  {AultONeal}, {Austin}, {Avendano}, {Avila-Alvarez}, {Babak}, {Bacon},
  {Badaracco}, {Bader}, {Bae}, {Baker}, {Baldaccini}, {Ballardin}, {Ballmer},
  {Banagiri}, {Barayoga}, {Barclay}, {Barish}, {Barker}, {Barkett}, {Barnum},
  {Barone}, {Barr}, {Barsotti}, {Barsuglia}, {Barta}, {Bartlett}, {Bartos},
  {Bassiri}, {Basti}, {Bawaj}, {Bayley}, {Bazzan}, {B{\'e}csy}, {Bejger},
  {Belahcene}, {Bell}, {Beniwal}, {Berger}, {Bergmann}, {Bernuzzi}, {Bero},
  {Berry}, {Bersanetti}, {Bertolini}, {Betzwieser}, {Bhandare}, {Bidler},
  {Bilenko}, {Bilgili}, {Billingsley}, {Birch}, {Birney}, {Birnholtz},
  {Biscans}, {Biscoveanu}, {Bisht}, {Bitossi}, {Bizouard}, {Blackburn},
  {Blackman}, {Blair}, {Blair}, {Blair}, {Bloemen}, {Bode}, {Boer}, {Boetzel},
  {Bogaert}, {Bondu}, {Bonilla}, {Bonnand}, {Booker}, {Boom}, {Booth}, {Bork},
  {Boschi}, {Bose}, {Bossie}, {Bossilkov}, {Bosveld}, {Bouffanais}, {Bozzi},
  {Bradaschia}, {Brady}, {Bramley}, {Branchesi}, {Brau}, {Briant}, {Briggs},
  {Brighenti}, {Brillet}, {Brinkmann}, {Brisson}, {Brockill}, {Brooks},
  {Brown}, {Brunett}, {Buikema}, {Bulik}, {Bulten}, {Buonanno}, {Buskulic},
  {Bustamante Rosell}, {Buy}, {Byer}, {Cabero}, {Cadonati}, {Cagnoli},
  {Cahillane}, {Calder{\'o}n Bustillo}, {Callister}, {Calloni}, {Camp},
  {Campbell}, {Canepa}, {Cannon}, {Cao}, {Cao}, {Capocasa}, {Carbognani},
  {Caride}, {Carney}, {Carullo}, {Casanueva Diaz}, {Casentini}, {Caudill},
  {Cavagli{\`a}}, {Cavalier}, {Cavalieri}, {Cella}, {Cerd{\'a}-Dur{\'a}n},
  {Cerretani}, {Cesarini}, {Chaibi}, {Chakravarti}, {Chamberlin}, {Chan},
  {Chao}, {Charlton}, {Chase}, {Chassande-Mottin}, {Chatterjee}, {Chaturvedi},
  {Chatziioannou}, {Cheeseboro}, {Chen}, {Chen}, {Chen}, {Cheng}, {Cheong},
  {Chia}, {Chincarini}, {Chiummo}, {Cho}, {Cho}, {Cho}, {Christensen}, {Chu},
  {Chua}, {Chung}, {Chung}, {Ciani}, {Ciobanu}, {Ciolfi}, {Cipriano}, {Cirone},
  {Clara}, {Clark}, {Clearwater}, {Cleva}, {Cocchieri}, {Coccia}, {Cohadon},
  {Cohen}, {Colgan}, {Colleoni}, {Collette}, {Collins}, {Cominsky},
  {Constancio}, {Conti}, {Cooper}, {Corban}, {Corbitt}, {Cordero-Carri{\'o}n},
  {Corley}, {Cornish}, {Corsi}, {Cortese}, {Costa}, {Cotesta}, {Coughlin},
  {Coughlin}, {Coulon}, {Countryman}, {Couvares}, {Covas}, {Cowan}, {Coward},
  {Cowart}, {Coyne}, {Coyne}, {Creighton}, {Creighton}, {Cripe}, {Croquette},
  {Crowder}, {Cullen}, {Cumming}, {Cunningham}, {Cuoco}, {Canton}, {D{\'a}lya},
  {Danilishin}, {D'Antonio}, {Danzmann}, {Dasgupta}, {Da Silva Costa},
  {Datrier}, {Dattilo}, {Dave}, {Davier}, {Davis}, {Daw}, {DeBra},
  {Deenadayalan}, {Degallaix}, {De Laurentis}, {Del{\'e}glise}, {Del Pozzo},
  {DeMarchi}, {Demos}, {Dent}, {De Pietri}, {Derby}, {De Rosa}, {De Rossi},
  {DeSalvo}, {de Varona}, {Dhurandhar}, {D{\'\i}az}, {Dietrich}, {Di Fiore},
  {Di Giovanni}, {Di Girolamo}, {Di Lieto}, {Ding}, {Di Pace}, {Di Palma}, {Di
  Renzo}, {Dmitriev}, {Doctor}, {Donovan}, {Dooley}, {Doravari}, {Dorrington},
  {Downes}, {Drago}, {Driggers}, {Du}, {Ducoin}, {Dupej}, {Dwyer}, {Easter},
  {Edo}, {Edwards}, {Effler}, {Ehrens}, {Eichholz}, {Eikenberry}, {Eisenmann},
  {Eisenstein}, {Essick}, {Estelles}, {Estevez}, {Etienne}, {Etzel}, {Evans},
  {Evans}, {Fafone}, {Fair}, {Fairhurst}, {Fan}, {Farinon}, {Farr}, {Farr},
  {Fauchon-Jones}, {Favata}, {Fays}, {Fazio}, {Fee}, {Feicht}, {Fejer}, {Feng},
  {Fernandez-Galiana}, {Ferrante}, {Ferreira}, {Ferreira}, {Ferrini},
  {Fidecaro}, {Fiori}, {Fiorucci}, {Fishbach}, {Fisher}, {Fishner},
  {Fitz-Axen}, {Flaminio}, {Fletcher}, {Flynn}, {Fong}, {Font}, {Forsyth},
  {Fournier}, {Frasca}, {Frasconi}, {Frei}, {Freise}, {Frey}, {Frey},
  {Fritschel}, {Frolov}, {Fulda}, {Fyffe}, {Gabbard}, {Gadre}, {Gaebel},
  {Gair}, {Gammaitoni}, {Ganija}, {Gaonkar}, {Garcia},
  {Garc{\'\i}a-Quir{\'o}s}, {Garufi}, {Gateley}, {Gaudio}, {Gaur}, {Gayathri},
  {Gemme}, {Genin}, {Gennai}, {George}, {George}, {Gergely}, {Germain},
  {Ghonge}, {Ghosh}, {Ghosh}, {Ghosh}, {Giacomazzo}, {Giaime}, {Giardina},
  {Giazotto}, {Gill}, {Giordano}, {Glover}, {Godwin}, {Goetz}, {Goetz},
  {Goncharov}, {Gonz{\'a}lez}, {Gonzalez Castro}, {Gopakumar}, {Gorodetsky},
  {Gossan}, {Gosselin}, {Gouaty}, {Grado}, {Graef}, {Granata}, {Grant}, {Gras},
  {Grassia}, {Gray}, {Gray}, {Greco}, {Green}, {Green}, {Gretarsson}, {Groot},
  {Grote}, {Grunewald}, {Gruning}, {Guidi}, {Gulati}, {Guo}, {Gupta}, {Gupta},
  {Gustafson}, {Gustafson}, {Haegel}, {Halim}, {Hall}, {Hall}, {Hamilton},
  {Hammond}, {Haney}, {Hanke}, {Hanks}, {Hanna}, {Hannam}, {Hannuksela},
  {Hanson}, {Hardwick}, {Haris}, {Harms}, {Harry}, {Harry}, {Haster},
  {Haughian}, {Hayes}, {Healy}, {Heidmann}, {Heintze}, {Heitmann}, {Hello},
  {Hemming}, {Hendry}, {Heng}, {Hennig}, {Heptonstall}, {Hernandez Vivanco},
  {Heurs}, {Hild}, {Hinderer}, {Hoak}, {Hochheim}, {Hofman}, {Holgado},
  {Holland}, {Holt}, {Holz}, {Hopkins}, {Horst}, {Hough}, {Howell}, {Hoy},
  {Hreibi}, {Huang}, {Huerta}, {Huet}, {Hughey}, {Hulko}, {Husa}, {Huttner},
  {Huynh-Dinh}, {Idzkowski}, {Iess}, {Ingram}, {Inta}, {Intini}, {Irwin},
  {Isa}, {Isac}, {Isi}, {Iyer}, {Izumi}, {Jacqmin}, {Jadhav}, {Jani},
  {Janthalur}, {Jaranowski}, {Jenkins}, {Jiang}, {Johnson}, {Johnson-McDaniel},
  {Jones}, {Jones}, {Jones}, {Jonker}, {Ju}, {Junker}, {Kalaghatgi},
  {Kalogera}, {Kamai}, {Kandhasamy}, {Kang}, {Kanner}, {Kapadia}, {Karki},
  {Karvinen}, {Kashyap}, {Kasprzack}, {Katsanevas}, {Katsavounidis}, {Katzman},
  {Kaufer}, {Kawabe}, {Keerthana}, {K{\'e}f{\'e}lian}, {Keitel}, {Kennedy},
  {Key}, {Khalili}, {Khan}, {Khan}, {Khan}, {Khan}, {Khazanov}, {Khursheed},
  {Kijbunchoo}, {Kim}, {Kim}, {Kim}, {Kim}, {Kim}, {Kim}, {Kimball}, {King},
  {King}, {Kinley-Hanlon}, {Kirchhoff}, {Kissel}, {Kleybolte}, {Klika},
  {Klimenko}, {Knowles}, {Koch}, {Koehlenbeck}, {Koekoek}, {Koley},
  {Kondrashov}, {Kontos}, {Koper}, {Korobko}, {Korth}, {Kowalska}, {Kozak},
  {Kringel}, {Krishnendu}, {Kr{\'o}lak}, {Kuehn}, {Kumar}, {Kumar}, {Kumar},
  {Kumar}, {Kuo}, {Kutynia}, {Kwang}, {Lackey}, {Lai}, {Lam}, {Landry}, {Lane},
  {Lang}, {Lange}, {Lantz}, {Lanza}, {Lartaux-Vollard}, {Lasky}, {Laxen},
  {Lazzarini}, {Lazzaro}, {Leaci}, {Leavey}, {Lecoeuche}, {Lee}, {Lee}, {Lee},
  {Lee}, {Lee}, {Lee}, {Lehmann}, {Lenon}, {Leroy}, {Letendre}, {Levin}, {Li},
  {Li}, {Li}, {Li}, {Lin}, {Linde}, {Linker}, {Littenberg}, {Liu}, {Liu}, {Lo},
  {Lockerbie}, {London}, {Longo}, {Lorenzini}, {Loriette}, {Lormand},
  {Losurdo}, {Lough}, {Lousto}, {Lovelace}, {Lower}, {L{\"u}ck}, {Lumaca},
  {Lundgren}, {Lynch}, {Ma}, {Macas}, {Macfoy}, {MacInnis}, {Macleod},
  {Macquet}, {Maga{\~n}a-Sandoval}, {Maga{\~n}a Zertuche}, {Magee}, {Majorana},
  {Maksimovic}, {Malik}, {Man}, {Mandic}, {Mangano}, {Mansell}, {Manske},
  {Mantovani}, {Marchesoni}, {Marion}, {M{\'a}rka}, {M{\'a}rka}, {Markakis},
  {Markosyan}, {Markowitz}, {Maros}, {Marquina}, {Marsat}, {Martelli},
  {Martin}, {Martin}, {Martynov}, {Mason}, {Massera}, {Masserot}, {Massinger},
  {Masso-Reid}, {Mastrogiovanni}, {Matas}, {Matichard}, {Matone}, {Mavalvala},
  {Mazumder}, {McCann}, {McCarthy}, {McClelland}, {McCormick}, {McCuller},
  {McGuire}, {McIver}, {McManus}, {McRae}, {McWilliams}, {Meacher}, {Meadors},
  {Mehmet}, {Mehta}, {Meidam}, {Melatos}, {Mendell}, {Mercer}, {Mereni},
  {Merilh}, {Merzougui}, {Meshkov}, {Messenger}, {Messick}, {Metzdorff},
  {Meyers}, {Miao}, {Michel}, {Middleton}, {Mikhailov}, {Milano}, {Miller},
  {Miller}, {Millhouse}, {Mills}, {Milovich-Goff}, {Minazzoli}, {Minenkov},
  {Mishkin}, {Mishra}, {Mistry}, {Mitra}, {Mitrofanov}, {Mitselmakher},
  {Mittleman}, {Mo}, {Moffa}, {Mogushi}, {Mohapatra}, {Montani}, {Moore},
  {Moraru}, {Moreno}, {Morisaki}, {Mours}, {Mow-Lowry}, {Mukherjee},
  {Mukherjee}, {Mukherjee}, {Mukund}, {Mullavey}, {Munch}, {Mu{\~n}iz},
  {Muratore}, {Murray}, {Nagar}, {Nardecchia}, {Naticchioni}, {Nayak},
  {Neilson}, {Nelemans}, {Nelson}, {Nery}, {Neunzert}, {Ng}, {Ng}, {Nguyen},
  {Nichols}, {Nielsen}, {Nissanke}, {Nitz}, {Nocera}, {North}, {Nuttall},
  {Obergaulinger}, {Oberling}, {O'Brien}, {O'Dea}, {Ogin}, {Oh}, {Oh}, {Ohme},
  {Ohta}, {Okada}, {Oliver}, {Oppermann}, {Oram}, {O'Reilly}, {Ormiston},
  {Ortega}, {O'Shaughnessy}, {Ossokine}, {Ottaway}, {Overmier}, {Owen}, {Pace},
  {Pagano}, {Page}, {Pai}, {Pai}, {Palamos}, {Palashov}, {Palomba},
  {Pal-Singh}, {Pan}, {Pang}, {Pang}, {Pankow}, {Pannarale}, {Pant},
  {Paoletti}, {Paoli}, {Papa}, {Parida}, {Parker}, {Pascucci}, {Pasqualetti},
  {Passaquieti}, {Passuello}, {Patil}, {Patricelli}, {Pearlstone}, {Pedersen},
  {Pedraza}, {Pedurand}, {Pele}, {Penn}, {Perego}, {Perez}, {Perreca},
  {Pfeiffer}, {Phelps}, {Phukon}, {Piccinni}, {Pichot}, {Piergiovanni},
  {Pillant}, {Pinard}, {Pirello}, {Pitkin}, {Poggiani}, {Pong}, {Ponrathnam},
  {Popolizio}, {Porter}, {Powell}, {Prajapati}, {Prasad}, {Prasai}, {Prasanna},
  {Pratten}, {Prestegard}, {Privitera}, {Prodi}, {Prokhorov}, {Puncken},
  {Punturo}, {Puppo}, {P{\"u}rrer}, {Qi}, {Quetschke}, {Quinonez}, {Quintero},
  {Quitzow-James}, {Raab}, {Radkins}, {Radulescu}, {Raffai}, {Raja}, {Rajan},
  {Rajbhandari}, {Rakhmanov}, {Ramirez}, {Ramos-Buades}, {Rana}, {Rao},
  {Rapagnani}, {Raymond}, {Razzano}, {Read}, {Regimbau}, {Rei}, {Reid},
  {Reitze}, {Ren}, {Ricci}, {Richardson}, {Richardson}, {Ricker},
  {Riemenschneider}, {Riles}, {Rizzo}, {Robertson}, {Robie}, {Robinet},
  {Rocchi}, {Rolland}, {Rollins}, {Roma}, {Romanelli}, {Romano}, {Romel},
  {Romie}, {Rose}, {Rosi{\'n}ska}, {Rosofsky}, {Ross}, {Rowan}, {R{\"u}diger},
  {Ruggi}, {Rutins}, {Ryan}, {Sachdev}, {Sadecki}, {Sakellariadou}, {Salafia},
  {Salconi}, {Saleem}, {Salemi}, {Samajdar}, {Sammut}, {Sanchez}, {Sanchez},
  {Sanchis-Gual}, {Sandberg}, {Sanders}, {Santiago}, {Sarin}, {Sassolas},
  {Sathyaprakash}, {Saulson}, {Sauter}, {Savage}, {Schale}, {Scheel},
  {Scheuer}, {Schmidt}, {Schnabel}, {Schofield}, {Sch{\"o}nbeck}, {Schreiber},
  {Schulte}, {Schutz}, {Schwalbe}, {Scott}, {Scott}, {Seidel}, {Sellers},
  {Sengupta}, {Sennett}, {Sentenac}, {Sequino}, {Sergeev}, {Setyawati},
  {Shaddock}, {Shaffer}, {Shahriar}, {Shaner}, {Shao}, {Sharma}, {Shawhan},
  {Shen}, {Shink}, {Shoemaker}, {Shoemaker}, {ShyamSundar}, {Siellez},
  {Sieniawska}, {Sigg}, {Silva}, {Singer}, {Singh}, {Singhal}, {Sintes},
  {Sitmukhambetov}, {Skliris}, {Slagmolen}, {Slaven-Blair}, {Smith}, {Smith},
  {Somala}, {Son}, {Sorazu}, {Sorrentino}, {Souradeep}, {Sowell}, {Spencer},
  {Srivastava}, {Srivastava}, {Staats}, {Stachie}, {Standke}, {Steer},
  {Steinke}, {Steinlechner}, {Steinlechner}, {Steinmeyer}, {Stevenson},
  {Stocks}, {Stone}, {Stops}, {Strain}, {Stratta}, {Strigin}, {Strunk},
  {Sturani}, {Stuver}, {Sudhir}, {Summerscales}, {Sun}, {Sunil}, {Suresh},
  {Sutton}, {Swinkels}, {Szczepa{\'n}czyk}, {Tacca}, {Tait}, {Talbot},
  {Talukder}, {Tanner}, {T{\'a}pai}, {Taracchini}, {Tasson}, {Taylor}, {Thies},
  {Thomas}, {Thomas}, {Thondapu}, {Thorne}, {Thrane}, {Tiwari}, {Tiwari},
  {Tiwari}, {Toland}, {Tonelli}, {Tornasi}, {Torres-Forn{\'e}}, {Torrie},
  {T{\"o}yr{\"a}}, {Travasso}, {Traylor}, {Tringali}, {Trovato}, {Trozzo},
  {Trudeau}, {Tsang}, {Tse}, {Tso}, {Tsukada}, {Tsuna}, {Tuyenbayev}, {Ueno},
  {Ugolini}, {Unnikrishnan}, {Urban}, {Usman}, {Vahlbruch}, {Vajente},
  {Valdes}, {van Bakel}, {van Beuzekom}, {van den Brand}, {Van Den Broeck},
  {Vander-Hyde}, {van Heijningen}, {van der Schaaf}, {van Veggel}, {Vardaro},
  {Varma}, {Vass}, {Vas{\'u}th}, {Vecchio}, {Vedovato}, {Veitch}, {Veitch},
  {Venkateswara}, {Venugopalan}, {Verkindt}, {Vetrano}, {Vicer{\'e}}, {Viets},
  {Vine}, {Vinet}, {Vitale}, {Vo}, {Vocca}, {Vorvick}, {Vyatchanin}, {Wade},
  {Wade}, {Wade}, {Walet}, {Walker}, {Wallace}, {Walsh}, {Wang}, {Wang},
  {Wang}, {Wang}, {Wang}, {Ward}, {Warden}, {Warner}, {Was}, {Watchi},
  {Weaver}, {Wei}, {Weinert}, {Weinstein}, {Weiss}, {Wellmann}, {Wen},
  {Wessel}, {We{\ss}els}, {Westhouse}, {Wette}, {Whelan}, {White}, {Whiting},
  {Whittle}, {Wilken}, {Williams}, {Williamson}, {Willis}, {Willke}, {Wimmer},
  {Winkler}, {Wipf}, {Wittel}, {Woan}, {Woehler}, {Wofford}, {Worden},
  {Wright}, {Wu}, {Wysocki}, {Xiao}, {Yamamoto}, {Yancey}, {Yang}, {Yap},
  {Yazback}, {Yeeles}, {Yu}, {Yu}, {Yuen}, {Yvert}, {Zadro{\.Z}ny}, {Zanolin},
  {Zappa}, {Zelenova}, {Zendri}, {Zevin}, {Zhang}, {Zhang}, {Zhang}, {Zhao},
  {Zhou}, {Zhou}, {Zhu}, {Zimmerman}, {Zlochower}, {Zucker}, {Zweizig}, {LIGO
  Scientific Collaboration}, \& {Virgo Collaboration}}]{GWTC1}
---. 2019, Physical Review X, 9, 031040

\bibitem[{{Abbott} {et~al.}(2020{\natexlab{a}}){Abbott}, {Abbott}, {Abbott},
  {Abraham}, {Acernese}, {Ackley}, {Adams}, {Adhikari}, {Adya}, {Affeldt},
  {Agathos}, {Agatsuma}, {Aggarwal}, {Aguiar}, {Aiello}, {Ain}, {Ajith},
  {Allen}, {Allocca}, {Aloy}, {Altin}, {Amato}, {Anand}, {Ananyeva},
  {Anderson}, {Anderson}, {Angelova}, {Antier}, {Appert}, {Arai}, {Araya},
  {Areeda}, {Ar{\`e}ne}, {Arnaud}, {Aronson}, {Arun}, {Ascenzi}, {Ashton},
  {Aston}, {Astone}, {Aubin}, {Aufmuth}, {AultONeal}, {Austin}, {Avendano},
  {Avila-Alvarez}, {Babak}, {Bacon}, {Badaracco}, {Bader}, {Bae}, {Baird},
  {Baker}, {Baldaccini}, {Ballardin}, {Ballmer}, {Bals}, {Banagiri},
  {Barayoga}, {Barbieri}, {Barclay}, {Barish}, {Barker}, {Barkett}, {Barnum},
  {Barone}, {Barr}, {Barsotti}, {Barsuglia}, {Barta}, {Bartlett}, {Bartos},
  {Bassiri}, {Basti}, {Bawaj}, {Bayley}, {Baylor}, {Bazzan}, {B{\'e}csy},
  {Bejger}, {Belahcene}, {Bell}, {Beniwal}, {Benjamin}, {Berger}, {Bergmann},
  {Bernuzzi}, {Berry}, {Bersanetti}, {Bertolini}, {Betzwieser}, {Bhandare},
  {Bidler}, {Biggs}, {Bilenko}, {Bilgili}, {Billingsley}, {Birney},
  {Birnholtz}, {Biscans}, {Bischi}, {Biscoveanu}, {Bisht}, {Bitossi},
  {Bizouard}, {Blackburn}, {Blackman}, {Blair}, {Blair}, {Blair}, {Bloemen},
  {Bobba}, {Bode}, {Boer}, {Boetzel}, {Bogaert}, {Bondu}, {Bonnand}, {Booker},
  {Boom}, {Bork}, {Boschi}, {Bose}, {Bossilkov}, {Bosveld}, {Bouffanais},
  {Bozzi}, {Bradaschia}, {Brady}, {Bramley}, {Branchesi}, {Brau}, {Breschi},
  {Briant}, {Briggs}, {Brighenti}, {Brillet}, {Brinkmann}, {Brockill},
  {Brooks}, {Brooks}, {Brown}, {Brunett}, {Buikema}, {Bulik}, {Bulten},
  {Buonanno}, {Buskulic}, {Buy}, {Byer}, {Cabero}, {Cadonati}, {Cagnoli},
  {Cahillane}, {Calder{\'o}n Bustillo}, {Callister}, {Calloni}, {Camp},
  {Campbell}, {Canepa}, {Cannon}, {Cao}, {Cao}, {Carapella}, {Carbognani},
  {Caride}, {Carney}, {Carullo}, {Casanueva Diaz}, {Casentini}, {Caudill},
  {Cavagli{\`a}}, {Cavalier}, {Cavalieri}, {Cella}, {Cerd{\'a}-Dur{\'a}n},
  {Cesarini}, {Chaibi}, {Chakravarti}, {Chamberlin}, {Chan}, {Chao},
  {Charlton}, {Chase}, {Chassande-Mottin}, {Chatterjee}, {Chaturvedi},
  {Chatziioannou}, {Cheeseboro}, {Chen}, {Chen}, {Chen}, {Cheng}, {Cheong},
  {Chia}, {Chiadini}, {Chincarini}, {Chiummo}, {Cho}, {Cho}, {Cho},
  {Christensen}, {Chu}, {Chua}, {Chung}, {Chung}, {Ciani}, {Cie{\'s}lar},
  {Ciobanu}, {Ciolfi}, {Cipriano}, {Cirone}, {Clara}, {Clark}, {Clearwater},
  {Cleva}, {Coccia}, {Cohadon}, {Cohen}, {Colleoni}, {Collette}, {Collins},
  {Colpi}, {Cominsky}, {Constancio}, {Conti}, {Cooper}, {Corban}, {Corbitt},
  {Cordero-Carri{\'o}n}, {Corezzi}, {Corley}, {Cornish}, {Corre}, {Corsi},
  {Cortese}, {Costa}, {Cotesta}, {Coughlin}, {Coughlin}, {Coulon},
  {Countryman}, {Couvares}, {Covas}, {Cowan}, {Coward}, {Cowart}, {Coyne},
  {Coyne}, {Creighton}, {Creighton}, {Cripe}, {Croquette}, {Crowder}, {Cullen},
  {Cumming}, {Cunningham}, {Cuoco}, {Dal Canton}, {D{\'a}lya}, {D'Angelo},
  {Danilishin}, {D'Antonio}, {Danzmann}, {Dasgupta}, {Da Silva Costa},
  {Datrier}, {Dattilo}, {Dave}, {Davier}, {Davis}, {Daw}, {DeBra},
  {Deenadayalan}, {Degallaix}, {De Laurentis}, {Del{\'e}glise}, {De Lillo},
  {Del Pozzo}, {DeMarchi}, {Demos}, {Dent}, {De Pietri}, {De Rosa}, {De Rossi},
  {DeSalvo}, {de Varona}, {Dhurandhar}, {D{\'\i}az}, {Dietrich}, {Di Fiore},
  {DiFronzo}, {Di Giorgio}, {Di Giovanni}, {Di Giovanni}, {Di Girolamo}, {Di
  Lieto}, {Ding}, {Di Pace}, {Di Palma}, {Di Renzo}, {Divakarla}, {Dmitriev},
  {Doctor}, {Donovan}, {Dooley}, {Doravari}, {Dorrington}, {Downes}, {Drago},
  {Driggers}, {Du}, {Ducoin}, {Dudi}, {Dupej}, {Durante}, {Dwyer}, {Easter},
  {Eddolls}, {Edo}, {Effler}, {Ehrens}, {Eichholz}, {Eikenberry}, {Eisenmann},
  {Eisenstein}, {Errico}, {Essick}, {Estelles}, {Estevez}, {Etienne}, {Etzel},
  {Evans}, {Evans}, {Fafone}, {Fairhurst}, {Fan}, {Farinon}, {Farr}, {Farr},
  {Fauchon-Jones}, {Favata}, {Fays}, {Fazio}, {Fee}, {Feicht}, {Fejer}, {Feng},
  {Fernandez-Galiana}, {Ferrante}, {Ferreira}, {Ferreira}, {Fidecaro}, {Fiori},
  {Fiorucci}, {Fishbach}, {Fisher}, {Fishner}, {Fittipaldi}, {Fitz-Axen},
  {Fiumara}, {Flaminio}, {Fletcher}, {Floden}, {Flynn}, {Fong}, {Font},
  {Forsyth}, {Fournier}, {Vivanco}, {Frasca}, {Frasconi}, {Frei}, {Freise},
  {Frey}, {Frey}, {Fritschel}, {Frolov}, {Fronz{\`e}}, {Fulda}, {Fyffe},
  {Gabbard}, {Gadre}, {Gaebel}, {Gair}, {Gamba}, {Gammaitoni}, {Gaonkar},
  {Garc{\'\i}a-Quir{\'o}s}, {Garufi}, {Gateley}, {Gaudio}, {Gaur}, {Gayathri},
  {Gemme}, {Genin}, {Gennai}, {George}, {George}, {George}, {Gergely},
  {Ghonge}, {Ghosh}, {Ghosh}, {Ghosh}, {Giacomazzo}, {Giaime}, {Giardina},
  {Gibson}, {Gill}, {Glover}, {Gniesmer}, {Godwin}, {Goetz}, {Goetz},
  {Goncharov}, {Gonz{\'a}lez}, {Castro}, {Gopakumar}, {Gossan}, {Gosselin},
  {Gouaty}, {Grace}, {Grado}, {Granata}, {Grant}, {Gras}, {Grassia}, {Gray},
  {Gray}, {Greco}, {Green}, {Green}, {Gretarsson}, {Grimaldi}, {Grimm},
  {Groot}, {Grote}, {Grunewald}, {Gruning}, {Guidi}, {Gulati}, {Guo}, {Gupta},
  {Gupta}, {Gupta}, {Gustafson}, {Gustafson}, {Haegel}, {Halim}, {Hall},
  {Hall}, {Hamilton}, {Hammond}, {Haney}, {Hanke}, {Hanks}, {Hanna}, {Hannam},
  {Hannuksela}, {Hansen}, {Hanson}, {Harder}, {Hardwick}, {Haris}, {Harms},
  {Harry}, {Harry}, {Hasskew}, {Haster}, {Haughian}, {Hayes}, {Healy},
  {Heidmann}, {Heintze}, {Heitmann}, {Hellman}, {Hello}, {Hemming}, {Hendry},
  {Heng}, {Hennig}, {Heurs}, {Hild}, {Hinderer}, {Ho}, {Hochheim}, {Hofman},
  {Holgado}, {Holland}, {Holt}, {Holz}, {Hopkins}, {Horst}, {Hough}, {Howell},
  {Hoy}, {Huang}, {H{\"u}bner}, {Huerta}, {Huet}, {Hughey}, {Hui}, {Husa},
  {Huttner}, {Huynh-Dinh}, {Idzkowski}, {Iess}, {Inchauspe}, {Ingram}, {Inta},
  {Intini}, {Irwin}, {Isa}, {Isac}, {Isi}, {Iyer}, {Jacqmin}, {Jadhav}, {Jani},
  {Janthalur}, {Jaranowski}, {Jariwala}, {Jenkins}, {Jiang}, {Johnson},
  {Johnson-McDaniel}, {Jones}, {Jones}, {Jones}, {Jones}, {Jonker}, {Ju},
  {Junker}, {Kalaghatgi}, {Kalogera}, {Kamai}, {Kandhasamy}, {Kang}, {Kanner},
  {Kapadia}, {Karki}, {Kashyap}, {Kasprzack}, {Kastaun}, {Katsanevas},
  {Katsavounidis}, {Katzman}, {Kaufer}, {Kawabe}, {Keerthana},
  {K{\'e}f{\'e}lian}, {Keitel}, {Kennedy}, {Key}, {Khalili}, {Khan}, {Khan},
  {Khazanov}, {Khetan}, {Khursheed}, {Kijbunchoo}, {Kim}, {Kim}, {Kim}, {Kim},
  {Kim}, {Kim}, {Kimball}, {King}, {Kinley-Hanlon}, {Kirchhoff}, {Kissel},
  {Kleybolte}, {Klika}, {Klimenko}, {Knowles}, {Koch}, {Koehlenbeck},
  {Koekoek}, {Koley}, {Kondrashov}, {Kontos}, {Koper}, {Korobko}, {Korth},
  {Kovalam}, {Kozak}, {Kr{\"a}mer}, {Kringel}, {Krishnendu}, {Kr{\'o}lak},
  {Krupinski}, {Kuehn}, {Kumar}, {Kumar}, {Kumar}, {Kumar}, {Kuo}, {Kutynia},
  {Kwang}, {Lackey}, {Laghi}, {Lai}, {Lam}, {Landry}, {Landry}, {Lane}, {Lang},
  {Lange}, {Lantz}, {Lanza}, {Lartaux-Vollard}, {Lasky}, {Laxen}, {Lazzarini},
  {Lazzaro}, {Leaci}, {Leavey}, {Lecoeuche}, {Lee}, {Lee}, {Lee}, {Lee}, {Lee},
  {Lee}, {Lehmann}, {Lenon}, {Leroy}, {Letendre}, {Levin}, {Li}, {Li}, {Li},
  {Li}, {Li}, {Lin}, {Linde}, {Linker}, {Littenberg}, {Liu}, {Liu},
  {Llorens-Monteagudo}, {Lo}, {London}, {Longo}, {Lorenzini}, {Loriette},
  {Lormand}, {Losurdo}, {Lough}, {Lousto}, {Lovelace}, {Lower}, {Lucaccioni},
  {L{\"u}ck}, {Lumaca}, {Lundgren}, {Lynch}, {Ma}, {Macas}, {Macfoy},
  {MacInnis}, {Macleod}, {Macquet}, {Maga{\~n}a Hernandez},
  {Maga{\~n}a-Sandoval}, {Magee}, {Majorana}, {Maksimovic}, {Malik}, {Man},
  {Mandic}, {Mangano}, {Mansell}, {Manske}, {Mantovani}, {Mapelli},
  {Marchesoni}, {Marion}, {M{\'a}rka}, {M{\'a}rka}, {Markakis}, {Markosyan},
  {Markowitz}, {Maros}, {Marquina}, {Marsat}, {Martelli}, {Martin}, {Martin},
  {Martinez}, {Martynov}, {Masalehdan}, {Mason}, {Massera}, {Masserot},
  {Massinger}, {Masso-Reid}, {Mastrogiovanni}, {Matas}, {Matichard}, {Matone},
  {Mavalvala}, {McCann}, {McCarthy}, {McClelland}, {McCormick}, {McCuller},
  {McGuire}, {McIsaac}, {McIver}, {McManus}, {McRae}, {McWilliams}, {Meacher},
  {Meadors}, {Mehmet}, {Mehta}, {Meidam}, {Mejuto Villa}, {Melatos}, {Mendell},
  {Mercer}, {Mereni}, {Merfeld}, {Merilh}, {Merzougui}, {Meshkov}, {Messenger},
  {Messick}, {Messina}, {Metzdorff}, {Meyers}, {Meylahn}, {Miani}, {Miao},
  {Michel}, {Middleton}, {Milano}, {Miller}, {Millhouse}, {Mills},
  {Milovich-Goff}, {Minazzoli}, {Minenkov}, {Mishkin}, {Mishra}, {Mistry},
  {Mitra}, {Mitrofanov}, {Mitselmakher}, {Mittleman}, {Mo}, {Moffa}, {Mogushi},
  {Mohapatra}, {Molina-Ruiz}, {Mondin}, {Montani}, {Moore}, {Moraru},
  {Morawski}, {Moreno}, {Morisaki}, {Mours}, {Mow-Lowry}, {Muciaccia},
  {Mukherjee}, {Mukherjee}, {Mukherjee}, {Mukherjee}, {Mukund}, {Mullavey},
  {Munch}, {Mu{\~n}iz}, {Muratore}, {Murray}, {Nagar}, {Nardecchia},
  {Naticchioni}, {Nayak}, {Neil}, {Neilson}, {Nelemans}, {Nelson}, {Nery},
  {Neunzert}, {Nevin}, {Ng}, {Ng}, {Nguyen}, {Nguyen}, {Nichols}, {Nichols},
  {Nissanke}, {Nocera}, {North}, {Nuttall}, {Obergaulinger}, {Oberling},
  {O'Brien}, {Oganesyan}, {Ogin}, {Oh}, {Oh}, {Ohme}, {Ohta}, {Okada},
  {Oliver}, {Oppermann}, {Oram}, {O'Reilly}, {Ormiston}, {Ortega},
  {O'Shaughnessy}, {Ossokine}, {Ottaway}, {Overmier}, {Owen}, {Pace}, {Pagano},
  {Page}, {Pagliaroli}, {Pai}, {Pai}, {Palamos}, {Palashov}, {Palomba}, {Pan},
  {Panda}, {Pang}, {Pankow}, {Pannarale}, {Pant}, {Paoletti}, {Paoli},
  {Parida}, {Parker}, {Pascucci}, {Pasqualetti}, {Passaquieti}, {Passuello},
  {Patil}, {Patricelli}, {Payne}, {Pearlstone}, {Pechsiri}, {Pedersen},
  {Pedraza}, {Pedurand}, {Pele}, {Penn}, {Perego}, {Perez}, {P{\'e}rigois},
  {Perreca}, {Petermann}, {Pfeiffer}, {Phelps}, {Phukon}, {Piccinni}, {Pichot},
  {Piergiovanni}, {Pierro}, {Pillant}, {Pinard}, {Pinto}, {Pirello}, {Pitkin},
  {Plastino}, {Poggiani}, {Pong}, {Ponrathnam}, {Popolizio}, {Porter},
  {Powell}, {Prajapati}, {Prasad}, {Prasai}, {Prasanna}, {Pratten},
  {Prestegard}, {Principe}, {Prodi}, {Prokhorov}, {Punturo}, {Puppo},
  {P{\"u}rrer}, {Qi}, {Quetschke}, {Quinonez}, {Raab}, {Raaijmakers},
  {Radkins}, {Radulesco}, {Raffai}, {Raja}, {Rajan}, {Rajbhandari},
  {Rakhmanov}, {Ramirez}, {Ramos-Buades}, {Rana}, {Rao}, {Rapagnani},
  {Raymond}, {Razzano}, {Read}, {Regimbau}, {Rei}, {Reid}, {Reitze},
  {Rettegno}, {Ricci}, {Richardson}, {Richardson}, {Ricker}, {Riemenschneider},
  {Riles}, {Rizzo}, {Robertson}, {Robinet}, {Rocchi}, {Rolland}, {Rollins},
  {Roma}, {Romanelli}, {Romano}, {Romel}, {Romie}, {Rose}, {Rose}, {Rose},
  {Rosell}, {Rosi{\'n}ska}, {Rosofsky}, {Ross}, {Rowan}, {Roy}, {R{\"u}diger},
  {Ruggi}, {Rutins}, {Ryan}, {Sachdev}, {Sadecki}, {Sakellariadou}, {Salafia},
  {Salconi}, {Saleem}, {Samajdar}, {Sammut}, {Sanchez}, {Sanchez},
  {Sanchis-Gual}, {Sanders}, {Santiago}, {Santos}, {Sarin}, {Sassolas},
  {Sathyaprakash}, {Sauter}, {Savage}, {Schale}, {Scheel}, {Scheuer},
  {Schmidt}, {Schnabel}, {Schofield}, {Sch{\"o}nbeck}, {Schreiber}, {Schulte},
  {Schutz}, {Scott}, {Scott}, {Seidel}, {Sellers}, {Sengupta}, {Sennett},
  {Sentenac}, {Sequino}, {Sergeev}, {Setyawati}, {Shaddock}, {Shaffer},
  {Shahriar}, {Shaner}, {Sharma}, {Sharma}, {Shawhan}, {Shen}, {Shink},
  {Shoemaker}, {Shoemaker}, {Shukla}, {ShyamSundar}, {Siellez}, {Sieniawska},
  {Sigg}, {Singer}, {Singh}, {Singh}, {Singhal}, {Sintes}, {Sitmukhambetov},
  {Skliris}, {Slagmolen}, {Slaven-Blair}, {Smith}, {Smith}, {Somala}, {Son},
  {Soni}, {Sorazu}, {Sorrentino}, {Souradeep}, {Sowell}, {Spencer}, {Spera},
  {Srivastava}, {Srivastava}, {Staats}, {Stachie}, {Standke}, {Steer},
  {Steinke}, {Steinlechner}, {Steinlechner}, {Steinmeyer}, {Stevenson},
  {Stocks}, {Stone}, {Stops}, {Strain}, {Stratta}, {Strigin}, {Strunk},
  {Sturani}, {Stuver}, {Sudhir}, {Summerscales}, {Sun}, {Sunil}, {Sur},
  {Suresh}, {Sutton}, {Swinkels}, {Szczepa{\'n}czyk}, {Tacca}, {Tait},
  {Talbot}, {Tanner}, {Tao}, {T{\'a}pai}, {Tapia}, {Tasson}, {Taylor},
  {Tenorio}, {Terkowski}, {Thomas}, {Thomas}, {Thondapu}, {Thorne}, {Thrane},
  {Tiwari}, {Tiwari}, {Tiwari}, {Toland}, {Tonelli}, {Tornasi},
  {Torres-Forn{\'e}}, {Torrie}, {T{\"o}yr{\"a}}, {Travasso}, {Traylor},
  {Tringali}, {Tripathee}, {Trovato}, {Trozzo}, {Tsang}, {Tse}, {Tso},
  {Tsukada}, {Tsuna}, {Tsutsui}, {Tuyenbayev}, {Ueno}, {Ugolini},
  {Unnikrishnan}, {Urban}, {Usman}, {Vahlbruch}, {Vajente}, {Valdes},
  {Valentini}, {van Bakel}, {van Beuzekom}, {van den Brand}, {Van Den Broeck},
  {Vander-Hyde}, {van der Schaaf}, {VanHeijningen}, {van Veggel}, {Vardaro},
  {Varma}, {Vass}, {Vas{\'u}th}, {Vecchio}, {Vedovato}, {Veitch}, {Veitch},
  {Venkateswara}, {Venugopalan}, {Verkindt}, {Vetrano}, {Vicer{\'e}}, {Viets},
  {Vinciguerra}, {Vine}, {Vinet}, {Vitale}, {Vo}, {Vocca}, {Vorvick},
  {Vyatchanin}, {Wade}, {Wade}, {Wade}, {Walet}, {Walker}, {Wallace}, {Walsh},
  {Wang}, {Wang}, {Wang}, {Wang}, {Ward}, {Warden}, {Warner}, {Was}, {Watchi},
  {Weaver}, {Wei}, {Weinert}, {Weinstein}, {Weiss}, {Wellmann}, {Wen},
  {Wessel}, {We{\ss}els}, {Westhouse}, {Wette}, {Whelan}, {White}, {Whiting},
  {Whittle}, {Wilken}, {Williams}, {Williamson}, {Willis}, {Willke}, {Winkler},
  {Wipf}, {Wittel}, {Woan}, {Woehler}, {Wofford}, {Wright}, {Wu}, {Wysocki},
  {Xiao}, {Xu}, {Yamamoto}, {Yancey}, {Yang}, {Yang}, {Yang}, {Yap}, {Yazback},
  {Yeeles}, {Yu}, {Yu}, {Yuen}, {Zadro{\.z}ny}, {Zadro{\.z}ny}, {Zanolin},
  {Zelenova}, {Zendri}, {Zevin}, {Zhang}, {Zhang}, {Zhang}, {Zhao}, {Zhao},
  {Zhou}, {Zhou}, {Zhu}, {Zimmerman}, {Zucker}, \&
  {Zweizig}}]{GW190425discovery}
---. 2020{\natexlab{a}}, \apjl, 892, L3

\bibitem[{{Abbott} {et~al.}(2020{\natexlab{b}}){Abbott}, {Abbott}, {Abbott},
  {Abraham}, {Acernese}, {Ackley}, {Adams}, {Adya}, {Affeldt}, {Agathos},
  {Agatsuma}, {Aggarwal}, {Aguiar}, {Aiello}, {Ain}, {Ajith}, {Alford},
  {Allen}, {Allocca}, {Aloy}, {Altin}, {Amato}, {Ananyeva}, {Anderson},
  {Anderson}, {Angelova}, {Antier}, {Appert}, {Arai}, {Araya}, {Areeda},
  {Ar{\`e}ne}, {Arnaud}, {Arun}, {Ascenzi}, {Ashton}, {Aston}, {Astone},
  {Aubin}, {Aufmuth}, {AultONeal}, {Austin}, {Avendano}, {Avila-Alvarez},
  {Babak}, {Bacon}, {Badaracco}, {Bader}, {Bae}, {Baker}, {Baldaccini},
  {Ballardin}, {Ballmer}, {Banagiri}, {Barayoga}, {Barclay}, {Barish},
  {Barker}, {Barkett}, {Barnum}, {Barone}, {Barr}, {Barsotti}, {Barsuglia},
  {Barta}, {Bartlett}, {Bartos}, {Bassiri}, {Basti}, {Bawaj}, {Bayley},
  {Bazzan}, {B{\'e}csy}, {Bejger}, {Belahcene}, {Bell}, {Beniwal}, {Berger},
  {Bergmann}, {Bernuzzi}, {Bero}, {Berry}, {Bersanetti}, {Bertolini},
  {Betzwieser}, {Bhandare}, {Bidler}, {Bilenko}, {Bilgili}, {Billingsley},
  {Birch}, {Birney}, {Birnholtz}, {Biscans}, {Biscoveanu}, {Bisht}, {Bitossi},
  {Bizouard}, {Blackburn}, {Blair}, {Blair}, {Blair}, {Bloemen}, {Bode},
  {Boer}, {Boetzel}, {Bogaert}, {Bondu}, {Bonilla}, {Bonnand}, {Booker},
  {Boom}, {Booth}, {Bork}, {Boschi}, {Bose}, {Bossie}, {Bossilkov}, {Bosveld},
  {Bouffanais}, {Bozzi}, {Bradaschia}, {Brady}, {Bramley}, {Branchesi}, {Brau},
  {Briant}, {Briggs}, {Brighenti}, {Brillet}, {Brinkmann}, {Brisson},
  {Brockill}, {Brooks}, {Brown}, {Brunett}, {Buikema}, {Bulik}, {Bulten},
  {Buonanno}, {Buskulic}, {Buy}, {Byer}, {Cabero}, {Cadonati}, {Cagnoli},
  {Cahillane}, {Calder{\'o}n Bustillo}, {Callister}, {Calloni}, {Camp},
  {Campbell}, {Canepa}, {Cannon}, {Cao}, {Cao}, {Capocasa}, {Carbognani},
  {Caride}, {Carney}, {Carullo}, {Casanueva Diaz}, {Casentini}, {Caudill},
  {Cavagli{\`a}}, {Cavalier}, {Cavalieri}, {Cella}, {Cerd{\'a}-Dur{\'a}n},
  {Cerretani}, {Cesarini}, {Chaibi}, {Chakravarti}, {Chamberlin}, {Chan},
  {Chao}, {Charlton}, {Chase}, {Chassande-Mottin}, {Chatterjee}, {Chaturvedi},
  {Chatziioannou}, {Cheeseboro}, {Chen}, {Chen}, {Chen}, {Cheng}, {Cheong},
  {Chia}, {Chincarini}, {Chiummo}, {Cho}, {Cho}, {Cho}, {Christensen}, {Chu},
  {Chua}, {Chung}, {Chung}, {Ciani}, {Ciobanu}, {Ciolfi}, {Cipriano}, {Cirone},
  {Clara}, {Clark}, {Clearwater}, {Cleva}, {Cocchieri}, {Coccia}, {Cohadon},
  {Cohen}, {Colgan}, {Colleoni}, {Collette}, {Collins}, {Cominsky},
  {Constancio}, {Conti}, {Cooper}, {Corban}, {Corbitt}, {Cordero-Carri{\'o}n},
  {Corley}, {Cornish}, {Corsi}, {Cortese}, {Costa}, {Cotesta}, {Coughlin},
  {Coughlin}, {Coulon}, {Countryman}, {Couvares}, {Covas}, {Cowan}, {Coward},
  {Cowart}, {Coyne}, {Coyne}, {Creighton}, {Creighton}, {Cripe}, {Croquette},
  {Crowder}, {Cullen}, {Cumming}, {Cunningham}, {Cuoco}, {Dal Canton},
  {D{\'a}lya}, {Danilishin}, {D'Antonio}, {Danzmann}, {Dasgupta}, {Da Silva
  Costa}, {Datrier}, {Dattilo}, {Dave}, {Davier}, {Davis}, {Daw}, {DeBra},
  {Deenadayalan}, {Degallaix}, {De Laurentis}, {Del{\'e}glise}, {Del Pozzo},
  {DeMarchi}, {Demos}, {Dent}, {De Pietri}, {Derby}, {De Rosa}, {De Rossi},
  {DeSalvo}, {de Varona}, {Dhurandhar}, {D{\'\i}az}, {Dietrich}, {Di Fiore},
  {Di Giovanni}, {Di Girolamo}, {Di Lieto}, {Ding}, {Di Pace}, {Di Palma}, {Di
  Renzo}, {Dmitriev}, {Doctor}, {Donovan}, {Dooley}, {Doravari}, {Dorrington},
  {Downes}, {Drago}, {Driggers}, {Du}, {Ducoin}, {Dupej}, {Dwyer}, {Easter},
  {Edo}, {Edwards}, {Effler}, {Ehrens}, {Eichholz}, {Eikenberry}, {Eisenmann},
  {Eisenstein}, {Essick}, {Estelles}, {Estevez}, {Etienne}, {Etzel}, {Evans},
  {Evans}, {Fafone}, {Fair}, {Fairhurst}, {Fan}, {Farinon}, {Farr}, {Farr},
  {Fauchon-Jones}, {Favata}, {Fays}, {Fazio}, {Fee}, {Feicht}, {Fejer}, {Feng},
  {Fernandez-Galiana}, {Ferrante}, {Ferreira}, {Ferreira}, {Ferrini},
  {Fidecaro}, {Fiori}, {Fiorucci}, {Fishbach}, {Fisher}, {Fishner},
  {Fitz-Axen}, {Flaminio}, {Fletcher}, {Flynn}, {Fong}, {Font}, {Forsyth},
  {Fournier}, {Frasca}, {Frasconi}, {Frei}, {Freise}, {Frey}, {Frey},
  {Fritschel}, {Frolov}, {Fulda}, {Fyffe}, {Gabbard}, {Gadre}, {Gaebel},
  {Gair}, {Gammaitoni}, {Ganija}, {Gaonkar}, {Garcia},
  {Garc{\'\i}a-Quir{\'o}s}, {Garufi}, {Gateley}, {Gaudio}, {Gaur}, {Gayathri},
  {Gemme}, {Genin}, {Gennai}, {George}, {George}, {Gergely}, {Germain},
  {Ghonge}, {Ghosh}, {Ghosh}, {Ghosh}, {Giacomazzo}, {Giaime}, {Giardina},
  {Giazotto}, {Gill}, {Giordano}, {Glover}, {Godwin}, {Goetz}, {Goetz},
  {Goncharov}, {Gonz{\'a}lez}, {Gonzalez Castro}, {Gopakumar}, {Gorodetsky},
  {Gossan}, {Gosselin}, {Gouaty}, {Grado}, {Graef}, {Granata}, {Grant}, {Gras},
  {Grassia}, {Gray}, {Gray}, {Greco}, {Green}, {Green}, {Gretarsson}, {Groot},
  {Grote}, {Grunewald}, {Gruning}, {Guidi}, {Gulati}, {Guo}, {Gupta}, {Gupta},
  {Gustafson}, {Gustafson}, {Haegel}, {Halim}, {Hall}, {Hall}, {Hamilton},
  {Hammond}, {Haney}, {Hanke}, {Hanks}, {Hanna}, {Hannam}, {Hannuksela},
  {Hanson}, {Hardwick}, {Haris}, {Harms}, {Harry}, {Harry}, {Haster},
  {Haughian}, {Hayes}, {Healy}, {Heidmann}, {Heintze}, {Heitmann}, {Hello},
  {Hemming}, {Hendry}, {Heng}, {Hennig}, {Heptonstall}, {Vivanco}, {Heurs},
  {Hild}, {Hinderer}, {Hoak}, {Hochheim}, {Hofman}, {Holgado}, {Holland},
  {Holt}, {Holz}, {Hopkins}, {Horst}, {Hough}, {Howell}, {Hoy}, {Hreibi},
  {Huerta}, {Huet}, {Hughey}, {Hulko}, {Husa}, {Huttner}, {Huynh-Dinh},
  {Idzkowski}, {Iess}, {Ingram}, {Inta}, {Intini}, {Irwin}, {Isa}, {Isac},
  {Isi}, {Iyer}, {Izumi}, {Jacqmin}, {Jadhav}, {Jani}, {Janthalur},
  {Jaranowski}, {Jenkins}, {Jiang}, {Johnson}, {Jones}, {Jones}, {Jones},
  {Jonker}, {Ju}, {Junker}, {Kalaghatgi}, {Kalogera}, {Kamai}, {Kandhasamy},
  {Kang}, {Kanner}, {Kapadia}, {Karki}, {Karvinen}, {Kashyap}, {Kasprzack},
  {Katsanevas}, {Katsavounidis}, {Katzman}, {Kaufer}, {Kawabe}, {Keerthana},
  {K{\'e}f{\'e}lian}, {Keitel}, {Kennedy}, {Key}, {Khalili}, {Khan}, {Khan},
  {Khan}, {Khan}, {Khazanov}, {Khursheed}, {Kijbunchoo}, {Kim}, {Kim}, {Kim},
  {Kim}, {Kim}, {Kim}, {Kim}, {Kimball}, {King}, {King}, {Kinley-Hanlon},
  {Kirchhoff}, {Kissel}, {Kleybolte}, {Klika}, {Klimenko}, {Knowles}, {Koch},
  {Koehlenbeck}, {Koekoek}, {Koley}, {Kondrashov}, {Kontos}, {Koper},
  {Korobko}, {Korth}, {Kowalska}, {Kozak}, {Kringel}, {Krishnendu},
  {Kr{\'o}lak}, {Kuehn}, {Kumar}, {Kumar}, {Kumar}, {Kumar}, {Kuo}, {Kutynia},
  {Kwang}, {Lackey}, {Lai}, {Lam}, {Landry}, {Lane}, {Lang}, {Lange}, {Lantz},
  {Lanza}, {Larson}, {Lartaux-Vollard}, {Lasky}, {Laxen}, {Lazzarini},
  {Lazzaro}, {Leaci}, {Leavey}, {Lecoeuche}, {Lee}, {Lee}, {Lee}, {Lee}, {Lee},
  {Lee}, {Lehmann}, {Lenon}, {Leroy}, {Letendre}, {Levin}, {Li}, {Li}, {Li},
  {Li}, {Lin}, {Linde}, {Linker}, {Littenberg}, {Liu}, {Liu}, {Lo},
  {Lockerbie}, {London}, {Longo}, {Lorenzini}, {Loriette}, {Lormand},
  {Losurdo}, {Lough}, {Lousto}, {Lovelace}, {Lower}, {L{\"u}ck}, {Lumaca},
  {Lundgren}, {Lynch}, {Ma}, {Macas}, {Macfoy}, {MacInnis}, {Macleod},
  {Macquet}, {Maga{\~n}a-Sandoval}, {Maga{\~n}a Zertuche}, {Magee}, {Majorana},
  {Maksimovic}, {Malik}, {Man}, {Mandic}, {Mangano}, {Mansell}, {Manske},
  {Mantovani}, {Marchesoni}, {Marion}, {M{\'a}rka}, {M{\'a}rka}, {Markakis},
  {Markosyan}, {Markowitz}, {Maros}, {Marquina}, {Marsat}, {Martelli},
  {Martin}, {Martin}, {Martynov}, {Mason}, {Massera}, {Masserot}, {Massinger},
  {Masso-Reid}, {Mastrogiovanni}, {Matas}, {Matichard}, {Matone}, {Mavalvala},
  {Mazumder}, {McCann}, {McCarthy}, {McClelland}, {McCormick}, {McCuller},
  {McGuire}, {McIver}, {McManus}, {McRae}, {McWilliams}, {Meacher}, {Meadors},
  {Mehmet}, {Mehta}, {Meidam}, {Melatos}, {Mendell}, {Mercer}, {Mereni},
  {Merilh}, {Merzougui}, {Meshkov}, {Messenger}, {Messick}, {Metzdorff},
  {Meyers}, {Miao}, {Michel}, {Middleton}, {Mikhailov}, {Milano}, {Miller},
  {Miller}, {Millhouse}, {Mills}, {Milovich-Goff}, {Minazzoli}, {Minenkov},
  {Mishkin}, {Mishra}, {Mistry}, {Mitra}, {Mitrofanov}, {Mitselmakher},
  {Mittleman}, {Mo}, {Moffa}, {Mogushi}, {Mohapatra}, {Montani}, {Moore},
  {Moraru}, {Moreno}, {Morisaki}, {Mours}, {Mow-Lowry}, {Mukherjee},
  {Mukherjee}, {Mukherjee}, {Mukund}, {Mullavey}, {Munch}, {Mu{\~n}iz},
  {Muratore}, {Murray}, {Nagar}, {Nardecchia}, {Naticchioni}, {Nayak},
  {Neilson}, {Nelemans}, {Nelson}, {Nery}, {Neunzert}, {Ng}, {Ng}, {Nguyen},
  {Nichols}, {Nissanke}, {Nocera}, {North}, {Nuttall}, {Obergaulinger},
  {Oberling}, {O'Brien}, {O'Dea}, {Ogin}, {Oh}, {Oh}, {Ohme}, {Ohta}, {Okada},
  {Oliver}, {Oppermann}, {Oram}, {O'Reilly}, {Ormiston}, {Ortega},
  {O'Shaughnessy}, {Ossokine}, {Ottaway}, {Overmier}, {Owen}, {Pace}, {Pagano},
  {Page}, {Pai}, {Pai}, {Palamos}, {Palashov}, {Palomba}, {Pal-Singh}, {Pan},
  {Pang}, {Pang}, {Pankow}, {Pannarale}, {Pant}, {Paoletti}, {Paoli}, {Parida},
  {Parker}, {Pascucci}, {Pasqualetti}, {Passaquieti}, {Passuello}, {Patil},
  {Patricelli}, {Pearlstone}, {Pedersen}, {Pedraza}, {Pedurand}, {Pele},
  {Penn}, {Perez}, {Perreca}, {Pfeiffer}, {Phelps}, {Phukon}, {Piccinni},
  {Pichot}, {Piergiovanni}, {Pillant}, {Pinard}, {Pirello}, {Pitkin},
  {Poggiani}, {Pong}, {Ponrathnam}, {Popolizio}, {Porter}, {Powell},
  {Prajapati}, {Prasad}, {Prasai}, {Prasanna}, {Pratten}, {Prestegard},
  {Privitera}, {Prodi}, {Prokhorov}, {Puncken}, {Punturo}, {Puppo},
  {P{\"u}rrer}, {Qi}, {Quetschke}, {Quinonez}, {Quintero}, {Quitzow-James},
  {Raab}, {Radkins}, {Radulescu}, {Raffai}, {Raja}, {Rajan}, {Rajbhandari},
  {Rakhmanov}, {Ramirez}, {Ramos-Buades}, {Rana}, {Rao}, {Rapagnani},
  {Raymond}, {Razzano}, {Read}, {Regimbau}, {Rei}, {Reid}, {Reitze}, {Ren},
  {Ricci}, {Richardson}, {Richardson}, {Ricker}, {Riles}, {Rizzo}, {Robertson},
  {Robie}, {Robinet}, {Rocchi}, {Rolland}, {Rollins}, {Roma}, {Romanelli},
  {Romano}, {Romel}, {Romie}, {Rose}, {Rosi{\'n}ska}, {Rosofsky}, {Ross},
  {Rowan}, {R{\"u}diger}, {Ruggi}, {Rutins}, {Ryan}, {Sachdev}, {Sadecki},
  {Sakellariadou}, {Salconi}, {Saleem}, {Samajdar}, {Sammut}, {Sanchez},
  {Sanchez}, {Sanchis-Gual}, {Sandberg}, {Sanders}, {Santiago}, {Sarin},
  {Sassolas}, {Sathyaprakash}, {Saulson}, {Sauter}, {Savage}, {Schale},
  {Scheel}, {Scheuer}, {Schmidt}, {Schnabel}, {Schofield}, {Sch{\"o}nbeck},
  {Schreiber}, {Schulte}, {Schutz}, {Schwalbe}, {Scott}, {Scott}, {Seidel},
  {Sellers}, {Sengupta}, {Sennett}, {Sentenac}, {Sequino}, {Sergeev},
  {Setyawati}, {Shaddock}, {Shaffer}, {Shahriar}, {Shaner}, {Shao}, {Sharma},
  {Shawhan}, {Shen}, {Shink}, {Shoemaker}, {Shoemaker}, {ShyamSundar},
  {Siellez}, {Sieniawska}, {Sigg}, {Silva}, {Singer}, {Singh}, {Singhal},
  {Sintes}, {Sitmukhambetov}, {Skliris}, {Slagmolen}, {Slaven-Blair}, {Smith},
  {Smith}, {Somala}, {Son}, {Sorazu}, {Sorrentino}, {Souradeep}, {Sowell},
  {Spencer}, {Srivastava}, {Srivastava}, {Staats}, {Stachie}, {Standke},
  {Steer}, {Steinke}, {Steinlechner}, {Steinlechner}, {Steinmeyer},
  {Stevenson}, {Stocks}, {Stone}, {Stops}, {Strain}, {Stratta}, {Strigin},
  {Strunk}, {Sturani}, {Stuver}, {Sudhir}, {Summerscales}, {Sun}, {Sunil},
  {Suresh}, {Sutton}, {Swinkels}, {Szczepa{\'n}czyk}, {Tacca}, {Tait},
  {Talbot}, {Talukder}, {Tanner}, {T{\'a}pai}, {Taracchini}, {Tasson},
  {Taylor}, {Thies}, {Thomas}, {Thomas}, {Thondapu}, {Thorne}, {Thrane},
  {Tiwari}, {Tiwari}, {Tiwari}, {Toland}, {Tonelli}, {Tornasi},
  {Torres-Forn{\'e}}, {Torrie}, {T{\"o}yr{\"a}}, {Travasso}, {Traylor},
  {Tringali}, {Trovato}, {Trozzo}, {Trudeau}, {Tsang}, {Tse}, {Tso}, {Tsukada},
  {Tsuna}, {Tuyenbayev}, {Ueno}, {Ugolini}, {Unnikrishnan}, {Urban}, {Usman},
  {Vahlbruch}, {Vajente}, {Valdes}, {van Bakel}, {van Beuzekom}, {van den
  Brand}, {Van Den Broeck}, {Vander-Hyde}, {van Heijningen}, {van der Schaaf},
  {van Veggel}, {Vardaro}, {Varma}, {Vass}, {Vas{\'u}th}, {Vecchio},
  {Vedovato}, {Veitch}, {Veitch}, {Venkateswara}, {Venugopalan}, {Verkindt},
  {Vetrano}, {Vicer{\'e}}, {Viets}, {Vine}, {Vinet}, {Vitale}, {Vo}, {Vocca},
  {Vorvick}, {Vyatchanin}, {Wade}, {Wade}, {Wade}, {Walet}, {Walker},
  {Wallace}, {Walsh}, {Wang}, {Wang}, {Wang}, {Wang}, {Wang}, {Ward}, {Warden},
  {Warner}, {Was}, {Watchi}, {Weaver}, {Wei}, {Weinert}, {Weinstein}, {Weiss},
  {Wellmann}, {Wen}, {Wessel}, {We{\ss}els}, {Westhouse}, {Wette}, {Whelan},
  {Whiting}, {Whittle}, {Wilken}, {Williams}, {Williamson}, {Willis}, {Willke},
  {Wimmer}, {Winkler}, {Wipf}, {Wittel}, {Woan}, {Woehler}, {Wofford},
  {Worden}, {Wright}, {Wu}, {Wysocki}, {Xiao}, {Yamamoto}, {Yancey}, {Yang},
  {Yap}, {Yazback}, {Yeeles}, {Yu}, {Yu}, {Yuen}, {Yvert}, {Zadro{\.z}ny},
  {Zanolin}, {Zelenova}, {Zendri}, {Zevin}, {Zhang}, {Zhang}, {Zhang}, {Zhao},
  {Zhou}, {Zhou}, {Zhu}, {Zucker}, {Zweizig}, {LIGO Scientific Collaboration},
  \& {the Virgo Collaboration}}]{GWdataguide}
---. 2020{\natexlab{b}}, Classical and Quantum Gravity, 37, 055002

\bibitem[{{Abbott} {et~al.}(2020{\natexlab{c}}){Abbott}, {Abbott}, {Abraham},
  {Acernese}, {Ackley}, {Adams}, {Adams}, {Adhikari}, {Adya}, {Affeldt},
  {Agathos}, {Agatsuma}, {Aggarwal}, {Aguiar}, {Aiello}, {Ain}, {Ajith},
  {Akcay}, {Allen}, {Allocca}, {Altin}, {Amato}, {Anand}, {Ananyeva},
  {Anderson}, {Anderson}, {Angelova}, {Ansoldi}, {Antelis}, {Antier}, {Appert},
  {Arai}, {Araya}, {Areeda}, {Ar{\`e}ne}, {Arnaud}, {Aronson}, {Arun}, {Asali},
  {Ascenzi}, {Ashton}, {Aston}, {Astone}, {Aubin}, {Aufmuth}, {AultONeal},
  {Austin}, {Avendano}, {Babak}, {Badaracco}, {Bader}, {Bae}, {Baer},
  {Bagnasco}, {Baird}, {Ball}, {Ballardin}, {Ballmer}, {Bals}, {Balsamo},
  {Baltus}, {Banagiri}, {Bankar}, {Bankar}, {Barayoga}, {Barbieri}, {Barish},
  {Barker}, {Barneo}, {Barnum}, {Barone}, {Barr}, {Barsotti}, {Barsuglia},
  {Barta}, {Bartlett}, {Bartos}, {Bassiri}, {Basti}, {Bawaj}, {Bayley},
  {Bazzan}, {Becher}, {B{\'e}csy}, {Bedakihale}, {Bejger}, {Belahcene},
  {Beniwal}, {Benjamin}, {Bennett}, {Bentley}, {Bergamin}, {Berger},
  {Bergmann}, {Bernuzzi}, {Berry}, {Bersanetti}, {Bertolini}, {Betzwieser},
  {Bhandare}, {Bhandari}, {Bhattacharjee}, {Bidler}, {Bilenko}, {Billingsley},
  {Birney}, {Birnholtz}, {Biscans}, {Bischi}, {Biscoveanu}, {Bisht}, {Bitossi},
  {Bizouard}, {Blackburn}, {Blackman}, {Blair}, {Blair}, {Blair}, {Blanch},
  {Bobba}, {Bode}, {Boer}, {Boetzel}, {Bogaert}, {Boldrini}, {Bondu},
  {Bonnand}, {Bonilla}, {Booker}, {Boom}, {Bork}, {Boschi}, {Bose},
  {Bossilkov}, {Boudart}, {Bouffanais}, {Bozzi}, {Bradaschia}, {Brady},
  {Bramley}, {Branchesi}, {Brau}, {Breschi}, {Briant}, {Briggs}, {Brighenti},
  {Brillet}, {Brinkmann}, {Brockill}, {Brooks}, {Brooks}, {Brown}, {Brunett},
  {Bruno}, {Bruntz}, {Buikema}, {Bulik}, {Bulten}, {Buonanno}, {Buscicchio},
  {Buskulic}, {Byer}, {Cabero}, {Cadonati}, {Caesar}, {Cagnoli}, {Cahillane},
  {Calder{\'o}n Bustillo}, {Callaghan}, {Callister}, {Calloni}, {Camp},
  {Canepa}, {Cannon}, {Cao}, {Cao}, {Carapella}, {Carbognani}, {Carney},
  {Carpinelli}, {Carullo}, {Carver}, {Casanueva Diaz}, {Casentini}, {Caudill},
  {Cavagli{\`a}}, {Cavalier}, {Cavalieri}, {Cella}, {Cerd{\'a}-Dur{\'a}n},
  {Cesarini}, {Chaibi}, {Chakravarti}, {Chan}, {Chan}, {Chandra}, {Chanial},
  {Chao}, {Charlton}, {Chase}, {Chassande-Mottin}, {Chatterjee},
  {Chattopadhyay}, {Chaturvedi}, {Chatziioannou}, {Chen}, {Chen}, {Chen},
  {Chen}, {Cheng}, {Cheong}, {Chia}, {Chiadini}, {Chierici}, {Chincarini},
  {Chiummo}, {Cho}, {Cho}, {Cho}, {Choate}, {Christensen}, {Chu}, {Chua},
  {Chung}, {Chung}, {Ciani}, {Ciecielag}, {Cie{\'s}lar}, {Cifaldi}, {Ciobanu},
  {Ciolfi}, {Cipriano}, {Cirone}, {Clara}, {Clark}, {Clark}, {Clarke},
  {Clearwater}, {Clesse}, {Cleva}, {Coccia}, {Cohadon}, {Cohen}, {Colleoni},
  {Collette}, {Collins}, {Colpi}, {Constancio}, {Conti}, {Cooper}, {Corban},
  {Corbitt}, {Cordero-Carri{\'o}n}, {Corezzi}, {Corley}, {Cornish}, {Corre},
  {Corsi}, {Cortese}, {Costa}, {Cotesta}, {Coughlin}, {Coughlin}, {Coulon},
  {Countryman}, {Cousins}, {Couvares}, {Covas}, {Coward}, {Cowart}, {Coyne},
  {Coyne}, {Creighton}, {Creighton}, {Croquette}, {Crowder}, {Cudell},
  {Cullen}, {Cumming}, {Cummings}, {Cunningham}, {Cuoco}, {Curylo}, {Dal
  Canton}, {D{\'a}lya}, {Dana}, {DaneshgaranBajastani}, {D'Angelo}, {Danila},
  {Danilishin}, {D'Antonio}, {Danzmann}, {Darsow-Fromm}, {Dasgupta}, {Datrier},
  {Dattilo}, {Dave}, {Davier}, {Davies}, {Davis}, {Daw}, {Dean}, {DeBra},
  {Deenadayalan}, {Degallaix}, {De Laurentis}, {Del{\'e}glise}, {Del Favero},
  {De Lillo}, {De Lillo}, {Del Pozzo}, {DeMarchi}, {De Matteis}, {D'Emilio},
  {Demos}, {Denker}, {Dent}, {Depasse}, {De Pietri}, {De Rosa}, {De Rossi},
  {DeSalvo}, {de Varona}, {Dhurandhar}, {D{\'\i}az}, {Diaz-Ortiz}, {Didio},
  {Dietrich}, {Di Fiore}, {DiFronzo}, {Di Giorgio}, {Di Giovanni}, {Di
  Giovanni}, {Di Girolamo}, {Di Lieto}, {Ding}, {Di Pace}, {Di Palma}, {Di
  Renzo}, {Divakarla}, {Dmitriev}, {Doctor}, {D'Onofrio}, {Donovan}, {Dooley},
  {Doravari}, {Dorrington}, {Downes}, {Drago}, {Driggers}, {Du}, {Ducoin},
  {Dupej}, {Durante}, {D'Urso}, {Duverne}, {Dwyer}, {Easter}, {Eddolls},
  {Edelman}, {Edo}, {Edy}, {Effler}, {Eichholz}, {Eikenberry}, {Eisenmann},
  {Eisenstein}, {Ejlli}, {Errico}, {Essick}, {Estell{\'e}s}, {Estevez},
  {Etienne}, {Etzel}, {Evans}, {Evans}, {Ewing}, {Fafone}, {Fair}, {Fairhurst},
  {Fan}, {Farah}, {Farinon}, {Farr}, {Farr}, {Fauchon-Jones}, {Favata}, {Fays},
  {Fazio}, {Feicht}, {Fejer}, {Feng}, {Fenyvesi}, {Ferguson},
  {Fernandez-Galiana}, {Ferrante}, {Ferreira}, {Fidecaro}, {Figura}, {Fiori},
  {Fiorucci}, {Fishbach}, {Fisher}, {Fishner}, {Fittipaldi}, {Fitz-Axen},
  {Fiumara}, {Flaminio}, {Floden}, {Flynn}, {Fong}, {Font}, {Forsyth},
  {Fournier}, {Frasca}, {Frasconi}, {Frei}, {Freise}, {Frey}, {Frey},
  {Fritschel}, {Frolov}, {Fronz{\'e}}, {Fulda}, {Fyffe}, {Gabbard}, {Gadre},
  {Gaebel}, {Gair}, {Gais}, {Galaudage}, {Gamba}, {Ganapathy}, {Ganguly},
  {Gaonkar}, {Garaventa}, {Garc{\'\i}a-Quir{\'o}s}, {Garufi}, {Gateley},
  {Gaudio}, {Gayathri}, {Gemme}, {Gennai}, {George}, {George}, {George},
  {Gergely}, {Ghonge}, {Ghosh}, {Ghosh}, {Ghosh}, {Giacomazzo}, {Giacoppo},
  {Giaime}, {Giardina}, {Gibson}, {Gier}, {Gill}, {Giri}, {Glanzer}, {Gleckl},
  {Godwin}, {Goetz}, {Goetz}, {Gohlke}, {Goncharov}, {Gonz{\'a}lez},
  {Gopakumar}, {Gossan}, {Gosselin}, {Gouaty}, {Grace}, {Grado}, {Granata},
  {Granata}, {Grant}, {Gras}, {Grassia}, {Gray}, {Gray}, {Greco}, {Green},
  {Green}, {Gretarsson}, {Griggs}, {Grignani}, {Grimaldi}, {Grimes}, {Grimm},
  {Grote}, {Grunewald}, {Gruning}, {Guerrero}, {Guidi}, {Guimaraes},
  {Guix{\'e}}, {Gulati}, {Guo}, {Gupta}, {Gupta}, {Gupta}, {Gustafson},
  {Gustafson}, {Guzman}, {Haegel}, {Halim}, {Hall}, {Hamilton}, {Hammond},
  {Haney}, {Hanke}, {Hanks}, {Hanna}, {Hannam}, {Hannuksela}, {Hannuksela},
  {Hansen}, {Hansen}, {Hanson}, {Harder}, {Hardwick}, {Haris}, {Harms},
  {Harry}, {Harry}, {Hartwig}, {Hasskew}, {Haster}, {Haughian}, {Hayes},
  {Healy}, {Heidmann}, {Heintze}, {Heinze}, {Heinzel}, {Heitmann}, {Hellman},
  {Hello}, {Helmling-Cornell}, {Hemming}, {Hendry}, {Heng}, {Hennes}, {Hennig},
  {Hennig}, {Hernandez Vivanco}, {Heurs}, {Hild}, {Hill}, {Hines}, {Hochheim},
  {Hofgard}, {Hofman}, {Hohmann}, {Holgado}, {Holland}, {Hollows}, {Holmes},
  {Holt}, {Holz}, {Hopkins}, {Horst}, {Hough}, {Howell}, {Hoy}, {Hoyland},
  {Huang}, {H{\"u}bner}, {Huddart}, {Huerta}, {Hughey}, {Hui}, {Husa},
  {Huttner}, {Hutzler}, {Huxford}, {Huynh-Dinh}, {Idzkowski}, {Iess},
  {Imperato}, {Inchauspe}, {Ingram}, {Intini}, {Isi}, {Iyer},
  {JaberianHamedan}, {Jacqmin}, {Jadhav}, {Jadhav}, {James}, {Jani},
  {Janssens}, {Janthalur}, {Jaranowski}, {Jariwala}, {Jaume}, {Jenkins},
  {Jeunon}, {Jiang}, {Johns}, {Johnson-McDaniel}, {Jones}, {Jones}, {Jones},
  {Jones}, {Jones}, {Jonker}, {Ju}, {Junker}, {Kalaghatgi}, {Kalogera},
  {Kamai}, {Kandhasamy}, {Kang}, {Kanner}, {Kapadia}, {Kapasi}, {Karathanasis},
  {Karki}, {Kashyap}, {Kasprzack}, {Kastaun}, {Katsanevas}, {Katsavounidis},
  {Katzman}, {Kawabe}, {K{\'e}f{\'e}lian}, {Keitel}, {Key}, {Khadka},
  {Khalili}, {Khan}, {Khan}, {Khazanov}, {Khetan}, {Khursheed}, {Kijbunchoo},
  {Kim}, {Kim}, {Kim}, {Kim}, {Kim}, {Kim}, {Kimball}, {King}, {Kinley-Hanlon},
  {Kirchhoff}, {Kissel}, {Kleybolte}, {Klimenko}, {Knowles}, {Knyazev}, {Koch},
  {Koehlenbeck}, {Koekoek}, {Koley}, {Kolstein}, {Komori}, {Kondrashov},
  {Kontos}, {Koper}, {Korobko}, {Korth}, {Kovalam}, {Kozak}, {Kr{\"a}mer},
  {Kringel}, {Krishnendu}, {Kr{\'o}lak}, {Kuehn}, {Kumar}, {Kumar}, {Kumar},
  {Kumar}, {Kuns}, {Kwang}, {Lackey}, {Laghi}, {Lalande}, {Lam}, {Lamberts},
  {Landry}, {Lane}, {Lang}, {Lange}, {Lantz}, {Lanza}, {La Rosa},
  {Lartaux-Vollard}, {Lasky}, {Laxen}, {Lazzarini}, {Lazzaro}, {Leaci},
  {Leavey}, {Lecoeuche}, {Lee}, {Lee}, {Lee}, {Lee}, {Lehmann}, {Leon},
  {Leroy}, {Letendre}, {Levin}, {Li}, {Li}, {Li}, {Li}, {Li}, {Linde},
  {Linker}, {Linley}, {Littenberg}, {Liu}, {Liu}, {Llorens-Monteagudo}, {Lo},
  {Lockwood}, {London}, {Longo}, {Lorenzini}, {Loriette}, {Lormand}, {Losurdo},
  {Lough}, {Lousto}, {Lovelace}, {L{\"u}ck}, {Lumaca}, {Lundgren}, {Ma},
  {Macas}, {MacInnis}, {Macleod}, {MacMillan}, {Macquet}, {Maga{\~n}a
  Hernandez}, {Maga{\~n}a-Sandoval}, {Magazz{\`u}}, {Magee}, {Majorana},
  {Maksimovic}, {Maliakal}, {Malik}, {Man}, {Mandic}, {Mangano}, {Mansell},
  {Manske}, {Mantovani}, {Mapelli}, {Marchesoni}, {Marion}, {M{\'a}rka},
  {M{\'a}rka}, {Markakis}, {Markosyan}, {Markowitz}, {Maros}, {Marquina},
  {Marsat}, {Martelli}, {Martin}, {Martin}, {Martinez}, {Martinez}, {Martynov},
  {Masalehdan}, {Mason}, {Massera}, {Masserot}, {Massinger}, {Masso-Reid},
  {Mastrogiovanni}, {Matas}, {Mateu-Lucena}, {Matichard}, {Matiushechkina},
  {Mavalvala}, {Maynard}, {McCann}, {McCarthy}, {McClelland}, {McCormick},
  {McCuller}, {McGuire}, {McIsaac}, {McIver}, {McManus}, {McRae}, {McWilliams},
  {Meacher}, {Meadors}, {Mehmet}, {Mehta}, {Melatos}, {Melchor}, {Mendell},
  {Menendez-Vazquez}, {Mercer}, {Mereni}, {Merfeld}, {Merilh}, {Merritt},
  {Merzougui}, {Meshkov}, {Messenger}, {Messick}, {Metzdorff}, {Meyers},
  {Meylahn}, {Mhaske}, {Miani}, {Miao}, {Michaloliakos}, {Michel}, {Middleton},
  {Milano}, {Miller}, {Millhouse}, {Mills}, {Milotti}, {Milovich-Goff},
  {Minazzoli}, {Minenkov}, {Mir}, {Mishkin}, {Mishra}, {Mistry}, {Mitra},
  {Mitrofanov}, {Mitselmakher}, {Mittleman}, {Mo}, {Mogushi}, {Mohapatra},
  {Mohite}, {Molina}, {Molina-Ruiz}, {Mondin}, {Montani}, {Moore}, {Moraru},
  {Morawski}, {Moreno}, {Morisaki}, {Mours}, {Mow-Lowry}, {Mozzon},
  {Muciaccia}, {Mukherjee}, {Mukherjee}, {Mukherjee}, {Mukherjee}, {Mukund},
  {Mullavey}, {Munch}, {Mu{\~n}iz}, {Murray}, {Nadji}, {Nagar}, {Nardecchia},
  {Naticchioni}, {Nayak}, {Neil}, {Neilson}, {Nelemans}, {Nelson}, {Nery},
  {Neunzert}, {Nitz}, {Ng}, {Ng}, {Nguyen}, {Nguyen}, {Nguyen}, {Nichols},
  {Nissanke}, {Nocera}, {Noh}, {North}, {Nothard}, {Nuttall}, {Oberling},
  {O'Brien}, {O'Dell}, {Oganesyan}, {Ogin}, {Oh}, {Oh}, {Ohme}, {Ohta},
  {Okada}, {Olivetto}, {Oppermann}, {Oram}, {O'Reilly}, {Ormiston}, {Ortega},
  {O'Shaughnessy}, {Ossokine}, {Osthelder}, {Ottaway}, {Overmier}, {Owen},
  {Pace}, {Pagano}, {Page}, {Pagliaroli}, {Pai}, {Pai}, {Palamos}, {Palashov},
  {Palomba}, {Pan}, {Panda}, {Pang}, {Pankow}, {Pannarale}, {Pant}, {Paoletti},
  {Paoli}, {Paolone}, {Parker}, {Pascucci}, {Pasqualetti}, {Passaquieti},
  {Passuello}, {Patel}, {Patricelli}, {Payne}, {Pechsiri}, {Pedraza},
  {Pegoraro}, {Pele}, {Penn}, {Perego}, {Perez}, {P{\'e}rigois}, {Perreca},
  {Perri{\`e}s}, {Petermann}, {Petterson}, {Pfeiffer}, {Pham}, {Phukon},
  {Piccinni}, {Pichot}, {Piendibene}, {Piergiovanni}, {Pierini}, {Pierro},
  {Pillant}, {Pilo}, {Pinard}, {Pinto}, {Piotrzkowski}, {Pirello}, {Pitkin},
  {Placidi}, {Plastino}, {Pluchar}, {Poggiani}, {Polini}, {Pong}, {Ponrathnam},
  {Popolizio}, {Porter}, {Poverman}, {Powell}, {Pracchia}, {Prajapati},
  {Prasai}, {Prasanna}, {Pratten}, {Prestegard}, {Principe}, {Prodi},
  {Prokhorov}, {Prosposito}, {Prudenzi}, {Puecher}, {Punturo}, {Puosi},
  {Puppo}, {P{\"u}rrer}, {Qi}, {Quetschke}, {Quinonez}, {Quitzow-James},
  {Raab}, {Raaijmakers}, {Radkins}, {Radulesco}, {Raffai}, {Rafferty}, {Rail},
  {Raja}, {Rajan}, {Rajbhandari}, {Rakhmanov}, {Ramirez}, {Ramirez},
  {Ramos-Buades}, {Rana}, {Rao}, {Rapagnani}, {Rapol}, {Ratto}, {Raymond},
  {Razzano}, {Read}, {Regimbau}, {Rei}, {Reid}, {Reitze}, {Rettegno}, {Ricci},
  {Richardson}, {Richardson}, {Richardson}, {Ricker}, {Riemenschneider},
  {Riles}, {Rizzo}, {Robertson}, {Robinet}, {Rocchi}, {Rocha}, {Rodriguez},
  {Rodriguez-Soto}, {Rolland}, {Rollins}, {Roma}, {Romanelli}, {Romano},
  {Romel}, {Romero}, {Romero-Shaw}, {Romie}, {Ronchini}, {Rose}, {Rose},
  {Rose}, {Rosell}, {Rosi{\'n}ska}, {Rosofsky}, {Ross}, {Rowan}, {Rowlinson},
  {Roy}, {Roy}, {Ruggi}, {Ryan}, {Sachdev}, {Sadecki}, {Sadiq},
  {Sakellariadou}, {Salafia}, {Salconi}, {Saleem}, {Samajdar}, {Sanchez},
  {Sanchez}, {Sanchez}, {Sanchis-Gual}, {Sanders}, {Sandles}, {Santiago},
  {Santos}, {Saravanan}, {Sarin}, {Sassolas}, {Sathyaprakash}, {Sauter},
  {Savage}, {Savant}, {Sawant}, {Sayah}, {Schaetzl}, {Schale}, {Scheel},
  {Scheuer}, {Schindler-Tyka}, {Schmidt}, {Schnabel}, {Schofield},
  {Sch{\"o}nbeck}, {Schreiber}, {Schulte}, {Schutz}, {Schwarm}, {Schwartz},
  {Scott}, {Scott}, {Seglar-Arroyo}, {Seidel}, {Sellers}, {Sengupta},
  {Sennett}, {Sentenac}, {Sequino}, {Sergeev}, {Setyawati}, {Shaffer},
  {Shahriar}, {Sharifi}, {Sharma}, {Sharma}, {Shawhan}, {Shen}, {Shikauchi},
  {Shink}, {Shoemaker}, {Shoemaker}, {Shukla}, {ShyamSundar}, {Sieniawska},
  {Sigg}, {Singer}, {Singh}, {Singh}, {Singha}, {Singhal}, {Sintes}, {Sipala},
  {Skliris}, {Slagmolen}, {Slaven-Blair}, {Smetana}, {Smith}, {Smith},
  {Somala}, {Son}, {Soni}, {Soni}, {Sorazu}, {Sordini}, {Sorrentino},
  {Sorrentino}, {Soulard}, {Souradeep}, {Sowell}, {Spencer}, {Spera},
  {Srivastava}, {Srivastava}, {Staats}, {Stachie}, {Steer}, {Steinhoff},
  {Steinke}, {Steinlechner}, {Steinlechner}, {Steinmeyer}, {Stevenson},
  {Stolle-McAllister}, {Stops}, {Stover}, {Strain}, {Stratta}, {Strunk},
  {Sturani}, {Stuver}, {S{\"u}dbeck}, {Sudhagar}, {Sudhir}, {Suh},
  {Summerscales}, {Sun}, {Sun}, {Sunil}, {Sur}, {Suresh}, {Sutton}, {Swinkels},
  {Szczepa{\'n}czyk}, {Tacca}, {Tait}, {Talbot}, {Tanasijczuk}, {Tanner},
  {Tao}, {Tapia}, {Tapia San Martin}, {Tasson}, {Taylor}, {Tenorio},
  {Terkowski}, {Thirugnanasambandam}, {Thomas}, {Thomas}, {Thomas}, {Thompson},
  {Thondapu}, {Thorne}, {Thrane}, {Tiwari}, {Tiwari}, {Tiwari}, {Toland},
  {Tolley}, {Tonelli}, {Tornasi}, {Torres-Forn{\'e}}, {Torrie}, {Melo},
  {T{\"o}yr{\"a}}, {Tran}, {Trapananti}, {Travasso}, {Traylor}, {Tringali},
  {Tripathee}, {Trovato}, {Trudeau}, {Tsai}, {Tsang}, {Tse}, {Tso}, {Tsukada},
  {Tsuna}, {Tsutsui}, {Turconi}, {Ubhi}, {Udall}, {Ueno}, {Ugolini},
  {Unnikrishnan}, {Urban}, {Usman}, {Utina}, {Vahlbruch}, {Vajente}, {Vajpeyi},
  {Valdes}, {Valentini}, {Valsan}, {van Bakel}, {van Beuzekom}, {van den
  Brand}, {Van Den Broeck}, {Vander-Hyde}, {van der Schaaf}, {van Heijningen},
  {Vardaro}, {Vargas}, {Varma}, {Vass}, {Vas{\'u}th}, {Vecchio}, {Vedovato},
  {Veitch}, {Veitch}, {Venkateswara}, {Venneberg}, {Venugopalan}, {Verkindt},
  {Verma}, {Veske}, {Vetrano}, {Vicer{\'e}}, {Viets}, {Vijaykumar},
  {Villa-Ortega}, {Vinet}, {Vitale}, {Vo}, {Vocca}, {Vorvick}, {Vyatchanin},
  {Wade}, {Wade}, {Wade}, {Walet}, {Walker}, {Wallace}, {Wallace}, {Walsh},
  {Wang}, {Wang}, {Wang}, {Wang}, {Ward}, {Warner}, {Was}, {Washington},
  {Watchi}, {Weaver}, {Wei}, {Weinert}, {Weinstein}, {Weiss}, {Wellmann},
  {Wen}, {We{\ss}els}, {Westhouse}, {Wette}, {Whelan}, {White}, {White},
  {Whiting}, {Whittle}, {Wilken}, {Williams}, {Williams}, {Williamson},
  {Willis}, {Willke}, {Wilson}, {Wimmer}, {Winkler}, {Wipf}, {Woan}, {Woehler},
  {Wofford}, {Wong}, {Wrangel}, {Wright}, {Wu}, {Wysocki}, {Xiao}, {Yamamoto},
  {Yang}, {Yang}, {Yang}, {Yap}, {Yeeles}, {Yoon}, {Yu}, {Yu}, {Yuen},
  {Zadro{\.z}ny}, {Zanolin}, {Zelenova}, {Zendri}, {Zevin}, {Zhang}, {Zhang},
  {Zhang}, {Zhang}, {Zhao}, {Zhao}, {Zheng}, {Zhou}, {Zhou}, {Zhu},
  {Zimmerman}, {Zlochower}, {Zucker}, \& {Zweizig}}]{GWTC2}
{Abbott}, R., {Abbott}, T.~D., {Abraham}, S., {et~al.} 2020{\natexlab{c}},
  arXiv e-prints, arXiv:2010.14527

\bibitem[{{Abbott} {et~al.}(2020{\natexlab{d}}){Abbott}, {Abbott}, {Abraham},
  {Acernese}, {Ackley}, {Adams}, {Adhikari}, {Adya}, {Affeldt}, {Agathos},
  {Agatsuma}, {Aggarwal}, {Aguiar}, {Aich}, {Aiello}, {Ain}, {Ajith}, {Akcay},
  {Allen}, {Allocca}, {Altin}, {Amato}, {Anand}, {Ananyeva}, {Anderson},
  {Anderson}, {Angelova}, {Ansoldi}, {Antier}, {Appert}, {Arai}, {Araya},
  {Areeda}, {Ar{\`e}ne}, {Arnaud}, {Aronson}, {Arun}, {Asali}, {Ascenzi},
  {Ashton}, {Aston}, {Astone}, {Aubin}, {Aufmuth}, {AultONeal}, {Austin},
  {Avendano}, {Babak}, {Bacon}, {Badaracco}, {Bader}, {Bae}, {Baer}, {Baird},
  {Baldaccini}, {Ballardin}, {Ballmer}, {Bals}, {Balsamo}, {Baltus},
  {Banagiri}, {Bankar}, {Bankar}, {Barayoga}, {Barbieri}, {Barish}, {Barker},
  {Barkett}, {Barneo}, {Barone}, {Barr}, {Barsotti}, {Barsuglia}, {Barta},
  {Bartlett}, {Bartos}, {Bassiri}, {Basti}, {Bawaj}, {Bayley}, {Bazzan},
  {B{\'e}csy}, {Bejger}, {Belahcene}, {Bell}, {Beniwal}, {Benjamin}, {Benkel},
  {Bentley}, {Bergamin}, {Berger}, {Bergmann}, {Bernuzzi}, {Berry},
  {Bersanetti}, {Bertolini}, {Betzwieser}, {Bhandare}, {Bhandari}, {Bidler},
  {Biggs}, {Bilenko}, {Billingsley}, {Birney}, {Birnholtz}, {Biscans},
  {Bischi}, {Biscoveanu}, {Bisht}, {Bissenbayeva}, {Bitossi}, {Bizouard},
  {Blackburn}, {Blackman}, {Blair}, {Blair}, {Blair}, {Bobba}, {Bode}, {Boer},
  {Boetzel}, {Bogaert}, {Bondu}, {Bonilla}, {Bonnand}, {Booker}, {Boom},
  {Bork}, {Boschi}, {Bose}, {Bossilkov}, {Bosveld}, {Bouffanais}, {Bozzi},
  {Bradaschia}, {Brady}, {Bramley}, {Branchesi}, {Brau}, {Breschi}, {Briant},
  {Briggs}, {Brighenti}, {Brillet}, {Brinkmann}, {Brito}, {Brockill}, {Brooks},
  {Brooks}, {Brown}, {Brunett}, {Bruno}, {Bruntz}, {Buikema}, {Bulik},
  {Bulten}, {Buonanno}, {Buskulic}, {Byer}, {Cabero}, {Cadonati}, {Cagnoli},
  {Cahillane}, {Bustillo}, {Callaghan}, {Callister}, {Calloni}, {Camp},
  {Canepa}, {Cannon}, {Cao}, {Cao}, {Carapella}, {Carbognani}, {Caride},
  {Carney}, {Carullo}, {Diaz}, {Casentini}, {Casta{\~n}eda}, {Caudill},
  {Cavagli{\`a}}, {Cavalier}, {Cavalieri}, {Cella}, {Cerd{\'a}-Dur{\'a}n},
  {Cesarini}, {Chaibi}, {Chakravarti}, {Chan}, {Chan}, {Chao}, {Charlton},
  {Chase}, {Chassande-Mottin}, {Chatterjee}, {Chaturvedi}, {Chatziioannou},
  {Chen}, {Chen}, {Chen}, {Cheng}, {Cheong}, {Chia}, {Chiadini}, {Chierici},
  {Chincarini}, {Chiummo}, {Cho}, {Cho}, {Cho}, {Christensen}, {Chu}, {Chua},
  {Chung}, {Chung}, {Ciani}, {Ciecielag}, {Cie{\'s}lar}, {Ciobanu}, {Ciolfi},
  {Cipriano}, {Cirone}, {Clara}, {Clark}, {Clearwater}, {Clesse}, {Cleva},
  {Coccia}, {Cohadon}, {Cohen}, {Colleoni}, {Collette}, {Collins}, {Colpi},
  {Constancio}, {Conti}, {Cooper}, {Corban}, {Corbitt}, {Cordero-Carri{\'o}n},
  {Corezzi}, {Corley}, {Cornish}, {Corre}, {Corsi}, {Cortese}, {Costa},
  {Cotesta}, {Coughlin}, {Coughlin}, {Coulon}, {Countryman}, {Couvares},
  {Covas}, {Coward}, {Cowart}, {Coyne}, {Coyne}, {Creighton}, {Creighton},
  {Cripe}, {Croquette}, {Crowder}, {Cudell}, {Cullen}, {Cumming}, {Cummings},
  {Cunningham}, {Cuoco}, {Curylo}, {Canton}, {D{\'a}lya}, {Dana},
  {Daneshgaran-Bajastani}, {D'Angelo}, {Danilishin}, {D'Antonio}, {Danzmann},
  {Darsow-Fromm}, {Dasgupta}, {Datrier}, {Dattilo}, {Dave}, {Davier}, {Davies},
  {Davis}, {Daw}, {DeBra}, {Deenadayalan}, {Degallaix}, {De Laurentis},
  {Del{\'e}glise}, {Delfavero}, {De Lillo}, {Del Pozzo}, {DeMarchi},
  {D'Emilio}, {Demos}, {Dent}, {De Pietri}, {De Rosa}, {De Rossi}, {DeSalvo},
  {de Varona}, {Dhurandhar}, {D{\'\i}az}, {Diaz-Ortiz}, {Dietrich}, {Di Fiore},
  {Di Fronzo}, {Di Giorgio}, {Di Giovanni}, {Di Giovanni}, {Di Girolamo}, {Di
  Lieto}, {Ding}, {Di Pace}, {Di Palma}, {Di Renzo}, {Divakarla}, {Dmitriev},
  {Doctor}, {Donovan}, {Dooley}, {Doravari}, {Dorrington}, {Downes}, {Drago},
  {Driggers}, {Du}, {Ducoin}, {Dupej}, {Durante}, {D'Urso}, {Dwyer}, {Easter},
  {Eddolls}, {Edelman}, {Edo}, {Edy}, {Effler}, {Ehrens}, {Eichholz},
  {Eikenberry}, {Eisenmann}, {Eisenstein}, {Ejlli}, {Errico}, {Essick},
  {Estelles}, {Estevez}, {Etienne}, {Etzel}, {Evans}, {Evans}, {Ewing},
  {Fafone}, {Fairhurst}, {Fan}, {Farinon}, {Farr}, {Farr}, {Fauchon-Jones},
  {Favata}, {Fays}, {Fazio}, {Feicht}, {Fejer}, {Feng}, {Fenyvesi}, {Ferguson},
  {Fernandez-Galiana}, {Ferrante}, {Ferreira}, {Ferreira}, {Fidecaro}, {Fiori},
  {Fiorucci}, {Fishbach}, {Fisher}, {Fittipaldi}, {Fitz-Axen}, {Fiumara},
  {Flaminio}, {Floden}, {Flynn}, {Fong}, {Font}, {Forsyth}, {Fournier},
  {Frasca}, {Frasconi}, {Frei}, {Freise}, {Frey}, {Frey}, {Fritschel},
  {Frolov}, {Fronz{\`e}}, {Fulda}, {Fyffe}, {Gabbard}, {Gadre}, {Gaebel},
  {Gair}, {Galaudage}, {Ganapathy}, {Ganguly}, {Gaonkar},
  {Garc{\'\i}a-Quir{\'o}s}, {Garufi}, {Gateley}, {Gaudio}, {Gayathri}, {Gemme},
  {Genin}, {Gennai}, {George}, {George}, {Gergely}, {Ghonge}, {Ghosh}, {Ghosh},
  {Ghosh}, {Giacomazzo}, {Giaime}, {Giardina}, {Gibson}, {Gier}, {Gill},
  {Glanzer}, {Gniesmer}, {Godwin}, {Goetz}, {Goetz}, {Gohlke}, {Goncharov},
  {Gonz{\'a}lez}, {Gopakumar}, {Gossan}, {Gosselin}, {Gouaty}, {Grace},
  {Grado}, {Granata}, {Grant}, {Gras}, {Grassia}, {Gray}, {Gray}, {Greco},
  {Green}, {Green}, {Gretarsson}, {Griggs}, {Grignani}, {Grimaldi}, {Grimm},
  {Grote}, {Grunewald}, {Gruning}, {Guidi}, {Guimaraes}, {Guix{\'e}}, {Gulati},
  {Guo}, {Gupta}, {Gupta}, {Gupta}, {Gustafson}, {Gustafson}, {Haegel},
  {Halim}, {Hall}, {Hamilton}, {Hammond}, {Haney}, {Hanke}, {Hanks}, {Hanna},
  {Hannam}, {Hannuksela}, {Hansen}, {Hanson}, {Harder}, {Hardwick}, {Haris},
  {Harms}, {Harry}, {Harry}, {Hasskew}, {Haster}, {Haughian}, {Hayes}, {Healy},
  {Heidmann}, {Heintze}, {Heinze}, {Heitmann}, {Hellman}, {Hello}, {Hemming},
  {Hendry}, {Heng}, {Hennes}, {Hennig}, {Heurs}, {Hild}, {Hinderer}, {Hoback},
  {Hochheim}, {Hofgard}, {Hofman}, {Holgado}, {Holland}, {Holt}, {Holz},
  {Hopkins}, {Horst}, {Hough}, {Howell}, {Hoy}, {Huang}, {H{\"u}bner},
  {Huerta}, {Huet}, {Hughey}, {Hui}, {Husa}, {Huttner}, {Huxford},
  {Huynh-Dinh}, {Idzkowski}, {Iess}, {Inchauspe}, {Ingram}, {Intini}, {Isac},
  {Isi}, {Iyer}, {Jacqmin}, {Jadhav}, {Jadhav}, {James}, {Jani}, {Janthalur},
  {Jaranowski}, {Jariwala}, {Jaume}, {Jenkins}, {Jiang}, {Johns},
  {Johnson-McDaniel}, {Jones}, {Jones}, {Jones}, {Jones}, {Jones}, {Jonker},
  {Ju}, {Junker}, {Kalaghatgi}, {Kalogera}, {Kamai}, {Kandhasamy}, {Kang},
  {Kanner}, {Kapadia}, {Karki}, {Kashyap}, {Kasprzack}, {Kastaun},
  {Katsanevas}, {Katsavounidis}, {Katzman}, {Kaufer}, {Kawabe},
  {K{\'e}f{\'e}lian}, {Keitel}, {Keivani}, {Kennedy}, {Key}, {Khadka},
  {Khalili}, {Khan}, {Khan}, {Khan}, {Khazanov}, {Khetan}, {Khursheed},
  {Kijbunchoo}, {Kim}, {Kim}, {Kim}, {Kim}, {Kim}, {Kim}, {Kim}, {Kimball},
  {King}, {Kinley-Hanlon}, {Kirchhoff}, {Kissel}, {Kleybolte}, {Klimenko},
  {Knowles}, {Knyazev}, {Koch}, {Koehlenbeck}, {Koekoek}, {Koley},
  {Kondrashov}, {Kontos}, {Koper}, {Korobko}, {Korth}, {Kovalam}, {Kozak},
  {Kringel}, {Krishnendu}, {Kr{\'o}lak}, {Krupinski}, {Kuehn}, {Kumar},
  {Kumar}, {Kumar}, {Kumar}, {Kumar}, {Kuo}, {Kutynia}, {Lackey}, {Laghi},
  {Lalande}, {Lam}, {Lamberts}, {Landry}, {Landry}, {Lane}, {Lang}, {Lange},
  {Lantz}, {Lanza}, {La Rosa}, {Lartaux-Vollard}, {Lasky}, {Laxen},
  {Lazzarini}, {Lazzaro}, {Leaci}, {Leavey}, {Lecoeuche}, {Lee}, {Lee}, {Lee},
  {Lee}, {Lee}, {Lehmann}, {Leroy}, {Letendre}, {Levin}, {Li}, {Li}, {li},
  {Li}, {Li}, {Linde}, {Linker}, {Linley}, {Littenberg}, {Liu}, {Liu},
  {Llorens-Monteagudo}, {Lo}, {Lockwood}, {London}, {Longo}, {Lorenzini},
  {Loriette}, {Lormand}, {Losurdo}, {Lough}, {Lousto}, {Lovelace}, {L{\"u}ck},
  {Lumaca}, {Lundgren}, {Ma}, {Macas}, {Macfoy}, {MacInnis}, {Macleod},
  {MacMillan}, {Macquet}, {Hernandez}, {Maga{\~n}a-Sandoval}, {Magee},
  {Majorana}, {Maksimovic}, {Malik}, {Man}, {Mandic}, {Mangano}, {Mansell},
  {Manske}, {Mantovani}, {Mapelli}, {Marchesoni}, {Marion}, {M{\'a}rka},
  {M{\'a}rka}, {Markakis}, {Markosyan}, {Markowitz}, {Maros}, {Marquina},
  {Marsat}, {Martelli}, {Martin}, {Martin}, {Martinez}, {Martynov},
  {Masalehdan}, {Mason}, {Massera}, {Masserot}, {Massinger}, {Masso-Reid},
  {Mastrogiovanni}, {Matas}, {Matichard}, {Mavalvala}, {Maynard}, {McCann},
  {McCarthy}, {McClelland}, {McCormick}, {McCuller}, {McGuire}, {McIsaac},
  {McIver}, {McManus}, {McRae}, {McWilliams}, {Meacher}, {Meadors}, {Mehmet},
  {Mehta}, {Villa}, {Melatos}, {Mendell}, {Mercer}, {Mereni}, {Merfeld},
  {Merilh}, {Merritt}, {Merzougui}, {Meshkov}, {Messenger}, {Messick},
  {Metzdorff}, {Meyers}, {Meylahn}, {Mhaske}, {Miani}, {Miao}, {Michaloliakos},
  {Michel}, {Middleton}, {Milano}, {Miller}, {Millhouse}, {Mills}, {Milotti},
  {Milovich-Goff}, {Minazzoli}, {Minenkov}, {Mishkin}, {Mishra}, {Mistry},
  {Mitra}, {Mitrofanov}, {Mitselmakher}, {Mittleman}, {Mo}, {Mogushi},
  {Mohapatra}, {Mohite}, {Molina-Ruiz}, {Mondin}, {Montani}, {Moore}, {Moraru},
  {Morawski}, {Moreno}, {Morisaki}, {Mours}, {Mow-Lowry}, {Mozzon},
  {Muciaccia}, {Mukherjee}, {Mukherjee}, {Mukherjee}, {Mukherjee}, {Mukund},
  {Mullavey}, {Munch}, {Mu{\~n}iz}, {Murray}, {Nagar}, {Nardecchia},
  {Naticchioni}, {Nayak}, {Neil}, {Neilson}, {Nelemans}, {Nelson}, {Nery},
  {Neunzert}, {Ng}, {Ng}, {Nguyen}, {Nguyen}, {Nichols}, {Nichols}, {Nissanke},
  {Nocera}, {Noh}, {North}, {Nothard}, {Nuttall}, {Oberling}, {O'Brien},
  {Oganesyan}, {Ogin}, {Oh}, {Oh}, {Ohme}, {Ohta}, {Okada}, {Oliver},
  {Olivetto}, {Oppermann}, {Oram}, {O'Reilly}, {Ormiston}, {Ortega},
  {O'Shaughnessy}, {Ossokine}, {Osthelder}, {Ottaway}, {Overmier}, {Owen},
  {Pace}, {Pagano}, {Page}, {Pagliaroli}, {Pai}, {Pai}, {Palamos}, {Palashov},
  {Palomba}, {Pan}, {Panda}, {Pang}, {Pankow}, {Pannarale}, {Pant}, {Paoletti},
  {Paoli}, {Parida}, {Parker}, {Pascucci}, {Pasqualetti}, {Passaquieti},
  {Passuello}, {Patricelli}, {Payne}, {Pearlstone}, {Pechsiri}, {Pedersen},
  {Pedraza}, {Pele}, {Penn}, {Perego}, {Perez}, {P{\'e}rigois}, {Perreca},
  {Perri{\`e}s}, {Petermann}, {Pfeiffer}, {Phelps}, {Phukon}, {Piccinni},
  {Pichot}, {Piendibene}, {Piergiovanni}, {Pierro}, {Pillant}, {Pinard},
  {Pinto}, {Piotrzkowski}, {Pirello}, {Pitkin}, {Plastino}, {Poggiani}, {Pong},
  {Ponrathnam}, {Popolizio}, {Porter}, {Powell}, {Prajapati}, {Prasai},
  {Prasanna}, {Pratten}, {Prestegard}, {Principe}, {Prodi}, {Prokhorov},
  {Punturo}, {Puppo}, {P{\"u}rrer}, {Qi}, {Quetschke}, {Quinonez}, {Raab},
  {Raaijmakers}, {Radkins}, {Radulesco}, {Raffai}, {Rafferty}, {Raja}, {Rajan},
  {Rajbhandari}, {Rakhmanov}, {Ramirez}, {Ramos-Buades}, {Rana}, {Rao},
  {Rapagnani}, {Raymond}, {Razzano}, {Read}, {Regimbau}, {Rei}, {Reid},
  {Reitze}, {Rettegno}, {Ricci}, {Richardson}, {Richardson}, {Ricker},
  {Riemenschneider}, {Riles}, {Rizzo}, {Robertson}, {Robinet}, {Rocchi},
  {Rodriguez-Soto}, {Rolland}, {Rollins}, {Roma}, {Romanelli}, {Romano},
  {Romel}, {Romero-Shaw}, {Romie}, {Rose}, {Rose}, {Rose}, {Rosi{\'n}ska},
  {Rosofsky}, {Ross}, {Rowan}, {Rowlinson}, {Roy}, {Roy}, {Roy}, {Ruggi},
  {Rutins}, {Ryan}, {Sachdev}, {Sadecki}, {Sakellariadou}, {Salafia},
  {Salconi}, {Saleem}, {Salemi}, {Samajdar}, {Sanchez}, {Sanchez},
  {Sanchis-Gual}, {Sanders}, {Santiago}, {Santos}, {Sarin}, {Sassolas},
  {Sathyaprakash}, {Sauter}, {Savage}, {Savant}, {Sawant}, {Sayah}, {Schaetzl},
  {Schale}, {Scheel}, {Scheuer}, {Schmidt}, {Schnabel}, {Schofield},
  {Sch{\"o}nbeck}, {Schreiber}, {Schulte}, {Schutz}, {Schwarm}, {Schwartz},
  {Scott}, {Scott}, {Seidel}, {Sellers}, {Sengupta}, {Sennett}, {Sentenac},
  {Sequino}, {Sergeev}, {Setyawati}, {Shaddock}, {Shaffer}, {Shahriar},
  {Sharma}, {Sharma}, {Shawhan}, {Shen}, {Shikauchi}, {Shink}, {Shoemaker},
  {Shoemaker}, {Shukla}, {ShyamSundar}, {Siellez}, {Sieniawska}, {Sigg},
  {Singer}, {Singh}, {Singh}, {Singha}, {Singhal}, {Sintes}, {Sipala},
  {Skliris}, {Slagmolen}, {Slaven-Blair}, {Smetana}, {Smith}, {Smith},
  {Somala}, {Son}, {Soni}, {Sorazu}, {Sordini}, {Sorrentino}, {Souradeep},
  {Sowell}, {Spencer}, {Spera}, {Srivastava}, {Srivastava}, {Staats},
  {Stachie}, {Standke}, {Steer}, {Steinhoff}, {Steinke}, {Steinlechner},
  {Steinlechner}, {Steinmeyer}, {Stevenson}, {Stocks}, {Stops}, {Stover},
  {Strain}, {Stratta}, {Strunk}, {Sturani}, {Stuver}, {Sudhagar}, {Sudhir},
  {Summerscales}, {Sun}, {Sunil}, {Sur}, {Suresh}, {Sutton}, {Swinkels},
  {Szczepa{\'n}czyk}, {Tacca}, {Tait}, {Talbot}, {Tanasijczuk}, {Tanner},
  {Tao}, {T{\'a}pai}, {Tapia}, {San Martin}, {Tasson}, {Taylor}, {Tenorio},
  {Terkowski}, {Thirugnanasambandam}, {Thomas}, {Thomas}, {Thompson},
  {Thondapu}, {Thorne}, {Thrane}, {Tinsman}, {Saravanan}, {Tiwari}, {Tiwari},
  {Tiwari}, {Toland}, {Tonelli}, {Tornasi}, {Torres-Forn{\'e}}, {Torrie},
  {Tosta e Melo}, {T{\"o}yr{\"a}}, {Trail}, {Travasso}, {Traylor}, {Tringali},
  {Tripathee}, {Trovato}, {Trudeau}, {Tsang}, {Tse}, {Tso}, {Tsukada}, {Tsuna},
  {Tsutsui}, {Turconi}, {Ubhi}, {Ueno}, {Ugolini}, {Unnikrishnan}, {Urban},
  {Usman}, {Utina}, {Vahlbruch}, {Vajente}, {Valdes}, {Valentini}, {van Bakel},
  {van Beuzekom}, {van den Brand}, {Van Den Broeck}, {Vander-Hyde}, {van der
  Schaaf}, {Van Heijningen}, {van Veggel}, {Vardaro}, {Varma}, {Vass},
  {Vas{\'u}th}, {Vecchio}, {Vedovato}, {Veitch}, {Veitch}, {Venkateswara},
  {Venugopalan}, {Verkindt}, {Veske}, {Vetrano}, {Vicer{\'e}}, {Viets},
  {Vinciguerra}, {Vine}, {Vinet}, {Vitale}, {Vivanco}, {Vo}, {Vocca},
  {Vorvick}, {Vyatchanin}, {Wade}, {Wade}, {Wade}, {Walet}, {Walker},
  {Wallace}, {Wallace}, {Walsh}, {Wang}, {Wang}, {Wang}, {Ward}, {Warden},
  {Warner}, {Was}, {Watchi}, {Weaver}, {Wei}, {Weinert}, {Weinstein}, {Weiss},
  {Wellmann}, {Wen}, {We{\ss}els}, {Westhouse}, {Wette}, {Whelan}, {Whiting},
  {Whittle}, {Wilken}, {Williams}, {Willis}, {Willke}, {Winkler}, {Wipf},
  {Wittel}, {Woan}, {Woehler}, {Wofford}, {Wong}, {Wright}, {Wu}, {Wysocki},
  {Xiao}, {Yamamoto}, {Yang}, {Yang}, {Yang}, {Yap}, {Yazback}, {Yeeles}, {Yu},
  {Yu}, {Yuen}, {Zadro{\.z}ny}, {Zadro{\.z}ny}, {Zanolin}, {Zelenova},
  {Zendri}, {Zevin}, {Zhang}, {Zhang}, {Zhang}, {Zhao}, {Zhao}, {Zhou}, {Zhou},
  {Zhu}, {Zimmerman}, {Zucker}, {Zweizig}, {LIGO Scientific Collaboration}, \&
  {Virgo Collaboration}}]{GW190814}
---. 2020{\natexlab{d}}, \apjl, 896, L44

\bibitem[{{Ackley} {et~al.}(2020){Ackley}, {Amati}, {Barbieri}, {Bauer},
  {Benetti}, {Bernardini}, {Bhirombhakdi}, {Botticella}, {Branchesi},
  {Brocato}, {Bruun}, {Bulla}, {Campana}, {Cappellaro}, {Castro-Tirado},
  {Chambers}, {Chaty}, {Chen}, {Ciolfi}, {Coleiro}, {Copperwheat}, {Covino},
  {Cutter}, {D'Ammando}, {D'Avanzo}, {De Cesare}, {D'Elia}, {Della Valle},
  {Denneau}, {De Pasquale}, {Dhillon}, {Dyer}, {Elias-Rosa}, {Evans},
  {Eyles-Ferris}, {Fiore}, {Fraser}, {Fruchter}, {Fynbo}, {Galbany}, {Gall},
  {Galloway}, {Getman}, {Ghirlanda}, {Gillanders}, {Gomboc}, {Gompertz},
  {Gonz{\'a}lez-Fern{\'a}ndez}, {Gonz{\'a}lez-Gait{\'a}n}, {Grado}, {Greco},
  {Gromadzki}, {Groot}, {Guti{\'e}rrez}, {Heikkil{\"a}}, {Heintz}, {Hjorth},
  {Hu}, {Huber}, {Inserra}, {Izzo}, {Japelj}, {Jerkstrand}, {Jin}, {Jonker},
  {Kankare}, {Kann}, {Kennedy}, {Kim}, {Klose}, {Kool}, {Kotak},
  {Kuncarayakti}, {Lamb}, {Leloudas}, {Levan}, {Longo}, {Lowe}, {Lyman},
  {Magnier}, {Maguire}, {Maiorano}, {Mandel}, {Mapelli}, {Mattila}, {McBrien},
  {Melandri}, {Micha{\l}owski}, {Milvang-Jensen}, {Moran}, {Nicastro},
  {Nicholl}, {Nicuesa Guelbenzu}, {Nuttal}, {Oates}, {O'Brien}, {Onori},
  {Palazzi}, {Patricelli}, {Perego}, {Torres}, {Perley}, {Pian}, {Pignata},
  {Piranomonte}, {Poshyachinda}, {Possenti}, {Pumo}, {Quirola-V{\'a}squez},
  {Ragosta}, {Ramsay}, {Rau}, {Rest}, {Reynolds}, {Rosetti}, {Rossi},
  {Rosswog}, {Sabha}, {Sagu{\'e}s Carracedo}, {Salafia}, {Salmon},
  {Salvaterra}, {Savaglio}, {Sbordone}, {Schady}, {Schipani}, {Schultz},
  {Schweyer}, {Smartt}, {Smith}, {Smith}, {Sollerman}, {Srivastav}, {Stanway},
  {Starling}, {Steeghs}, {Stratta}, {Stubbs}, {Tanvir}, {Testa}, {Thrane},
  {Tonry}, {Turatto}, {Ulaczyk}, {van der Horst}, {Vergani}, {Walton},
  {Watson}, {Wiersema}, {Wiik}, {Wyrzykowski}, {Yang}, {Yi}, \&
  {Young}}]{ENGRAVE_190814}
{Ackley}, K., {Amati}, L., {Barbieri}, C., {et~al.} 2020, \aap, 643, A113

\bibitem[{{Anand} {et~al.}(2020){Anand}, {Coughlin}, {Kasliwal}, {Bulla},
  {Ahumada}, {Sagu{\'e}s Carracedo}, {Almualla}, {Andreoni}, {Stein},
  {Foucart}, {Singer}, {Sollerman}, {Bellm}, {Bolin}, {Caballero-Garc{\'\i}a},
  {Castro-Tirado}, {Cenko}, {De}, {Dekany}, {Duev}, {Feeney}, {Fremling},
  {Goldstein}, {Golkhou}, {Graham}, {Guessoum}, {Hankins}, {Hu}, {Kong},
  {Kool}, {Kulkarni}, {Kumar}, {Laher}, {Masci}, {Mr{\'o}z}, {Nissanke},
  {Porter}, {Reusch}, {Riddle}, {Rosnet}, {Rusholme}, {Serabyn},
  {S{\'a}nchez-Ram{\'\i}rez}, {Rigault}, {Shupe}, {Smith}, {Soumagnac},
  {Walters}, \& {Valeev}}]{Anand2020}
{Anand}, S., {Coughlin}, M.~W., {Kasliwal}, M.~M., {et~al.} 2020, Nature
  Astronomy, arXiv:2009.07210

\bibitem[{{Andreoni} {et~al.}(2020){Andreoni}, {Goldstein}, {Kasliwal},
  {Nugent}, {Zhou}, {Newman}, {Bulla}, {Foucart}, {Hotokezaka}, {Nakar},
  {Nissanke}, {Raaijmakers}, {Bloom}, {De}, {Jencson}, {Ward}, {Ahumada},
  {Anand}, {Buckley}, {Caballero-Garc{\'\i}a}, {Castro-Tirado}, {Copperwheat},
  {Coughlin}, {Cenko}, {Gromadzki}, {Hu}, {Karambelkar}, {Perley}, {Sharma},
  {Valeev}, {Cook}, {Fremling}, {Kumar}, {Taggart}, {Bagdasaryan}, {Cooke},
  {Dahiwale}, {Dhawan}, {Dobie}, {Gatkine}, {Golkhou}, {Goobar}, {Chaves},
  {Hankins}, {Kaplan}, {Kong}, {Kool}, {Mohite}, {Sollerman}, {Tzanidakis},
  {Webb}, \& {Zhang}}]{Andreoni20}
{Andreoni}, I., {Goldstein}, D.~A., {Kasliwal}, M.~M., {et~al.} 2020, \apj,
  890, 131

\bibitem[{{Antier} {et~al.}(2020){Antier}, {Agayeva}, {Almualla}, {Awiphan},
  {Baransky}, {Barynova}, {Beradze}, {Bla{\v{z}}ek}, {Bo{\"e}r}, {Burkhonov},
  {Christensen}, {Coleiro}, {Corre}, {Coughlin}, {Crisp}, {Dietrich}, {Ducoin},
  {Duverne}, {Marchal-Duval}, {Gendre}, {Gokuldass}, {Eggenstein}, {Eymar},
  {Hello}, {Howell}, {Ismailov}, {Kann}, {Karpov}, {Klotz}, {Kochiashvili},
  {Lachaud}, {Leroy}, {Lin}, {Li}, {Ma{\v{s}}ek}, {Mo}, {Menard}, {Morris},
  {Noysena}, {Orange}, {Prouza}, {Rattanamala}, {Sadibekova}, {Saint-Gelais},
  {Serrau}, {Simon}, {Stachie}, {Th{\"o}ne}, {Tillayev}, {Turpin}, {de Ugarte
  Postigo}, {Vasylenko}, {Vidadi}, {Was}, {Wang}, {Zhang}, {Zhang}, \&
  {Zhang}}]{GRANDMA_o3}
{Antier}, S., {Agayeva}, S., {Almualla}, M., {et~al.} 2020, \mnras, 497, 5518

\bibitem[{{Baiotti} \& {Rezzolla}(2017)}]{baiotti16}
{Baiotti}, L., \& {Rezzolla}, L. 2017, Rep. Prog. Phys., 80, 096901

\bibitem[{{Barbieri} {et~al.}(2020{\natexlab{a}}){Barbieri}, {Salafia},
  {Colpi}, {Ghirlanda}, \& {Perego}}]{Barbieri20b}
{Barbieri}, C., {Salafia}, O.~S., {Colpi}, M., {Ghirlanda}, G., \& {Perego}, A.
  2020{\natexlab{a}}, arXiv e-prints, arXiv:2002.09395

\bibitem[{{Barbieri} {et~al.}(2019){Barbieri}, {Salafia}, {Perego}, {Colpi}, \&
  {Ghirlanda}}]{Barbieri19a}
{Barbieri}, C., {Salafia}, O.~S., {Perego}, A., {Colpi}, M., \& {Ghirlanda}, G.
  2019, \aap, 625, A152

\bibitem[{{Barbieri} {et~al.}(2020{\natexlab{b}}){Barbieri}, {Salafia},
  {Perego}, {Colpi}, \& {Ghirlanda}}]{Barbieri20}
---. 2020{\natexlab{b}}, European Physical Journal A, 56, 8

\bibitem[{Bardeen {et~al.}(1972)Bardeen, Press, \& Teukolsky}]{Bardeen1972}
Bardeen, J.~M., Press, W.~H., \& Teukolsky, S.~A. 1972, Astrophys.\ J., 178,
  347

\bibitem[{{Barnes} {et~al.}(2020){Barnes}, {Zhu}, {Lund}, {Sprouse}, {Vassh},
  {McLaughlin}, {Mumpower}, \& {Surman}}]{Barnes20}
{Barnes}, J., {Zhu}, Y.~L., {Lund}, K.~A., {et~al.} 2020, arXiv e-prints,
  arXiv:2010.11182

\bibitem[{{Behnel} {et~al.}(2011){Behnel}, {Bradshaw}, {Citro}, {Dalcin},
  {Seljebotn}, \& {Smith}}]{cython2011}
{Behnel}, S., {Bradshaw}, R., {Citro}, C., {et~al.} 2011, Computing in Science
  Engineering, 13, 31

\bibitem[{Binnington \& Poisson(2009)}]{Binnington09}
Binnington, T., \& Poisson, E. 2009, Physical Review D, 80,
  doi:10.1103/physrevd.80.084018.
\newblock \url{http://dx.doi.org/10.1103/PhysRevD.80.084018}

\bibitem[{{Bloemen} {et~al.}(2016){Bloemen}, {Groot}, {Woudt}, {Klein Wolt},
  {McBride}, {Nelemans}, {K{\"o}rding}, {Pretorius}, {Roelfsema}, {Bettonvil},
  {Balster}, {Bakker}, {Dolron}, {van Elteren}, {Elswijk}, {Engels}, {Fender},
  {Fokker}, {de Haan}, {Hagoort}, {de Hoog}, {ter Horst}, {van der Kevie},
  {Koz{\l}owski}, {Kragt}, {Lech}, {Le Poole}, {Lesman}, {Morren}, {Navarro},
  {Paalberends}, {Paterson}, {Paw{\l}aszek}, {Pessemier}, {Raskin}, {Rutten},
  {Scheers}, {Schuil}, \& {Sybilski}}]{BlackGEM}
{Bloemen}, S., {Groot}, P., {Woudt}, P., {et~al.} 2016, in Society of
  Photo-Optical Instrumentation Engineers (SPIE) Conference Series, Vol. 9906,
  Ground-based and Airborne Telescopes VI, ed. H.~J. {Hall}, R.~{Gilmozzi}, \&
  H.~K. {Marshall}, 990664

\bibitem[{Breschi {et~al.}(2021{\natexlab{a}})Breschi, Gamba, \&
  Bernuzzi}]{Breschi21a}
Breschi, M., Gamba, R., \& Bernuzzi, S. 2021{\natexlab{a}}, arXiv:2102.00017

\bibitem[{Breschi {et~al.}(2021{\natexlab{b}})Breschi, Perego, Bernuzzi,
  Del~Pozzo, Nedora, Radice, \& Vescovi}]{Breschi21b}
Breschi, M., Perego, A., Bernuzzi, S., {et~al.} 2021{\natexlab{b}},
  arXiv:2101.01201

\bibitem[{{Buchner} {et~al.}(2014){Buchner}, {Georgakakis}, {Nandra}, {Hsu},
  {Rangel}, {Brightman}, {Merloni}, {Salvato}, {Donley}, \&
  {Kocevski}}]{Buchner14}
{Buchner}, J., {Georgakakis}, A., {Nandra}, K., {et~al.} 2014, \aap, 564, A125

\bibitem[{{Bulla}(2019)}]{Bulla19}
{Bulla}, M. 2019, \mnras, 489, 5037

\bibitem[{Burbidge {et~al.}(1957)Burbidge, Burbidge, Fowler, \&
  Hoyle}]{Burbidge57}
Burbidge, E.~M., Burbidge, G.~R., Fowler, W.~A., \& Hoyle, F. 1957, Rev. Mod.
  Phys., 29, 547.
\newblock \url{https://link.aps.org/doi/10.1103/RevModPhys.29.547}

\bibitem[{Cameron(1957)}]{Cameron57}
Cameron, A. G.~W. 1957, Publications of the Astronomical Society of the
  Pacific, 69, 201.
\newblock \url{https://doi.org/10.1086/127051}

\bibitem[{{Capano} {et~al.}(2020){Capano}, {Tews}, {Brown}, {Margalit}, {De},
  {Kumar}, {Brown}, {Krishnan}, \& {Reddy}}]{Capano20}
{Capano}, C.~D., {Tews}, I., {Brown}, S.~M., {et~al.} 2020, Nature Astronomy,
  4, 625

\bibitem[{Chaurasia {et~al.}(2020)Chaurasia, Dietrich, Ujevic, Hendriks, Dudi,
  Fabbri, Tichy, \& Br\"ugmann}]{Chaurasia20}
Chaurasia, S.~V., Dietrich, T., Ujevic, M., {et~al.} 2020, Phys. Rev. D, 102,
  024087

\bibitem[{{Chia}(2020)}]{Chia20}
{Chia}, H.~S. 2020, arXiv e-prints, arXiv:2010.07300

\bibitem[{{Chirenti} {et~al.}(2020){Chirenti}, {Posada}, \&
  {Guedes}}]{Chirenti20}
{Chirenti}, C., {Posada}, C., \& {Guedes}, V. 2020, Classical and Quantum
  Gravity, 37, 195017

\bibitem[{{Chornock} {et~al.}(2019){Chornock}, {Cowperthwaite}, {Margutti},
  {Milisavljevic}, {Alexander}, {Andreoni}, {Arcavi}, {Baldeschi}, {Barnes},
  {Bellm}, {Beniamini}, {Berger}, {Berry}, {Bianco}, {Blanchard}, {Bloom},
  {Burke-Spolaor}, {Burns}, {Carbone}, {Cenko}, {Coppejans}, {Corsi},
  {Coughlin}, {Drout}, {Eftekhari}, {Foley}, {Fong}, {Fox}, {Frail},
  {Giannios}, {Golkhou}, {Gomez}, {Graham}, {Graur}, {Hajela}, {Hallinan},
  {Hanna}, {Hotokezaka}, {Kalogera}, {Kasen}, {Kasliwal}, {Kathirgamaraju},
  {Kerzendorf}, {Kilpatrick}, {Laskar}, {Levesque}, {MacFadyen}, {Macias},
  {Margalit}, {Matheson}, {Metzger}, {Miller}, {Modjaz}, {Murase}, {Murguia},
  {Nissanke}, {Palmese}, {Pankow}, {Paterson}, {Patton}, {Perna}, {Radice},
  {Ramirez-Ruiz}, {Rest}, {Rho}, {Rojas-Bravo}, {Roth}, {Safarzadeh}, {Sand},
  {Sbarufatti}, {Siegel}, {Sironi}, {Soares-Santos}, {Sravan}, {Starrfield},
  {Street}, {Stringfellow}, {Tchekhovskoy}, {Terreran}, {Valenti}, {Villar},
  {Wang}, {Wheeler}, {Williams}, \& {Zrake}}]{Chornock19}
{Chornock}, R., {Cowperthwaite}, P.~S., {Margutti}, R., {et~al.} 2019, \baas,
  51, 237

\bibitem[{{Christie} {et~al.}(2019){Christie}, {Lalakos}, {Tchekhovskoy},
  {Fern{\'a}ndez}, {Foucart}, {Quataert}, \& {Kasen}}]{Christie19}
{Christie}, I.~M., {Lalakos}, A., {Tchekhovskoy}, A., {et~al.} 2019, \mnras,
  490, 4811

\bibitem[{Ciolfi(2020)}]{Ciolfi20}
Ciolfi, R. 2020, Front. Astron. Space Sci., 7, 27

\bibitem[{{Coughlin} \& {Dietrich}(2019)}]{Coughlin19c}
{Coughlin}, M.~W., \& {Dietrich}, T. 2019, \prd, 100, 043011

\bibitem[{{Coughlin} {et~al.}(2019{\natexlab{a}}){Coughlin}, {Dietrich},
  {Margalit}, \& {Metzger}}]{Coughlin19}
{Coughlin}, M.~W., {Dietrich}, T., {Margalit}, B., \& {Metzger}, B.~D.
  2019{\natexlab{a}}, \mnras, 489, L91

\bibitem[{Coughlin {et~al.}(2018)Coughlin, Dietrich, Doctor, Kasen, Coughlin,
  Jerkstrand, Leloudas, McBrien, Metzger, O’Shaughnessy, \&
  et~al.}]{Coughlin18}
Coughlin, M.~W., Dietrich, T., Doctor, Z., {et~al.} 2018, Monthly Notices of
  the Royal Astronomical Society, 480, 3871–3878.
\newblock \url{http://dx.doi.org/10.1093/mnras/sty2174}

\bibitem[{{Coughlin} {et~al.}(2019{\natexlab{b}}){Coughlin}, {Ahumada},
  {Anand}, {De}, {Hankins}, {Kasliwal}, {Singer}, {Bellm}, {Andreoni}, {Cenko},
  {Cooke}, {Copperwheat}, {Dugas}, {Jencson}, {Perley}, {Yu}, {Bhalerao},
  {Kumar}, {Bloom}, {Anupama}, {Ashley}, {Bagdasaryan}, {Biswas}, {Buckley},
  {Burdge}, {Cook}, {Cromer}, {Cunningham}, {D'A{\`\i}}, {Dekany}, {Delacroix},
  {Dichiara}, {Duev}, {Dutta}, {Feeney}, {Frederick}, {Gatkine}, {Ghosh},
  {Goldstein}, {Golkhou}, {Goobar}, {Graham}, {Hanayama}, {Horiuchi}, {Hung},
  {Jha}, {Kong}, {Giomi}, {Kaplan}, {Karambelkar}, {Kowalski}, {Kulkarni},
  {Kupfer}, {Masci}, {Mazzali}, {Moore}, {Mogotsi}, {Neill}, {Ngeow},
  {Mart{\'\i}nez-Palomera}, {La Parola}, {Pavana}, {Ofek}, {Patil}, {Riddle},
  {Rigault}, {Rusholme}, {Serabyn}, {Shupe}, {Sharma}, {Singh}, {Sollerman},
  {Soon}, {Staats}, {Taggart}, {Tan}, {Travouillon}, {Troja}, {Waratkar}, \&
  {Yatsu}}]{Coughlin19b}
{Coughlin}, M.~W., {Ahumada}, T., {Anand}, S., {et~al.} 2019{\natexlab{b}},
  \apjl, 885, L19

\bibitem[{{Coughlin} {et~al.}(2019{\natexlab{c}}){Coughlin}, {Ahumada},
  {Anand}, {De}, {Hankins}, {Kasliwal}, {Singer}, {Bellm}, {Andreoni}, {Cenko},
  {Cooke}, {Copperwheat}, {Dugas}, {Jencson}, {Perley}, {Yu}, {Bhalerao},
  {Kumar}, {Bloom}, {Anupama}, {Ashley}, {Bagdasaryan}, {Biswas}, {Buckley},
  {Burdge}, {Cook}, {Cromer}, {Cunningham}, {D'A{\`\i}}, {Dekany}, {Delacroix},
  {Dichiara}, {Duev}, {Dutta}, {Feeney}, {Frederick}, {Gatkine}, {Ghosh},
  {Goldstein}, {Golkhou}, {Goobar}, {Graham}, {Hanayama}, {Horiuchi}, {Hung},
  {Jha}, {Kong}, {Giomi}, {Kaplan}, {Karambelkar}, {Kowalski}, {Kulkarni},
  {Kupfer}, {Masci}, {Mazzali}, {Moore}, {Mogotsi}, {Neill}, {Ngeow},
  {Mart{\'\i}nez-Palomera}, {La Parola}, {Pavana}, {Ofek}, {Patil}, {Riddle},
  {Rigault}, {Rusholme}, {Serabyn}, {Shupe}, {Sharma}, {Singh}, {Sollerman},
  {Soon}, {Staats}, {Taggart}, {Tan}, {Travouillon}, {Troja}, {Waratkar}, \&
  {Yatsu}}]{GROWTH190425}
---. 2019{\natexlab{c}}, \apjl, 885, L19

\bibitem[{{Coughlin} {et~al.}(2020{\natexlab{a}}){Coughlin}, {Dietrich},
  {Antier}, {Bulla}, {Foucart}, {Hotokezaka}, {Raaijmakers}, {Hinderer}, \&
  {Nissanke}}]{Coughlin20a}
{Coughlin}, M.~W., {Dietrich}, T., {Antier}, S., {et~al.} 2020{\natexlab{a}},
  \mnras, 492, 863

\bibitem[{{Coughlin} {et~al.}(2020{\natexlab{b}}){Coughlin}, {Dietrich},
  {Antier}, {Almualla}, {Anand}, {Bulla}, {Foucart}, {Guessoum}, {Hotokezaka},
  {Kumar}, {Raaijmakers}, \& {Nissanke}}]{Coughlin20b}
---. 2020{\natexlab{b}}, \mnras, 497, 1181

\bibitem[{{Coulter} {et~al.}(2017){Coulter}, {Foley}, {Kilpatrick}, {Drout},
  {Piro}, {Shappee}, {Siebert}, {Simon}, {Ulloa}, {Kasen}, {Madore},
  {Murguia-Berthier}, {Pan}, {Prochaska}, {Ramirez-Ruiz}, {Rest}, \&
  {Rojas-Bravo}}]{GW170817swope}
{Coulter}, D.~A., {Foley}, R.~J., {Kilpatrick}, C.~D., {et~al.} 2017, Science,
  358, 1556

\bibitem[{{Cowperthwaite} {et~al.}(2019){Cowperthwaite}, {Chen}, {Margalit},
  {Margutti}, {May}, {Metzger}, \& {Pankow}}]{Cowperthwaite19}
{Cowperthwaite}, P.~S., {Chen}, H.-Y., {Margalit}, B., {et~al.} 2019, arXiv
  e-prints, arXiv:1904.02718

\bibitem[{Dalc\'{i}n {et~al.}(2008)Dalc\'{i}n, Paz, Storti, \&
  D'El\'{i}a}]{mpi4py}
Dalc\'{i}n, L., Paz, R., Storti, M., \& D'El\'{i}a, J. 2008, Journal of
  Parallel and Distributed Computing, 68, 655

\bibitem[{{Darbha} \& {Kasen}(2020)}]{Darbha20}
{Darbha}, S., \& {Kasen}, D. 2020, \apj, 897, 150

\bibitem[{{De} \& {Siegel}(2020)}]{De20}
{De}, S., \& {Siegel}, D. 2020, arXiv e-prints, arXiv:2011.07176

\bibitem[{Dekany {et~al.}(2020)Dekany, Smith, Riddle, Feeney, Porter, Hale,
  Zolkower, Belicki, Kaye, Henning, \& et~al.}]{ZTF}
Dekany, R., Smith, R.~M., Riddle, R., {et~al.} 2020, Publications of the
  Astronomical Society of the Pacific, 132, 038001.
\newblock \url{http://dx.doi.org/10.1088/1538-3873/ab4ca2}

\bibitem[{{Dhawan} {et~al.}(2020){Dhawan}, {Bulla}, {Goobar}, {Sagu{\'e}s
  Carracedo}, \& {Setzer}}]{Dhawan20}
{Dhawan}, S., {Bulla}, M., {Goobar}, A., {Sagu{\'e}s Carracedo}, A., \&
  {Setzer}, C.~N. 2020, \apj, 888, 67

\bibitem[{{Dietrich} {et~al.}(2020){Dietrich}, {Coughlin}, {Pang}, {Bulla},
  {Heinzel}, {Issa}, {Tews}, \& {Antier}}]{Dietrich20}
{Dietrich}, T., {Coughlin}, M.~W., {Pang}, P. T.~H., {et~al.} 2020, arXiv
  e-prints, arXiv:2002.11355

\bibitem[{Dietrich {et~al.}(2020)Dietrich, Hinderer, \& Samajdar}]{Dietrich20a}
Dietrich, T., Hinderer, T., \& Samajdar, A. 2020, arXiv:2004.02527

\bibitem[{Dietrich {et~al.}(2019)Dietrich, Samajdar, Khan, Johnson-McDaniel,
  Dudi, \& Tichy}]{Dietrich19}
Dietrich, T., Samajdar, A., Khan, S., {et~al.} 2019, Physical Review D, 100,
  doi:10.1103/physrevd.100.044003.
\newblock \url{http://dx.doi.org/10.1103/PhysRevD.100.044003}

\bibitem[{Dietrich \& Ujevic(2017)}]{Dietrich17}
Dietrich, T., \& Ujevic, M. 2017, Classical and Quantum Gravity, 34, 105014.
\newblock \url{https://doi.org/10.1088/1361-6382/aa6bb0}

\bibitem[{Droettboom {et~al.}(2018)Droettboom, Caswell, Hunter, Firing,
  Nielsen, Lee, de~Andrade, Varoquaux, Stansby, Root, Elson, Dale, Lee, May,
  Seppänen, Klymak, McDougall, Straw, Hobson, cgohlke, Yu, Ma, Vincent,
  Silvester, Moad, Katins, Kniazev, Hoffmann, Ariza, \& Würtz}]{matplotlibv2}
Droettboom, M., Caswell, T.~A., Hunter, J., {et~al.} 2018,
  matplotlib/matplotlib v2.2.2, , , doi:10.5281/zenodo.1202077.
\newblock \url{https://doi.org/10.5281/zenodo.1202077}

\bibitem[{{East} {et~al.}(2019){East}, {Paschalidis}, {Pretorius}, \&
  {Tsokaros}}]{East19}
{East}, W.~E., {Paschalidis}, V., {Pretorius}, F., \& {Tsokaros}, A. 2019,
  \prd, 100, 124042

\bibitem[{{Fahlman} \& {Fern{\'a}ndez}(2018)}]{fahlman2018}
{Fahlman}, S., \& {Fern{\'a}ndez}, R. 2018, \apjl, 869, L3

\bibitem[{{Fern{\'a}ndez} {et~al.}(2020){Fern{\'a}ndez}, {Foucart}, \&
  {Lippuner}}]{Fernandez20}
{Fern{\'a}ndez}, R., {Foucart}, F., \& {Lippuner}, J. 2020, \mnras, 497, 3221

\bibitem[{{Fern{\'a}ndez} {et~al.}(2019){Fern{\'a}ndez}, {Tchekhovskoy},
  {Quataert}, {Foucart}, \& {Kasen}}]{Fernandez19}
{Fern{\'a}ndez}, R., {Tchekhovskoy}, A., {Quataert}, E., {Foucart}, F., \&
  {Kasen}, D. 2019, \mnras, 482, 3373

\bibitem[{{Feroz} {et~al.}(2009){Feroz}, {Hobson}, \& {Bridges}}]{Feroz09}
{Feroz}, F., {Hobson}, M.~P., \& {Bridges}, M. 2009, \mnras, 398, 1601

\bibitem[{{Feroz} {et~al.}(2013){Feroz}, {Hobson}, {Cameron}, \&
  {Pettitt}}]{Feroz13}
{Feroz}, F., {Hobson}, M.~P., {Cameron}, E., \& {Pettitt}, A.~N. 2013, ArXiv
  e-prints, arXiv:1306.2144

\bibitem[{{Finstad} \& {Brown}(2020)}]{Finstad20}
{Finstad}, D., \& {Brown}, D.~A. 2020, arXiv e-prints, arXiv:2009.13759

\bibitem[{Forum(1994)}]{MPI}
Forum, M.~P. 1994, MPI: A Message-Passing Interface Standard, Tech. rep.,
  Knoxville, TN, USA

\bibitem[{Foucart(2020)}]{Foucart20}
Foucart, F. 2020, Front. Astron. Space Sci., 7, 46

\bibitem[{{Foucart} {et~al.}(2018){Foucart}, {Hinderer}, \&
  {Nissanke}}]{Foucart18}
{Foucart}, F., {Hinderer}, T., \& {Nissanke}, S. 2018, \prd, 98, 081501

\bibitem[{{Foucart} {et~al.}(2017){Foucart}, {Desai}, {Brege}, {Duez}, {Kasen},
  {Hemberger}, {Kidder}, {Pfeiffer}, \& {Scheel}}]{Foucart17}
{Foucart}, F., {Desai}, D., {Brege}, W., {et~al.} 2017, Classical and Quantum
  Gravity, 34, 044002

\bibitem[{{Fujibayashi} {et~al.}(2018){Fujibayashi}, {Kiuchi}, {Nishimura},
  {Sekiguchi}, \& {Shibata}}]{fujibayashi2018}
{Fujibayashi}, S., {Kiuchi}, K., {Nishimura}, N., {Sekiguchi}, Y., \&
  {Shibata}, M. 2018, \apj, 860, 64

\bibitem[{{Fujibayashi} {et~al.}(2020){Fujibayashi}, {Wanajo}, {Kiuchi},
  {Kyutoku}, {Sekiguchi}, \& {Shibata}}]{fujibayashi2020}
{Fujibayashi}, S., {Wanajo}, S., {Kiuchi}, K., {et~al.} 2020, \apj, 901, 122

\bibitem[{Goldstein {et~al.}(2017)Goldstein, Veres, Burns, Briggs, Hamburg,
  Kocevski, Wilson-Hodge, Preece, Poolakkil, Roberts, \& et~al.}]{Goldstein17}
Goldstein, A., Veres, P., Burns, E., {et~al.} 2017, The Astrophysical Journal,
  848, L14.
\newblock \url{http://dx.doi.org/10.3847/2041-8213/aa8f41}

\bibitem[{{Goldstein} {et~al.}(2019){Goldstein}, {Andreoni}, {Nugent},
  {Kasliwal}, {Coughlin}, {Anand}, {Bloom}, {Mart{\'\i}nez-Palomera}, {Zhang},
  {Ahumada}, {Bagdasaryan}, {Cooke}, {De}, {Duev}, {Fremling}, {Gatkine},
  {Graham}, {Ofek}, {Singer}, \& {Yan}}]{Goldstein19}
{Goldstein}, D.~A., {Andreoni}, I., {Nugent}, P.~E., {et~al.} 2019, \apjl, 881,
  L7

\bibitem[{{Gompertz} {et~al.}(2020){Gompertz}, {Cutter}, {Steeghs}, {Galloway},
  {Lyman}, {Ulaczyk}, {Dyer}, {Ackley}, {Dhillon}, {O'Brien}, {Ramsay},
  {Poshyachinda}, {Kotak}, {Nuttall}, {Breton}, {Pall{\'e}}, {Pollacco},
  {Thrane}, {Aukkaravittayapun}, {Awiphan}, {Brown}, {Burhanudin}, {Chote},
  {Chrimes}, {Daw}, {Duffy}, {Eyles-Ferris}, {Heikkil{\"a}}, {Irawati},
  {Kennedy}, {Killestein}, {Levan}, {Littlefair}, {Makrygianni}, {Marsh}, {Mata
  S{\'a}nchez}, {Mattila}, {Maund}, {McCormac}, {Mkrtichian}, {Mong},
  {Mullaney}, {M{\"u}ller}, {Obradovic}, {Rol}, {Sawangwit}, {Stanway},
  {Starling}, {Str{\o}m}, {Tooke}, {West}, \& {Wiersema}}]{GOTO_o3}
{Gompertz}, B.~P., {Cutter}, R., {Steeghs}, D., {et~al.} 2020, \mnras, 497, 726

\bibitem[{Gompertz {et~al.}(2020)Gompertz, Cutter, Steeghs, Galloway, Lyman,
  Ulaczyk, Dyer, Ackley, Dhillon, O’Brien, \& et~al.}]{GOTO}
Gompertz, B.~P., Cutter, R., Steeghs, D., {et~al.} 2020, Monthly Notices of the
  Royal Astronomical Society, 497, 726–738.
\newblock \url{http://dx.doi.org/10.1093/mnras/staa1845}

\bibitem[{{Graham} {et~al.}(2020){Graham}, {Ford}, {McKernan}, {Ross}, {Stern},
  {Burdge}, {Coughlin}, {Djorgovski}, {Drake}, {Duev}, {Kasliwal}, {Mahabal},
  {van Velzen}, {Belecki}, {Bellm}, {Burruss}, {Cenko}, {Cunningham}, {Helou},
  {Kulkarni}, {Masci}, {Prince}, {Reiley}, {Rodriguez}, {Rusholme}, {Smith}, \&
  {Soumagnac}}]{Graham20}
{Graham}, M.~J., {Ford}, K.~E.~S., {McKernan}, B., {et~al.} 2020, \prl, 124,
  251102

\bibitem[{Guillochon {et~al.}(2018)Guillochon, Nicholl, Villar, Mockler,
  Narayan, Mandel, Berger, \& Williams}]{Guillochon18}
Guillochon, J., Nicholl, M., Villar, V.~A., {et~al.} 2018, The Astrophysical
  Journal Supplement Series, 236, 6.
\newblock \url{http://dx.doi.org/10.3847/1538-4365/aab761}

\bibitem[{{Hallinan} {et~al.}(2017){Hallinan}, {Corsi}, {Mooley}, {Hotokezaka},
  {Nakar}, {Kasliwal}, {Kaplan}, {Frail}, {Myers}, {Murphy}, {De}, {Dobie},
  {Allison}, {Bannister}, {Bhalerao}, {Chandra}, {Clarke}, {Giacintucci}, {Ho},
  {Horesh}, {Kassim}, {Kulkarni}, {Lenc}, {Lockman}, {Lynch}, {Nichols},
  {Nissanke}, {Palliyaguru}, {Peters}, {Piran}, {Rana}, {Sadler}, \&
  {Singer}}]{Hallinan17}
{Hallinan}, G., {Corsi}, A., {Mooley}, K.~P., {et~al.} 2017, Science, 358, 1579

\bibitem[{{Heinzel} {et~al.}(2021){Heinzel}, {Coughlin}, {Dietrich}, {Bulla},
  {Antier}, {Christensen}, {Coulter}, {Foley}, {Issa}, \& {Khetan}}]{Heinzel21}
{Heinzel}, J., {Coughlin}, M.~W., {Dietrich}, T., {et~al.} 2021, \mnras,
  arXiv:2010.10746

\bibitem[{{Hinderer} {et~al.}(2010){Hinderer}, {Lackey}, {Lang}, \&
  {Read}}]{Hinderer10}
{Hinderer}, T., {Lackey}, B.~D., {Lang}, R.~N., \& {Read}, J.~S. 2010, \prd,
  81, 123016

\bibitem[{{Hinderer} {et~al.}(2019){Hinderer}, {Nissanke}, {Foucart},
  {Hotokezaka}, {Vincent}, {Kasliwal}, {Schmidt}, {Williamson}, {Nichols},
  {Duez}, {Kidder}, {Pfeiffer}, \& {Scheel}}]{Hinderer19}
{Hinderer}, T., {Nissanke}, S., {Foucart}, F., {et~al.} 2019, \prd, 100, 063021

\bibitem[{{Hotokezaka} \& {Nakar}(2020)}]{Hotokezaka19}
{Hotokezaka}, K., \& {Nakar}, E. 2020, \apj, 891, 152

\bibitem[{{Hotokezaka} {et~al.}(2019){Hotokezaka}, {Nakar}, {Gottlieb},
  {Nissanke}, {Masuda}, {Hallinan}, {Mooley}, \& {Deller}}]{Hotokezaka19a}
{Hotokezaka}, K., {Nakar}, E., {Gottlieb}, O., {et~al.} 2019, Nature Astronomy,
  3, 940

\bibitem[{{Hotokezaka} {et~al.}(2016){Hotokezaka}, {Nissanke}, {Hallinan},
  {Lazio}, {Nakar}, \& {Piran}}]{Hotokezaka16}
{Hotokezaka}, K., {Nissanke}, S., {Hallinan}, G., {et~al.} 2016, \apj, 831, 190

\bibitem[{{Howlett} \& {Davis}(2020)}]{Howlett20}
{Howlett}, C., \& {Davis}, T.~M. 2020, \mnras, 492, 3803

\bibitem[{Hunter(2007)}]{Hunter:2007}
Hunter, J.~D. 2007, Computing in Science \& Engineering, 9, 90.
\newblock \url{http://dx.doi.org/10.1109/MCSE.2007.55}

\bibitem[{{Ivezi{\'c}} {et~al.}(2019){Ivezi{\'c}}, {Kahn}, {Tyson}, {Abel},
  {Acosta}, {Allsman}, {Alonso}, {AlSayyad}, {Anderson}, {Andrew}, \&
  et~al.}]{LSST}
{Ivezi{\'c}}, {\v Z}., {Kahn}, S.~M., {Tyson}, J.~A., {et~al.} 2019, \apj, 873,
  111

\bibitem[{Jones {et~al.}(2001--)Jones, Oliphant, Peterson, {et~al.}}]{Scipy}
Jones, E., Oliphant, T., Peterson, P., {et~al.} 2001--, {SciPy}: Open source
  scientific tools for {Python}, , , [Online; accessed 21.06.2019].
\newblock \url{http://www.scipy.org/}

\bibitem[{{Just} {et~al.}(2015){Just}, {Bauswein}, {Pulpillo}, {Goriely}, \&
  {Janka}}]{just2015}
{Just}, O., {Bauswein}, A., {Pulpillo}, R.~A., {Goriely}, S., \& {Janka}, H.-T.
  2015, MNRAS, 448, 541

\bibitem[{{Kasen} \& {Barnes}(2019)}]{Kasen19}
{Kasen}, D., \& {Barnes}, J. 2019, \apj, 876, 128

\bibitem[{{Kasliwal} {et~al.}(2017){Kasliwal}, {Nakar}, {Singer}, {Kaplan},
  {Cook}, {Van Sistine}, {Lau}, {Fremling}, {Gottlieb}, {Jencson}, {Adams},
  {Feindt}, {Hotokezaka}, {Ghosh}, {Perley}, {Yu}, {Piran}, {Allison},
  {Anupama}, {Balasubramanian}, {Bannister}, {Bally}, {Barnes}, {Barway},
  {Bellm}, {Bhalerao}, {Bhattacharya}, {Blagorodnova}, {Bloom}, {Brady},
  {Cannella}, {Chatterjee}, {Cenko}, {Cobb}, {Copperwheat}, {Corsi}, {De},
  {Dobie}, {Emery}, {Evans}, {Fox}, {Frail}, {Frohmaier}, {Goobar}, {Hallinan},
  {Harrison}, {Helou}, {Hinderer}, {Ho}, {Horesh}, {Ip}, {Itoh}, {Kasen},
  {Kim}, {Kuin}, {Kupfer}, {Lynch}, {Madsen}, {Mazzali}, {Miller}, {Mooley},
  {Murphy}, {Ngeow}, {Nichols}, {Nissanke}, {Nugent}, {Ofek}, {Qi}, {Quimby},
  {Rosswog}, {Rusu}, {Sadler}, {Schmidt}, {Sollerman}, {Steele}, {Williamson},
  {Xu}, {Yan}, {Yatsu}, {Zhang}, \& {Zhao}}]{Kasliwal17}
{Kasliwal}, M.~M., {Nakar}, E., {Singer}, L.~P., {et~al.} 2017, Science, 358,
  1559

\bibitem[{{Kawaguchi} {et~al.}(2016){Kawaguchi}, {Kyutoku}, {Shibata}, \&
  {Tanaka}}]{Kawaguchi16}
{Kawaguchi}, K., {Kyutoku}, K., {Shibata}, M., \& {Tanaka}, M. 2016, \apj, 825,
  52

\bibitem[{Kawaguchi {et~al.}(2020)Kawaguchi, Shibata, \& Tanaka}]{Kawaguchi20}
Kawaguchi, K., Shibata, M., \& Tanaka, M. 2020, The Astrophysical Journal, 889,
  171.
\newblock \url{http://dx.doi.org/10.3847/1538-4357/ab61f6}

\bibitem[{{Klion} {et~al.}(2021){Klion}, {Duffell}, {Kasen}, \&
  {Quataert}}]{Klion21}
{Klion}, H., {Duffell}, P.~C., {Kasen}, D., \& {Quataert}, E. 2021, \mnras,
  502, 865

\bibitem[{Kluyver {et~al.}(2016)Kluyver, Ragan-Kelley, P{\'e}rez, Granger,
  Bussonnier, Frederic, Kelley, Hamrick, Grout, Corlay, Ivanov, Avila, Abdalla,
  \& Willing}]{Kluyver:2016aa}
Kluyver, T., Ragan-Kelley, B., P{\'e}rez, F., {et~al.} 2016, in Positioning and
  Power in Academic Publishing: Players, Agents and Agendas, ed. F.~Loizides \&
  B.~Schmidt, IOS Press, 87 -- 90

\bibitem[{{Korobkin} {et~al.}(2012){Korobkin}, {Rosswog}, {Arcones}, \&
  {Winteler}}]{Korobkin12}
{Korobkin}, O., {Rosswog}, S., {Arcones}, A., \& {Winteler}, C. 2012, \mnras,
  426, 1940

\bibitem[{{Korobkin} {et~al.}(2020){Korobkin}, {Wollaeger}, {Fryer},
  {Hungerford}, {Rosswog}, {Fontes}, {Mumpower}, {Chase}, {Even}, {Miller},
  {Misch}, \& {Lippuner}}]{Korobkin20}
{Korobkin}, O., {Wollaeger}, R., {Fryer}, C., {et~al.} 2020, arXiv e-prints,
  arXiv:2004.00102

\bibitem[{{Kr{\"u}ger} \& {Foucart}(2020)}]{Kruger20}
{Kr{\"u}ger}, C.~J., \& {Foucart}, F. 2020, \prd, 101, 103002

\bibitem[{{Kulkarni}(2005)}]{Kulkarni05}
{Kulkarni}, S.~R. 2005, arXiv e-prints, astro

\bibitem[{{Kyutoku} {et~al.}(2020){Kyutoku}, {Fujibayashi}, {Hayashi},
  {Kawaguchi}, {Kiuchi}, {Shibata}, \& {Tanaka}}]{Kyutoku20}
{Kyutoku}, K., {Fujibayashi}, S., {Hayashi}, K., {et~al.} 2020, \apjl, 890, L4

\bibitem[{Lattimer \& Prakash(2001)}]{Lattimer01}
Lattimer, J.~M., \& Prakash, M. 2001, The Astrophysical Journal, 550, 426.
\newblock \url{https://doi.org/10.1086%2F319702}

\bibitem[{{Lattimer} \& {Schramm}(1974)}]{Lattimer74}
{Lattimer}, J.~M., \& {Schramm}, D.~N. 1974, \apjl, 192, L145

\bibitem[{Lewis(2019)}]{GetDist}
Lewis, A. 2019, arXiv:1910.13970.
\newblock \url{https://getdist.readthedocs.io}

\bibitem[{{Li} \& {Paczy{\'n}ski}(1998)}]{Li98}
{Li}, L.-X., \& {Paczy{\'n}ski}, B. 1998, \apjl, 507, L59

\bibitem[{{Lippuner} {et~al.}(2017){Lippuner}, {Fern{\'a}ndez}, {Roberts},
  {Foucart}, {Kasen}, {Metzger}, \& {Ott}}]{lippuner2017}
{Lippuner}, J., {Fern{\'a}ndez}, R., {Roberts}, L.~F., {et~al.} 2017, MNRAS,
  472, 904

\bibitem[{Maggiore {et~al.}(2020)Maggiore, Broeck, Bartolo, Belgacem, Bertacca,
  Bizouard, Branchesi, Clesse, Foffa, García-Bellido, \&
  et~al.}]{EinsteinTelescope}
Maggiore, M., Broeck, C. V.~D., Bartolo, N., {et~al.} 2020, Journal of
  Cosmology and Astroparticle Physics, 2020, 050–050.
\newblock \url{http://dx.doi.org/10.1088/1475-7516/2020/03/050}

\bibitem[{{Martin} {et~al.}(2015){Martin}, {Perego}, {Arcones}, {Thielemann},
  {Korobkin}, \& {Rosswog}}]{martin2015}
{Martin}, D., {Perego}, A., {Arcones}, A., {et~al.} 2015, ApJ, 813, 2

\bibitem[{Matas {et~al.}(2020)Matas, Dietrich, Buonanno, Hinderer, Pürrer,
  Foucart, Boyle, Duez, Kidder, Pfeiffer, \& et~al.}]{Matas20}
Matas, A., Dietrich, T., Buonanno, A., {et~al.} 2020, Physical Review D, 102,
  doi:10.1103/physrevd.102.043023.
\newblock \url{http://dx.doi.org/10.1103/PhysRevD.102.043023}

\bibitem[{{Metzger} \& {Fern{\'a}ndez}(2014)}]{metzger2014}
{Metzger}, B.~D., \& {Fern{\'a}ndez}, R. 2014, \mnras, 441, 3444

\bibitem[{{Metzger} {et~al.}(2010){Metzger}, {Mart{\'\i}nez-Pinedo}, {Darbha},
  {Quataert}, {Arcones}, {Kasen}, {Thomas}, {Nugent}, {Panov}, \&
  {Zinner}}]{Metzger10}
{Metzger}, B.~D., {Mart{\'\i}nez-Pinedo}, G., {Darbha}, S., {et~al.} 2010,
  \mnras, 406, 2650

\bibitem[{{Miller} {et~al.}(2019){Miller}, {Ryan}, {Dolence}, {Burrows},
  {Fontes}, {Fryer}, {Korobkin}, {Lippuner}, {Mumpower}, \&
  {Wollaeger}}]{miller2019}
{Miller}, J.~M., {Ryan}, B.~R., {Dolence}, J.~C., {et~al.} 2019, \prd, 100,
  023008

\bibitem[{{Most} {et~al.}(2020){Most}, {Papenfort}, {Tootle}, \&
  {Rezzolla}}]{Most20}
{Most}, E.~R., {Papenfort}, L.~J., {Tootle}, S., \& {Rezzolla}, L. 2020, arXiv
  e-prints, arXiv:2012.03896

\bibitem[{{Most} {et~al.}(2019){Most}, {Papenfort}, {Tsokaros}, \&
  {Rezzolla}}]{Most19}
{Most}, E.~R., {Papenfort}, L.~J., {Tsokaros}, A., \& {Rezzolla}, L. 2019,
  \apj, 884, 40

\bibitem[{{M{\"o}sta} {et~al.}(2020){M{\"o}sta}, {Radice}, {Haas}, {Schnetter},
  \& {Bernuzzi}}]{Moesta20}
{M{\"o}sta}, P., {Radice}, D., {Haas}, R., {Schnetter}, E., \& {Bernuzzi}, S.
  2020, \apjl, 901, L37

\bibitem[{{Mukherjee} {et~al.}(2019){Mukherjee}, {Lavaux}, {Bouchet}, {Jasche},
  {Wandelt}, {Nissanke}, {Leclercq}, \& {Hotokezaka}}]{Mukherjee19}
{Mukherjee}, S., {Lavaux}, G., {Bouchet}, F.~R., {et~al.} 2019, arXiv e-prints,
  arXiv:1909.08627

\bibitem[{{Nativi} {et~al.}(2021){Nativi}, {Bulla}, {Rosswog}, {Lundman},
  {Kowal}, {Gizzi}, {Lamb}, \& {Perego}}]{Nativi21}
{Nativi}, L., {Bulla}, M., {Rosswog}, S., {et~al.} 2021, \mnras, 500, 1772

\bibitem[{Nedora {et~al.}(2021)Nedora, Bernuzzi, Radice, Daszuta, Endrizzi,
  Perego, Prakash, Safarzadeh, Schianchi, \& Logoteta}]{Nedora21}
Nedora, V., Bernuzzi, S., Radice, D., {et~al.} 2021, The Astrophysical Journal,
  906, 98.
\newblock \url{http://dx.doi.org/10.3847/1538-4357/abc9be}

\bibitem[{Nicholl {et~al.}(2021)Nicholl, Margalit, Schmidt, Smith, Ridley, \&
  Nuttall}]{Nicholl21}
Nicholl, M., Margalit, B., Schmidt, P., {et~al.} 2021, arXiv:2102.02229

\bibitem[{Nicolaou {et~al.}(2020)Nicolaou, Lahav, Lemos, Hartley, \&
  Braden}]{Nicolaou20}
Nicolaou, C., Lahav, O., Lemos, P., Hartley, W., \& Braden, J. 2020, Monthly
  Notices of the Royal Astronomical Society, 495, 90–97.
\newblock \url{http://dx.doi.org/10.1093/mnras/staa1120}

\bibitem[{{Oliphant}(2007)}]{python2007}
{Oliphant}, T.~E. 2007, Computing in Science Engineering, 9, 10

\bibitem[{{Page} {et~al.}(2020){Page}, {Evans}, {Tohuvavohu}, {Kennea},
  {Klingler}, {Cenko}, {Oates}, {Ambrosi}, {Barthelmy}, {Beardmore},
  {Bernardini}, {Breeveld}, {Brown}, {Burrows}, {Campana}, {Caputo},
  {Cusumano}, {D'A{\`\i}}, {D'Avanzo}, {D'Elia}, {De Pasquale}, {Emery},
  {Giommi}, {Gronwall}, {Hartmann}, {Krimm}, {Kuin}, {Malesani}, {Marshall},
  {Melandri}, {Nousek}, {O'Brien}, {Osborne}, {Pagani}, {Page}, {Palmer},
  {Perri}, {Racusin}, {Sakamoto}, {Sbarufatti}, {Schlieder}, {Siegel},
  {Tagliaferri}, \& {Troja}}]{SWIFT_o3}
{Page}, K.~L., {Evans}, P.~A., {Tohuvavohu}, A., {et~al.} 2020, \mnras, 499,
  3459

\bibitem[{{Palmese} {et~al.}(2019){Palmese}, {Graur}, {Annis}, {BenZvi}, {Di
  Valentino}, {Garcia-Bellido}, {Gontcho}, {Keeley}, {Kim}, {Lahav},
  {Nissanke}, {Paterson}, {Sako}, {Shafieloo}, \& {Tsai}}]{Palmese19}
{Palmese}, A., {Graur}, O., {Annis}, J.~T., {et~al.} 2019, \baas, 51, 310

\bibitem[{{Papenfort} {et~al.}(2018){Papenfort}, {Gold}, \&
  {Rezzolla}}]{Papenfort18}
{Papenfort}, L.~J., {Gold}, R., \& {Rezzolla}, L. 2018, \prd, 98, 104028

\bibitem[{Perego {et~al.}(2017)Perego, Radice, \& Bernuzzi}]{Perego17}
Perego, A., Radice, D., \& Bernuzzi, S. 2017, The Astrophysical Journal, 850,
  L37.
\newblock \url{http://dx.doi.org/10.3847/2041-8213/aa9ab9}

\bibitem[{{Perez} \& {Granger}(2007)}]{IPython2007}
{Perez}, F., \& {Granger}, B.~E. 2007, Computing in Science Engineering, 9, 21

\bibitem[{{Piro} \& {Kollmeier}(2018)}]{Piro18}
{Piro}, A.~L., \& {Kollmeier}, J.~A. 2018, \apj, 855, 103

\bibitem[{{Raaijmakers} {et~al.}(2020){Raaijmakers}, {Greif}, {Riley},
  {Hinderer}, {Hebeler}, {Schwenk}, {Watts}, {Nissanke}, {Guillot}, {Lattimer},
  \& {Ludlam}}]{Raaijmakers20}
{Raaijmakers}, G., {Greif}, S.~K., {Riley}, T.~E., {et~al.} 2020, \apjl, 893,
  L21

\bibitem[{{Radice} \& {Dai}(2019)}]{Radice19}
{Radice}, D., \& {Dai}, L. 2019, European Physical Journal A, 55, 50

\bibitem[{{Radice} {et~al.}(2018{\natexlab{a}}){Radice}, {Perego},
  {Hotokezaka}, {Bernuzzi}, {Fromm}, \& {Roberts}}]{Radice18a}
{Radice}, D., {Perego}, A., {Hotokezaka}, K., {et~al.} 2018{\natexlab{a}},
  \apjl, 869, L35

\bibitem[{{Radice} {et~al.}(2018{\natexlab{b}}){Radice}, {Perego},
  {Hotokezaka}, {Fromm}, {Bernuzzi}, \& {Roberts}}]{Radice18b}
---. 2018{\natexlab{b}}, \apj, 869, 130

\bibitem[{{Radice} {et~al.}(2018{\natexlab{c}}){Radice}, {Perego}, {Zappa}, \&
  {Bernuzzi}}]{Radice18}
{Radice}, D., {Perego}, A., {Zappa}, F., \& {Bernuzzi}, S. 2018{\natexlab{c}},
  \apjl, 852, L29

\bibitem[{{Reitze} {et~al.}(2019){Reitze}, {Adhikari}, {Ballmer}, {Barish},
  {Barsotti}, {Billingsley}, {Brown}, {Chen}, {Coyne}, {Eisenstein}, {Evans},
  {Fritschel}, {Hall}, {Lazzarini}, {Lovelace}, {Read}, {Sathyaprakash},
  {Shoemaker}, {Smith}, {Torrie}, {Vitale}, {Weiss}, {Wipf}, \&
  {Zucker}}]{CosmicExplorer}
{Reitze}, D., {Adhikari}, R.~X., {Ballmer}, S., {et~al.} 2019, in Bulletin of
  the American Astronomical Society, Vol.~51, 35

\bibitem[{Rosswog {et~al.}(1999)Rosswog, Liebendoerfer, Thielemann, Davies,
  Benz, \& Piran}]{Rosswog98}
Rosswog, S., Liebendoerfer, M., Thielemann, F.~K., {et~al.} 1999, Astron.
  Astrophys., 341, 499

\bibitem[{{Ruiz} {et~al.}(2020){Ruiz}, {Paschalidis}, {Tsokaros}, \&
  {Shapiro}}]{Ruiz20}
{Ruiz}, M., {Paschalidis}, V., {Tsokaros}, A., \& {Shapiro}, S.~L. 2020, \prd,
  102, 124077

\bibitem[{{Ruiz} {et~al.}(2021){Ruiz}, {Shapiro}, \& {Tsokaros}}]{Ruiz21}
{Ruiz}, M., {Shapiro}, S.~L., \& {Tsokaros}, A. 2021, arXiv e-prints,
  arXiv:2102.03366

\bibitem[{Ruiz {et~al.}(2021)Ruiz, Shapiro, \& Tsokaros}]{Ruiz21a}
Ruiz, M., Shapiro, S.~L., \& Tsokaros, A. 2021, arXiv:2102.03366

\bibitem[{{Ryan} {et~al.}(2020){Ryan}, {van Eerten}, {Piro}, \&
  {Troja}}]{Ryan20}
{Ryan}, G., {van Eerten}, H., {Piro}, L., \& {Troja}, E. 2020, \apj, 896, 166

\bibitem[{Sekiguchi {et~al.}(2016)Sekiguchi, Kiuchi, Kyutoku, Shibata, \&
  Taniguchi}]{Sekiguchi16}
Sekiguchi, Y., Kiuchi, K., Kyutoku, K., Shibata, M., \& Taniguchi, K. 2016,
  Phys. Rev. D, 93, 124046.
\newblock \url{https://link.aps.org/doi/10.1103/PhysRevD.93.124046}

\bibitem[{Shibata \& Hotokezaka(2019)}]{Shibata19}
Shibata, M., \& Hotokezaka, K. 2019, Ann. Rev. Nucl. Part. Sci., 69, 41

\bibitem[{{Siegel} \& {Metzger}(2017)}]{Siegel17}
{Siegel}, D.~M., \& {Metzger}, B.~D. 2017, \prl, 119, 231102

\bibitem[{{Tanaka} \& {Hotokezaka}(2013)}]{Tanaka13}
{Tanaka}, M., \& {Hotokezaka}, K. 2013, \apj, 775, 113

\bibitem[{{Tanaka} {et~al.}(2020){Tanaka}, {Kato}, {Gaigalas}, \&
  {Kawaguchi}}]{Tanaka20}
{Tanaka}, M., {Kato}, D., {Gaigalas}, G., \& {Kawaguchi}, K. 2020, \mnras, 496,
  1369

\bibitem[{{Tews} {et~al.}(2020){Tews}, {Pang}, {Dietrich}, {Coughlin},
  {Antier}, {Bulla}, {Heinzel}, \& {Issa}}]{Tews20}
{Tews}, I., {Pang}, P. T.~H., {Dietrich}, T., {et~al.} 2020, arXiv e-prints,
  arXiv:2007.06057

\bibitem[{{The LIGO Scientific Collaboration} {et~al.}(2019){The LIGO
  Scientific Collaboration}, {the Virgo Collaboration}, {Abbott}, {Abbott},
  {Abbott}, {Abraham}, {Acernese}, {Ackley}, {Adams}, {Adhikari}, {Adya},
  {Affeldt}, {Agathos}, {Agatsuma}, {Aggarwal}, {Aguiar}, {Aiello}, {Ain},
  {Ajith}, {Allen}, {Allocca}, {Aloy}, {Altin}, {Amato}, {Anand}, {Ananyeva},
  {Anderson}, {Anderson}, {Angelova}, {Antier}, {Appert}, {Arai}, {Araya},
  {Areeda}, {Ar{\`e}ne}, {Arnaud}, {Aronson}, {Arun}, {Ascenzi}, {Ashton},
  {Aston}, {Astone}, {Aubin}, {Aufmuth}, {AultONeal}, {Austin}, {Avendano},
  {Avila-Alvarez}, {Babak}, {Bacon}, {Badaracco}, {Bader}, {Bae}, {Baird},
  {Baker}, {Baldaccini}, {Ballardin}, {Ballmer}, {Bals}, {Banagiri},
  {Barayoga}, {Barbieri}, {Barclay}, {Barish}, {Barker}, {Barkett}, {Barnum},
  {Barone}, {Barr}, {Barsotti}, {Barsuglia}, {Barta}, {Bartlett}, {Bartos},
  {Bassiri}, {Basti}, {Bawaj}, {Bayley}, {Bazzan}, {B{\'e}csy}, {Bejger},
  {Belahcene}, {Bell}, {Beniwal}, {Benjamin}, {Berger}, {Bergmann}, {Bernuzzi},
  {Berry}, {Bersanetti}, {Bertolini}, {Betzwieser}, {Bhandare}, {Bidler},
  {Biggs}, {Bilenko}, {Bilgili}, {Billingsley}, {Birney}, {Birnholtz},
  {Biscans}, {Bischi}, {Biscoveanu}, {Bisht}, {Bitossi}, {Bizouard},
  {Blackburn}, {Blackman}, {Blair}, {Blair}, {Blair}, {Bloemen}, {Bobba},
  {Bode}, {Boer}, {Boetzel}, {Bogaert}, {Bondu}, {Bonnand}, {Booker}, {Boom},
  {Bork. Boschi}, {Bose}, {Bossilkov}, {Bosveld}, {Bouffanais}, {Bozzi},
  {Bradaschia}, {Brady}, {Bramley}, {Branchesi}, {Brau}, {Breschi}, {Briant},
  {Briggs}, {Brighenti}, {Brillet}, {Brinkmann}, {Brockill}, {Brooks},
  {Brooks}, {Brown}, {Brunett}, {Buikema}, {Bulik}, {Bulten}, {Buonanno},
  {Buskulic}, {Buy}, {Byer}, {Cabero}, {Cadonati}, {Cagnoli}, {Cahillane},
  {Calder{\'o}n Bustillo}, {Callister}, {Calloni}, {Camp}, {Campbell},
  {Canepa}, {Cannon}, {Cao}, {Cao}, {Carapella}, {Carbognani}, {Caride},
  {Carney}, {Carullo}, {Casanueva Diaz}, {Casentini}, {Caudill},
  {Cavagli{\`a}}, {Cavalier}, {Cavalieri}, {Cella}, {Cerd{\'a}-Dur{\'a}n},
  {Cesarini}, {Chaibi}, {Chakravarti}, {Chamberlin}, {Chan}, {Chao},
  {Charlton}, {Chase}, {Chassande-Mottin}, {Chatterjee}, {Chaturvedi},
  {Cheeseboro}, {Chen}, {Chen}, {Chen}, {Cheng}, {Cheong}, {Chia}, {Chiadini},
  {Chincarini}, {Chiummo}, {Cho}, {Cho}, {Cho}, {Christensen}, {Chu}, {Chua},
  {Chung}, {Chung}, {Ciani}, {Cie{\'s}lar}, {Ciobanu}, {Ciolfi}, {Cipriano},
  {Cirone}, {Clara}, {Clark}, {Clearwater}, {Cleva}, {Coccia}, {Cohadon},
  {Cohen}, {Colleoni}, {Collette}, {Collins}, {Colpi}, {Cominsky},
  {Constancio}, {Conti}, {Cooper}, {Corban}, {Corbitt}, {Cordero-Carri{\'o}n},
  {Corezzi}, {Corley}, {Cornish}, {Corre}, {Corsi}, {Cortese}, {Costa},
  {Cotesta}, {Coughlin}, {Coughlin}, {Coulon}, {Countryman}, {Couvares},
  {Covas}, {Cowan}, {Coward}, {Cowart}, {Coyne}, {Coyne}, {Creighton},
  {Creighton}, {Cripe}, {Croquette}, {Crowder}, {Cullen}, {Cumming},
  {Cunningham}, {Cuoco}, {Dal Canton}, {D{\'a}lya}, {D'Angelo}, {Danilishin},
  {D'Antonio}, {Danzmann}, {Dasgupta}, {Da Silva Costa}, {Datrier}, {Dattilo},
  {Dave}, {Davier}, {Davis}, {Daw}, {DeBra}, {Deenadayalan}, {Degallaix}, {De
  Laurentis}, {Del{\'e}glise}, {Del Pozzo}, {DeMarchi}, {Demos}, {Dent}, {De
  Pietri}, {De Rosa}, {De Rossi}, {DeSalvo}, {de Varona}, {Dhurandhar},
  {D{\'\i}az}, {Dietrich}, {Di Fiore}, {DiFronzo}, {Di Giorgio}, {Di Giovanni},
  {Di Giovanni}, {Di Girolamo}, {Di Lieto}, {Ding}, {Di Pace}, {Di Palma}, {Di
  Renzo}, {Divakarla}, {Dmitriev}, {Doctor}, {Donovan}, {Dooley}, {Doravari},
  {Dorrington}, {Downes}, {Drago}, {Driggers}, {Du}, {Ducoin}, {Dupej},
  {Durante}, {Dwyer}, {Easter}, {Eddolls}, {Edo}, {Effler}, {Ehrens},
  {Eichholz}, {Eikenberry}, {Eisenmann}, {Eisenstein}, {Errico}, {Essick},
  {Estelles}, {Estevez}, {Etienne}, {Etzel}, {Evans}, {Evans}, {Fafone},
  {Fairhurst}, {Fan}, {Farinon}, {Farr}, {Farr}, {Fauchon-Jones}, {Favata},
  {Fays}, {Fazio}, {Fee}, {Feicht}, {Fejer}, {Feng}, {Fernandez-Galiana},
  {Ferrante}, {Ferreira}, {Ferreira}, {Fidecaro}, {Fiori}, {Fiorucci},
  {Fishbach}, {Fisher}, {Fishner}, {Fittipaldi}, {Fitz-Axen}, {Fiumara},
  {Flaminio}, {Fletcher}, {Floden}, {Flynn}, {Fong}, {Font}, {Forsyth},
  {Fournier}, {Hernandez Vivanco}, {Frasca}, {Frasconi}, {Frei}, {Freise},
  {Frey}, {Frey}, {Fritschel}, {Frolov}, {Fronz{\`e}}, {Fulda}, {Fyffe},
  {Gabbard}, {Gadre}, {Gaebel}, {Gair}, {Gammaitoni}, {Gaonkar},
  {Garc{\'\i}a-Quir{\'o}s}, {Garufi}, {Gateley}, {Gaudio}, {Gaur}, {Gayathri},
  {Gemme}, {Genin}, {Gennai}, {George}, {George}, {Gergely}, {Ghonge}, {Ghosh},
  {Ghosh}, {Ghosh}, {Giacomazzo}, {Giaime}, {Giardina}, {Gibson}, {Gill},
  {Glover}, {Gniesmer}, {Godwin}, {Goetz}, {Goetz}, {Goncharov},
  {Gonz{\'a}lez}, {Gonzalez Castro}, {Gopakumar}, {Gossan}, {Gosselin},
  {Gouaty}, {Grace}, {Grado}, {Granata}, {Grant}, {Gras}, {Grassia}, {Gray},
  {Gray}, {Greco}, {Green}, {Green}, {Gretarsson}, {Grimaldi}, {Grimm},
  {Groot}, {Grote}, {Grunewald}, {Gruning}, {Guidi}, {Gulati}, {Guo}, {Gupta},
  {Gupta}, {Gupta}, {Gustafson}, {Gustafson}, {Haegel}, {Halim}, {Hall},
  {Hall}, {Hamilton}, {Hammond}, {Haney}, {Hanke}, {Hanks}, {Hanna}, {Hannam},
  {Hannuksela}, {Hansen}, {Hanson}, {Harder}, {Hardwick}, {Haris}, {Harms},
  {Harry}, {Harry}, {Hasskew}, {Haster}, {Haughian}, {Hayes}, {Healy},
  {Heidmann}, {Heintze}, {Heitmann}, {Hellman}, {Hello}, {Hemming}, {Hendry},
  {Heng}, {Hennig}, {Heurs}, {Hild}, {Hinderer}, {Hochheim}, {Hofman},
  {Holgado}, {Holland}, {Holt}, {Holz}, {Hopkins}, {Horst}, {Hough}, {Howell},
  {Hoy}, {Huang}, {H{\"u}bner}, {Huerta}, {Huet}, {Hughey}, {Hui}, {Husa},
  {Huttner}, {Huynh-Dinh}, {Idzkowski}, {Iess}, {Inchauspe}, {Ingram}, {Inta},
  {Intini}, {Irwin}, {Isa}, {Isac}, {Isi}, {Iyer}, {Jacqmin}, {Jadhav}, {Jani},
  {Janthalur}, {Jaranowski}, {Jariwala}, {Jenkins}, {Jiang}, {Johnson},
  {Jones}, {Jones}, {Jones}, {Jones}, {Jonker}, {Ju}, {Junker}, {Kalaghatgi},
  {Kalogera}, {Kamai}, {Kandhasamy}, {Kang}, {Kanner}, {Kapadia}, {Karki},
  {Kashyap}, {Kasprzack}, {Katsanevas}, {Katsavounidis}, {Katzman}, {Kaufer},
  {Kawabe}, {Keerthana}, {K{\'e}f{\'e}lian}, {Keitel}, {Kennedy}, {Key},
  {Khalili}, {Khan}, {Khan}, {Khazanov}, {Khetan}, {Khursheed}, {Kijbunchoo},
  {Kim}, {Kim}, {Kim}, {Kim}, {Kim}, {Kim}, {Kimball}, {King}, {Kinley-Hanlon},
  {Kirchhoff}, {Kissel}, {Kleybolte}, {Klika}, {Klimenko}, {Knowles}, {Koch},
  {Koehlenbeck}, {Koekoek}, {Koley}, {Kondrashov}, {Kontos}, {Koper},
  {Korobko}, {Korth}, {Kovalam}, {Kozak}, {Kr{\"a}mer}, {Kringel},
  {Krishnendu}, {Kr{\'o}lak}, {Krupinski}, {Kuehn}, {Kumar}, {Kumar}, {Kumar},
  {Kumar}, {Kuo}, {Kutynia}, {Kwang}, {Lackey}, {Laghi}, {Lai}, {Lam},
  {Landry}, {Lane}, {Lang}, {Lange}, {Lantz}, {Lanza}, {Lartaux-Vollard},
  {Lasky}, {Laxen}, {Lazzarini}, {Lazzaro}, {Leaci}, {Leavey}, {Lecoeuche},
  {Lee}, {Lee}, {Lee}, {Lee}, {Lee}, {Lee}, {Lehmann}, {Lenon}, {Leroy},
  {Letendre}, {Levin}, {Li}, {Li}, {Li}, {Li}, {Li}, {Lin}, {Linde}, {Linker},
  {Littenberg}, {Liu}, {Liu}, {Llorens-Monteagudo}, {Lo}, {London}, {Longo},
  {Lorenzini}, {Loriette}, {Lormand}, {Losurdo}, {Lough}, {Lousto}, {Lovelace},
  {Lower}, {L{\"u}ck}, {Lumaca}, {Lundgren}, {Lynch}, {Ma}, {Macas}, {Macfoy},
  {MacInnis}, {Macleod}, {Macquet}, {Maga{\~n}a Hernandez},
  {Maga{\~n}a-Sandoval}, {Magee}, {Majorana}, {Maksimovic}, {Malik}, {Man},
  {Mandic}, {Mangano}, {Mansell}, {Manske}, {Mantovani}, {Mapelli},
  {Marchesoni}, {Marion}, {M{\'a}rka}, {M{\'a}rka}, {Markakis}, {Markosyan},
  {Markowitz}, {Maros}, {Marquina}, {Marsat}, {Martelli}, {Martin}, {Martin},
  {Martinez}, {Martynov}, {Masalehdan}, {Mason}, {Massera}, {Masserot},
  {Massinger}, {Masso-Reid}, {Mastrogiovanni}, {Matas}, {Matichard}, {Matone},
  {Mavalvala}, {McCann}, {McCarthy}, {McClelland}, {McCormick}, {McCuller},
  {McGuire}, {McIsaac}, {McIver}, {McManus}, {McRae}, {McWilliams}, {Meacher},
  {Meadors}, {Mehmet}, {Mehta}, {Meidam}, {Mejuto Villa}, {Melatos}, {Mendell},
  {Mercer}, {Mereni}, {Merfeld}, {Merilh}, {Merzougui}, {Meshkov}, {Messenger},
  {Messick}, {Messina}, {Metzdorff}, {Meyers}, {Meylahn}, {Miani}, {Miao},
  {Michel}, {Middleton}, {Milano}, {Miller}, {Millhouse}, {Mills},
  {Milovich-Goff}, {Minazzoli}, {Minenkov}, {Mishkin}, {Mishra}, {Mistry},
  {Mitra}, {Mitrofanov}, {Mitselmakher}, {Mittleman}, {Mo}, {Moffa}, {Mogushi},
  {Mohapatra}, {Molina-Ruiz}, {Mondin}, {Montani}, {Moore}, {Moraru},
  {Morawski}, {Moreno}, {Morisaki}, {Mours}, {Mow-Lowry}, {Muciaccia},
  {Mukherjee}, {Mukherjee}, {Mukherjee}, {Mukherjee}, {Mukund}, {Mullavey},
  {Munch}, {Mu{\~n}iz}, {Muratore}, {Murray}, {Nagar}, {Nardecchia},
  {Naticchioni}, {Nayak}, {Neil}, {Neilson}, {Nelemans}, {Nelson}, {Nery},
  {Neunzert}, {Nevin}, {Ng}, {Ng}, {Nguyen}, {Nguyen}, {Nichols}, {Nichols},
  {Nissanke}, {Nocera}, {North}, {Nuttall}, {Obergaulinger}, {Oberling},
  {O'Brien}, {Oganesyan}, {Ogin}, {Oh}, {Oh}, {Ohme}, {Ohta}, {Okada},
  {Oliver}, {Oppermann}, {Oram}, {O'Reilly}, {Ormiston}, {Ortega},
  {O'Shaughnessy}, {Ossokine}, {Ottaway}, {Overmier}, {Owen}, {Pace}, {Pagano},
  {Page}, {Pagliaroli}, {Pai}, {Pai}, {Palamos}, {Palashov}, {Palomba}, {Pan},
  {Panda}, {Pang}, {Pankow}, {Pannarale}, {Pant}, {Paoletti}, {Paoli},
  {Parida}, {Parker}, {Pascucci}, {Pasqualetti}, {Passaquieti}, {Passuello},
  {Patil}, {Patricelli}, {Payne}, {Pearlstone}, {Pechsiri}, {Pedersen},
  {Pedraza}, {Pedurand}, {Pele}, {Penn}, {Perego}, {Perez}, {P{\'e}rigois},
  {Perreca}, {Petermann}, {Pfeiffer}, {Phelps}, {Phukon}, {Piccinni}, {Pichot},
  {Piergiovanni}, {Pierro}, {Pillant}, {Pinard}, {Pinto}, {Pirello}, {Pitkin},
  {Plastino}, {Poggiani}, {Pong}, {Ponrathnam}, {Popolizio}, {Porter},
  {Powell}, {Prajapati}, {Prasad}, {Prasai}, {Prasanna}, {Pratten},
  {Prestegard}, {Principe}, {Prodi}, {Prokhorov}, {Punturo}, {Puppo},
  {P{\"u}rrer}, {Qi}, {Quetschke}, {Quinonez}, {Raab}, {Raaijmakers},
  {Radkins}, {Radulesco}, {Raffai}, {Raja}, {Rajan}, {Rajbhandari},
  {Rakhmanov}, {Ramirez}, {Ramos-Buades}, {Rana}, {Rao}, {Rapagnani},
  {Raymond}, {Razzano}, {Read}, {Regimbau}, {Rei}, {Reid}, {Reitze},
  {Rettegno}, {Ricci}, {Richardson}, {Richardson}, {Ricker}, {Riemenschneider},
  {Riles}, {Rizzo}, {Robertson}, {Robinet}, {Rocchi}, {Rolland}, {Rollins},
  {Roma}, {Romanelli.}, {Romano}, {Romano}, {Romel}, {Romie}, {Rose}, {Rose},
  {Rose}, {Rosi{\'n}ska}, {Rosofsky}, {Ross}, {Rowan}, {R{\"u}diger}, {Ruggi},
  {Rutins}, {Ryan}, {Sachdev}, {Sadecki}, {Sakellariadou}, {Salafia},
  {Salconi}, {Saleem}, {Samajdar}, {Sammut}, {Sanchez}, {Sanchez},
  {Sanchis-Gual}, {Sanders}, {Santiago}, {Santos}, {Sarin}, {Sassolas},
  {Sathyaprakash}, {Sauter}, {Savage}, {Schale}, {Scheel}, {Scheuer},
  {Schmidt}, {Schnabel}, {Schofield}, {Sch{\"o}nbeck}, {Schreiber}, {Schulte},
  {Schutz}, {Scott}, {Scott}, {Seidel}, {Sellers}, {Sengupta}, {Sennett},
  {Sentenac}, {Sequino}, {Sergeev}, {Setyawati}, {Shaddock}, {Shaffer},
  {Shahriar}, {Shaner}, {Sharma}, {Sharma}, {Shawhan}, {Shen}, {Shink},
  {Shoemaker}, {Shoemaker}, {Shukla}, {ShyamSundar}, {Siellez}, {Sieniawska},
  {Sigg}, {Singer}, {Singh}, {Singh}, {Singhal}, {Sintes}, {Sitmukhambetov},
  {Skliris}, {Slagmolen}, {Slaven-Blair}, {Smith}, {Smith}, {Somala}, {Son},
  {Soni}, {Sorazu}, {Sorrentino}, {Souradeep}, {Sowell}, {Spencer}, {Spera},
  {Srivastava}, {Srivastava}, {Staats}, {Stachie}, {Standke}, {Steer},
  {Steinke}, {Steinlechner}, {Steinlechner}, {Steinmeyer}, {Stevenson},
  {Stocks}, {Stone}, {Stops}, {Strain}, {Stratta}, {Strigin}, {Strunk},
  {Sturani}, {Stuver}, {Sudhir}, {Summerscales}, {Sun}, {Sunil}, {Sur},
  {Suresh}, {Sutton}, {Swinkels}, {Szczepa{\'n}czyk}, {Tacca}, {Tait},
  {Talbot}, {Tanner}, {Tao}, {T{\'a}pai}, {Tapia}, {Tasson}, {Taylor},
  {Tenorio}, {Terkowski}, {Thomas}, {Thomas}, {Thondapu}, {Thorne}, {Thrane},
  {Tiwari}, {Tiwari}, {Tiwari}, {Toland}, {Tonelli}, {Tornasi},
  {Torres-Forn{\'e}}, {Torrie}, {T{\"o}yr{\"a}}, {Travasso}, {Traylor},
  {Tringali}, {Tripathee}, {Trovato}, {Trozzo}, {Tsang}, {Tse}, {Tso},
  {Tsukada}, {Tsuna}, {Tsutsui}, {Tuyenbayev}, {Ueno}, {Ugolini},
  {Unnikrishnan}, {Urban}, {Usman}, {Vahlbruch}, {Vajente}, {Valdes},
  {Valentini}, {van Bakel}, {van Beuzekom}, {van den Brand}, {Van Den Broeck},
  {Vander-Hyde}, {van der Schaaf}, {VanHeijningen}, {van Veggel}, {Vardaro},
  {Varma}, {Vass}, {Vas{\'u}th}, {Vecchio}, {Vedovato}, {Veitch}, {Veitch},
  {Venkateswara}, {Venugopalan}, {Verkindt}, {Vetrano}, {Vicer{\'e}}, {Viets},
  {Vinciguerra}, {Vine}, {Vinet}, {Vitale}, {Vo}, {Vocca}, {Vorvick},
  {Vyatchanin}, {Wade}, {Wade}, {Wade}, {Walet}, {Walker}, {Wallace}, {Walsh},
  {Wang}, {Wang}, {Wang}, {Wang}, {Wang}, {Ward}, {Warden}, {Warner}, {Was},
  {Watchi}, {Weaver}, {Wei}, {Weinert}, {Weinstein}, {Weiss}, {Wellmann},
  {Wen}, {Wessel}, {We{\ss}els}, {Westhouse}, {Wette}, {Whelan}, {Whiting},
  {Whittle}, {Wilken}, {Williams}, {Williamson}, {Willis}, {Willke}, {Winkler},
  {Wipf}, {Wittel}, {Woan}, {Woehler}, {Wofford}, {Wright}, {Wu}, {Wysocki},
  {Xiao}, {Xu}, {Yamamoto}, {Yancey}, {Yang}, {Yang}, {Yang}, {Yap}, {Yazback},
  {Yeeles}, {Yu}, {Yu}, {Yuen}, {Zadro{\.z}ny}, {Zadro{\.z}ny}, {Zanolin},
  {Zelenova}, {Zendri}, {Zevin}, {Zhang}, {Zhang}, {Zhang}, {Zhao}, {Zhao},
  {Zhou}, {Zhou}, {Zhu}, {Zimmerman}, {Zucker}, \& {Zweizig}}]{LVChubble}
{The LIGO Scientific Collaboration}, {the Virgo Collaboration}, {Abbott},
  B.~P., {et~al.} 2019, arXiv e-prints, arXiv:1908.06060

\bibitem[{Thompson {et~al.}(2020)Thompson, Fauchon-Jones, Khan, Nitoglia,
  Pannarale, Dietrich, \& Hannam}]{Thompson20}
Thompson, J.~E., Fauchon-Jones, E., Khan, S., {et~al.} 2020, Physical Review D,
  101, doi:10.1103/physrevd.101.124059.
\newblock \url{http://dx.doi.org/10.1103/PhysRevD.101.124059}

\bibitem[{{van der Walt} {et~al.}(2011){van der Walt}, {Colbert}, \&
  {Varoquaux}}]{Numpy2011}
{van der Walt}, S., {Colbert}, S.~C., \& {Varoquaux}, G. 2011, Computing in
  Science Engineering, 13, 22

\bibitem[{{Waxman} {et~al.}(2019){Waxman}, {Ofek}, \& {Kushnir}}]{Waxman19}
{Waxman}, E., {Ofek}, E.~O., \& {Kushnir}, D. 2019, \apj, 878, 93

\bibitem[{{Wu} {et~al.}(2016){Wu}, {Fern{\'a}ndez}, {Mart{\'{\i}}nez-Pinedo},
  \& {Metzger}}]{wu2016}
{Wu}, M.-R., {Fern{\'a}ndez}, R., {Mart{\'{\i}}nez-Pinedo}, G., \& {Metzger},
  B.~D. 2016, MNRAS, 463, 2323

\bibitem[{{Yagi} \& {Yunes}(2017)}]{Yagi17}
{Yagi}, K., \& {Yunes}, N. 2017, \physrep, 681, 1

\bibitem[{{Zackay} {et~al.}(2018){Zackay}, {Dai}, \& {Venumadhav}}]{Zackay18}
{Zackay}, B., {Dai}, L., \& {Venumadhav}, T. 2018, arXiv e-prints,
  arXiv:1806.08792

\bibitem[{{Zhu} {et~al.}(2020){Zhu}, {Yang}, {Liu}, {Huang}, {Zhang}, {Li},
  {Yu}, \& {Gao}}]{Zhu20}
{Zhu}, J.-P., {Yang}, Y.-P., {Liu}, L.-D., {et~al.} 2020, \apj, 897, 20

\bibitem[{{Zhu} {et~al.}(2018){Zhu}, {Thrane}, {Os{\l}owski}, {Levin}, \&
  {Lasky}}]{Zhu18}
{Zhu}, X., {Thrane}, E., {Os{\l}owski}, S., {Levin}, Y., \& {Lasky}, P.~D.
  2018, \prd, 98, 043002

\end{thebibliography}
\end{document}